\documentclass [12pt, twoside] {uwthesis}[2012/04/02]
 
\usepackage{alltt}  %
\usepackage{amsmath, amsthm, amssymb}
 \usepackage{graphicx,  subfigure} 
 \numberwithin{equation}{section}

\newcommand{\be}{\begin{equation}}
\newcommand{\ee}{\end{equation}}
\newcommand{\bea}{\begin{eqnarray}}
\newcommand{\eea}{\end{eqnarray}}
\newcommand{\eq}[1]{Eq.~(\ref{#1})}

\begin{document}

%
%

\prelimpages
 
%
%
\Title{Nuclei as Probes of Meson-Nucleon Interactions at High and Low Energy}
\Author{Gary Howell}
\Year{2013}
\Program{UW Department of Physics}

\Chair{Gerald A. Miller}{Professor}{Department of Physics}
\Signature{Henry J. Lubatti}
\Signature{Martin J. Savage}

\copyrightpage



{\Degreetext{A dissertation%
  \\
  submitted in partial fulfillment of the\\ requirements for the degree of}
 \def\thefootnote{\fnsymbol{footnote}}
 \let\footnoterule\relax
 \titlepage
 }
\setcounter{footnote}{0}

 
%
%

%
%

\setcounter{page}{-1}

\abstract{%

This dissertation explores two main topics:  1)  Color Transparency and quasi-elastic knockout reactions involving pions and $\rho$ mesons; and 2)  determination of the $J/\psi$-nucleon scattering amplitude and scattering length via $J/\psi$ electroproduction on the deuteron.  It is shown that at the energies available at the COMPASS experiment at CERN, Color Transparency should be detectable in the  reaction $\pi+A\to \pi+p+(A-1)^*$ (proton knockout).  It is also shown that Color Transparency should be detectable in the electroproduction reaction $\gamma^*+A\to\rho+p+(A-1)^*$ at small $Q^2$ (where $Q^2$ is the virtuality of the photon) but large $t$ (4-momentum transfer squared to the knocked out proton), which represents an as-yet unexplored kinematic region in the search for CT effects in electroproduction of vector mesons.   Calculations are also presented for the reaction $\gamma^*+D\to J/\psi+p+n$ at JLab energies in order to determine the feasibility of measuring the elastic $J/\psi$-nucleon scattering amplitude and/or scattering length.  It is found that it may be possible to measure the $J/\psi$-nucleon scattering amplitude at lower energies than previous measurements, but the scattering length cannot be measured.

}
 
%
%
\tableofcontents
\listoffigures
 
%
%
%
%
%
%
%
\acknowledgments{
  Thanks to my advisor, Gerald Miller, for all his help in my research, without which this dissertation wouldn't exist.  
}

%
%
\dedication{\begin{center}
To my parents.
\end{center}}

%

%
%

\textpages
 
 
\chapter {Introduction}
 
This thesis explores two main topics:  1)  Color Transparency and quasi-elastic knockout reactions involving pions and $\rho$ mesons; and 2)  determination of the $J/\psi$-nucleon scattering amplitude and scattering length via $J/\psi$ electroproduction on the deuteron.  

The thesis is organized as follows:

Ch. 2 discusses Color Transparency (CT) and calculation of the cross-section and transparency for the reaction $\pi+A\to \pi+p+(A-1)^*$ (i.e. pion scattering from a nucleus of nucleon number $A$ in which a proton is knocked-out).  In short, Color Transparency (CT) is the  vanishing of Final-State Interactions (e.g. scattering of the knocked-out proton by other nucleons in the nucleus) in large-momentum-transfer elastic or quasi-elastic nuclear reactions, and is a prediction of Quantum Chromodynamics (QCD).  Many experiments have been done in order to search for the predicted effects of CT; references to these experiments are provided in Ch. 2 and Ch. 3.   The quantity that measures the influence of Final-State Interactions is called the ``nuclear transparency" $T$, which is defined as the ratio of two cross-sections:
\be
T\equiv \frac{\sigma}{\sigma_{PWIA}}
\ee
where $\sigma$ is the actual measured cross-section (either total or differential) for the reaction occuring in a nucleus, and $\sigma_{PWIA}$ is the corresponding cross-section calculated in the Plane Wave Impulse Approximation.  Complete vanishing of Final-State Interactions would give $T=1$.   Calculations are done here both neglecting any effects of possible Color Transparency, and including the effects of Color Transparency.  The calculations are performed within the Glauber model of high-energy scattering from a composite target, suitably modified to account for Color Transparency effects.   It is shown that at the energies available at the COMPASS experiment at CERN, Color Transparency should be detectable in this reaction.  

Ch. 3 discusses Color Transparency and vector meson electroproduction, specifically the reaction $\gamma^*+A\to\rho+p+(A-1)^*$.  Electroproduction of the $\rho$ provides another means of detecting the effects of Color Transparency.  In contrast to the purely elastic pion scattering considered in Ch. 2, for electroproduction there are more parameters that may be varied, namely the virtual photon energy $\nu$ and virtuality $Q^2$ in addition to $t$ (4-momentum-transfer-squared to the knocked out proton).  These quantities, as well as a combination of them called the coherence length, $l_c=\frac{2\nu}{Q^2+m_V^2}$, can all affect the observed transparency.  The coherence length plays an especially important role, since by varying its value the nuclear transparency $T$ will vary even in the absence of any Color Transparency effects.  Thus to observe an actual CT effect, one must keep the coherence length fixed.   The calculations presented here show that CT effects may be observed in the reaction $\gamma^*+A\to\rho+p+(A-1)^*$.  So far experiments have concentrated on the large-$Q^2$ regime, but the results presented here show that CT may be observable in the small-$Q^2$ regime, as long as $t$ is large enough.  Calculations are presented here neglecting CT effects and including CT effects, and are performed within the Glauber model of high-energy scattering from a composite target modified to account for particle production and to inclued Color Transparency effects.    

Ch. 4 discusses the feasibilty of measuring the $J/\psi$-nucleon scattering length and/or scattering amplitude in a proposed experiment at JLab, in the reaction $\gamma^*+D\to J/\psi+p+n$ (where $D$ stands for deuteron).  With the mass of the $J/\psi$ being $3.097\;GeV$, the threshold photon energy for photoproduction on a single nucleon is $8.2\;GeV$, and is thus accessible with a $12\;GeV$ electron beam.  Most of the existing data on $J/\psi$ photo- and electroproduction is at much higher energy.  The upcoming $12\;GeV$ upgrade at JLab provides the opportunity to measure $J/\psi$ production near threshold~\cite{jlab12}.   The motivation for the work in the first part of this chapter (Secs. 4.2 - 4.4) was a proposal at JLab~\cite{jlab10} to measure the $J/\psi $-nucleon scattering length by the reaction $\gamma^*+D\to J/\psi +p+n$, where the $J/\psi$ is produced on one nucleon in the deuteron and then re-scatters from the other nucleon.  The reason the $J/\psi $-nucleon scattering length is of interest is that several authors have argued that a nuclear bound state of the $J/\psi$ may exist~\cite{savage92,brodsky97}.  They propose that the force between a $J/\psi$ and a nucleon is purely gluonic in nature, and therefore is the analogue in QCD of the van der Waals force in electrodynamics, since the hadrons are  color neutral objects (analogous to electrically neutral atoms in electrodynamics).  A $J/\psi$-nucleus bound state would represent a state of matter different from ``ordinary" nuclei, i.e. nuclei composed of protons and neutrons interacting by exchange of mesons, and would allow investigation of an aspect of QCD (namely, gluon exchange) within the nuclear environment different from the usual meson exchange aspects.  A diagrammatic model of $\gamma^*+D\to J/\psi +p+n$ is used for the calculations presented here, and it is determined that the kinematic conditions in the proposed JLab experiment and the small size of the contribution to the cross-section from $J/\psi$-nucleon rescattering do not allow the scattering length to be determined.  However, it may be possible to measure the $J/\psi$-nucleon scattering amplitude, at higher energy, in the same experiment; this is discussed in Sec. \ref{sec:intermedenergy}.  The energy of the $J/\psi$-nucleon elastic rescattering is high enough that many partial waves will be involved, in this case, and hence it is not sensitive to the value of the scattering length.  The energy of the $J/\psi$-nucleon elastic rescattering would be in a range for which $J/\psi$-nucleon elastic scattering has not been measured previously (it would be significantly smaller than in the only measurements so far performed).


\chapter{Color transparency and the reaction\\ $\pi+A\to\pi+p+(A-1)^*$}

\section{Introduction}

In this chapter, the ideas of Color Transparency are introduced, and the Glauber theory of high-energy scattering from nuclei is used to calculate properties of the reaction $\pi+A\to\pi+p+(A-1)^*$ for two cases.  The first case ignores any possible effects of Color Transparency, while the second case includes these effects.  The quantity of most interest here is called the nuclear transparency, which is defined as the ratio of two cross-sections:
\be
T\equiv \frac{\sigma}{\sigma_{PWIA}}
\ee
where $\sigma$ is the actual measured cross-section for the reaction occuring in a nucleus, and $\sigma_{PWIA}$ is the cross-section calculated in the Plane Wave Impulse Approximation.  In the PWIA, all interaction of incoming and outgoing particles with nucleons in the nucleus are neglected, except for the interaction which is responsible for the reaction in the first place (e.g. $\pi+p\to\pi+p$ for the reaction $\pi+A\to\pi+p+(A-1)^*$).  The actual cross-section $\sigma$ includes interactions of the incoming and outgoing particles.  These interactions will lead in general to a value for $\sigma$ which is smaller than $\sigma_{PWIA}$, and therefore $T<1$.  (In the above expression for $T$, the cross-sections can in general be total cross-sections or differential cross-sections.)  But a remarkable prediction of perturbative QCD (pQCD) is that under certain kinematic condtions, the outgoing particles from a reaction inside a nucleus will undergo no interaction at all with the other nucleons, and so the nucleus will appear ``transparent" to these outgoing particles.  The requirement for this to occur is that the reaction be a very-large-momentum-transfer elastic or quasi-elastic reaction.  For reasons discussed in the following, the reaction $\pi+A\to\pi+p+(A-1)^*$ should be a good candidate for observing the effects of Color Transparency.

This chapter is organized as follows.  In Sec. \ref{sec:ct}, the basic ideas of Color Transparency (CT) are discussed, along with discussion of the experimental searches for CT which have been performed.  In Sec. \ref{sec:glaubermodel} the Glauber model of high-energy hadron-nucleus scattering is presented.  The basic results of the Glauber theory which will be used are presented in this section.  In Sec. \ref{sec:pionreaction} and \ref{sec:integratedT} , the Glauber model is used to calculate the cross-section and the transparency $T$ for the reaction $\pi+A\to\pi+p+(A-1)^*$ at large pion incident momentum (200 GeV, which is the momentum available at the COMPASS experiment at CERN).  Sec. \ref{sec:pionconclusion} presents our conclusion, which is that at the energies available at COMPASS the effects of Color Transparency should be very evident.

\section{Color Transparency}
\label{sec:ct}

\subsection{Color Transparency basics}
Color Transparency is a prediction of perturbative Quantum Chromodynamics which asserts that when a hadron undergoes a high-momentum-transfer elastic or quasi-elastic reaction inside a nucleus, the outgoing hadron experiences reduced interactions with the nucleons of the nucleus, compared to their interaction in free-space~\cite{miller07}.  In the limit of very large momentum transfer, the outgoing hadron experiences no interaction at all with the rest of the nucleus (it passes through without interacting with any of the nucleons); this is termed the vanishing of Final State Interactions.  Thus the reason for using the word "transparency":  the nucleus appears transparent to the outgoing hadron.  For example, in the quasielastic scattering of an electron from a nucleus accompanied by proton knockout, $A(e,e'p)(A-1)$, perturbative QCD predicts that if the momentum transfer from the electron to the proton is large enough, the knocked-out proton will experience reduced interactions with the rest of the nucleons on its way out.  For very large momentum transfer, the fast moving proton would not interact with the other nucleons at all.  This is in contradistinction to what would happen if we just sent a fast moving proton impinging on a nucleus:  it certainly would not just pass through completely unaffected.  In fact in this case, if the nucleus is large enough (i.e. $A$ large enough) the proton would almost certainly undergo an inelastic collision with a nucleon.  This can be seen from the classical result that the mean-free-path of a particle passing through a system of scatterers is $l=\frac{1}{\sigma\rho}$ where $\sigma$ is the total cross-section of interaction of the particle with an individual scatterer and $\rho$ is the number density of the scatterers.  For a typical nuclear density $\rho\simeq0.2\;fm^(-3)$, and proton-nucleon total cross-section $\sigma=40\;mb$ (for proton momentum greater than a few $GeV$), the mean-free path is $l\simeq1.25\;fm$.  Thus for a nucleus of radius ~$3\;fm$ the incident proton would have a large probability of interacting.  But for the fast-moving proton knocked-out of a nucleus by a hard collision, pQCD predicts that the probability of its interacting with the other nucleons on its way out is much less, and zero in the limit of very large momentum transfer.  The reason why this is so is that pQCD predicts that the outgoing ``proton" is not in fact a usual proton at all, but instead a system of quarks in what is called a ``small-size configuration", where the 3 valence quarks in the proton are much closer together than they usually are in the proton.  And pQCD predicts that the cross-section of interaction of a small-size color singlet with another hadron decreases the closer together the quarks are.  In the limit of zero separation (a ``point-like configuration"), the cross-section of interaction is zero, and to such an object the nucleus appears ``transparent".  This is analogous to what happens to a classical physical electric dipole (which of course has zero net electric charge):  the closer together the two charges are, the smaller is the force exerted on the dipole by any external electric fields.  In the limit as the separation goes to zero, the net force on the dipole goes to zero.

It's important to note that the occurrence of the small-size configuration doesn't have anything to do with the proton being inside a nucleus.  For the free-space elastic reaction $e+p\to e+p$ with large momentum transfer, the outgoing proton will be in a small-size configuration.  But this small-size configuration is not detectable unless the outgoing proton has another particle to interact with.  The role of the nucleus in this is that it serves as a laboratory to study the spatial configuration and spacetime evolution of a hadron produced in a hard exclusive reaction, by observing its interactions with the other nucleons.  It is, however, important that the reaction be elastic or quasi-elastic; if the struck proton breaks apart, then it is not necessary for the outgoing particles to start in point-like configurations.  The reason is that for a large-momentum transfer elastic reaction, all of the quarks involved have to be in a small region of space at the time of interaction in order for the quarks of a given particle to remain together after the collision (see Sec. \ref{formation} for more discussion of this).  For an inelastic reaction, the outgoing quarks aren't required to remain together, and so they don't need to be close together at the time of the collision.

Another feature of the large momentum transfer elastic reaction, e.g. $e+p\to e+p$, is that the produced small-size object (called an ``ejectile") eventually must expand and become a normal-size proton, since a proton is what is detected.  The small-size object is not an eigenstate of the strong Hamiltonian, and so evolves in time.  For the reaction inside a nucleus, if the outgoing proton expands too quickly to its normal size, then it will experience normal-strength interactions with the other nucleons and no transparency will be observed.  Therefore to observe Color Transparency, the outgoing proton should be moving fast enough that it has left the nucleus by the time significant expansion occurs.  High velocity helps in two ways, the first of course being that it leaves the nucleus in a shorter time, and the second being that time-dilation slows down the rate of expansion as observed in the Lab compared to the rate in the proton's rest frame.

Other reactions for which the same ideas hold include quasi-elastic proton scattering ($p+A\to p+p+(A-1)$), pion photoproduction ($\gamma+n\to \pi^-+p$), pion electroproduction, and the two which are explored in this thesis:  pion elastic scattering with proton knockout ($\pi+A\to \pi+p+(A-1)$), and vector meson electroproduction with proton knockout ($\gamma^*+A\to V+p+(A-1)$, $V$ representing a vector meson, which in this thesis is the $\rho$).

\subsection{Formation of Pointlike Configuration}
\label{formation}

The idea of why a large momentum-transfer elastic reaction involves a small size configuration of quarks can be seen heuristically from the example of the pion form factor, i.e. the reaction $e+\pi\to e+\pi$, with large momentum transfer~\cite{dok91}.  To lowest order in the electromagnetic interaction this occurs through emission by the electron of a virtual photon which is absorbed by one of the two valence quarks in the pion, the pion remaining intact.  The 4-momentum of the virtual photon is denoted by $q$, with $q^2=-Q^2$.  Loosely speaking, in the frame where the pion is moving fast, the two quarks are essentially moving parallel to each other at the same speed.  When one of the quarks absorbs the large-momentum virtual photon (large $Q$) the direction of its motion is changed.  If the pion is to remain intact, the momentum of the other quark must also change so that the two quarks are again moving in the same direction.  This occurs by exchange of a gluon between the quarks.  The 4-momentum-squared of this gluon is of order $-Q^2$ also, and hence the gluon is far off its mass-shell.  By the uncertainty principle it can only exist for a time which is of order $~1/Q$.  But in order to be absorbed by the other quark it must traverse the distance between the 2 quarks, and so this distance must be less than order $~1/Q$.  Thus in order for the pion to remain a pion, i.e. in order for the reaction  $e+\pi\to e+\pi$ to proceed, at the time the photon is absorbed the two quarks need to be closer together in space than a distance of order $~1/Q$.  If they are then the reaction may proceed, and the outgoing "pion" will in fact consist of 2 valence quarks in a small-size configuration, with their separation being of order $~1/Q$.  If the 2 quarks are farther apart than this, then they will separate from each other, with each quark eventually hadronizing, yielding an inelastic reaction $e+\pi\to e+X$.  Since the wavefunction of the quarks in the pion has some amplitude for the 2 quarks to be close together, the elastic reaction may occur.  The outgoing ejectile then evolves over time, becoming the observed pion.

Nonperturbative studies of realistic hadron models also show the formation of a small-sized configuration during a large-momentum transfer reaction~\cite{miller93}. 

\subsection{Expansion from PLC}

Once a pointlike configuration is formed in a large-momentum transfer reaction, the system will expand until it reaches the ``normal" size of the hadron; once it has reached its normal size the expansion ceases.  In order to account for this, models must be used.  The model used in the analysis in this thesis is called the ``quantum diffusion model"~\cite{liu88,dok91}.  In this model, the interaction cross-section of the outgoing object with the nucleons increases linearly with distance from the interaction point where the hard scatter occurred which produced the pointlike configuration (see Eq. \ref{sigeff}).  This model is derived from perturbative QCD~\cite{dok91}:  for a quark-antiquark system starting from a transverse size of zero, gluon exchange between the quark and antiquark proceeds until the system reaches the normal meson size.  It is shown in~\cite{dok91} that the transverse area of the system  (and hence its cross-section) increases linearly with distance traveled.  The ``naive" model of expansion would correspond to free quarks expanding from zero transverse size in both directions transverse to the momentum of the system.  In this case the transverse area of the system would increase as the square of the distance traveled~\cite{liu88}.  Since the quantum diffusion model is derived from QCD (albeit perturbatively), it is the model used in the calculations presented here.  

\subsection{Experimental searches for CT}

The first dedicated experiment to search for effects of color transparency was in 1988 at Brookhaven National Laboratory~\cite{carroll88}.  Quasi-elastic scattering of protons (i.e. the reaction $p+A\to p+p+(A-1)$, where a proton is knocked out of the nucleus by the incoming projectile) in various nuclei was observed, at incident proton momentum of from 6 to 12 GeV.  The transparency, as a function of the 4-momentum transfer squared $t$, was observed to increase as $\vert t\vert$ increased, up to a point, but then the transparency decreased after that as $\vert t\vert$ was increased further.  This behavior did not appear to agree with the predictions of color transparency, as the transparency should increase as $\vert t\vert$ is increased.  However, their may be other factors at work in the elementary $pp$ scattering cross-section, and several models were proposed to try to explain this behavior~\cite{miller07}.  Another experiment was later performed~\cite{mardor98,leksanov01} wherein the momenta of both outgoing protons was measured (in contrast to the first experiment where only one of the outgoing proton's momentum was analyzed).  Similar results were obtained as in the earlier experiment, with the transparency first rising and then falling with $\vert t\vert$.

In the $(p,2p)$ reactions, in order for a small-sized configuration to be formed it is necessary to have 6 quarks all localized in a small region, which may have a very small probability.  The formation of a small-sized configuration may be more likely if fewer quarks are involved.  Thus quasi-elastic electron scattering ($e+A\to e+p+(A-1)$) may be a better candidate to observe color transparency.  In this case, the elementary reaction $e+p\to e+p$ is better understood also, being an electromagnetic interaction rather than a strong interaction.  This experiment has been performed at SLAC~\cite{makins94,oneill95,garrow02} with a range of momentum-transfer squared $Q^2$ from $1$ to $8.1\;GeV^2$.  The results did not show any indication of color transparency.  The observations agreed with the standard calculation which assumes that the outgoing object is a normal-sized proton with the usual free-space value of its cross-section of interaction with the other nucleons.

There has been one experiment that can be said to show unambiguous evidence of color transparency.  This was the diffractive dissociation into dijets of pions scattered from carbon and platinum nuclei~\cite{aitala01}.  In this process a high-energy incident pion strikes a nucleus, with the minimal quark configuration $q\bar{q}$ scattering coherently from the nucleus.  The individual $q$ and $\bar{q}$ then each form a jet of hadrons.  Observation of two jets with large transverse momentum (transverse to the pion beam direction) indicates that the $q$ and $\bar{q}$ had large relative transverse momentum and hence small transverse spatial separation.  If the $q\bar{q}$ are in a pointlike configuration, then for forward scattering $t\simeq 0$, since the scattering is coherent, the amplitude for scattering from a nucleus would be $\sim A\;{\cal M}$, where ${\cal M}$ is the amplitude for scattering from a single nucleon; hence the forward differential cross-section would depend on $A$ as $A^2$.  In an experiment, what is measured is the integrated cross-section, $\sigma(A)=\int \frac{d\sigma}{dt}dt$.  For the coherent reaction, ${d\sigma}{dt}$ is also proportional to the form factor of the nucleus,  $~e^{t R_A^2/3}$ where $R_A$ is the radius of the nucleus; this then gives an $A$-dependence of $\sigma(A)\sim A^{4/3}$~\cite{miller07}.  This is to be contrasted with the expectation that a normal-size incident pion would undergo strong absorption from a nucleus, and so essentially only the nucleons on the surface would participate in the reaction; thus the $A$-dependence of the cross-section on a large nucleus would go like $\sim A^{2/3}$ .  The result of the experiment~\cite{aitala01} was a cross-section depending on $A$ as $A^{1.55}$, a clear indication of the effects of color transparency.

One further experiment that has been performed, with somewhat inconclusive results, is the reaction $\gamma+n\to \pi^- + p$ in $^4He$~\cite{dutta03}.  The results show a momentum-transfer dependence that seems to indicate CT, with the transparency rising with $\vert t\vert$.  However, better statistical precision is needed in order to be conclusive.

Other candidate reactions are those involving production of vector mesons.  These are discussed in the next chapter.

As it would seem more likely to observe Color Transparency in reactions involving mesons, it would be of interest to measure the quasielastic scattering of pions from nuclei at large momentum transfer, i.e. $\pi +A\to \pi +p +(A-1)$.  This reaction is the subject of this chapter.  In the COMPASS experiment at CERN, pions with momenta of $200$ GeV are produced.  At this large momentum, the expansion of the produced point-like configuration does not occur (due to time-dilation) before the pion escapes the nucleus.  COMPASS should therefore be able to observe the effect of Color Transparency~\cite{miller2010}.   In order to calculate the cross-section, a formalism is needed which takes into account the initial- and final-state interactions of the pion, and the final-state interaction of the proton, with the spectator nucleons.  As we are interested in high incident pion energy, the Glauber model provides such a formalism.  In the following, the cross-section for the above reaction is calculated in the Glauber model.  The result is the same as obtained in the usual Distored Wave Impulse Approximation, which has been used extensively to analyze proton knockout reactions~\cite{jacob66}.  The Glauber model can be easily extended to account for particle production processes, such as vector meson production, e.g. $\gamma^*+A\to V+p+(A-1)$.  In the following chapter this reaction is analyzed in the Glauber model.  This represents a new result.

\section{Glauber model of high-energy hadron-nucleus scattering}
\label{sec:glaubermodel}

In order to calculate the scattering cross-section for a projectile incident on a nucleus, one needs a formalism which accounts for interaction of the projectile with more than one nucleon during its passage through the nucleus.  The Glauber model of high-energy hadron-nucleus scattering~\cite{glaub59} is a multiple-scattering model which is valid under certain conditions.  The conditions are:  1) high energy of the incident particle, compared to the binding energy of the nucleons in the target nucleus; 2)  small angle scattering of the projectile.  Under these conditions the momentum transfer is mostly transverse, and so the longitudinal momentum transfer is neglected; the energy transfer from the projectile is also small, and so the energy transfer is neglected also.  In order to calculate the scattering cross-section in the Glauber model, only knowledge of the free-space hadron-nucleon scattering amplitude and the wavefunctions of the target system is required.  The Glauber model does not take into account the Fermi motion of the nucleons; for a projectile of high energy the Fermi motion should matter little.

The Glauber model takes advantage of the fact that high-energy elastic hadron-hadron scattering occurs at mostly small scattering angles; see e.g. ~\cite{perl74}.  For example, for proton-proton scattering at LAB incident momentum of 25 GeV,  essentially all of the scattering events occur with $\vert t\vert\leq 1\; GeV^2$, which corresponds to a center-of-mass scattering angle of $6^{\circ}$.  This is also true for the usual eikonal approximation in potential scattering; the Glauber model is an extension of the eikonal approximation to include scattering from multiple scatterers in the target.  In the Glauber picture, the nucleons' positions are fixed in place during the time that the projectile traverses the nucleus (the "frozen" approximation).  Also, the projectile is assumed to scatter at most once from any individual nucleon.  In between scattering events the projectile travels in a straight line.  The Glauber result for the scattering amplitude is a sum of terms representing the possible multiple-scatterings of the projectile.  The first term represents one elastic scatter, the second term represents two elastic scatters (from different nucleons), and so on up to a maximum of $A$ elastic scatters, $A$ being the nucleon number of the nucleus.

\begin{figure}[bth]
     \begin{center}

            \includegraphics[width=5in,height=2.5in]{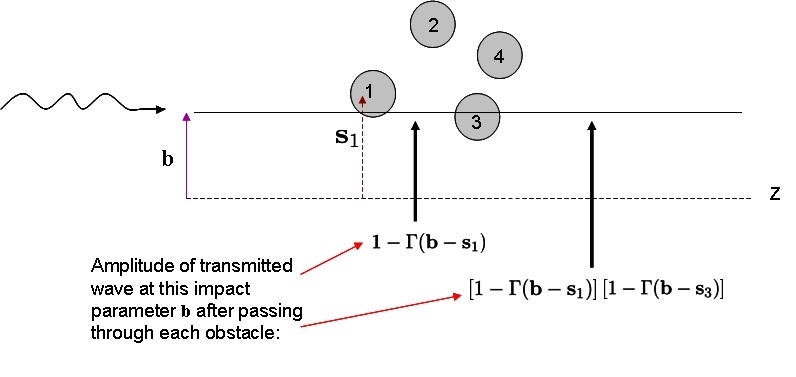}

    \end{center}
    \caption{%
        Diffraction from multiple scatterers.
     }%
   \label{fig:diff2}
\end{figure}

The Glauber scattering amplitude is very similar to the Fraunhofer diffraction amplitude in optics, and can be interpreted in terms of diffraction~\cite{glaub67}.  For 2-body elastic scattering, with initial and final momenta $\mathbf{k}$, $\mathbf{k}'$, we define the momentum transfer $\mathbf{q}\equiv\mathbf{k}-\mathbf{k}'$, and for high-energy scattering we have $\mathbf{q}\simeq\mathbf{q}_{\perp}$ where $\mathbf{q}_{\perp}$ is the component of $\mathbf{q}$ perpendicular to $\mathbf{k}$.
 The scattering amplitude is then given by (see Appendix A)
\be
f(\mathbf{q})=\frac{ik}{2\pi} \int d^2b\;e^{i\mathbf{q}\cdot \mathbf{b}}\Gamma(\mathbf{b})
\ee
which is the same as the expression for the scattering amplitude in Fraunhofer diffraction, for scattering of an incident wave from an obstacle.
\begin{figure}[bth]
     \begin{center}

            \includegraphics[width=3.5in,height=1.5in]{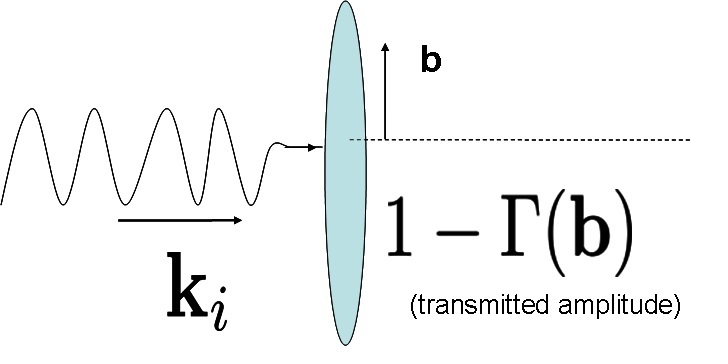}

    \end{center}
    \caption{%
        Diffraction of an incident wave by an obstacle.
     }%
   \label{fig:diffraction}
\end{figure}
In that case (see Fig. \ref{fig:diffraction}) the "profile function" $\Gamma(\mathbf{b})$ is related to the amplitude transmitted through the obstacle (for incident amplitude 1) by $t(\mathbf{b})=1-\Gamma(\mathbf{b})$.  In the Glauber model, a wave incident on a system of $n$ scatterers, each with profile function $\Gamma(\mathbf{b})$, undergoes absorption and transmission through each scatterer (see Fig. \ref{fig:diff2}).  After it has passed through all of the scatterers, the transmitted wave has amplitude (at a given transverse position $\mathbf{b}$)
\be
t_{tot}(\mathbf{b})=\prod_{j=1}^n (1-\Gamma(\mathbf{b}-\mathbf{s}_j))
\ee
where $\mathbf{s}_j$ is the transverse position of the center of the $j$th scatterer.  The scattering amplitude is then given by the 2-dimensional Fourier transform of $1-t_{tot}(\mathbf{b})\equiv\Gamma_{tot}(\lbrace\mathbf{b}-\mathbf{s}_j\rbrace)$.  This is for scatterers at given positions.  For a quantum system of scatterers (e.g. $A$ nucleons in a nucleus) the scattering amplitude for the target system to remain in its initial state is given by the expectation value of the scattering amplitude in the initial state:
\be
F_{ii}=\int d^3r_1 d^3r_2\ldots d^3r_A 
\vert\psi(\mathbf{r}_1,\mathbf{r}_2,\ldots,\mathbf{r}_A)\vert^2 
\frac{ik}{2\pi} \int d^2b\;e^{i\mathbf{q}\cdot \mathbf{b}}\; \Gamma_{tot}(\mathbf{b},\mathbf{s}_1,...,\mathbf{s}_A)
\ee
\be
=\frac{ik}{2\pi} \int d^2b\;e^{i\mathbf{q}\cdot \mathbf{b}}\;\langle i\vert \Gamma_{tot}(\mathbf{b},\mathbf{s}_1,...,\mathbf{s}_A)\vert i\rangle.
\ee
For transition of the target system to a final state $\vert f\rangle$ the amplitude is
\be\label{glauberamplitude}\boxed{
F_{fi}(\mathbf{q})=\frac{ik}{2\pi} \int d^2b\;e^{i\mathbf{q}\cdot \mathbf{b}}\;\langle f\vert \Gamma_{tot}(\mathbf{b},\mathbf{s}_1,...,\mathbf{s}_A)\vert i\rangle}\;
\ee
with the total profile function given by
\be\label{totgamma}\boxed{
\Gamma_{tot}(\mathbf{b},\mathbf{s}_1,...,\mathbf{s}_A)=1-\prod_{j=1}^A (1-\Gamma(\mathbf{b}-\mathbf{s}_j))}\;.
\ee
It is important to note that the states $\vert i\rangle$ and $\vert f\rangle$ are \emph{internal} states of the target nucleus.  (For a derivation of \eq{glauberamplitude} for the case of a projectile scattering from a system of scatterers bound by a potential, starting from the Schrodinger equation for the projectile-target system, see~\cite{glaub59}).

\section{Quasi-elastic scattering:  $\pi+A\to\pi+p+(A-1)^*$}
\label{sec:pionreaction}

The amplitude for a transition from the target (A-nucleon) state $\vert i\rangle$ to state $\vert f\rangle$ is given in Glauber theory by 
\begin{align}
\label{glauberamp}
F_{fi}=\frac{ik}{2\pi} \int d^2b\;e^{i\mathbf{q}\cdot \mathbf{b}}\;\int d^3r_1...d^3r_A\Phi_f^*\Phi_i \Gamma_{tot}(\mathbf{b},\mathbf{s}_1,...,\mathbf{s}_A)\; \delta^3\bigl(\frac{1}{A}\sum_{j=1}^A \mathbf{r}_j\bigr)
\end{align}
where \begin{equation}\Gamma_{tot}=1-\prod_{j=1}^A\bigl [1-\Gamma(\mathbf{b}-\mathbf{s}_j)\bigr]\end{equation} and the profile function $\Gamma(\mathbf{b})$ is related to the $\pi$-nucleon elastic scattering amplitude $f(\mathbf{q})$ by $f(\mathbf{q})=\frac{ik}{2\pi} \int d^2b\;e^{i\mathbf{q}\cdot \mathbf{b}}\;\Gamma(\mathbf{b})$.  Here $\mathbf{q}=\mathbf{k}-\mathbf{k}'$ with $\mathbf{k}=$ pion initial momentum and $\mathbf{k}'=$ pion final momentum.  And we take $\mathbf{k}$ along the positive $z$-axis, with $\mathbf{b}$ and $\mathbf{s}_j$ perpendicular to the $z$-axis. $\mathbf{r}_j=(\mathbf{s}_j,z_j)$ is the coordinate vector of the $j$th nucleon.  $\Phi_i(\mathbf{r}_1,\mathbf{r}_2,...,\mathbf{r}_A)$ and $\Phi_f(\mathbf{r}_1,\mathbf{r}_2,...,\mathbf{r}_A)$ are the initial and final wavefunctions of the $A$-nucleon system; these are in principle exact eigenstates of the $A$-nucleon Hamiltonian. The final $A$-nucleon state $\vert f\rangle$ can be either a bound or continuum state.  The delta function in the above equation enforces that the coordinates $\mathbf{r}_j$ of the nucleons are relative coordinates; hence the target wavefunctions above are internal wavefunctions.  For large $A$ we can neglect the delta function.   Note that for $f(0)$ pure imaginary, the optical theorem in terms of $\Gamma$ is $\int d^2b \Gamma(\mathbf{b})=\frac{1}{2}\sigma_{tot}^{\pi N}$.

For the case of proton knockout, the final state of the target nucleus consists of a continuum state, with one unbound proton.  We will be using shell-model wavefunctions for the initial target state.  For the final $A$-nucleon state, we will assume a wavefunction of the form 
\be
\phi_A(\mathbf{r}_1,\mathbf{r}_2,\ldots,\mathbf{r}_A)=\chi_{\mathbf{p}}(\mathbf{r}_1)\phi_{A-1}^f(\mathbf{r}_2,\ldots,\mathbf{r}_A)
\ee
where $\chi_{\mathbf{p}}(\mathbf{r}_1)$ is a scattering wavefunction for the proton of momentum $\mathbf{p}$.  Because of the (relatively) high energy of the outgoing proton, it is appropriate to use an eikonal wavefunction for the proton:
\be
\chi_{\mathbf{p}}(\mathbf{r}_1)=e^{i\mathbf{p}\cdot\mathbf{r}_1}e^{-\frac{1}{2}\int_0^{\infty}ds\; \sigma_{tot}^{pN}\rho(\mathbf{r}_1+s\hat{\mathbf{p}})}\equiv e^{i\mathbf{p}\cdot\mathbf{r}_1} e^{-\frac{1}{2}\alpha_{\mathbf{p}}(\mathbf{r}_1)} .
\ee
This represents scattering of the outgoing proton in the optical potential due to the other $A-1$ nucleons.  Therefore $\rho$ in the exponential should be the nucleon density of the residual nucleus and hence would depend on the final state $f$ of the residual nucleus.  We will assume, however, that $\rho$ is the same as the nucleon density of the initial nucleus, which should be valid for final states which are one-hole states or small excitations thereof.

In an actual experiment, the outgoing pion and proton are detected, while the recoiling residual nucleus is usually not detected.  Therefore the cross-section of interest is obtained by summing over all final states of the residual nucleus.  In calculating the differential cross-section, however, the phase space factors depend in principle on $f$ since the internal energy of the residual nucleus depends on $f$.  For the case of a high-energy projectile, and proton knockout, it is legitimate to neglect dependence of the phase-space factors on $f$.  Therefore we may simply square the amplitude  \eq{glauberamp}, and then sum over all final states $f$, and we can use closure on the residual nucleus states:  $\sum_f \phi_{A-1}^f(\mathbf{r}_2',\ldots,\mathbf{r}_A') \phi_{A-1}^f(\mathbf{r}_2,\ldots,\mathbf{r}_A) =\prod_{j=2}^A \delta^{(3)}(\mathbf{r}_j - \mathbf{r}_j')$.  The result gives a multiple scattering expansion, where the first term represents one hard scatter of the pion, with momentum transfer $\mathbf{q}$, together with multiple soft re-scatterings as the pion travels through the nucleus; the second term represents two scatterings of the pion with momentum transfers $\mathbf{q}_1$ and $\mathbf{q}_2$ such that $\mathbf{q}_1+\mathbf{q}_2=\mathbf{q}$, etc.  For the case of quasi-elastic kinematics, where we have $\mathbf{p}\simeq\mathbf{q}$, the outgoing proton has received almost all the momentum transferred from the pion.  Therefore the higher-order terms (representing multiple hard scatters of the pion) should be negligible, and only the first term should be appreciable.  Since in this term the pion only undergoes soft re-scatterings with the other $A-1$ nucleons, and by assumption the outgoing proton also only undergoes soft re-scatterings (inherent in the eikonal form of the proton wavefunction), the final state of the residual nucleus should be a one-hole state of the target nucleus.  And indeed this first term is identical to what is obtained if instead of summing over all final states of the residual nucleus we only sum over one-hole states.  So let us evaluate that case.       

If we only sum over final states of the residual nucleus which are one-hole states, i.e. obtained from the initial $A$-nucleon state by deleting one single-particle state, then the initial wavefunction can be written (in the shell model)
\be
\phi_A(\mathbf{r}_1,\ldots,\mathbf{r}_A)=\phi_n(\mathbf{r}_1) \phi_{A-1}^f(\mathbf{r}_2,\ldots,\mathbf{r}_A)
\ee
with $\phi_n$ being a shell-model single-particle wavefunction, and the Glauber amplitude is
\be
F_{fi}^{(n)}=\frac{ik}{2\pi}\int d^2b e^{i\mathbf{q}\cdot\mathbf{b}}\int d^3r_1\ldots d^3r_A \chi_p^*(\mathbf{r}_1)\phi_n(\mathbf{r}_1) \Bigl\vert\phi_{A-1}^f(\mathbf{r}_2,\ldots,\mathbf{r}_A)\Bigr\vert^2\Gamma_{tot}(\mathbf{b},\lbrace \mathbf{r}_j \rbrace)
\ee
Separating out the terms in $\Gamma_{tot}$ which are independent of $\mathbf{r}_1$, we have:
\be
\Gamma_{tot}=1-\prod_{j=2}^A (1-\Gamma_{bj}) +\Gamma_{b1}\prod_{j=2}^A (1-\Gamma_{bj}) 
\ee
\be
=1-\prod_{j=2}^A (1-\Gamma_{bj}) +\Gamma_{eff}(\mathbf{b},\mathbf{r}_1,\ldots,\mathbf{r}_A)
\ee
where
\be
\Gamma_{eff}(\mathbf{b},\mathbf{r}_1,\ldots,\mathbf{r}_A)\equiv\Gamma_{b1}\prod_{j=2}^A (1-\Gamma_{bj}) .
\ee
Because of the orthogonality of the single-particle wavefunctions $\chi_p$ and $\phi_n$, the terms in $\Gamma_{tot}$ that are independent of $\mathbf{r}_1$ contribute zero to $F_{fi}$.  Hence only $\Gamma_{eff}$ contributes, and we have:
\be
F_{fi}^{(n)}=\frac{ik}{2\pi}\int d^2b e^{i\mathbf{q}\cdot\mathbf{b}}\int d^3r_1\ldots d^3r_A \chi_p^*(\mathbf{r}_1)\phi_n(\mathbf{r}_1) \Bigl\vert\phi_{A-1}^f(\mathbf{r}_2,\ldots,\mathbf{r}_A)\Bigr\vert^2\Gamma_{eff}(\mathbf{b},\lbrace \mathbf{r}_j \rbrace)
\ee

For $\vert\phi_{A-1}\vert^2$, we assume an independent particle model, and write
\be
 \Bigl\vert\phi_{A-1}^f(\mathbf{r}_2,\ldots,\mathbf{r}_A)\Bigr\vert^2=\prod_{j=2}^A \rho_1(\mathbf{r}_j)
\ee
with the single-particle density $\rho_1$ normalized to 1.  The nucleon density of the state $\phi_{A-1}^f$ is then $\rho(\mathbf{r})=(A-1)\rho_1(\mathbf{r})$.  We then have
\be
\int d^3r_2\ldots d^3r_A \Bigl\vert\phi_{A-1}^f(\mathbf{r}_2,\ldots,\mathbf{r}_A)\Bigr\vert^2\Gamma_{eff}=\Gamma_{b1}\int d^3r_2\ldots d^3r_A \prod_{j=2}^A \rho_1(\mathbf{r}_j) (1-\Gamma_{bj})
\ee
\be
=\Gamma_{b1}\Biggl[\int d^3r_2 \rho_1(\mathbf{r}_2)(1-\Gamma(\mathbf{b}-\mathbf{s}_2))\Biggr]^{A-1}
\ee
\be
\equiv \Gamma_{b1}\; g(\mathbf{b})
\ee
and hence
\be
\label{amp1}
F_{fi}^{(n)}=\frac{ik}{2\pi}\int d^2b e^{i\mathbf{q}\cdot\mathbf{b}}\int d^3r_1 \chi_p^*(\mathbf{r}_1)\phi_n(\mathbf{r}_1)  \Gamma_{b1}\; g(\mathbf{b})
\ee

Since we are interested in large $A$, we can approximate $g(\mathbf{b})$ by an exponential, as follows.  We have
\be
\begin{split}
\int d^3r_2 \rho_1(\mathbf{r}_2)(1-\Gamma(\mathbf{b}-\mathbf{s}_2))&=1-\int d^3r_2 \rho_1(\mathbf{r}_2)\Gamma(\mathbf{b}-\mathbf{s}_2)\\
&=1-\frac{1}{A-1}\int d^2s_2 dz_2 \rho(\mathbf{r}_2)\Gamma(\mathbf{b}-\mathbf{s}_2)
\end{split}.
\ee
Now the profile function $\Gamma(\mathbf{b}-\mathbf{s}_2)$ is in general a sharply peaked function of its argument, peaked at $\mathbf{s}_2=\mathbf{b}$, while in contrast the nucleon density $\rho(\mathbf{r}_2)$ is a much more slowly varying function.  Hence we may approximate
\be
\int d^2s_2 dz_2 \rho(\mathbf{s}_2,z_2)\Gamma(\mathbf{b}-\mathbf{s}_2)\simeq\int  dz_2 \rho(\mathbf{b},z_2)\int d^2s_2\Gamma(\mathbf{b}-\mathbf{s}_2)=\frac{2\pi}{ik} f(0)\;T(\mathbf{b})
\ee
where $T(\mathbf{b})=\int_{-\infty}^{\infty}  dz \rho(\mathbf{b},z)$ is called the ``thickness function''~\cite{yennie78}, and $f(0)$ is the forward pion-nucleon elastic scattering amplitude.  For high-energy scattering, $f(0)$ is almost pure imaginary, and so using the optical theorem we obtain 
\be
\int d^2s_2 dz_2 \rho(\mathbf{s}_2,z_2)\Gamma(\mathbf{b}-\mathbf{s}_2)\simeq \frac{1}{2}\sigma_{tot}^{\pi N}\;T(\mathbf{b})
\ee
Hence
\be
\label{gofb1}
 g(\mathbf{b})\simeq \Bigl[1-\frac{1}{A-1}\frac{\sigma_{tot}^{\pi N}}{2}\;T(\mathbf{b})\Bigr]^{A-1}\simeq e^{-\frac{1}{2}\sigma_{tot}^{\pi N}\;T(\mathbf{b})}
\ee

Inserting this in the expression for $F_{fi}$, we have
\be
F_{fi}^{(n)}=\frac{ik}{2\pi}\int d^2b e^{i\mathbf{q}\cdot\mathbf{b}}\int d^3r_1 \chi_p^*(\mathbf{r}_1)\phi_n(\mathbf{r}_1)  \Gamma_{b1}\;e^{-\frac{1}{2}\sigma_{tot}^{\pi N}\;T(\mathbf{b})}
\ee
We may evaluate the integral over $\mathbf{b}$ at this point by using again the property of $\Gamma(\mathbf{b}-\mathbf{s}_1)$ that it is very sharply peaked at $\mathbf{b}=\mathbf{s}_1$, and write
\be
\frac{ik}{2\pi}\int d^2b \Gamma_{b1}\;e^{-\frac{1}{2}\sigma_{tot}^{\pi N}\;T(\mathbf{b})}\simeq e^{-\frac{1}{2}\sigma_{tot}^{\pi N}\;T(\mathbf{s}_1)}\frac{ik}{2\pi}\int d^2b \Gamma(\mathbf{b}-\mathbf{s}_1)
\ee
\be
=e^{i\mathbf{q}\cdot\mathbf{s}_1}\; f(\mathbf{q}) \;e^{-\frac{1}{2}\sigma_{tot}^{\pi N}\;T(\mathbf{s}_1)}
\ee
and therefore
 \be
F_{fi}^{(n)}= f(\mathbf{q})\int d^3r_1 e^{i\mathbf{q}\cdot\mathbf{s}_1} \chi_p^*(\mathbf{r}_1)\phi_n(\mathbf{r}_1) e^{-\frac{1}{2}\sigma_{tot}^{\pi N}\;T(\mathbf{s}_1)}.
\ee
Finally, writing $\chi_p^*(\mathbf{r}_1)=e^{-i\mathbf{p}\cdot\mathbf{r}_1}\;e^{-\frac{1}{2}\alpha_p(\mathbf{r}_1)}$, we can write $F_{fi}^{(n)}$ in terms of the missing momentum $\mathbf{p}_m\equiv \mathbf{p}-\mathbf{q}$ as
 \be
\label{pionresult}
F_{fi}^{(n)}= f(\mathbf{q})\int d^3r_1 e^{-i\mathbf{p}_m\cdot\mathbf{r}_1}\phi_n(\mathbf{r}_1)e^{-\frac{1}{2}\alpha_p(\mathbf{r}_1)} e^{-\frac{1}{2}\sigma_{tot}^{\pi N}\;T(\mathbf{s}_1)}.
\ee
and
\be
\label{pionressquared}
\begin{split}
\vert F_{fi}^{(n)}\vert^2=\vert  f(\mathbf{q})\vert^2 \int & d^3r_1 d^3r_1' e^{-i\mathbf{p}_m\cdot(\mathbf{r}_1-\mathbf{r}_1')}\phi_n^*(\mathbf{r}_1')\phi_n(\mathbf{r}_1)\\
&\times e^{-\frac{1}{2}\alpha_p(\mathbf{r}_1)} e^{-\frac{1}{2}\alpha_p(\mathbf{r}_1')}    e^{-\frac{1}{2}\sigma_{tot}^{\pi N}\;T(\mathbf{s}_1)}  e^{-\frac{1}{2}\sigma_{tot}^{\pi N}\;T(\mathbf{s}_1')}
\end{split}
\ee
Note that above we have used $\mathbf{q}\cdot\mathbf{s}_1\simeq\mathbf{q}\cdot\mathbf{r}_1$ which is valid since $q_z\ll\vert\mathbf{q}_{\perp}\vert$.
This result agrees with the usual Distorted Wave Impulse Approximation result for the amplitude of the $(p,2p)$ reaction~\cite{jacob66} if we identify the distortion factor for the incoming projectile as $D(\mathbf{r})=e^{-\frac{1}{2}\sigma\int_{-\infty}^z dz'\rho(\mathbf{s},z')}$ and the distortion factor for the outgoing projectile as  $D(\mathbf{r})=e^{-\frac{1}{2}\sigma\int_{z}^{\infty} dz'\rho(\mathbf{s},z')}$, which is valid here since the scattering angle of the projectile is very small and so both integrals in the exponentials are along the same straight-line path.  Note that these distortion factors are the same as one obtains in the eikonal approximation to the scattering wavefunction using an optical potential $V_{opt}(\mathbf{r})=-\frac{1}{2}\sigma\rho(\mathbf{r})$~\cite{joachain75}.

So now squaring and summing over all one-hole final states $f$, which is equivalent to summing over all occupied states $n$ of the initial nucleus, we obtain
\be
\label{sumsquared}
\begin{split}
\sum_{n=1}^A \vert F_{fi}^{(n)}\vert^2=\vert f(\mathbf{q})\vert^2 \int & d^3r_1  d^3r_1'  e^{-i\mathbf{p}_m\cdot(\mathbf{r}_1-\mathbf{r}_1')}     \;A \rho(\mathbf{r}_1,\mathbf{r}_1')\\
&\times e^{-\frac{1}{2}\alpha_p(\mathbf{r}_1)}e^{-\frac{1}{2}\alpha_p(\mathbf{r}_1')} e^{-\frac{1}{2}\sigma_{tot}^{\pi N}\;T(\mathbf{s}_1)} e^{-\frac{1}{2}\sigma_{tot}^{\pi N}\;T(\mathbf{s}_1')} 
\end{split}
\ee
where
\be
\rho(\mathbf{r}_1,\mathbf{r}_1')=\frac{1}{A}\sum_{n=1}^A \phi_n^*(\mathbf{r}_1')\;\phi_n(\mathbf{r}_1)
\ee
is the shell-model one-body density matrix.

To evaluate this using shell-model wavefunctions, it's easiest to write it as
\be
\label{sumsquared2}
\sum_{n=1}^A \vert F_{fi}^{(n)}\vert^2=\vert f(\mathbf{q})\vert^2 \sum_{n=1}^A \Bigl\vert g^{(n)}(\mathbf{p}_m)\Bigr\vert^2
\ee
where $\Bigl\vert g^{(n)}(\mathbf{p}_m)\Bigr\vert^2$ is called the distorted momentum distribution in the shell-model state $\phi_n$ ~\cite{jacob66}, with
\be
\label{gnpion}
g^{(n)}(\mathbf{p}_m)\equiv \int d^3r_1 e^{-i\mathbf{p}_m\cdot\mathbf{r}_1}\phi_n(\mathbf{r}_1)e^{-\frac{1}{2}\alpha_p(\mathbf{r}_1)} e^{-\frac{1}{2}\sigma_{tot}^{\pi N}\;T(\mathbf{s}_1)}.
\ee

For the proton knockout reaction, the transparency $T$ is defined as the ratio of the measured 5-fold differential cross-section to the differential cross-section calculated in the Plane Wave Impulse Approximation (PWIA)~\cite{garrow02, oneill95, makins94,benhar96}.  This can be evaluated at a specific kinematic point, i.e. a particular value of the missing momentum $\mathbf{p}_m$, or it can be the ratio of the integrated cross-sections, integrated over some domain $\cal{D}$ of $\mathbf{p}_m$.  Thus
\be
\label{transpdef}
T(\mathbf{p}_m)=\frac{\frac{d\sigma}{dE'd\Omega'  d\Omega_p}}{\frac{d\sigma_{PWIA}}{dE'd\Omega'  d\Omega_p}}
\ee 
or
\be
\label{integratedT}
T_{\cal{D}}=\frac{\int_{\cal{D}}d^3p_m\frac{d\sigma}{dE'd\Omega'  d\Omega_p}}{\int_{\cal{D}}d^3p_m \frac{d\sigma_{PWIA}}{dE'd\Omega'  d\Omega_p}}.
\ee 
We call the latter the ``integrated transparency".
At a given value of $\mathbf{p}_m$, the kinematic factors in the cross-sections cancel in the ratio Eq. \ref{transpdef}.  The 5-fold differential cross-section is proportional to $\sum_{n=1}^A \vert F_{fi}^{(n)}\vert^2$, and so we have
\be
\label{transp}
T(\mathbf{p}_m)=\frac{\sum_{n=1}^A \vert F_{fi}^{(n)}\vert^2}{\sum_{n=1}^A \vert F_{fi}^{(n)}\vert_{PWIA}^2}=\frac{ \sum_{n=1}^A \Bigl\vert g^{(n)}(\mathbf{p}_m)\Bigr\vert^2}{ \sum_{n=1}^A \Bigl\vert g^{(n)}(\mathbf{p}_m)\Bigr\vert_{PWIA}^2}
\ee
The PWIA value of $g^{(n)}$ is obtained from \eq{gnpion} by setting the attenuation factors equal to 1, which gives
\be
g^{(n)}_{PWIA}(\mathbf{p}_m)=\int d^3r_1 e^{-i\mathbf{p}_m\cdot\mathbf{r}_1}\phi_n(\mathbf{r}_1)=\sqrt{(2\pi)^3}\;\tilde{\phi}_n(\mathbf{p}_m)
\ee
i.e. just the momentum space wavefunction of the $n$th state.  Thus $ \sum_{n=1}^A \Bigl\vert g^{(n)}(\mathbf{p}_m)\Bigr\vert_{PWIA}^2$ is the momentum distribution of the initial nucleus.

To incorporate effects of Color Transparency into our result, we note that the expression for the amplitude $F_{fi}^{(n)}$,  \eq{pionresult} can be interpreted as follows:  the incoming pion strikes the proton in the nucleus at the position $\mathbf{r}_1$, which knocks the proton out; the proton suffers attenuation on its way out of the nucleus, beginning at the point $\mathbf{r}_1$, as represented by the factor
\be
e^{-\frac{1}{2}\alpha_p(\mathbf{r}_1)}=e^{-\frac{1}{2}\sigma_{tot}^{pN}\int_{0}^{\infty}\rho(\mathbf{r}_1+s\hat{\mathbf{p}})\;ds},
\ee
while the pion suffers attenuation on its way in (before the collision with the proton) from $z=-\infty$ up until the point $z=z_1$ and on its way out (after the collision with the proton) starting at $z=z_1$ until $z=\infty$, as represented by
\be
e^{-\frac{1}{2}\sigma_{tot}^{\pi N}\;T(\mathbf{s}_1)}  =e^{-\frac{1}{2}\sigma_{tot}^{\pi N}\;
\int_{-\infty}^{z_1}\rho(\mathbf{s}_1,z')\;dz'}\; e^{-\frac{1}{2}\sigma_{tot}^{\pi N}\;
\int_{z_1}^{\infty}\rho(\mathbf{s}_1,z')\;dz'}.
\ee
Because the scattering angle of the outgoing pion is very small, it's legitimate to approximate its entire trajectory as being a straight line parallel to the $z$-axis.

So now to include Color Transparency in the above result, we allow $\sigma_{tot}^{\pi N}$ and $\sigma_{tot}^{p N}$ to depend on the distance of the given particle from the point where the pion struck the proton.  Since the hard scatter occurs at the point $\mathbf{r}_1=(\mathbf{s}_1,z_1)$, in the above formula we make the replacements

\be
\label{ctsigma1}
\sigma_{tot}^{pN}  \int_0^{\infty}    \rho(\mathbf{r}_1+s\;\hat{\mathbf{p}})ds \;\to   \int_0^{\infty}\sigma_{tot}^{pN}(s)\rho(\mathbf{r}_1+s\;\hat{\mathbf{p}})ds
\ee
\be
\label{ctsigma2}
\sigma_{tot}^{\pi N}\int_{-\infty}^{\infty} dz^{\prime}\rho(\mathbf{s}_1,z^{\prime})                  \to\int_{-\infty}^z dz^{\prime}\sigma_{in}^{\pi N}(z^{\prime})\rho(\mathbf{s}_1,z^{\prime})+\int_{z}^{\infty} dz^{\prime}\sigma_{out}^{\pi N}(z^{\prime})\rho(\mathbf{s}_1,z^{\prime})
\ee
where the general form of the position dependent $\sigma$'s is ~\cite{liu88}
\be
\label{sigeff}
\sigma^{eff}_{hN}(z)=\sigma^{tot}_{hN}\Biggl[ \theta(l_h-z)\; \Bigl[\frac{z}{l_h}+\frac{n^2\langle k_t^2\rangle}{\vert t\vert}\Bigl (1-\frac{z}{l_h}\Bigr) \Bigr]  +\theta(z-l_h)\Biggr]   
\ee
where $\sigma_{tot}$ is the free-space total cross-section (this model of the expansion from the pointlike configuration is called the ``quantum diffusion model").  In this equation, $n$ is the number of valence quarks of the hadron, while $\langle k_t^2\rangle^{1/2}$ is the average transverse momentum of the quark in the hadron (taken to be  $\langle k_t^2\rangle^{1/2}=0.35$ GeV).  Thus $\frac{\langle k_t^2\rangle}{\vert t\vert}\sigma^{tot}_{hN}$ is a measure of the transverse size of the hadron at the time of collision.  The parameter $l_h$ (called the formation length; see Sec. \ref{sec:coherlength} for more discussion of this) is the distance the hadron travels after the collision until it reaches its normal size.  This is estimated as $l_h\simeq\frac{1}{E_n-E_h}\simeq\frac{2 p_h}{M_n^2-M_h^2}$, where $M_n$ is the mass of a typical intermediate state $n$ of the hadron~\cite{liu88}.  In principle the quantities $l_{\pi}$ and $l_p$ can be different from each other, but since the relation $l_h\simeq\frac{1}{E_n-E_h}\simeq\frac{2 p_h}{M_n^2-M_h^2}$ is only an estimate, we take here $M_n^2-M_N^2=M_n^2-M_{\pi}^2=0.7\;GeV^2$ for both $l_{\pi}$ and $l_p$~\cite{miller06}.                  

Making the replacements Eqs. \ref{ctsigma1} and \ref{ctsigma2} in \eq{gnpion}, we have
\be
\label{gnpion2}
\begin{split}
g^{(n)}_{CT}(\mathbf{p}_m)= \int & d^3r_1 e^{-i\mathbf{p}_m\cdot\mathbf{r}_1}\phi_n(\mathbf{r}_1)
e^{-\frac{1}{2} \int_{0}^{\infty} \sigma_{tot}^{pN}(s)  \rho(\mathbf{r}_1+s\hat{\mathbf{p}})\;ds}  \\
& \times e^{-\frac{1}{2}\int_{-\infty}^{z_1}  \sigma_{tot}^{\pi N}(z')   \rho(\mathbf{s}_1,z')\;dz'}\; e^{-\frac{1}{2}
\int_{z_1}^{\infty} \sigma_{tot}^{\pi N}(z')   \rho(\mathbf{s}_1,z')\;dz'}
\end{split}
\ee

For the case of $\pi+A\to \pi +p+(A-1)^*$ with incident pion momentum $\vert\mathbf{k}\vert=200\;GeV$, the transparency $T$ (\eq{transp}) was evaluated at $\mathbf{p}_m=0$ for a range of $t$ from $t=-1.5$ to $-10$ GeV$^2$, for the nuclei $^{12}$C and $^{40}$Ca.  The result when the effects of Color Transparency are not included is obtained using \eq{gnpion} for $g^{(n)}$ in the numerator of $T$, while the result that includes effects of Color Transparency are obtained using  \eq{gnpion2} for $g^{(n)}$ in the numerator.  We shall call the former the ``Glauber result" while the latter is the ``CT result".  The denominator of $T$ is of course the same for both.  The values of the free-space cross-sections used were $\sigma_{tot}^{pN}=40\;mb$ and $\sigma_{tot}^{\pi N}=25\;mb$ which are valid for proton LAB momenta $\gtrsim 0.6\;GeV$ and pion LAB momenta $\gtrsim 10\;GeV$~\cite{perl74}. 

For the wavefunctions $\phi_n$, harmonic oscillator wavefunctions were used.  The oscillator length $b=\sqrt{\frac{\hbar}{\mu\omega}}$ was chosen so that the mean-square radius $\bar{R^2}$ as calculated using the density from the wavefunctions, $\rho(\mathbf{r})=\sum_n \vert \phi_n(\mathbf{r}) \vert^2$, was equal to the mean-square radius $\bar{R^2}$ as calculated using the Woods-Saxon form of the nuclear number density:
\be
\rho(r)=\frac{\rho_0}{1+e^{\frac{r-R}{a}}}
\ee
where $R=1.1\;A^{1/3}\;fm$ and $a=0.56\;fm$; $\rho_0$ is determined by normalizing $\int d^3r \rho(r)$ to $A$.  The values obtained were $b=8.67 \;GeV^{-1}$ for $^{12}$C and  $b=10.48 \;GeV^{-1}$ for $^{40}$Ca.

\begin{figure}[bth]
     \begin{center}
        \subfigure[$A=12$]{%
            \label{fig:A=12pion2}
            \includegraphics[width=0.5\textwidth,height=2in]{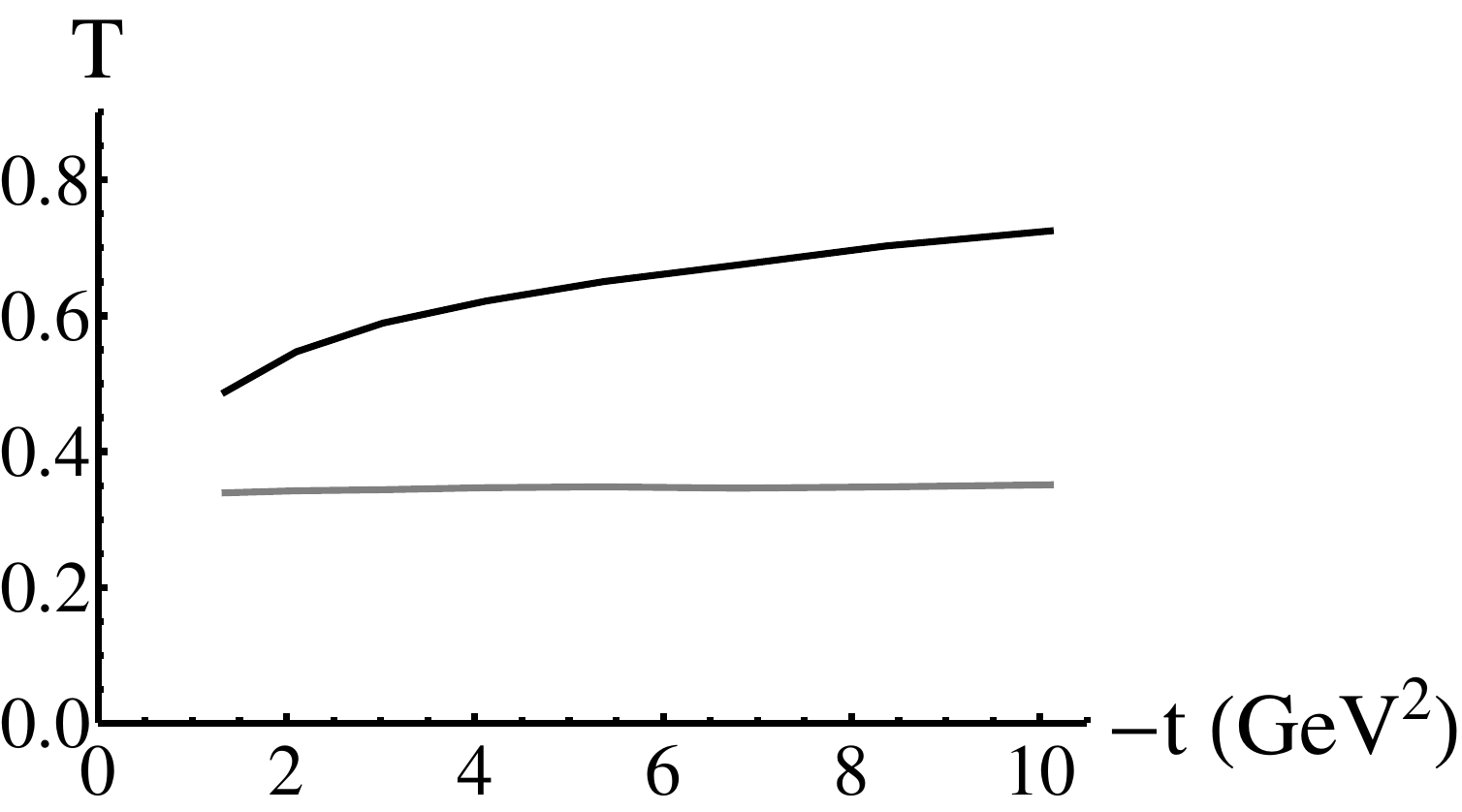}
        }%
        \subfigure[ $A=40$]{%
           \label{fig:A=40pion2}
           \includegraphics[width=0.5\textwidth,height=2in]{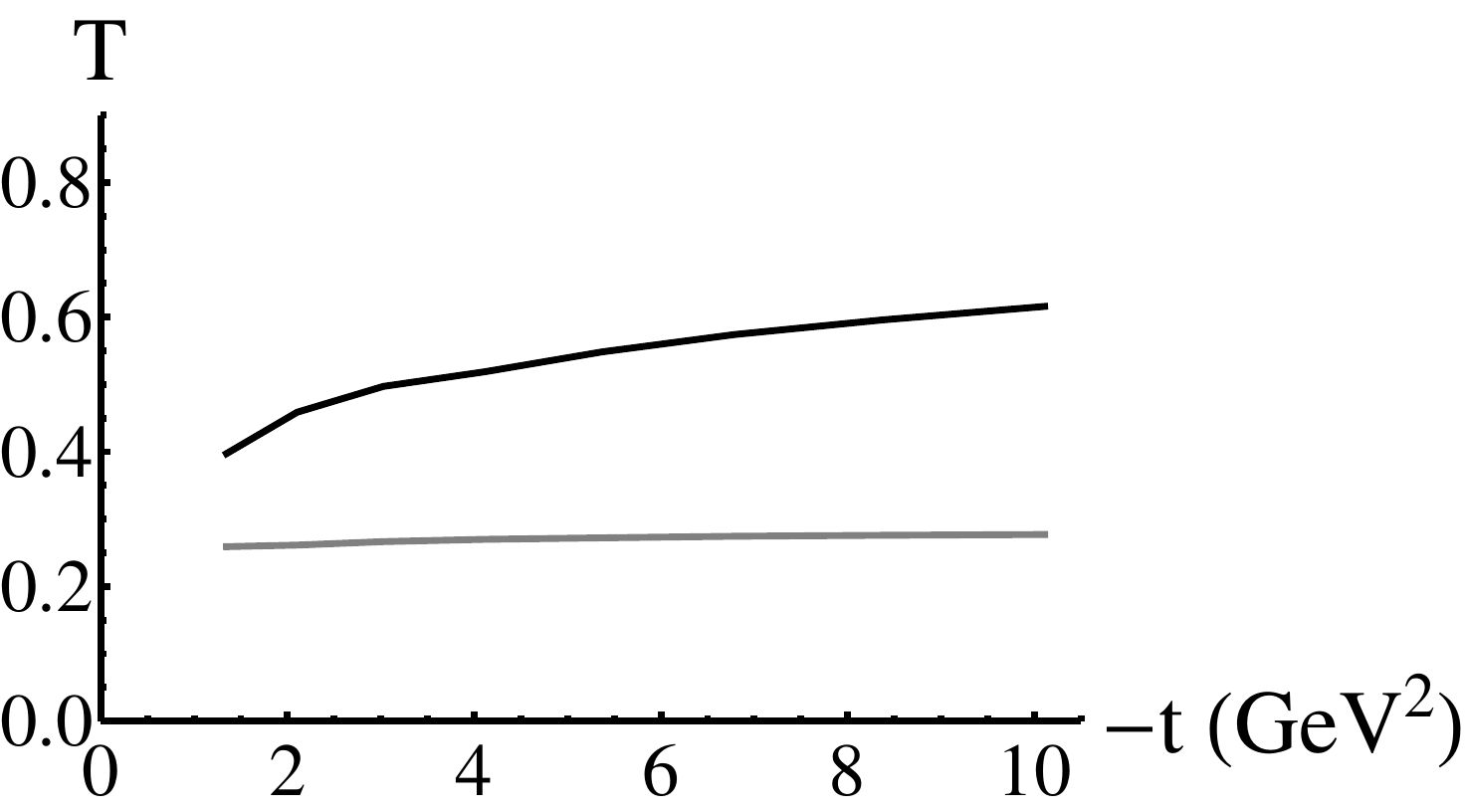}
        }\\ 

%
    \end{center}
    \caption{%
        Transparency $T(\mathbf{p}_m)$ for $A=12$ and $A=40$, for $\mathbf{p}_m=0$.  The bottom curves are the Glauber result; the top curves are the CT result.
     }%
   \label{fig:T1}
\end{figure}
\begin{figure}[htb]
     \begin{center}

            \includegraphics[width=4.8in,height=2.5in]{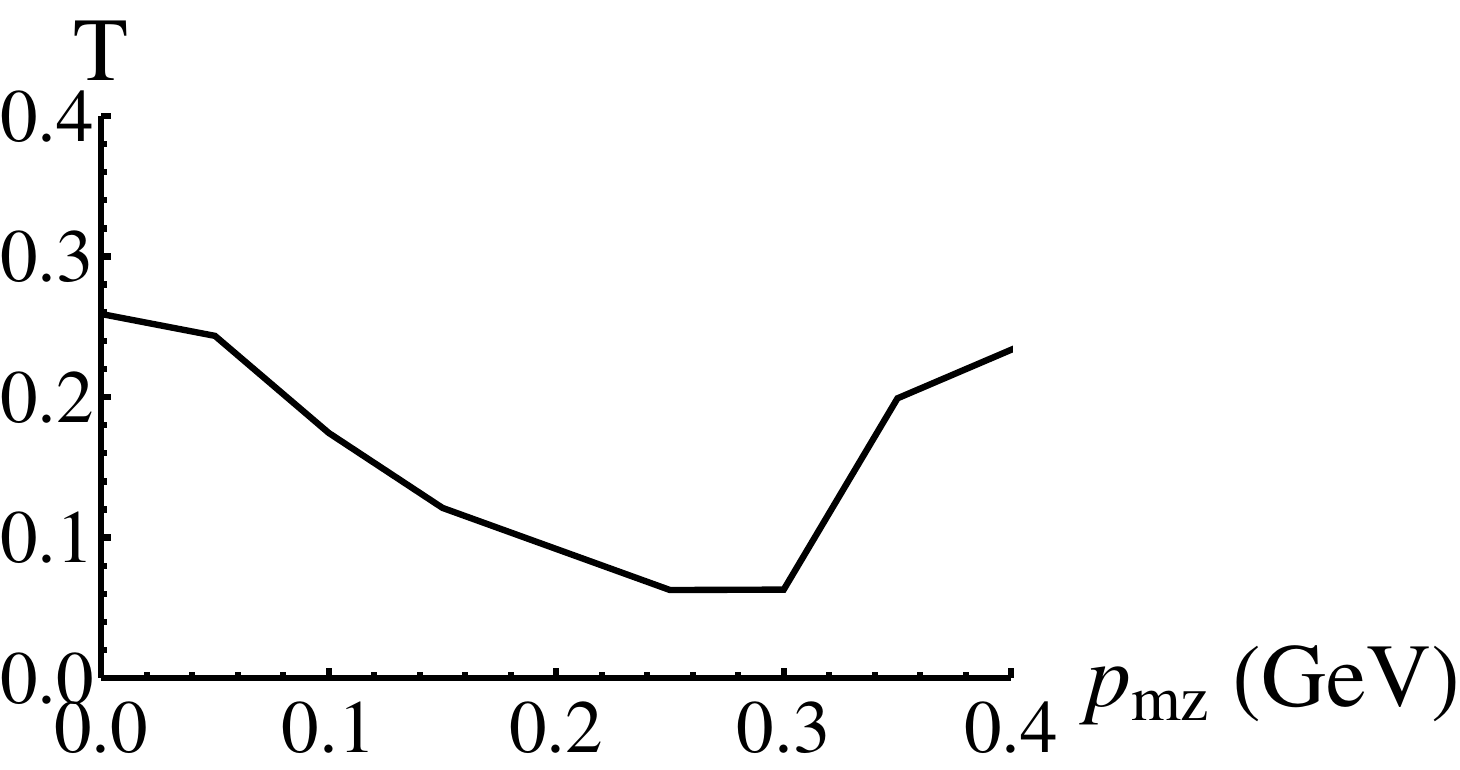}

    \end{center}
    \caption{%
        Transparency for $A=40$ as a function of $\mathbf{p}_{mz}$ for $\mathbf{p}_{mx}=\mathbf{p}_{my}=0$.   
     }%
   \label{fig:A=40missmom}
\end{figure}

Fig. \ref{fig:T1} shows the calculated transparency at $\mathbf{p}_m=0$.  The effects of Color Transparency are very apparent.  However, the value of the transparency is very sensitive to the value of $\mathbf{p}_m$.  Fig. \ref{fig:A=40missmom} shows the transparency, for $A=40$, as a function of $p_{mz}$ for $\mathbf{p}_{mx}=\mathbf{p}_{my}=0$, for $p_{mz}$ between $0$ and $400\;MeV$, for the Glauber case (i.e., not including effects of Color Transparency).

\section{Integrated transparency}
\label{sec:integratedT}

\begin{figure}[htb]
     \begin{center}
      
            \includegraphics[width=4.8in,height=2.5in]{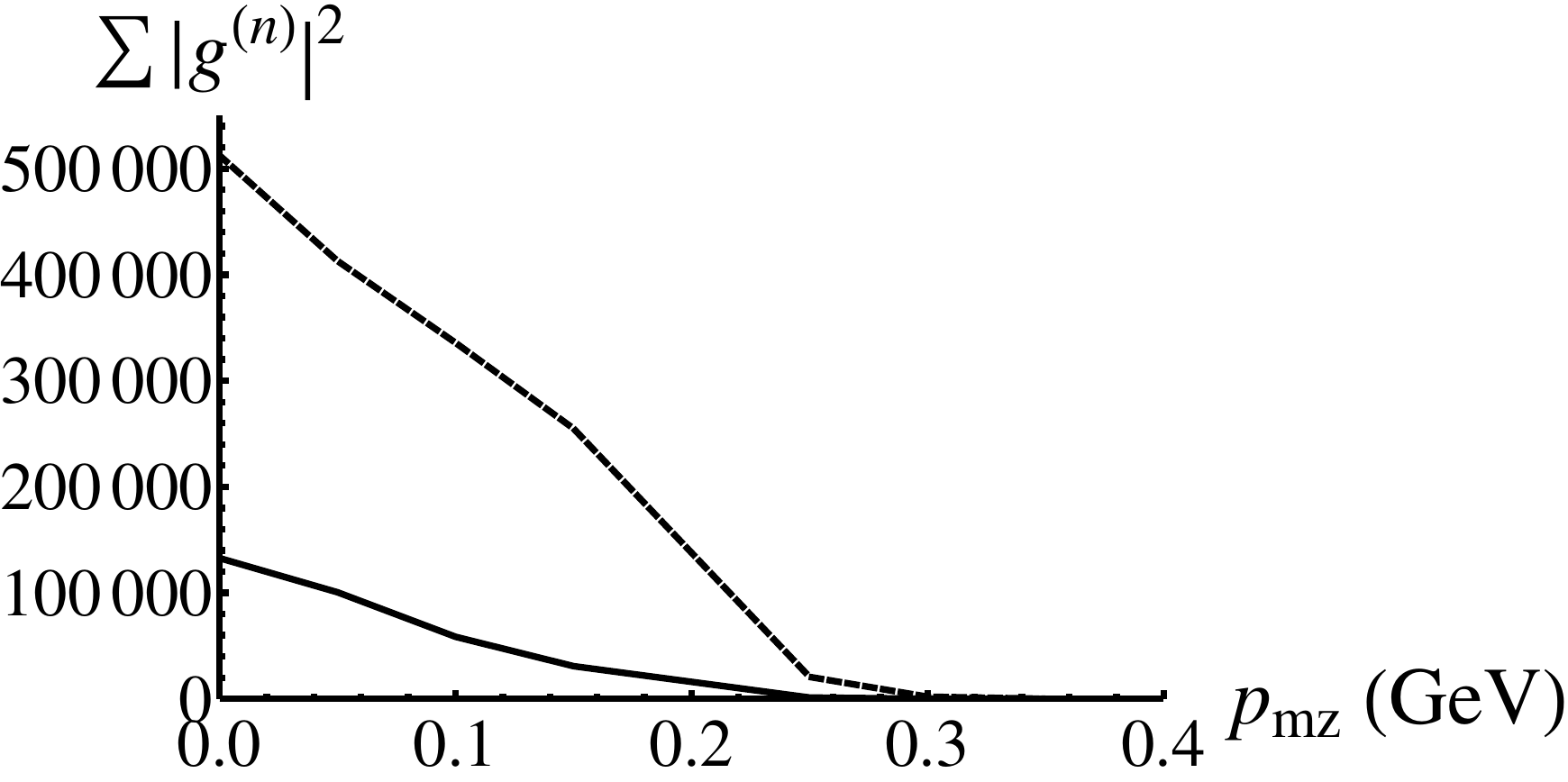}

    \end{center}
    \caption{%
        Numerator and denominator of $T(\mathbf{p}_{m})$ versus $p_{mz}$, for $p_{mx}=p_{my}=0$.  The dashed curve is the PWIA value (denominator) while the solid curve is the DWIA (numerator).  These are both without inclusion of Color Transparency effects.   
     }%
   \label{fig:gnvalues}
\end{figure}

In the work of Benhar~\cite{benhar96}, the integrated transparency $T_{\cal{D}}$, \eq{integratedT}, is calculated by integrating over the entire region of $\mathbf{p}_m$ such that the integrand is non-negligible.  As seen from Fig. \ref{fig:gnvalues}, the distorted momentum distribution $\sum_{n=1}^A\vert g^{(n)}(\mathbf{p}_m)\vert^2$ is only significant for $\vert\mathbf{p}_m\vert\leq 300\;MeV$.  The PWIA value of this, which is the actual momentum distribution, is negligible for $\mathbf{p}_m>300\;MeV$, as the Fermi momentum of the nucleons in a nucleus does not much exceed this.  The result in~\cite{benhar96}  for the integrated transparency, given without derivation, is 
\be
T_{\cal{D}}=\frac{1}{A}\int d^2s dz\rho(\mathbf{r})e^{-\alpha_p(\mathbf{r})}e^{-\sigma_{tot}^{\pi N}T(\mathbf{s})}
\ee
where the path of integration in $\alpha_p(\mathbf{r})$ is along direction of the outgoing proton's momentum $\mathbf{p}$.  In this section I derive this result, as well as the conditions under which it is valid.

We are interested in calculating the integrated transparency, which is (\eq{integratedT}):
\be
\label{intT1}
T_{\cal{D}}=\frac{\int_{\cal{D}}d^3p_m\frac{d\sigma}{dE'd\Omega'  d\Omega_p}}{\int_{\cal{D}}d^3p_m \frac{d\sigma_{PWIA}}{dE'd\Omega'  d\Omega_p}}=\frac{ \sum_{n=1}^A\int_{\cal{D}}d^3p_m \Bigl\vert g^{(n)}(\mathbf{p}_m)\Bigr\vert^2}{ \sum_{n=1}^A\int_{\cal{D}}d^3p_m \Bigl\vert g^{(n)}(\mathbf{p}_m)\Bigr\vert_{PWIA}^2}.
\ee 
The second equality above is valid in the case where the phase-space factors in the differential cross-section are approximately constant over the domain ${\cal D}$ that is integrated over; in that case they factor out of the integral and cancel in the ratio.
We wish to integrate over a region of $\mathbf{p}_m$ corresponding to the entire range of momentum that the proton in the nucleus has.  Therefore we integrate over all $\mathbf{p}_m$ such that $\vert\mathbf{p}_m\vert\le p_{max}$ with $p_{max}\simeq 300\;MeV$.  For any $\vert\mathbf{p}_m\vert$ larger than this, the integrand is negligible.  Thus the numerator in the above equation is 
\be
\label{intT2}
\begin{split}I_{num}\equiv \sum_{n=1}^A\int\limits_{p_m<p_{max}}\,d^3p_m \,
\int d^3r_1 & d^3r_1'  e^{-i\mathbf{p}_m\cdot(\mathbf{r}_1-\mathbf{r}_1')}     \;  \phi_n^*(\mathbf{r}_1')\;\phi_n(\mathbf{r}_1)\\
&\times e^{-\frac{1}{2}\alpha_p(\mathbf{r}_1)}e^{-\frac{1}{2}\alpha_p(\mathbf{r}_1')} e^{-\frac{1}{2}\sigma_{tot}^{\pi N}\;T(\mathbf{s}_1)} e^{-\frac{1}{2}\sigma_{tot}^{\pi N}\;T(\mathbf{s}_1')}
\end{split} 
\ee
We have $\mathbf{p}=\mathbf{q}+\mathbf{p}_m$, and so if $\vert\mathbf{q}\vert\gg p_{max}$, then over the entire domain of integration of $\mathbf{p}_m$ we have $\mathbf{p}\simeq\mathbf{q}$ and we can re-write \eq{intT2} as
\be
\label{intT3}
\begin{split}
I_{num}\simeq \sum_{n=1}^A\int d^3r_1' & \phi_n^*(\mathbf{r}_1') e^{-\frac{1}{2}\sigma_{tot}^{\pi N}\;T(\mathbf{s}_1')}e^{-\frac{1}{2}\alpha_q(\mathbf{r}_1')}\\ 
&\times\int\limits_{p_m<p_{max}}d^3p_m  e^{i\mathbf{p}_m\cdot\mathbf{r}_1'}  \int d^3r_1 \phi_n(\mathbf{r}_1)e^{-\frac{1}{2}\sigma_{tot}^{\pi N}\;T(\mathbf{s}_1)} e^{-\frac{1}{2}\alpha_q(\mathbf{r}_1)} e^{-i\mathbf{p}_m\cdot\mathbf{r}_1}
\end{split}
\ee
where we have set $\mathbf{p}=\mathbf{q}$ in the $\alpha_p$'s.  
The last integral, over $\mathbf{r}_1$, is almost the distorted momentum distribution at $\mathbf{p}_m$ for the state $n$; it differs from it in that it has $\alpha_q(\mathbf{r}_1)$ instead of  $\alpha_p(\mathbf{r}_1)$.  We may therefore assume that this integral vanishes for $\mathbf{p}_m>p_{max}$; this is certainly true of the actual momentum distribution, which is just $ \int d^3r_1 \phi_n(\mathbf{r}_1)e^{-i\mathbf{p}_m\cdot\mathbf{r}_1}$.  Therefore we may extend the upper limit on $\vert\mathbf{p}_m\vert$ to infinity, with exact equality:
\be
\begin{split}
I_{num}\simeq \sum_{n=1}^A\int d^3r_1' & \phi_n^*(\mathbf{r}_1') e^{-\frac{1}{2}\sigma_{tot}^{\pi N}\;T(\mathbf{s}_1')}e^{-\frac{1}{2}\alpha_q(\mathbf{r}_1')}\\
&\times\int d^3p_m  e^{i\mathbf{p}_m\cdot\mathbf{r}_1'}  \int d^3r_1 \phi_n(\mathbf{r}_1)e^{-\frac{1}{2}\sigma_{tot}^{\pi N}\;T(\mathbf{s}_1)} e^{-\frac{1}{2}\alpha_q(\mathbf{r}_1)} e^{-i\mathbf{p}_m\cdot\mathbf{r}_1}
\end{split}
\ee
where we now have integration over all $\mathbf{p}_m\in\Re^3$. Integrating over $\mathbf{p}_m$ now gives a delta function, $\delta^{(3)}(\mathbf{r}_1-\mathbf{r}_1')$, and so finally we have, summing over $n=1,\ldots,A$
\be
I_{num}\simeq (2\pi)^3  \int   d^3r_1 \rho(\mathbf{r}_1)e^{-\sigma_{tot}^{\pi N}\;T(\mathbf{s}_1)} e^{-\alpha_q(\mathbf{r}_1)}. 
\ee
Then the denominator of \eq{intT1} is just the PWIA value of the above expresion, which is $(2\pi)^3\int d^3r_1 \rho(\mathbf{r}_1)=(2\pi)^3\;A$.  Thus we have for the integrated transparency:
\be
\label{simplifiedT}
\boxed{
T_{\cal{D}}=\frac{1}{A} \int   d^3r_1 \rho(\mathbf{r}_1)e^{-\sigma_{tot}^{\pi N}\;T(\mathbf{s}_1)} e^{-\alpha_q(\mathbf{r}_1)} }
\ee 
where the domain $\cal{D}$ of missing momentum integrated over is all $\mathbf{p}_m$ such that the distorted momentum distribution $g^{(n)}(\mathbf{p}_m)$ for all states $n$
is non-zero (or at least non-negligible).  

It is important to note the essential assumption behind the preceding derivation:  the momentum transfer $\vert\mathbf{q}\vert\gg p_{max}$ where $p_{max}$ is the maximum momentum present in the distorted momentum distribution.  This allowed us to go from \eq{intT2} to  \eq{intT3}, which removed the dependence on $\mathbf{p}_m$ from the $\alpha_p$'s.  If $\vert\mathbf{q}\vert\gg p_{max}$ does not hold, then it's certainly not the case that $\alpha_p\simeq\alpha_q$.  If, for example, $\vert\mathbf{q}\vert=p_{max}$, then in integrating over $\mathbf{p}_m$ the direction of integration along the path of the outgoing proton in $\alpha_p(\mathbf{r})$ varies drastically.  For $\mathbf{r}$ near the edge of the nucleus, that could make the difference between $\alpha_p(\mathbf{r})$ being zero (if $\mathbf{p}=\mathbf{q}+\mathbf{p}_m$ points radially outward) and $\alpha_p(\mathbf{r})$ being significant (if $\mathbf{p}=\mathbf{q}+\mathbf{p}_m$ points radially inward).  Therefore in order for the expression  \eq{simplifiedT} to be valid, we need to have the momentum transfer be much larger than the Fermi momentum of the nucleons in the nucleus.

The result \ref{simplifiedT} is the result which is given in~\cite{liu88} as the semiclassical result for the transparency.  We see that it does indeed have a semiclassical interpretation.  The hard scatter with momentum transfer $\mathbf{q}$ occurs on a nucleon at the point $(\mathbf{s},z)$, which knocks out the nucleon.  The nucleon then propagates out of the nucleus.  The incoming and outgoing pion, and the outgoing nucleon both suffer attenuation along their paths, which is given by the classical result for the attenuation of the intensity of a beam of particles passing through a material composed of scatterers of number density $\rho(\mathbf{r})$.  The position-dependent mean-free path of the particles in the material is then $L(\mathbf{r})=( \sigma\rho(\mathbf{r}))^{-1}$ where $\sigma$ is the cross-section of interaction, and so the attenuation factor starting from a given point $\mathbf{r}$ is just $e^{-\int_0^{\infty} ds [L(\mathbf{r}+s\hat{p})]^{-1}}$.

The integrated transparency \eq{simplifiedT} was calculated for $A=12$, $A=40$, and $A=208$, for both the Glauber case (no CT effects included) and the CT case.  In the CT case, the position dependent cross-sections Eq. \ref{sigeff} are used, as was done for the transparency $T(\mathbf{p}_m=0)$.  The results are shown below in Fig. \ref{fig:Tpion}.  It can be seen that the integrated transparency, for a given $A$ and $t$, is smaller than the transparency $T(\mathbf{p}_m=0)$.  In all cases the integrated transparency for the Glauber case is essentially independent of $t$, while for the CT case the transparency is much larger than for the Glauber case and increases markedly with $\vert t\vert$.

\section{Conclusion}
\label{sec:pionconclusion}

We have calculated the transparency and integrated transparency for the proton knockout reaction $\pi+A\to \pi + p+(A-1)^*$ within the Glauber theory, for the incident pion momentum of $200$ GeV which is available at the COMPASS experiment.  With the estimated values of the parameters that enter in the position-dependent cross-section Eq. \ref{sigeff}, both the transparency $T(\mathbf{p}_m=0)$ and the integrated transparency $T$ show large effects due to Color Transparency.  In particular, for $A=208$, even for modest values of $\vert t\vert$ the integrated transparency $T$ is larger in the CT case than in the Glauber case by a factor of $\sim 3 - 4$, and increases substantially as $\vert t\vert$ increases.

\begin{figure}[hbt]
     \begin{center}
        \subfigure[ $A=12$]{%
            \label{fig:A=12piT}
            \includegraphics[width=0.6\textwidth]{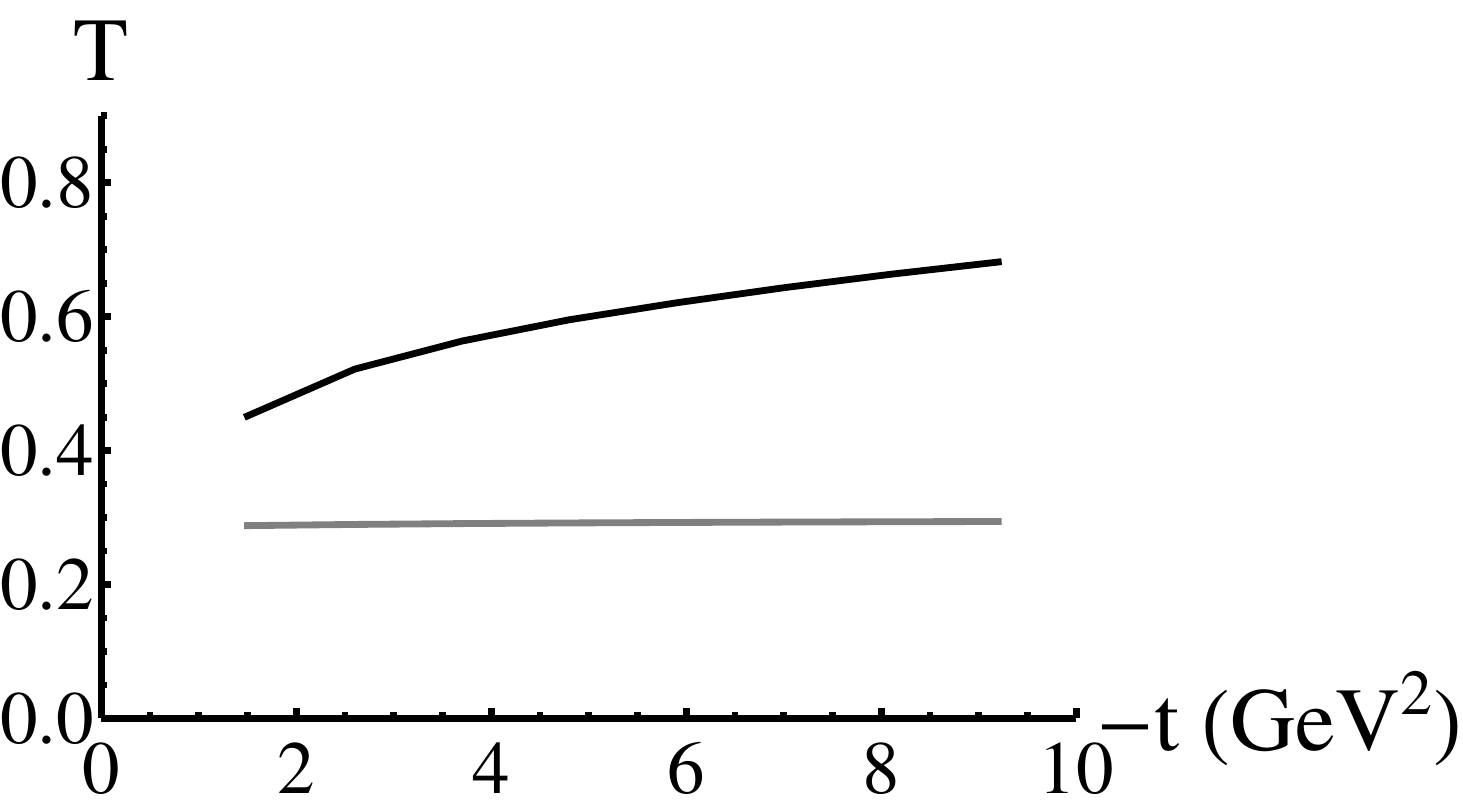}
        }%
         \hspace{0.5in}
        \subfigure[  $A=40$]{%
           \label{fig:A=40piT}
           \includegraphics[width=0.6\textwidth]{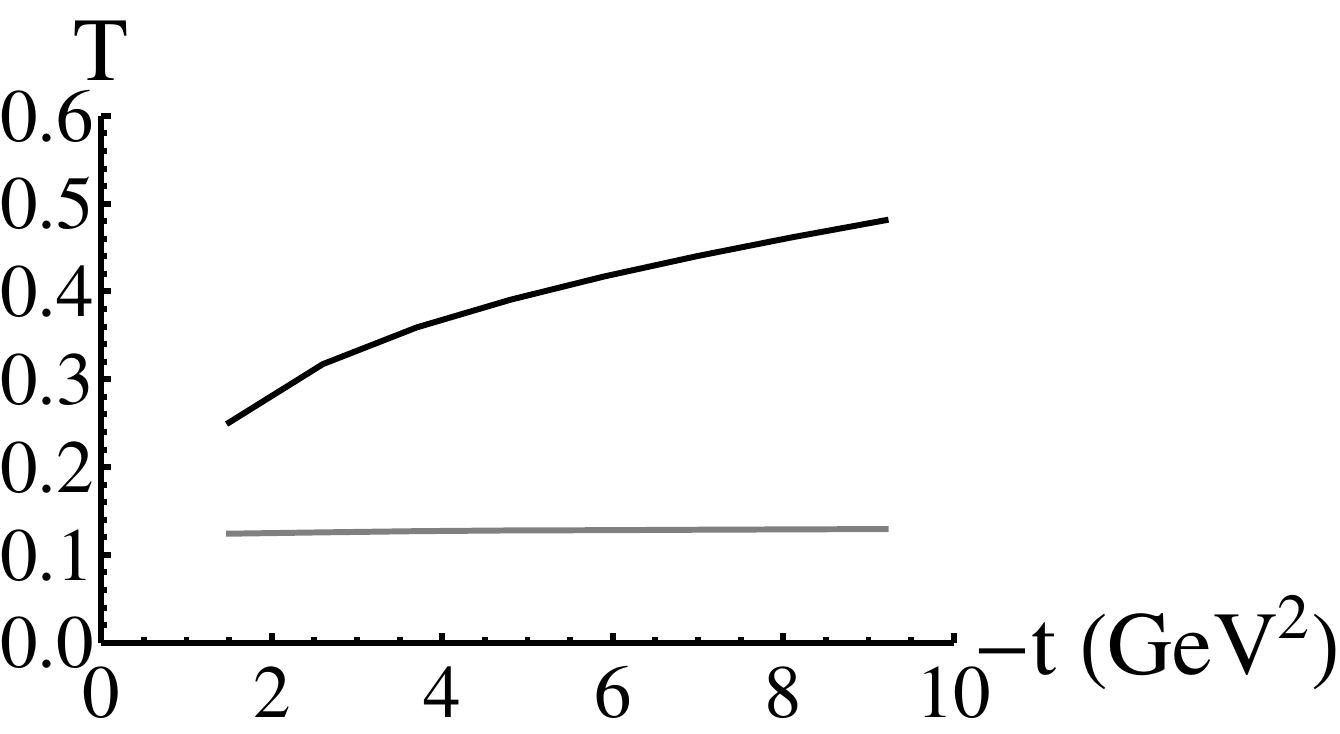}
        }\\ 

        \subfigure[  $A=208$]{%
            \label{fig:A=208piT}
            \includegraphics[width=0.6\textwidth,height=2.2in]{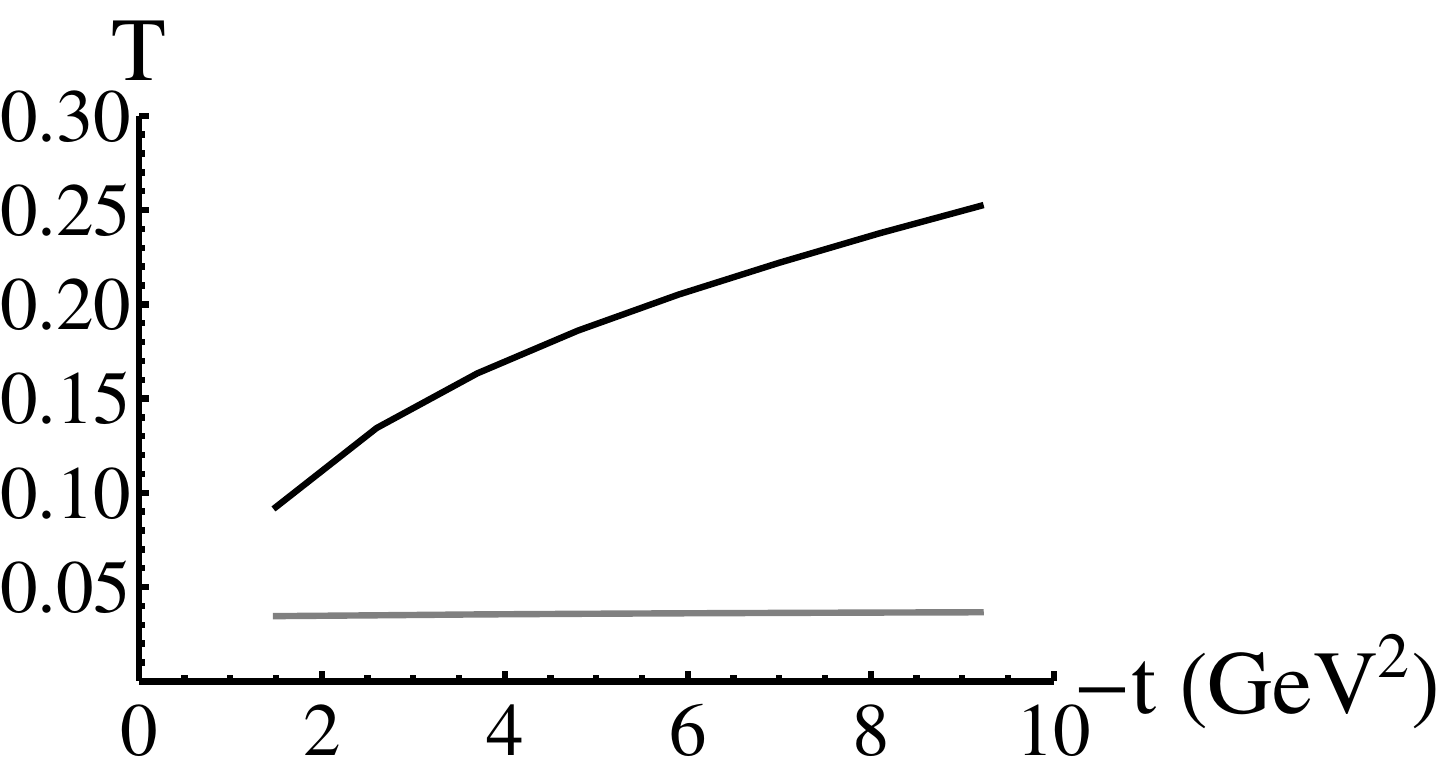}
        }%

    \end{center}
    \caption{%
        Integrated transparency $T$.  The bottom curves are the Glauber result; the top curves are the CT result.
     }%
   \label{fig:Tpion}
\end{figure}





\chapter{Color transparency and the reaction\\ $\gamma^*+A\to\rho+p+(A-1)^*$ }

\section{Introduction}

In this chapter we calculate the transparency $T(\mathbf{p}_m)$ and integrated transparency $T$ for the case of electroproduction of the $\rho$ meson with proton knockout, $\gamma^*+A\to\rho+p+(A-1)^*$.  At the photon energies we are interested in, Glauber theory, modified to account for particle production, is valid, as it was for the case of pion scattering considered in the previous chapter.  Electroproduction of the $\rho$ provides another means of detecting the effects of Color Transparency.  In contrast to the purely elastic pion scattering considered in Ch. 2, for electroproduction there are more parameters that may be varied, namely the virtual photon energy $\nu$ and virtuality $Q^2$.  These quantities, as well as a combination of them called the coherence length, $l_c=\frac{2\nu}{Q^2+m_V^2}$, can all affect the observed transparency.  The coherence length plays an especially important role, since by varying its value the transparency $T$ will vary even in the absence of any Color Transparency effects.  Thus to observe an actual CT effect, one must keep the coherence length fixed.

This chapter is organized as follows.  In Sec. \ref{sec:first}, electroproduction of vector mesons on a single nucleon is discussed.  The Vector Meson Dominance model is introduced, and the coherence length and formation time for vector meson production are discussed.  Existing experimental results in the search for CT involving vector mesons is discussed as well. In Sec. \ref{sec:second}, the Glauber formalism for particle production is presented.  The amplitude for the reaction  $\gamma^*+A\to\rho+p+(A-1)^*$ is derived.  Two limiting cases are analyzed, one for $l_c\to 0$ and one for $l_c\to\infty$, and it is shown that for the case of $l_c\to\infty$ the result reduces to the result for the pion elastic scattering case discussed in the previous chapter.  The results for the values of the transparency $T(\mathbf{p}_m=0)$ are presented for several different values of $A$ and $l_c$.  In Sec. \ref{sec:third}, the integrated transparency $T$ is calculated, for several different $A$ and $l_c$ values.  Sec. \ref{sec:conclusion} summarizes our results.

\section{Electroproduction of a vector meson on a single nucleon}
\label{sec:first}

There are several pictures of electroproduction of vector mesons.  In the Vector Meson Dominance model (VMD)~\cite{feyn72}, the interaction of a real or virtual photon with a nucleon proceeds with the photon first fluctuating into a (virtual) neutral vector meson (i.e. a meson with the same quantum numbers as the photon), followed by the virtual vector meson scattering elastically from the nucleon.  The elastic scattering of the virtual meson on the nucleon puts the meson on its mass shell.  The amplitude for the production process $\gamma^*+N\to N+V$ is then proportional to the elastic scattering amplitude for $V+N\to V+N$.  In this picture, the physical photon is a superposition of a bare photon state and vector meson states (the bare photon state would be the real photon state in the absence of the strong interaction). Thus at a given photon energy and $q^2$, and a given momentum transfer $t=(p_{\gamma}-p_{\rho})^2$, the production amplitude is simply proportional to the elastic scattering amplitude.

\begin{figure}[htb]
     \begin{center}
      
            \includegraphics[width=5in,height=2.5in]{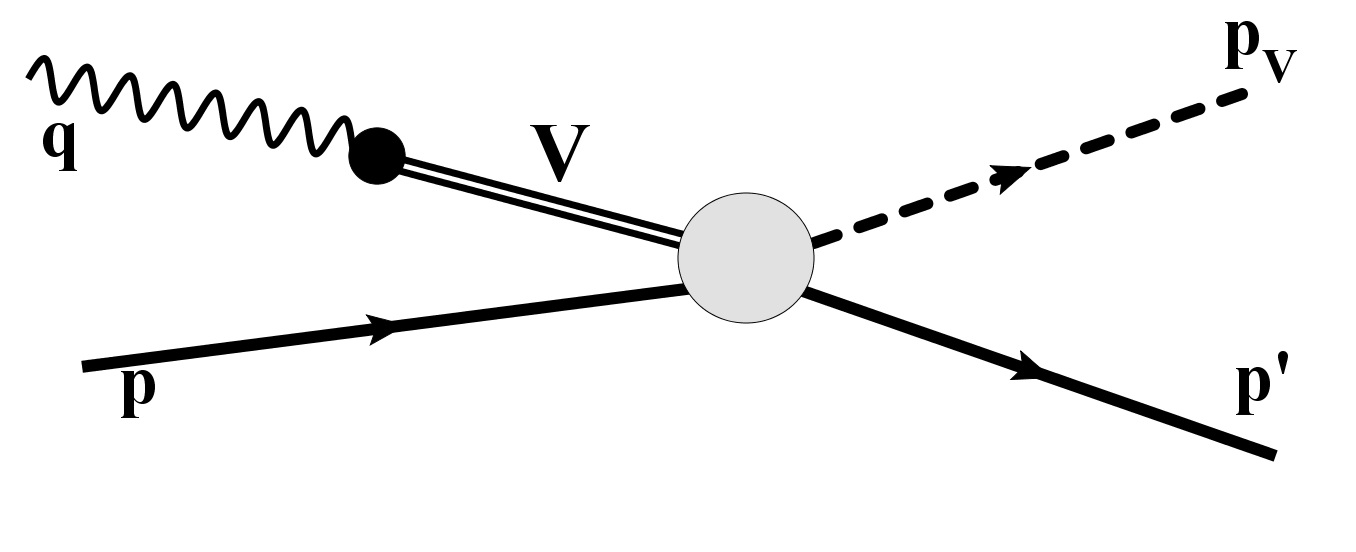}

    \end{center}
    \caption{%
        Vector meson production in the VMD model.  The incoming photon fluctuates into a virtual vector meson (V) which then scatters elastically from the nucleon.  
     }%
   \label{fig:vmd}
\end{figure}

Electroproduction of vector mesons can also be described in terms of quarks, using QCD.  The virtual photon fluctuates into a virtual $q\bar{q}$ pair, which propagates over a distance called the coherence length (determined by the energy-time uncertainty principle) before scattering elastically from the nucleon, which puts the $q\bar{q}$ pair on the mass-shell of the vector meson.  The $q\bar{q}$ state then evolves over time to form the final real vector meson state.  The transverse size of the $q\bar{q}$ that is produced by the virtual photon goes as $r_{\perp}\simeq 1/Q$~\cite{miller07}, so the larger $Q$ is, the smaller is the size of the produced $q\bar{q}$.  In the limit of $Q\to\infty$ the size goes to zero:  a point-like configuration.  Thus for large $Q^2$ the produced object should have vanishing interactions with the other nucleons and the transparency should approach $1$.

\subsection{Coherence length and formation time}
\label{sec:coherlength}

\begin{figure}[htb]
     \begin{center}
      
            \includegraphics[width=4.8in,height=2.5in]{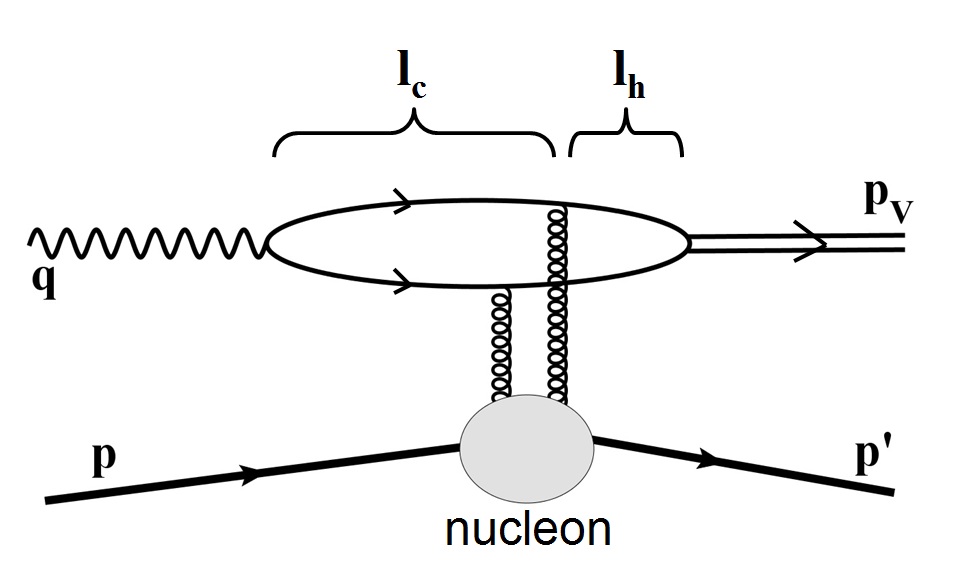}

    \end{center}
    \caption{%
        Coherence length ($l_c$) and formation length ($l_f$) for vector meson production.  The incoming photon dissociates into a $q\bar{q}$ pair which then interact with the nucleon by exchanging gluons.  
     }%
   \label{fig:coherlength}
\end{figure}

There are two length scales (or time scales) of relevance to vector meson production, the coherence length and the formation time (see Fig. \ref{fig:coherlength}).  The distance that the virtual hadronic fluctuation of the photon can travel in the LAB frame (target nucleon or nucleus at rest) is known as the coherence length~\cite{miller07}.  The energy-time uncertainty relation is used to determine this distance.  For a photon and virtual meson with the same momentum $\mathbf{k}$, the difference in energy between the photon and the virtual meson is
\be
\Delta\;E=\sqrt{\mathbf{k}^2+m_V^2}-\nu=\sqrt{\mathbf{k}^2+m_V^2}-\sqrt{\mathbf{k}^2-Q^2}\simeq\frac{Q^2+m_V^2}{2\mathbf{k}}\simeq\frac{Q^2+m_V^2}{2\nu}
\ee
where we've assumed $\nu\gg Q,\;m_V$.  For this high-energy case, the velocity of the vector meson is essentially $c$, and so the energy-time uncertainty relation gives the coherence length as
\be
l_c=\frac{2\nu}{Q^2+m_V^2}.
\ee

For vector meson production in a nucleus, while the virtual hadron or $q\bar{q}$ is propagating over the distance $l_c$ it may interact with nucleons and be absorbed, before it has a chance to undergo the elastic scatter which puts it on mass-shell.  These Initial State Interactions (ISI) therefore affect the measured production cross-section in the nucleus.  In general, as $l_c$ increases, the probability of absorption increases and so the measured production cross-section in a given nucleus should decrease.  Thus the production cross-section at low energy (small $\nu$) should be larger than the production cross-section at high energy (large $\nu$), for a given $Q^2$.  Or conversely, for a given $\nu$, as $Q^2$ is increased, $l_c$ will decrease and therefore the measured production cross-section should increase.  This effect mimics the effect of Color Transparency.  Therefore in order to detect effects of CT, the coherence length should be kept fixed in a given experiment.

The formation time is the time scale over which the virtual meson or $q\bar{q}$ pair develops into the final real vector meson state, after scattering from the nucleon.  The scattering with the nucleon puts the virtual meson or $q\bar{q}$ pair onto the mass shell of the vector meson.  At the time of scattering the transverse size of the $q\bar{q}$ is small, and as it propagates away it evolves into the final meson state.  This time can be estimated by considering the on-mass-shell small-size $q\bar{q}$ pair as a superposition of hadron states, namely the final real vector meson state and the next higher-mass meson state~\cite{miller07}.  Then the energy-time uncertainty principle in the rest frame of the outgoing meson gives
\be
\Delta t=\frac{1}{m_{V'}-m_V},
\ee
while in the LAB frame this is time-dilated by the factor $\gamma=\frac{E_V}{\frac{1}{2}(m_{V'}+m_V)}$, where $\frac{1}{2}(m_{V'}+m_V)$ is the average mass of the two states, and so the formation time or length in the LAB (assuming $\beta\simeq c$) is
\be
l_h=\frac{2\;p_V}{m_{V'}^2-m_V^2}.
\ee

\subsection{Experimental results for $\rho$ electroproduction in nuclei}

There have been several searches for evidence of Color Transparency in electroproduction of $\rho$ mesons in nuclei.  At Fermilab in 1995~\cite{adams95}, high energy muons were scattered from nuclei to produce $\rho$'s.  It was thought that CT was observed because the transparency, for a given $A$, increased as $Q^2$ was increased.  However, in this experiment the coherence length was not held constant as $Q^2$ was increased, so it is difficult to draw conclusions from their data.  A later experiment at DESY was conducted to explicitly measure the coherence length effect~\cite{ackerstaff99}.  It was observed, as expected, that the transparency decreased as $l_c$ was increased, in $\rho$ electroproduction in $^{14}N$.  The $Q^2$ values for this experiment were such that no CT effects should occur, i.e. the produced object would interact with the full $\rho$-nucleon cross-section.  Hence any dependence of the transparency on $l_c$ was not an indication of CT.  This was a clear indication that any attempt to detect CT in vector meson electroproduction must look for effects while holding $l_c$ constant.  Another experiment at DESY~\cite{airapetian03} was performed,  where the transparency as a function of $Q^2$ was measured for different values of $l_c$.  There appeared to be an increase in the transparency as $Q^2$ increased, although the number of events at each fixed value of $l_c$ was not large, and so better statistics are needed.  Finally, the most recent experiment to search for CT in $\rho$ production was at JLAB~\cite{fassi12}.  In this experiment, the coherence length varied from $\sim 0.5\;fm$ to $\sim 0.85\;fm$.  For this range of coherence length, the $q\bar{q}$ is produced essentially right at the location of the nucleon that it scatters from, and so there are no Initial State Interactions.  The transparencies on $^{12}C$ and $^{56}Fe$ were measured for $Q^2$ from $1.0$ to $2.3\;GeV^2$.  The transparencies appeared to show an increase with $Q^2$, although statistics again were low.

\section{The Glauber formalism for particle production}
\label{sec:second}

For the case of particle production, the profile operator $\Gamma_{tot}$ now depends on longitudinal momentum transfer~\cite{marg68}.  The reason is that for forward production on a single nucleon there is necessarily non-zero longitudinal momentum transfer due to the difference in mass between the incident particle and the outgoing particle.  For the case of $\gamma^*+N\to \rho +N$ at high energy, the energy transfer from the photon to the nucleon can be neglected, and so for an incident photon of momentum $\mathbf{k}$, energy $\nu$, 4-momentum squared $k^2=\nu^2-\mathbf{k}^2=-Q^2$, and an outgoing particle of mass $M$, energy $\nu$, and momentum $\mathbf{k}'$ with momentum parallel to $\mathbf{k}$ (i.e. forward production) conservation of energy gives
\be
\vert\mathbf{k}'\vert^2+M^2=\vert\mathbf{k}\vert^2-Q^2.
\ee
With the longitudinal momentum transfer $q_L\equiv\vert\mathbf{k}\vert-\vert\mathbf{k}'\vert$, we have 
\be
M^2+Q^2=\vert\mathbf{k}\vert^2-\vert\mathbf{k}'\vert^2=2\vert\mathbf{k}\vert\;q_L+{\cal O}(q_L^2)
\ee
and so for $\nu\gg Q$ we have
\be
q_L=\frac{Q^2+M^2}{2\nu}.
\ee

This longitudinal momentum transfer modifies the profile function $\Gamma(\mathbf{b})$~\cite{yennie78}, due to the phase difference between the incident (photon) wave and the outgoing (meson) wave.  Consider vector meson production on a nucleon located at $z_j$ (with the incident photon along the $z$-direction).  The phase of the transmitted wave at a point $z>z_j$ equals the phase of the incident (photon) wave at $z_j$ plus the change in phase of the transmitted wave as it propagates from $z_j$ to $z$.  Thus the transmitted wave at the point $z$ is $e^{ikz_j}e^{ik'(z-z_j)}=e^{ik'z}e^{iz_j(k-k')}=e^{ik'z}e^{iq_L z_j}$.  For elastic scattering of a projectile, the wave at point $z$ would just be $e^{ik'z}$.  Therefore the phase difference of the incident and transmitted waves is just $q_L z_j$, and so the profile function for production on a nucleon at $(\mathbf{s}_j,z_j)$ is $\Gamma^{\gamma}(\mathbf{b}-\mathbf{s}_j)e^{iq_L z_j}\equiv \Gamma^{\gamma}_{bj}e^{iq_Lz_j}$.  (The notation here is the same as in the previous chapter: $\mathbf{b}$ and $\mathbf{s}_j$ are two-dimensional vectors perpendicular to the incident photon's momentum direction, and $z$ and $z_j$ are coordinates along the $z$-axis which is parallel to the incident photon's momentum).   Here $\Gamma^{\gamma}$ is related to the production amplitude $f^{\gamma V}(\mathbf{q})$ for $\gamma^*+N\to V+N$  (where $\mathbf{q}$ is the transverse momentum transfer) by 
\be
\label{ampgammaV}
f^{\gamma V}(\mathbf{q})=\frac{ik}{2\pi}\int d^2b e^{i\mathbf{q}\cdot\mathbf{b}}\;\Gamma^{\gamma}(\mathbf{b})
\ee 
which is the same as for the elastic scattering case.  So we have also
\be
\Gamma^{\gamma}(\mathbf{b})=\frac{1}{2\pi ik}\int d^2q e^{-i\mathbf{q}\cdot\mathbf{b}}f^{\gamma V}(\mathbf{q}),
\ee 
giving $\Gamma^{\gamma}$ in terms of $f^{\gamma V}$.

Taking into account $q_L$, the total profile operator $\Gamma_{tot}$ represents production of the vector meson on a nucleon at $(\mathbf{s}_j,z_j)$, followed by any number of re-scatterings of the produced meson on the other nucleons.  Hence $\Gamma_{tot}$ has the form~\cite{yennie78,kopel96}:
\be
\label{Vgamma}
\Gamma_{tot}(\mathbf{b},\mathbf{r}_1,\ldots,\mathbf{r}_A)=\sum_{j=1}^{A} \Gamma^{\gamma}_{bj}e^{iq_Lz_j} \prod_{m\neq j}\bigl [1-\Gamma_{bm}\theta_{mj}\bigr]  
\ee
where $\Gamma_{bm}\equiv\Gamma(\mathbf{b}-\mathbf{s}_m)$ is the profile function for elastic meson-nucleon scattering, and $\theta_{mj}\equiv\theta(z_m-z_j)$ ensures that any elastic scattering of the produced vector meson occurs \underline{after} the meson has been produced (for high-energy scattering, the waves are all ``moving forward", which is along the z-direction, and so ``later in time" is equivalent to ``farther along in the $z$-direction") .

As in the pion case (Ch. 2), we will sum over the residual nucleus final states which are one-hole states of the initial nucleus, and so the scattering amplitude is again
\be
F_{fi}^{(n)}=\frac{ik}{2\pi}\int d^2b e^{i\mathbf{q}\cdot\mathbf{b}}\int d^3r_1\ldots d^3r_A \chi_p^*(\mathbf{r}_1)\phi_n(\mathbf{r}_1) \Bigl\vert\phi_{A-1}^f(\mathbf{r}_2,\ldots,\mathbf{r}_A)\Bigr\vert^2\Gamma_{tot}(\mathbf{b},\lbrace \mathbf{r}_j \rbrace)
\ee
Expanding out $\Gamma_{tot}$ into terms that depend on $\mathbf{r}_1$ and terms that don't, we have
\be
\Gamma_{tot}= \Gamma^{\gamma}_{b1}e^{iq_Lz_1} \prod_{k=2}^A\bigl [1-\Gamma_{bk}\theta_{k1}\bigr]-\Gamma_{b1}\sum_{j=2}^{A} \Gamma^{\gamma}_{bj}e^{iq_Lz_j}\theta_{1j} \prod_{k\neq1,\; j}\bigl [1-\Gamma_{bk}\theta_{kj}\bigr] 
\ee
\be
+ \sum_{j=2}^{A} \Gamma^{\gamma}_{bj}e^{iq_Lz_j} \prod_{k\neq 1,\; j}\bigl [1-\Gamma_{bk}\theta_{kj}\bigr]  
\ee
and note that the third term is independent of $\mathbf{r}_1$ and so contributes zero to $F_{fi}$ due to orthogonality of $\chi_p$ and $\phi_n$.  Therefore we have
\be
\label{Feff}
F_{fi}^{(n)}=\frac{ik}{2\pi}\int d^2b e^{i\mathbf{q}\cdot\mathbf{b}}\int d^3r_1\ldots d^3r_A \chi_p^*(\mathbf{r}_1)\phi_n(\mathbf{r}_1) \Bigl\vert\phi_{A-1}^f(\mathbf{r}_2,\ldots,\mathbf{r}_A)\Bigr\vert^2\Gamma_{eff}(\mathbf{b},\lbrace \mathbf{r}_j \rbrace)
\ee
where
\be
\label{gammaeff}
\Gamma_{eff}= \Gamma^{\gamma}_{b1}e^{iq_Lz_1} \prod_{k=2}^A\bigl [1-\Gamma_{bk}\theta_{k1}\bigr]-\Gamma_{b1}\sum_{j=2}^{A} \Gamma^{\gamma}_{bj}e^{iq_Lz_j}\theta_{1j} \prod_{k\neq1,\; j}\bigl [1-\Gamma_{bk}\theta_{kj}\bigr] 
\ee
Taking again an independent particle model for the residual nucleus, so that 
\be
 \Bigl\vert\phi_{A-1}^f(\mathbf{r}_2,\ldots,\mathbf{r}_A)\Bigr\vert^2=\prod_{j=2}^A \rho_1(\mathbf{r}_j),
\ee
the second term in \eq{gammaeff} contributes $A-1$ equal terms to $F_{fi}$.  Performing the integral over $\mathbf{r}_2,\ldots,\mathbf{r}_A$ we obtain
\be
\int d^3r_2\ldots d^3r_A \Bigl\vert\phi_{A-1}^f(\mathbf{r}_2,\ldots,\mathbf{r}_A)\Bigr\vert^2\Gamma_{eff}(\mathbf{b},\lbrace \mathbf{r}_j \rbrace)
\ee
\be
\label{exactGammaint}
=\Gamma^{\gamma}_{b1}e^{iq_Lz_1}g_1(\mathbf{b})-\Gamma_{b1}\int d^3r_2 \rho(\mathbf{r}_2) \Gamma^{\gamma}_{b2}e^{iq_Lz_2}\;\theta_{12}\; g_2(\mathbf{b})
\ee
where
\be
g_1(\mathbf{b})\equiv \Bigl[1-\int d^2s \int_{z_1}^{\infty}dz \rho_1(\mathbf{s},z)\Gamma(\mathbf{b}-\mathbf{s})\Bigr]^{A-1}
\ee
and
\be
g_2(\mathbf{b})\equiv \Bigl[1-\int d^2s \int_{z_2}^{\infty}dz \rho_1(\mathbf{s},z)\Gamma(\mathbf{b}-\mathbf{s})\Bigr]^{A-2}.
\ee
In the large-$A$ limit, we have
\be
g_1(\mathbf{b})\simeq e^{-\frac{1}{2}\sigma_{tot}^{\pi N}\;T_1(\mathbf{b})}
\ee
\be
g_2(\mathbf{b})\simeq e^{-\frac{1}{2}\sigma_{tot}^{\pi N}\;T_2(\mathbf{b})}
\ee
where $T_j(\mathbf{b})\equiv \int_{z_j}^{\infty} dz' \rho(\mathbf{b},z')$ is the ``partial thickness function".

As in the pion case, we can utilize the fact that the profile functions are sharply peaked and the other factors are relatively slowly varying. For a slowly varying function $f(\mathbf{r})$ we thus have, to good approximation, 
\be
\label{approxGammaint}
\int d^2s f(\mathbf{s},z)\Gamma(\mathbf{s}-\mathbf{a})\simeq f(\mathbf{a},z)\int d^2s \Gamma(\mathbf{s}-\mathbf{a})
\ee
and similarly for $\Gamma^{\gamma}(\mathbf{s}-\mathbf{a})$.

Using the above approximation, we may integrate over $s_2$ in Eq. \ref{exactGammaint}, using 
\be
\int d^2s_2 \Gamma^{\gamma}(\mathbf{b}-\mathbf{s}_2)=\frac{2\pi}{ik}f^{\gamma V}(0),
\ee
and then we may integrate over 
$\mathbf{b}$ in Eq. \ref{Feff} by setting $\mathbf{b}=\mathbf{s}_1$ everywhere except in the profile functions and $e^{i\mathbf{q}\cdot\mathbf{b}}$, with the result:
\begin{equation}
\label{Vresult}
\begin{split}F_{fi}^{(n)}=& \int d^2s_1  d z_1  e^{-i\mathbf{p}_m\cdot\mathbf{r}_1}   e^{-\frac{1}{2}\alpha_p(\mathbf{r}_1)}  \phi_n(\mathbf{r}_1)\\
& \times\Bigl(f^{\gamma V}(\mathbf{q}) e^{-\frac{1}{2}\sigma_{tot}^{V N}\;T_1(\mathbf{s}_1)} - \frac{2\pi}{ik}f^{\gamma V}(0)  \int_{-\infty}^{z_1} dz_2\; \rho(\mathbf{s}_1 ,z_2) \;e^{iq_L(z_2-z_1)}\;  e^{-\frac{1}{2}\sigma_{tot}^{V N}\;T_2(\mathbf{s}_1)} f(\mathbf{q})\Bigr)\end{split}.
\ee
Here we have written the result for $F_{fi}^{(n)}$ in terms of the missing momentum $\mathbf{p}_m$, which is defined by 
\be
\label{pmdef}
\mathbf{p}_m\equiv \mathbf{p} -\mathbf{k}+\mathbf{k}'=\mathbf{p}_{\perp}-\mathbf{q} + (p_z-q_L) \hat{\mathbf{z}}
\ee
where $\mathbf{p}$ is the momentum of the outgoing proton. 

\begin{figure}[tbp]
     \begin{center}
        \subfigure[First term of Eq. \ref{Vresult} ]{%
            \label{fig:pictorialrep1}
            \includegraphics[width=0.5\textwidth,height=1.6in]{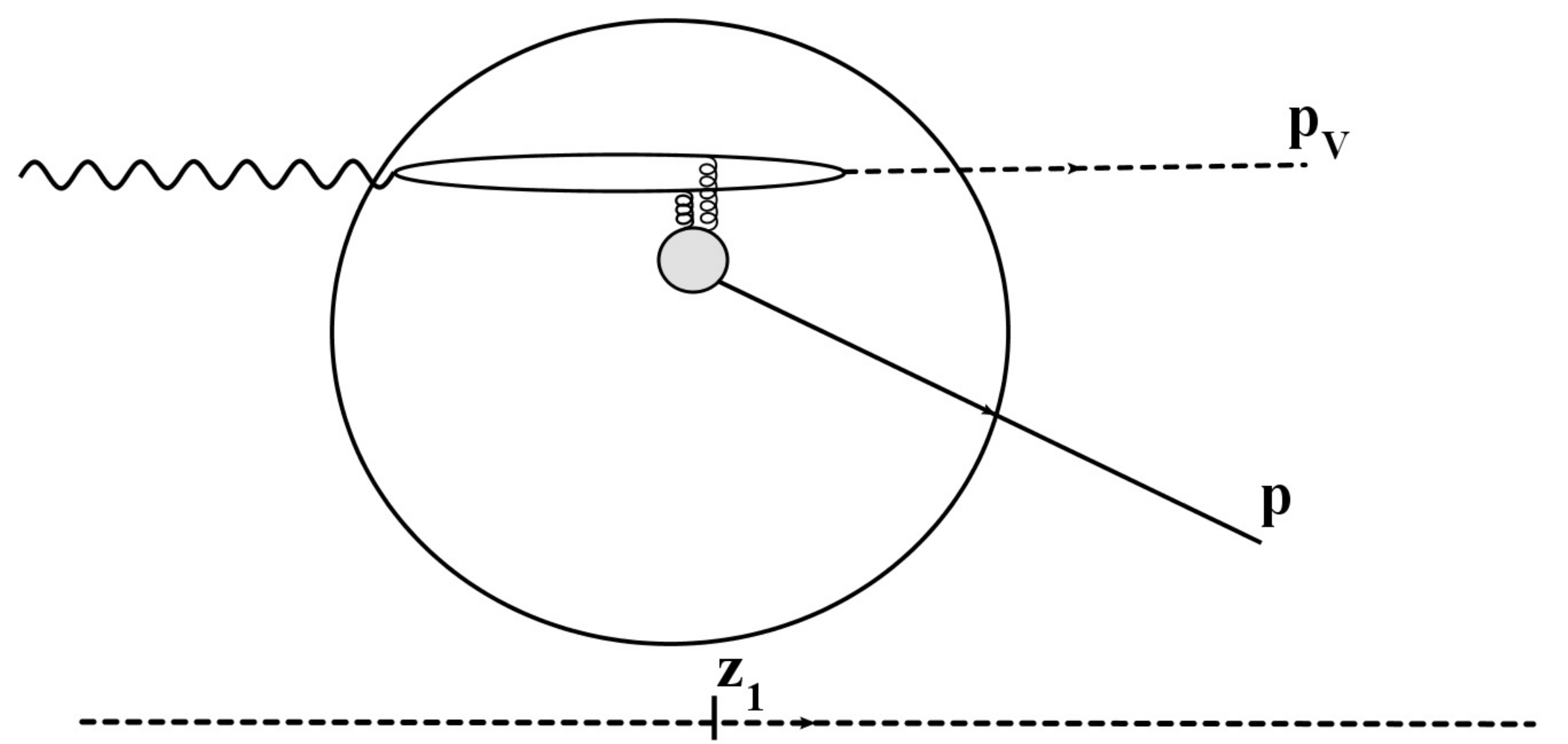}
        }%
        \subfigure[Second term of Eq. \ref{Vresult}]{%
           \label{fig:pictorialrep2}
           \includegraphics[width=0.5\textwidth,height=1.6in]{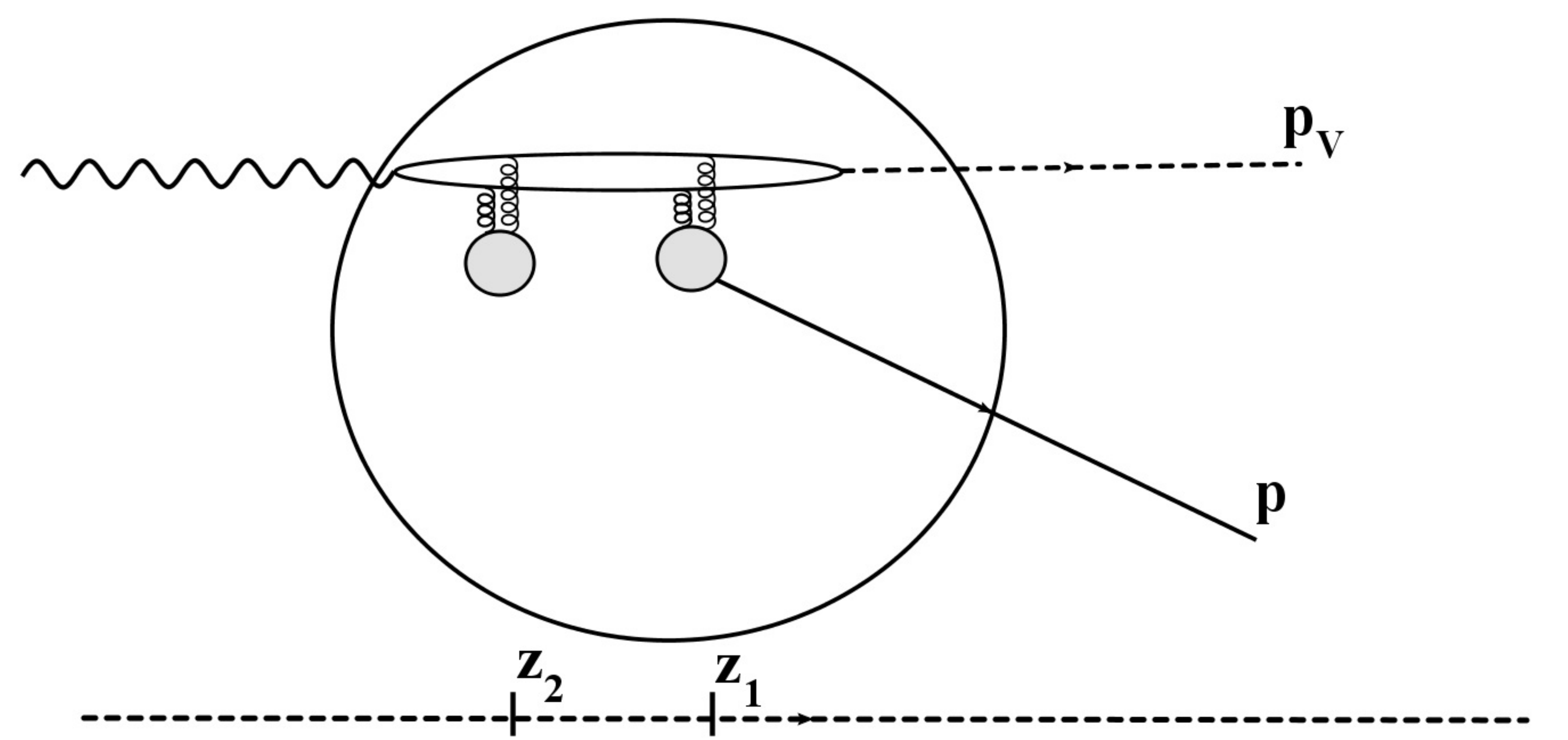}
        }\\ 

    \end{center}
    \caption{%
        Pictorial representation of the two terms in the amplitude of Eq. \ref{Vresult}.  
     }%
   \label{fig:pictorialrep}
\end{figure}

The physical interpretation of the two terms in \ref{Vresult} is as follows.  The first term in parentheses corresponds to production of the vector meson on nucleon 1 at position $(\mathbf{s}_1,z_1)$ with transverse momentum transfer $\mathbf{q}$, nucleon 1 being therefore knocked out. The second term in parentheses corresponds to forward production of the vector meson on nucleon 2 at position $(\mathbf{s}_1,z_2)$; the produced meson then propagates in the $z$-direction until the point $(\mathbf{s}_1,z_1)$  where it scatters elastically from nucleon 1 with transverse momentum transfer $\mathbf{q}$ to nucleon 1, nucleon 1 being knocked out.  In both cases the vector meson suffers attenuation beginning at the point where it is created as a physical meson through interaction with a nucleon (either at $(\mathbf{s}_1,z_1)$ for the first term or at $(\mathbf{s}_1,z_2)$ for the second term), while the proton suffers attenuation beginning at the point $\mathbf{r}_1=(\mathbf{s}_1,z_1)$ where it was located when the vector meson struck it.  The total amplitude is the sum of these two amplitudes; hence the square of the amplitude contains interference between the two amplitudes.

The result Eq. \ref{Vresult} is the Glauber theory result for the scattering amplitude for $\gamma^*+A\to\rho+p+(A-1)^*$, for the case where the final state of the residual nucleus is a one-hole state of the initial nucleus.  To obtain the differential cross-section, summed over all one-hole states, we would square $F_{fi}^{(n)}$, multiply by the appropriate phase-space and flux factors, and then sum over $n=1$ to $A$.  For the high-energy case we are considering, we may consider the energies of the outgoing particles to be essentially independent of $n$.  In that case, the phase-space and flux factors are independent of $n$, and so we may just sum $\vert F_{fi}^{(n)}\vert^2$ over $n$.   No inclusion of Color Transparency effects has been made up to this point, since the cross-sections $\sigma_{tot}$ that appear in it are the measured free-space cross-sections, and the elastic $\rho$-nucleon rescattering amplitude $f(\mathbf{q})$ that appears is also the free-space elastic amplitude.  Hence the outgoing meson or proton interacts with the other nucleons with the full free-space interaction cross-section.  To include CT effects, these cross-sections, and the elastic amplitude $f(\mathbf{q})$, must be modified to account for the smaller size of the outgoing hadrons compared to their usual sizes.

The result for $\sum_n \vert F_{fi}^{(n)}\vert^2$, where $n$ is summed only over one-hole states, is identical to the result one would obtain if instead one summed over \underline{all} final states of the residual nucleus (the incoherent cross-section) but only kept the terms corresponding to a single rescattering of the produced vector meson on a proton, and neglected terms where the vector meson rescatters two or more times on different nucleons.  The experimental situation, wherein the recoiling nucleus is not detected, corresponds to summing over all final states of the residual nucleus.  However, because of the exclusive nature of the reaction, if $\mathbf{p}_m$ is small, then the outgoing proton's momentum $\mathbf{p}\simeq\mathbf{q}$ and so only a single rescattering of the $\rho$ can have occurred, where the entire momentum transfer $\mathbf{q}$ was delivered to the detected proton.  Multiple rescattering terms in this case should be negligible, and so we need only sum $ \vert F_{fi}^{(n)}\vert^2$ over one-hole final states.  This implies that the transparency $T$ using the result Eq. \ref{Vresult}, which neglects any Color Transparency effects, will show very little dependence on the 4-momentum-transfer-squared $t\simeq -\mathbf{q}^2$.  Any significant variation of $T$ with $t$ will be due to Color Transparency.

Two limiting cases of the above result are of interest.  For the case of large $q_L$, which corresponds to a small coherence length $l_c$, the above result simplifies.  Taking $q_L\to\infty$, the second term in \eq{Vresult} is zero because of the oscillating exponential.  So in that case, 
\be
F_{fi}^{(n)}=f^{\gamma V}(\mathbf{q})  \int d^2s_1  d z_1  e^{-i\mathbf{p}_m\cdot\mathbf{r}_1}   e^{-\frac{1}{2}\alpha_p(\mathbf{r}_1)}  \phi_n(\mathbf{r}_1) e^{-\frac{1}{2}\sigma_{tot}^{V N}\;T_1(\mathbf{s}_1)}.
\ee
This is similar to the result for the pion case, \eq{pionresult}, except that the vector meson only undergoes attenuation (the factor $e^{-\frac{1}{2}\sigma_{tot}^{V N}\;T_1(\mathbf{s}_1)}$) starting at the point $z_1$, which is also the point where the initial proton was when it got knocked out.  There is no attenuation before this point; this agrees with the small coherence length, which means that the photon fluctuates into the vector meson essentially at the same point where it interacts with the proton with momentum transfer $\mathbf{q}$.  For the case of the pion (Ch. 2), the incoming pion can of course interact all along its incoming trajectory, and hence its attenuation factor $e^{-\frac{1}{2}\sigma_{tot}^{V N}\;T(\mathbf{s}_1)}$ includes integration from $z=-\infty$ to $z=\infty$.

The other limiting case of interest is for $q_L=0$.  In this case the profile operator $\Gamma_{tot}$ (\eq{Vgamma}) is equal to the $\Gamma_{tot}$ for the pion case, \eq{totgamma}, if we take $\Gamma^{\gamma}=\Gamma$.  This is easily shown for $A=2$ (and then proved for arbitrary $A$ by induction on $A$).  For $A=2$ we have:
\be
\begin{split}
\Gamma_{tot}(\mathbf{b},\mathbf{r}_1,\mathbf{r}_2)&= \Gamma_{b1}\bigl [1-\Gamma_{b2}\theta_{21}\bigr]+\Gamma_{b2}\bigl [1-\Gamma_{b1}\theta_{12}\bigr]  \\
&=\Gamma_{b1}+\Gamma_{b2}-\Gamma_{b1}\Gamma_{b2}\theta_{21}-\Gamma_{b2}\Gamma_{b1}\theta_{12}\\
&=\Gamma_{b1}+\Gamma_{b2}-\Gamma_{b1}\Gamma_{b2}=1-(1-\Gamma_{b1})(1-\Gamma_{b2})\;\surd
\end{split}
\ee
Therefore the result \eq{Vresult} should also reduce to the result for the pion case, \eq{pionresult}, when we set $q_L=0$ and $\Gamma^{\gamma}=\Gamma$, and indeed it does:  for $\Gamma^{\gamma}=\Gamma$ we have $f^{\gamma V}(\mathbf{q})=f^{}(\mathbf{q})$, and for $q_L=0$ the second term in parentheses in \eq{Vresult} becomes
\be
\begin{split}
& - \frac{1}{2}\sigma_{tot}^{VN} f(\mathbf{q})  \int_{-\infty}^{z_1} dz_2 \rho(\mathbf{s}_1 ,z_2)\;  e^{-\frac{1}{2}\sigma_{tot}^{V N}\;\int_{z_2}^{\infty} dz' \rho(\mathbf{s}_1,z')}\\
&= - \frac{1}{2}\sigma_{tot}^{VN}  f(\mathbf{q})   \int_{-\infty}^{z_1} dz_2 \Bigl(\frac{2}{\sigma_{tot}^{VN}}\Bigr)\frac{d}{dz_2} e^{-\frac{1}{2}\sigma_{tot}^{V N}\;\int_{z_2}^{\infty} dz' \rho(\mathbf{s}_1,z')}\\
&= f(\mathbf{q}) (- e^{-\frac{1}{2}\sigma_{tot}^{V N}\;\int_{z_1}^{\infty} dz' \rho(\mathbf{s}_1,z')}+e^{-\frac{1}{2}\sigma_{tot}^{V N}\;\int_{-\infty}^{\infty} dz' \rho(\mathbf{s}_1,z')})\\
&= f(\mathbf{q}) (-e^{-\frac{1}{2}\sigma_{tot}^{V N}\;T_1(\mathbf{s}_1)}+e^{-\frac{1}{2}\sigma_{tot}^{V N}\;T(\mathbf{s}_1)}).
\end{split}
\ee
Note that the optical theorem was used to relate $f^{\gamma V}(0)=f(0)$ to $\sigma_{tot}^{VN}$.  Thus we have
\be
F_{fi}^{(n)}=f^{\gamma V}(\mathbf{q})  \int d^2s_1  d z_1  e^{-i\mathbf{p}_m\cdot\mathbf{r}_1}   e^{-\frac{1}{2}\alpha_p(\mathbf{r}_1)}  \phi_n(\mathbf{r}_1) e^{-\frac{1}{2}\sigma_{tot}^{V N}\;T(\mathbf{s}_1)},
\ee
in agreement with the pion result.

\subsection{Inclusion of Color Transparency effects}

Effects of Color Transparency can be incorporated into the result \eq{Vresult} by including position dependent cross-sections from the quantum diffusion model~\cite{liu88,dok91}.  In this model, the total cross-section of interaction of the outgoing hadrons with a nucleon in the nucleus is~\cite{liu88}
\be
\label{sigeff}
\sigma^{eff}_{hN}(z,t)=\sigma^{tot}_{hN}\Biggl[ \theta(l_h-z)\; \Bigl[\frac{z}{l_h}+\frac{n^2\langle k_t^2\rangle}{\vert t\vert}\Bigl (1-\frac{z}{l_h}\Bigr) \Bigr]  +\theta(z-l_h)\Biggr].  
\ee
Here $z$ is the distance the hadron has traveled from the point where the hard hadron-nucleon interaction (with 4-momentum-transfer-squared $t$) occurred (Fig. \ref{fig:hardscatter}), 
$\sigma^{tot}_{hN}$ is the free-space total hadron-nucleon cross-section,  $n$ is the number of valence quarks of the hadron, and $\langle k_t^2\rangle^{1/2}$ is the average transverse momentum of the quark in the hadron (taken to be  $\langle k_t^2\rangle^{1/2}=0.35$ GeV).  Thus $\frac{\langle k_t^2\rangle}{\vert t\vert}\sigma^{tot}_{hN}$ is a measure of the transverse size of the hadron at the time of collision.  The parameter $l_h$ (the formation length) is the distance the hadron travels after the collision until it reaches its normal size.  This is estimated as $l_h\simeq\frac{1}{E_n-E_h}\simeq\frac{2 p_h}{M_n^2-M_h^2}$, where $M_n$ is the mass of a typical intermediate state $n$ of the hadron~\cite{liu88}.  In principle the quantity $l_h$ can be different for the pion and the proton, but since the relation $l_h\simeq\frac{1}{E_n-E_h}\simeq\frac{2 p_h}{M_n^2-M_h^2}$ is only an estimate, we take here $M_n^2-M_N^2=M_n^2-M_{\pi}^2=0.7\;GeV^2$ for both $l_{\pi}$ and $l_p$~\cite{miller06}.      

\begin{figure}[htb]
     \begin{center}
      
            \includegraphics[width=4.8in,height=2.5in]{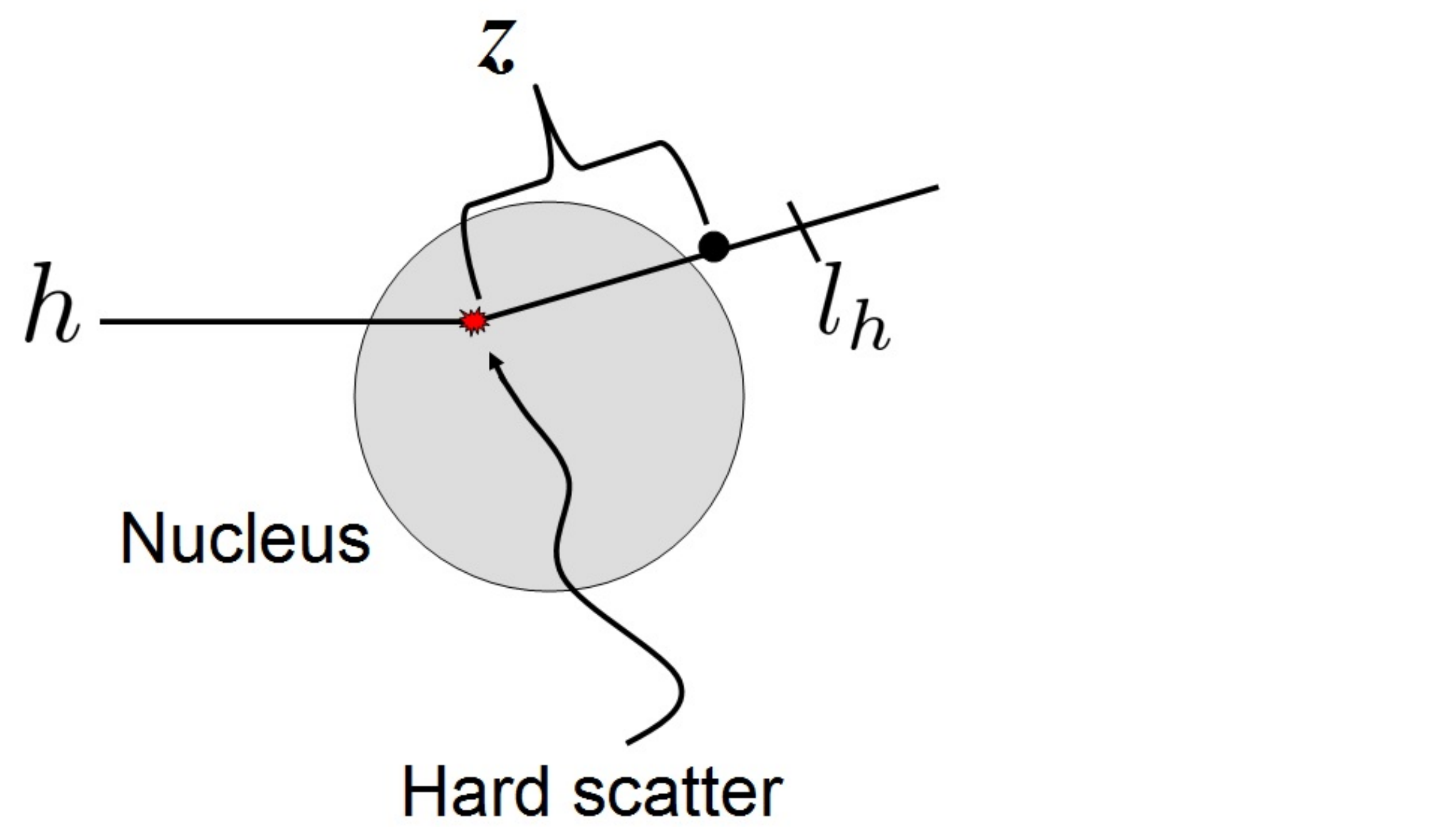}

    \end{center}
    \caption{%
        Formation length ($l_h$) for vector meson production. $z$ is the distance of the outgoing hadron from the point where the hard scattering occurred.
     }%
   \label{fig:hardscatter}
\end{figure}

The expression Eq. \ref{sigeff} is used for the cross-sections that appear in the exponentials in Eq. \ref{Vresult}.  The amplitudes $f^{\gamma V}(\mathbf{q})$ and $f^{\gamma V}(0)$ that appear in Eq. \ref{Vresult} are the same as the measured free-space production amplitudes.  However, the elastic rescattering amplitude $f(\mathbf{q})$ must be modified to include the effects of Color Transparency.  For large enough $Q^2$, the $q\bar{q}$ pair produced at the point $(\mathbf{s}_1,z_2)$ will be in a pointlike configuration; it will then expand as it propagates, and scatter elastically from a nucleon at $z_2$; if $z_2$ is close enough to $z_1$, the scattering amplitude $f(\mathbf{q})$ of the $q\bar{q}$ pair on the nucleon will be smaller than that of a normal $\rho$ meson.   Therefore the scattering amplitude $f(\mathbf{q})$ in Eq. \ref{Vresult} should be replaced by~\cite{Frankfurt:1994kk}
\be
\label{eq:fplc}
f^{PLC}(z_1-z_2,\mathbf{q},Q^2)=f(\mathbf{q})\frac{\sigma^{eff}_{VN}(z_1-z_2,Q^2)}{\sigma^{tot}_{VN}}\frac{G_V\Bigl(t\;\frac{\sigma^{eff}_{VN}(z_1-z_2,Q^2)}{\sigma^{tot}_{VN}}\Bigr)}{G_V(t)}
\ee
where $G_V(t)$ is the $\rho$-meson form factor, and $t\simeq -\mathbf{q}^2$, and $f(\mathbf{q})$ is the measured free-space elastic $\rho$-nucleon scattering amplitude.  This form for $f^{PLC}$ is derived using the optical theorem (and assuming $f(0)$ is pure imaginary) together with the empirical result~\cite{collins84,Frankfurt:1994kk} that the differential cross-section for hadron-nucleon scattering satisfies 
\be
\frac{d\sigma^{hN\to hN}}{dt}\sim G_h^2(t)G_N^2(t).
\ee 
in terms of the form factors of the $h$ and $N$.

Thus the result for the scattering amplitude including Color Transparency effects is
\begin{equation}
\label{VresultCT}
\begin{split}&F_{fi}^{(n)}=\int d^2s_1  d z_1  e^{-i\mathbf{p}_m\cdot\mathbf{r}_1}   e^{-\frac{1}{2}\alpha_{\mathbf{p}}(\mathbf{r}_1)}  \phi_n(\mathbf{r}_1)\\
& \times\Bigl(f^{\gamma V}(\mathbf{q}) e^{-\frac{1}{2}\alpha_V(\mathbf{s}_1,z_1)} - \frac{2\pi}{ik}f^{\gamma V}(0)  \int_{-\infty}^{z_1} dz_2\; \rho(\mathbf{s}_1 ,z_2)\; e^{iq_L(z_2-z_1)}\;  e^{-\frac{1}{2}\alpha_V(\mathbf{s}_1,z_2)} f^{PLC}(z_1,z_2,\mathbf{q},Q^2)\Bigr)\end{split}.
\ee
where
\be
\alpha_{\mathbf{p}}(\mathbf{r}_1)= \int_0^{\infty}\sigma^{eff}_{pN}(s,t)\rho(\mathbf{r}_1+s\;\hat{\mathbf{p}})ds
\ee
\be
\alpha_V(\mathbf{s}_1,z_1)=\int_{z_1}^{\infty} dz^{\prime}\sigma^{eff}_{V N}(z^{\prime}-z_1,t)\rho(\mathbf{s}_1,z^{\prime})
\ee
\be
\alpha_V(\mathbf{s}_1,z_2)=\int_{z_2}^{\infty} dz^{\prime}\sigma^{eff}_{V N}(z^{\prime}-z_2,Q^2)\rho(\mathbf{s}_1,z^{\prime}).
\ee
These expressions for $\alpha_V$ reflect the fact that the transverse size of the initial $q\bar{q}$ (at $z_2$) is determined by $1/Q^2$, while the transverse size of the outgoing  $q\bar{q}$ and proton, after the hard scatter from the proton at $(\mathbf{s}_1,z_1)$, is determined by $1/\vert t\vert$.

The transparency $T$ was calculated for $^{12}C$ and $^{40}Ca$ at $\mathbf{p}_m=0$ for kinematics corresponding roughly to those in the JLAB proposal for electroproduction of $\rho$ in nuclei~\cite{jlab06}.  The same harmonic oscillator nuclear wavefunctions were used as were used for the pion scattering case in Ch. 2.  The free-space cross-sections used were $\sigma_{tot}^{pN}=40\;mb$, and $\sigma_{tot}^{V N}=25\;mb$~\cite{anderson71}. 

\begin{figure}[tbp]
     \begin{center}
        \subfigure[VMD ]{%
            \label{fig:A=12rho1b}
            \includegraphics[width=0.4\textwidth,height=1.4in]{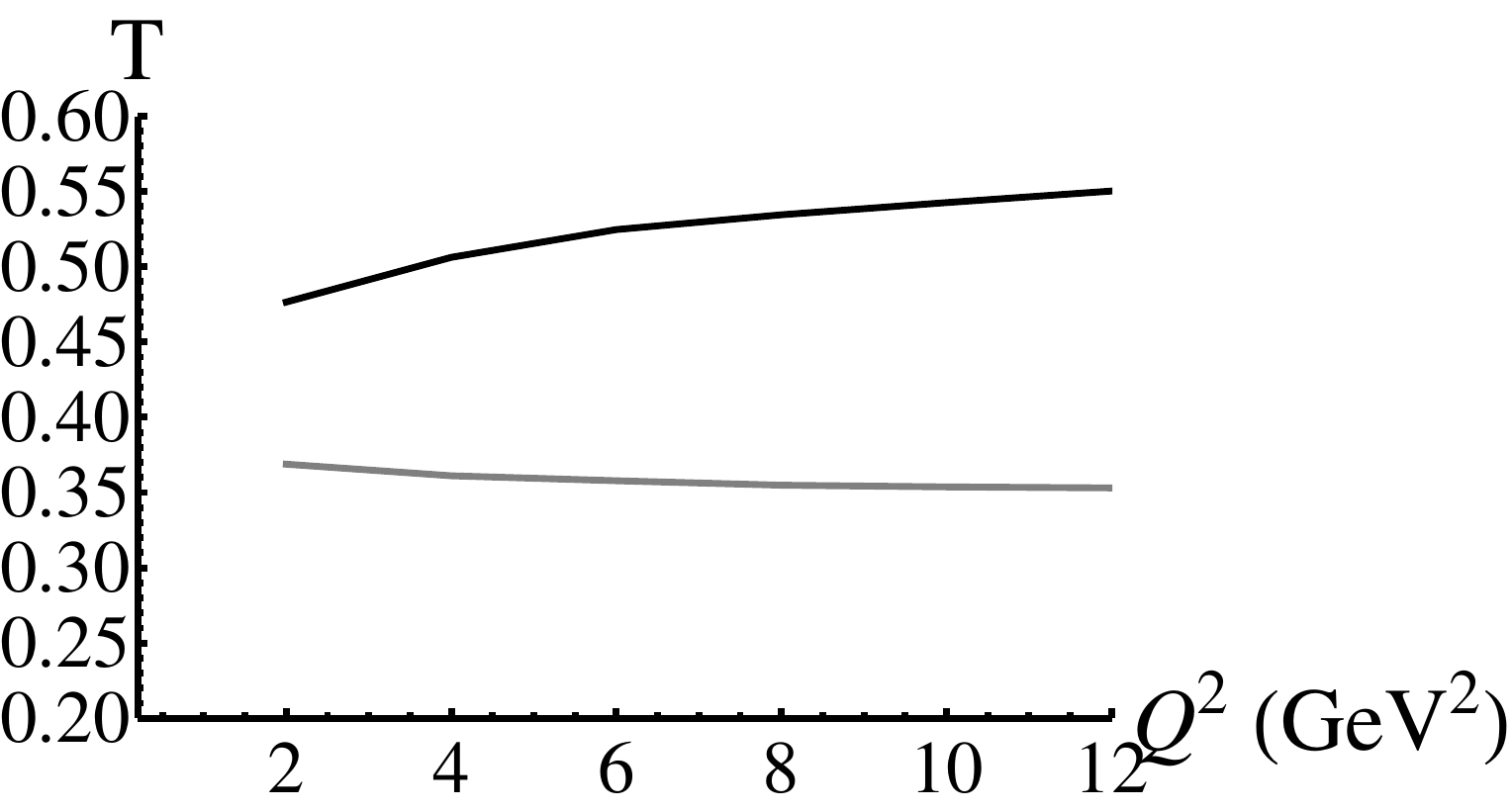}
}%
        \subfigure[$b=7$ GeV$^{-2}$ ]{%
           \label{fig:A=12rho2}
           \includegraphics[width=0.4\textwidth,height=1.4in]{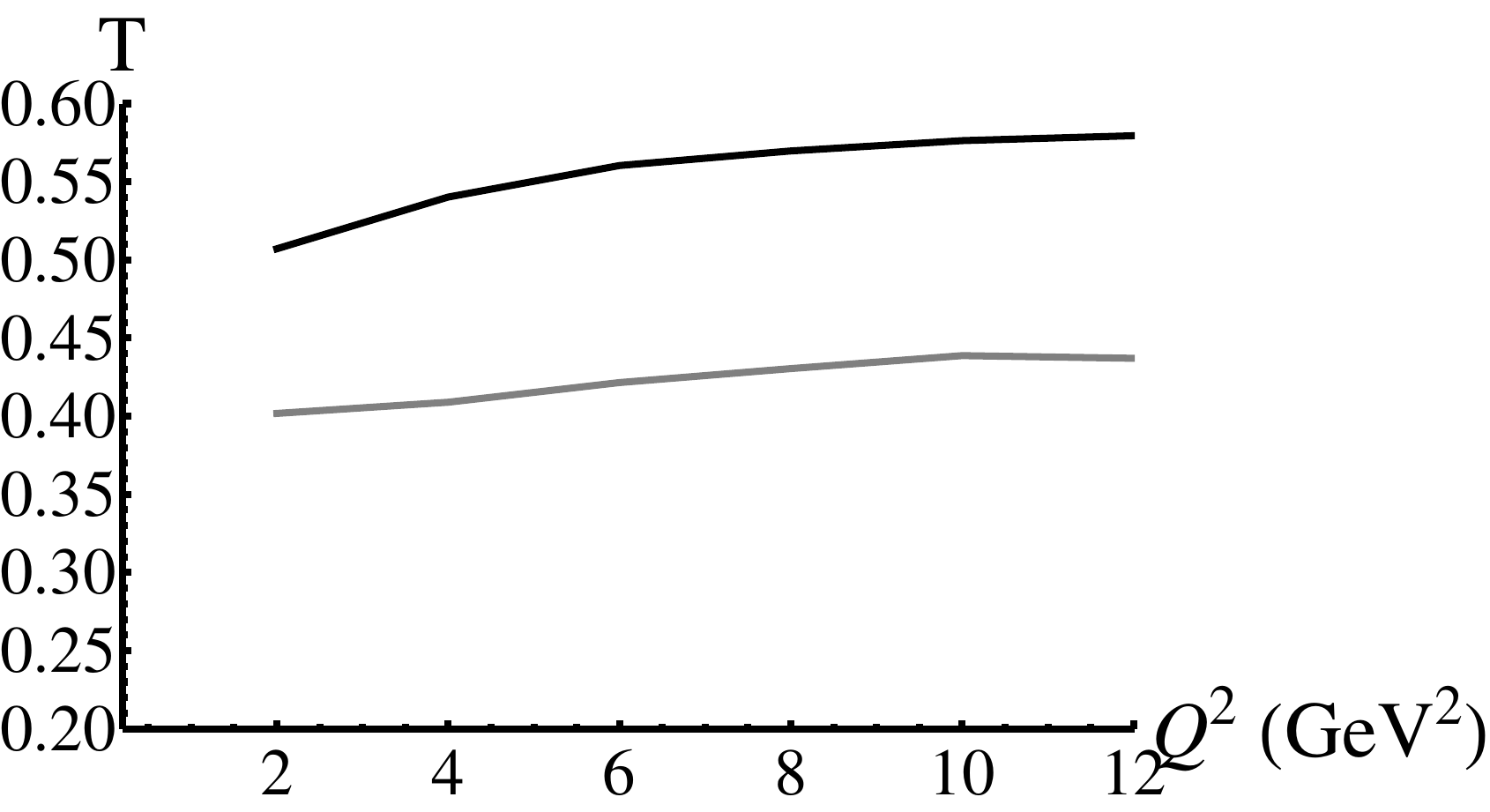}
        }\\ 

        \subfigure[ $b=8$ GeV$^{-2}$]{%
            \label{fig:A=12rho3}
            \includegraphics[width=0.4\textwidth,height=1.4in]{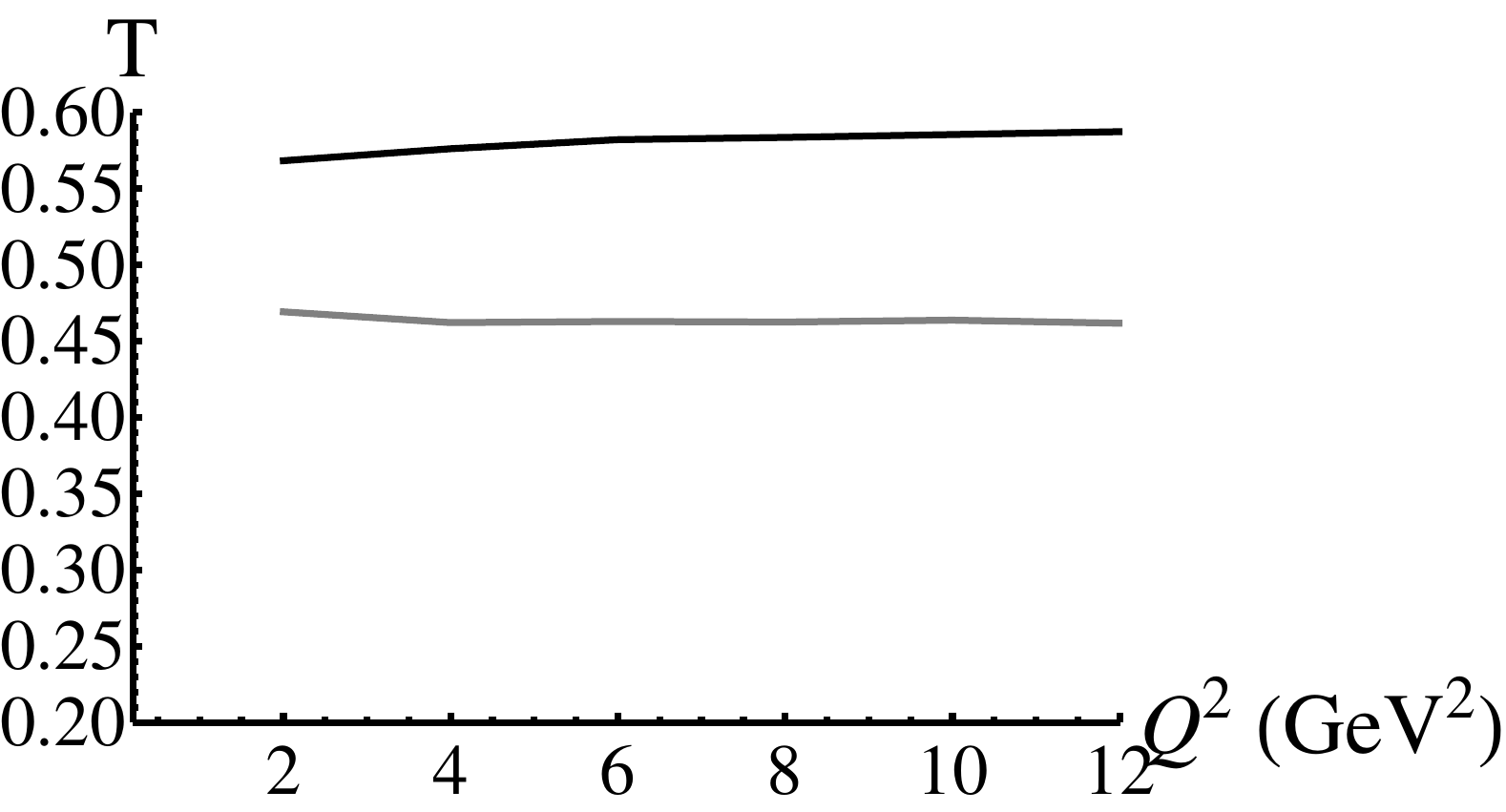}
        }%

    \end{center}
    \caption{%
        Transparency $T(\mathbf{p}_m)$ for $\mathbf{p}_m=0$ for $A=12$, $t=-2$ GeV$^2$, $l_c=5$ fm .  The bottom curves (gray) are the Glauber result; the top curves (black) are the CT result.  The value of the elastic $\rho$-nucleon $t$-slope parameter $b$ used in the calculation is indicated for each graph; VMD corresponds to $b_{\gamma V}=b$.
     }%
   \label{fig:pmzeroTnew}
\end{figure}

\begin{figure}[tbp]
     \begin{center}
        \subfigure[VMD ]{%
            \label{fig:A=40rho1b}
            \includegraphics[width=0.4\textwidth,height=1.4in]{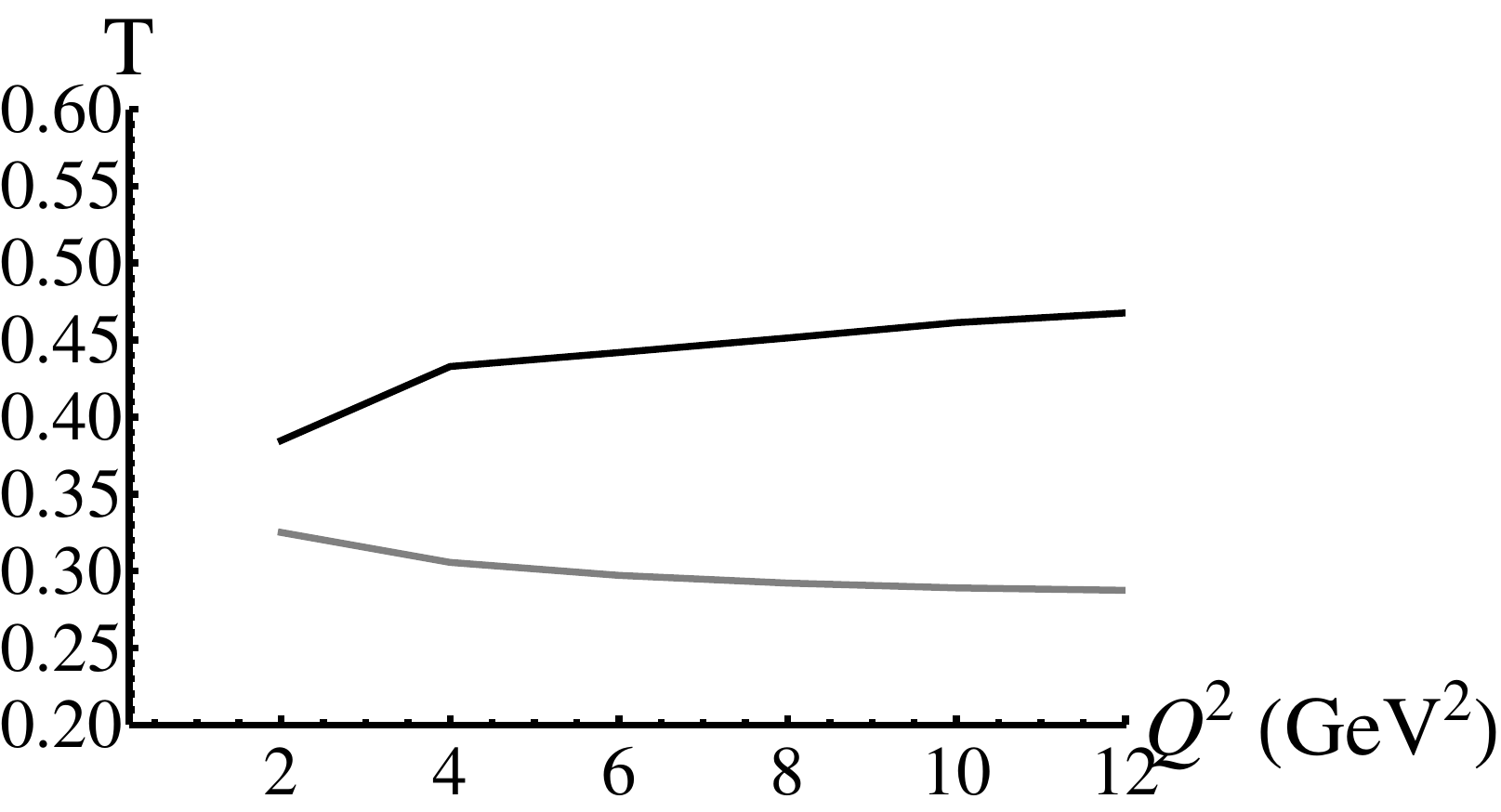}
}%
        \subfigure[$b=7$ GeV$^{-2}$ ]{%
           \label{fig:A=40rho2}
           \includegraphics[width=0.4\textwidth,height=1.4in]{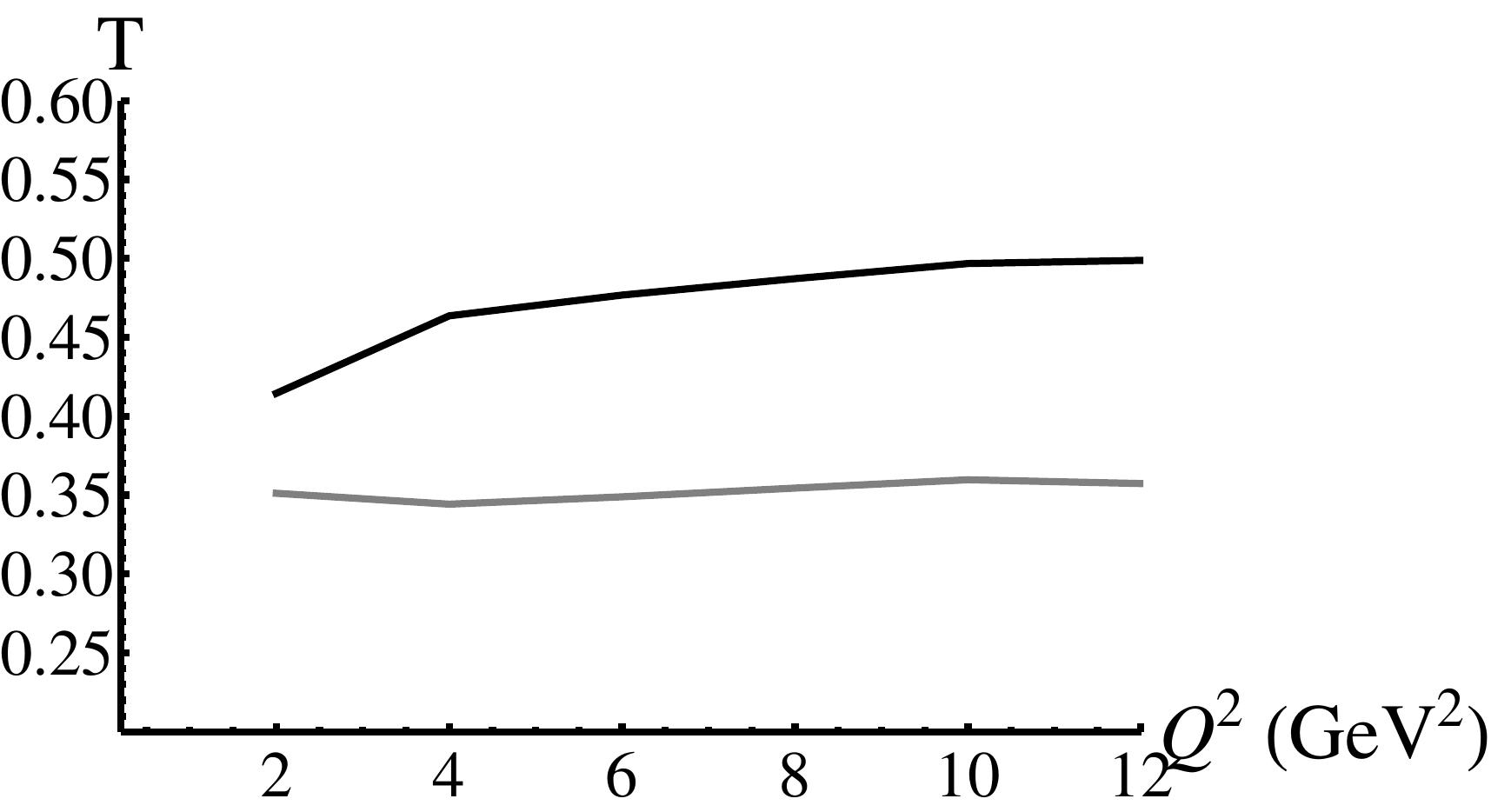}
        }\\ 

        \subfigure[ $b=8$ GeV$^{-2}$]{%
            \label{fig:A=40rho3}
            \includegraphics[width=0.4\textwidth,height=1.4in]{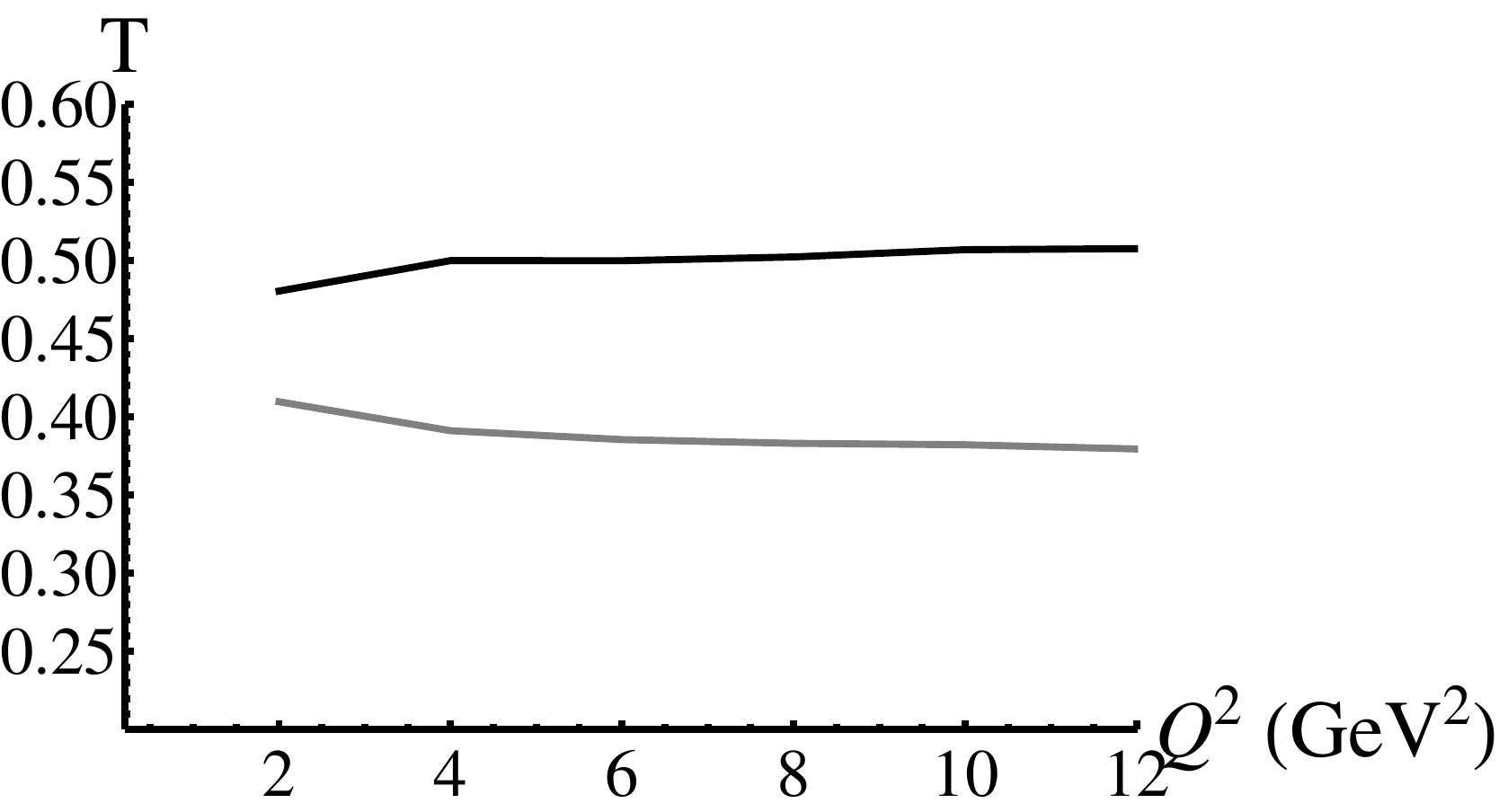}
        }%

    \end{center}
    \caption{%
        Transparency $T(\mathbf{p}_m)$ for $\mathbf{p}_m=0$ for $A=40$, $t=-2$ GeV$^2$, $l_c=5$ fm .  The bottom curves (gray) are the Glauber result; the top curves (black) are the CT result.  The value of the elastic $\rho$-nucleon $t$-slope parameter $b$ used in the calculation is indicated for each graph; VMD corresponds to $b_{\gamma V}=b$.
     }%
   \label{fig:pmzeroTA=40new}
\end{figure}

 Graphs of $T(\mathbf{p}_m=0)$ vs. $Q^2$ are shown in Figs. \ref{fig:pmzeroTnew} and \ref{fig:pmzeroTA=40new}.  It is important to note that the transparency as a function of $t$ is calculated for fixed $\nu$ and $Q^2$, so that the coherence length $l_c$ is held constant.  If the coherence length varied, this could mimic Color Transparency because as $l_c$ gets smaller the attenuation due the Initial State Interaction of the vector meson (before the hard scatter) decreases since the vector meson propagates a smaller distance before undergoing the hard scatter; this would cause the value of the transparency $T$ to increase as $l_c$ decreases.  The production and elastic scattering amplitudes $f^{\gamma V}(\mathbf{q})$ and $f(\mathbf{q})$ in \eq{VresultCT} and \eq{eq:fplc} were taken to be of the form $f^{\gamma V}(\mathbf{q})=A_{\gamma V}e^{\frac{1}{2}b_{\gamma V}t}$ and $f^{\gamma V}(\mathbf{q})=Ae^{\frac{1}{2}bt}$ (where $t=-\mathbf{q}^2$) with the parameters  $A_{\gamma V}$, $b_{\gamma V}$, $A$ and $b$ taken from experimental data.  The t-slope $b$ for elastic $\rho$-nucleon scattering has been measured to be between $7$ and $8$ GeV$^{-2}$~\cite{anderson71}, while the t-slope for the production amplitude varies with $Q^2$.  The available electroproduction data~\cite{Chekanov:2007zr} are at higher virtual photon energies than are considered in this paper, but the values of $b_{\gamma V}(Q^2)$ measured in that experiment were what were used in our calculations.  Calculations were done for $b=7$ GeV$^{-2}$ and for $b=8$ GeV$^{-2}$ with $b_{\gamma V}$ depending on $Q^2$.  For comparison, calculations were also done assuming the validity of Vector Meson Dominance, in which case $b_{\gamma V}=b$ and the transparency $T(\mathbf{p}_m)$, \eq{transp}, is independent of the value of $b$ since both numerator and denominator are proportional to $e^{bt}$. 
    The expected properties of the transparency are evident in  Figs. \ref{fig:pmzeroTnew} and \ref{fig:pmzeroTA=40new}.  For a given value of $Q^2$, the transparency (both Glauber and CT results) decreases with increasing $A$.    For a given $A$, as $Q^2$ increases the transparency in the CT case increases, which is also expected.  However, for the Glauber case, the behavior of $T$ as $Q^2$ varies is more sensitive to the values of $b$ and $b_{\gamma V}$ that are used.  Some of the dependence of $T$ on $Q^2$ is also due to the dependence of $\alpha_{\mathbf{p}} (\mathbf{r})$ on kinematics through the relation \eq{pmdef}.

\section{Integrated transparency}
\label{sec:third}

As in Sec. 2.4.2, the experimental situation corresponds to a range of values of the missing momentum $\mathbf{p}_m$.  The integrated transparency is again
\be
\label{intT1}
T_{\cal{D}}=\frac{\int_{\cal{D}}d^3p_m\frac{d\sigma}{dE'd\Omega'  d\Omega_p}}{\int_{\cal{D}}d^3p_m \frac{d\sigma_{PWIA}}{dE'd\Omega'  d\Omega_p}}=\frac{ \sum_{n=1}^A\int_{\cal{D}}d^3p_m \Bigl\vert F^{(n)}(\mathbf{p}_m)\Bigr\vert^2}{ \sum_{n=1}^A\int_{\cal{D}}d^3p_m \Bigl\vert F^{(n)}(\mathbf{p}_m)\Bigr\vert_{PWIA}^2}.
\ee 
Following the same steps as in the case of pion scattering, Sec. 2.4.2,  if we integrate over $\mathbf{p}_m$ up to $p_{max}\simeq 300\; MeV$, we may set $\mathbf{p}=\mathbf{q}$ in $\alpha_{\mathbf{p}}$; then assuming that the momentum distribution is zero for $\mathbf{p}_m>p_{max}$, we may extend the upper limit to infinity, $p_{max}\to\infty$.   For the denominator we obtain simply $(2\pi)^3\;A\;\vert f^{\gamma V}(\mathbf{q})\vert^2$.       For the numerator we obtain 3 terms:
\be
\label{eq:numerator}
(2\pi)^3\vert f^{\gamma V}(\mathbf{q})\vert^2  \int d^2s_1  d z_1 \rho(\mathbf{r}_1) e^{-\alpha_p(\mathbf{r}_1)} (h_1(\mathbf{r}_1)+h_2(\mathbf{r}_1)+h_3(\mathbf{r}_1))
\ee
where
\begin{flalign}
&h_1(\mathbf{r}_1) =   e^{-\alpha_V(\mathbf{s}_1,z_1)}&
\end{flalign}
\be
h_2(\mathbf{r}_1)=\frac{4\pi}{ik} \frac{f^{\gamma V}(\mathbf{q})f^{\gamma V}(0)}{\vert f^{\gamma V}(\mathbf{q})\vert^2}  e^{-\frac{1}{2}\alpha_V(\mathbf{s}_1,z_1)}\int_{-\infty}^{z_1} dz_2 \rho(\mathbf{s}_1,z_2) e^{-\frac{1}{2}\alpha_V(\mathbf{s}_1,z_2)}\cos{q_L(z_1-z_2)  f^{PLC}(z_1,z_2,\mathbf{q})}
\ee
\be
\begin{split}
h_3(\mathbf{r}_1)= \Bigl(\frac{2\pi}{k}\Bigr)^2 \frac{\vert f^{\gamma V}(0)\vert^2}{\vert f^{\gamma V}(\mathbf{q})\vert^2}        \int_{-\infty}^{z_1} dz_2 \int_{-\infty}^{z_1} dz_3  & \rho(\mathbf{s}_1,z_2) \rho(\mathbf{s}_1,z_3)  e^{-\frac{1}{2}\alpha_V(\mathbf{s}_1,z_2)} e^{-\frac{1}{2}\alpha_V(\mathbf{s}_1,z_3)}  \cos{q_L(z_2-z_3)}\\
& \times f^{PLC}(z_1,z_2,\mathbf{q}) f^{*PLC}(z_1,z_3,\mathbf{q})  
\end{split}
\ee
Thus we have for the integrated transparency
\be
\label{eq:integratedT}
T_{{\cal D}}=\frac{1}{A}\int d^2s_1  d z_1 \rho(\mathbf{r}_1) e^{-\alpha_p(\mathbf{r}_1)} (h_1(\mathbf{r}_1)+h_2(\mathbf{r}_1)+h_3(\mathbf{r}_1)).
\ee
This simplifies somewhat if we assume the validity of Vector Meson Dominance for the relation between the free-space production amplitude $f^{\gamma V}(\mathbf{q})$ and the free-space elastic scattering amplitude $f(\mathbf{q})$ (which appears inside $f^{PLC}$; see Eq.\ref{eq:fplc}).  Assuming that the high-energy amplitudes are purely imaginary, use of the optical theorem then gives:

\be
\label{eq:h2}
\begin{split}
h_2(\mathbf{r}_1)=- \frac{\sigma^{tot}_{VN}}{G_V(t)} e^{-\frac{1}{2}\alpha_V(\mathbf{s}_1,z_1)}\int_{-\infty}^{z_1} dz_2 \rho(\mathbf{s}_1,z_2)& e^{-\frac{1}{2}\alpha_V(\mathbf{s}_1,z_2)}\cos{q_L(z_1-z_2)}\\
& \times h(z_1-z_2)G_V\Bigl(t\;h(z_1-z_2)\Bigr)
\end{split}
\ee
\be
\label{eq:h3}
\begin{split}
h_3(\mathbf{r}_1)= \frac{1}{4}\Bigl(\frac{\sigma^{tot}_{VN}}{G_V(t)} \Bigr)^2       \int_{-\infty}^{z_1} &dz_2 \int_{-\infty}^{z_1} dz_3   \rho(\mathbf{s}_1,z_2) \rho(\mathbf{s}_1,z_3)  e^{-\frac{1}{2}\alpha_V(\mathbf{s}_1,z_2)} e^{-\frac{1}{2}\alpha_V(\mathbf{s}_1,z_3)}  \cos{q_L(z_2-z_3)}\\
& \times h(z_1-z_2) h(z_1-z_3) G_V\Bigl(t\;h(z_1-z_2)\Bigr) G_V\Bigl(t\;h(z_1-z_3)\Bigr)
\end{split}
\ee
where
\be
h(z)\equiv \frac{\sigma^{eff}_{VN}(z,Q^2)}{\sigma^{tot}_{VN}}=\Biggl[ \theta(l_h-z)\; \Bigl[\frac{z}{l_h}+\frac{n^2\langle k_t^2\rangle}{Q^2}\Bigl (1-\frac{z}{l_h}\Bigr) \Bigr]  +\theta(z-l_h)\Biggr].
\ee
The form factor $G_V$ used in evaluating Eq. \ref{eq:integratedT}  was taken to be the same form factor as for the pion: \be
G_V(t)=\frac{1}{1-t/0.59},
\ee
for $t$ in GeV$^2$.

The 3 terms of Eq. \ref{eq:numerator} or Eq. \ref{eq:integratedT} are represented pictorially by the same diagrams as in Fig. \ref{fig:pictorialrep}.  The term with $h_1$ is the square of the diagram in Fig. \ref{fig:pictorialrep1} and represents  incoherent production from nucleon 1; the term with $h_2$ represents interference between the diagrams of Fig. \ref{fig:pictorialrep1} and Fig. \ref{fig:pictorialrep2}, with interference between production on nucleon 1 and nucleon 2; and the term with $h_3$ is the square of the diagram in Fig. \ref{fig:pictorialrep2}, which represents interference between production on nucleon 2 and production on a different nucleon 3, with incoherent scattering from nucleon 1.

\begin{figure}[tbp]
     \begin{center}

        \subfigure[ VMD]{%
            \label{fig:A=56rho1}
            \includegraphics[width=0.4\textwidth,height=1.5in]{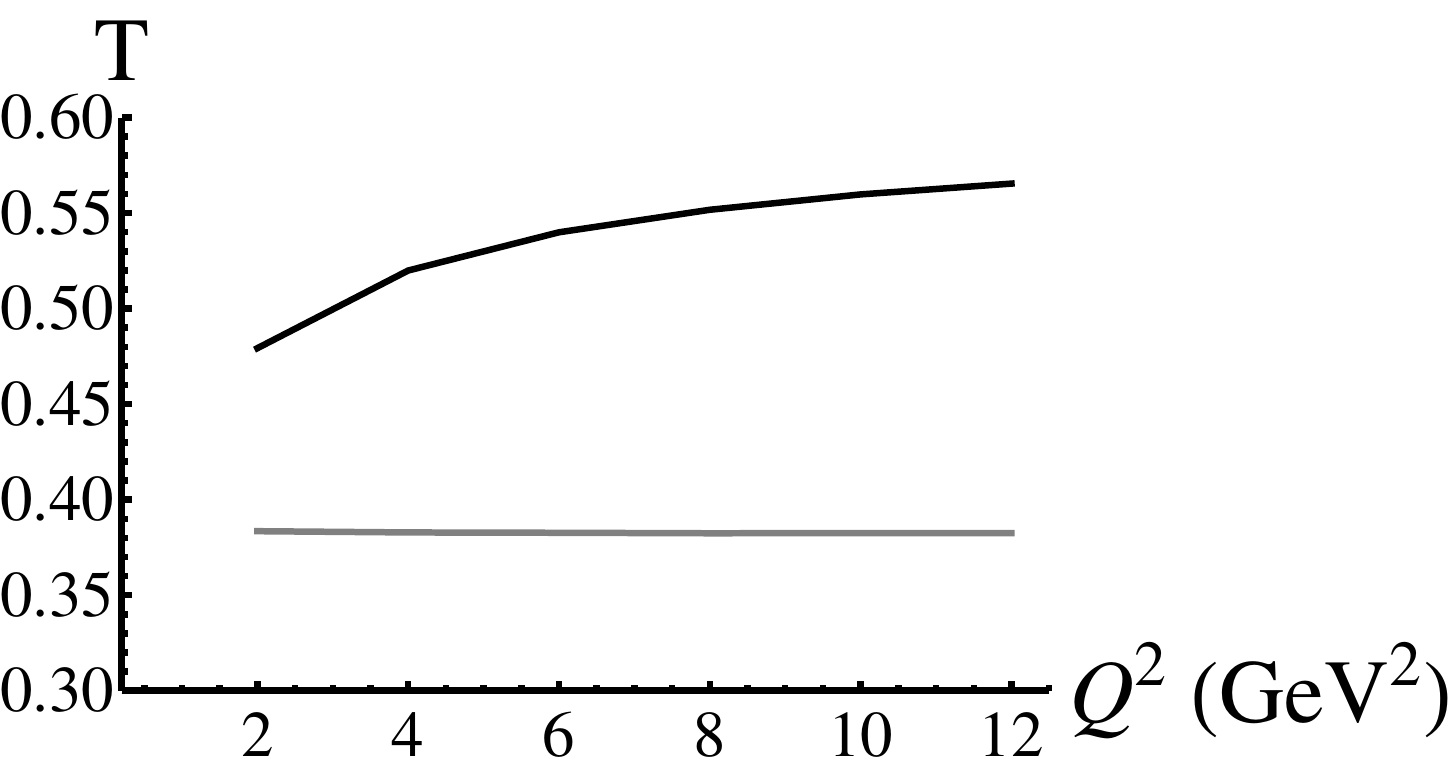}
        }%
    \hspace{0.5in}
        \subfigure[  $b=7$ GeV$^{-2}$]{%
           \label{fig:A=56rho2}
           \includegraphics[width=0.4\textwidth,height=1.5in]{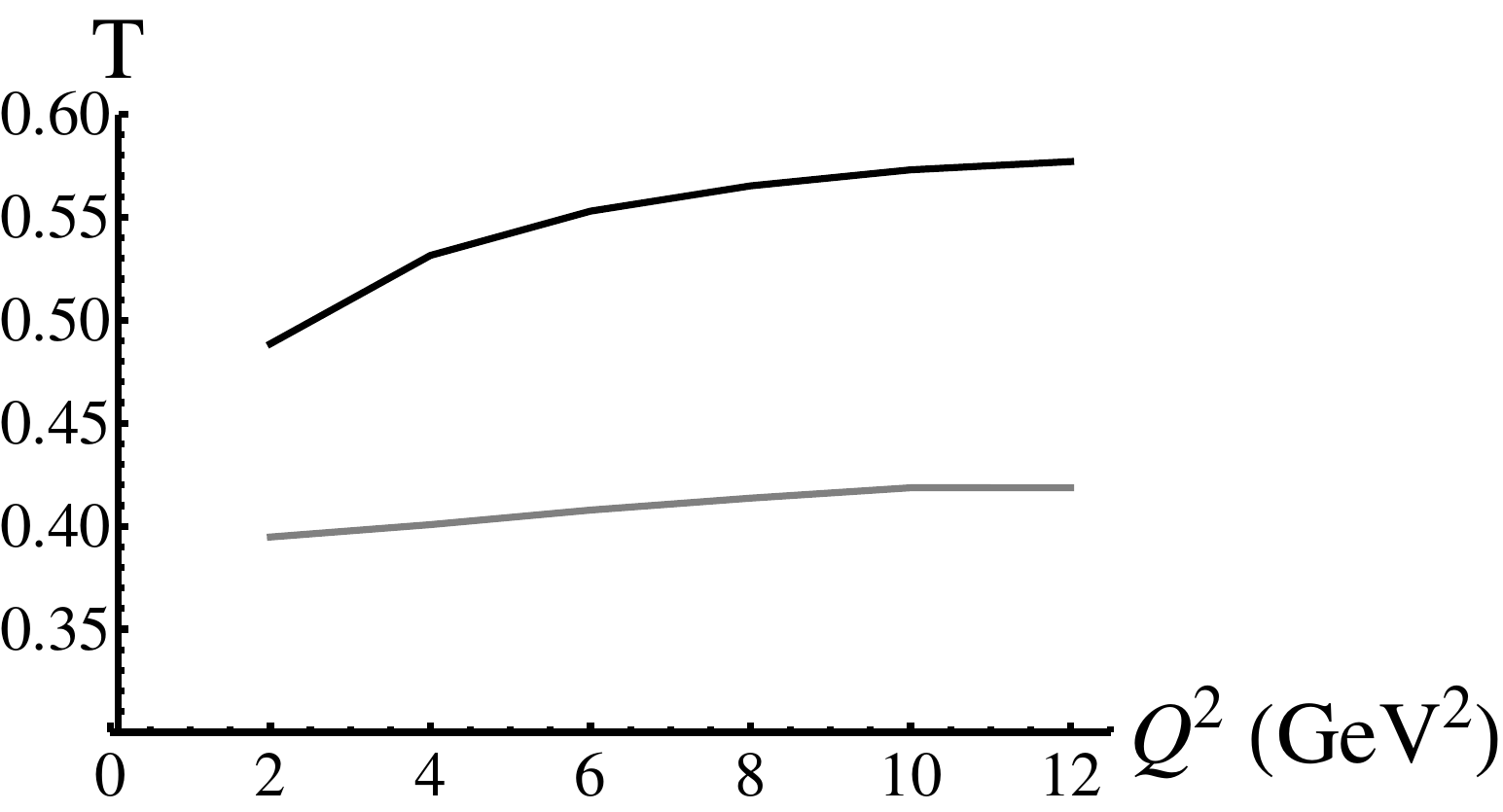}
        }\\ 
        \subfigure[ $b=8$ GeV$^{-2}$]{%
            \label{fig:A=12rho3}
            \includegraphics[width=0.4\textwidth,height=1.5in]{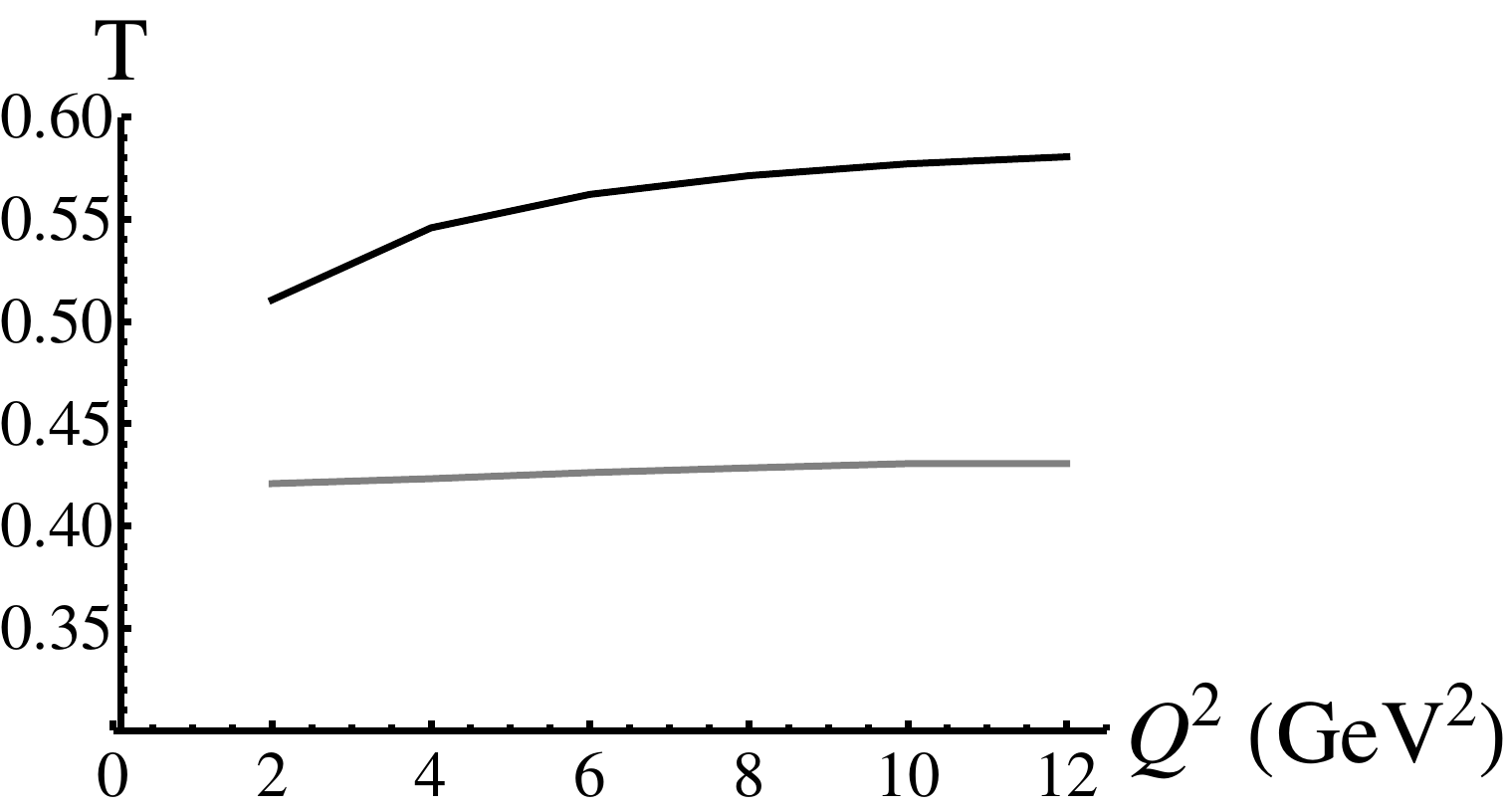}
        }%
    \end{center}
    \caption{%
        Integrated transparency $T$ for  $A=12$, $t=-2$ GeV$^2$, $l_c=2$ fm.  
        The bottom curves (gray) are the Glauber result; the top curves (black) are the CT result.  The value of the elastic $\rho$-nucleon $t$-slope parameter $b$ used in the calculation is indicated for each graph; VMD corresponds to $b_{\gamma V}=b$.
     }%
   \label{fig:intTA=12a}
\end{figure}

\begin{figure}[tbp]
     \begin{center}

        \subfigure[ VMD]{%
            \label{fig:A=56rho1}
            \includegraphics[width=0.4\textwidth,height=1.5in]{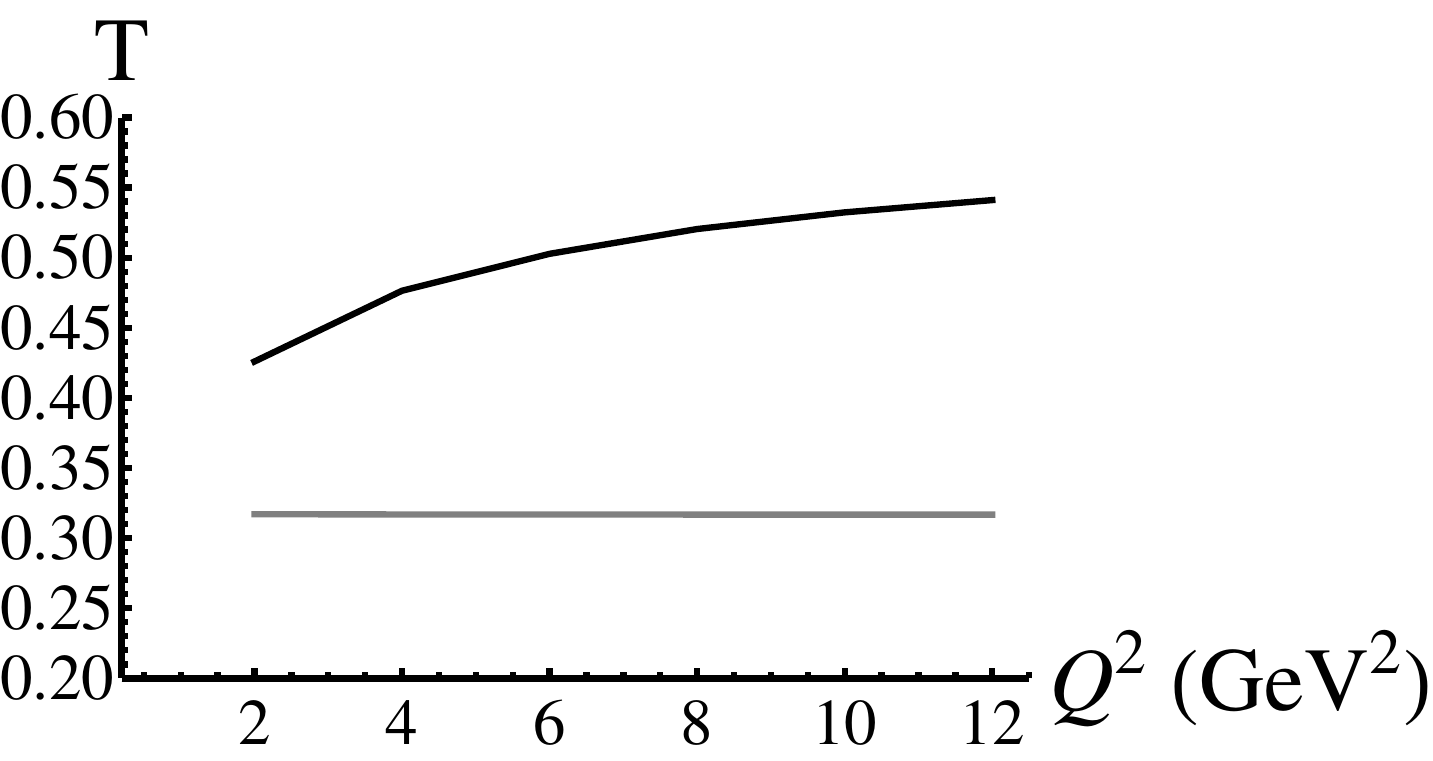}
        }%
    \hspace{0.5in}
        \subfigure[  $b=7$ GeV$^{-2}$]{%
           \label{fig:A=56rho2}
           \includegraphics[width=0.4\textwidth,height=1.5in]{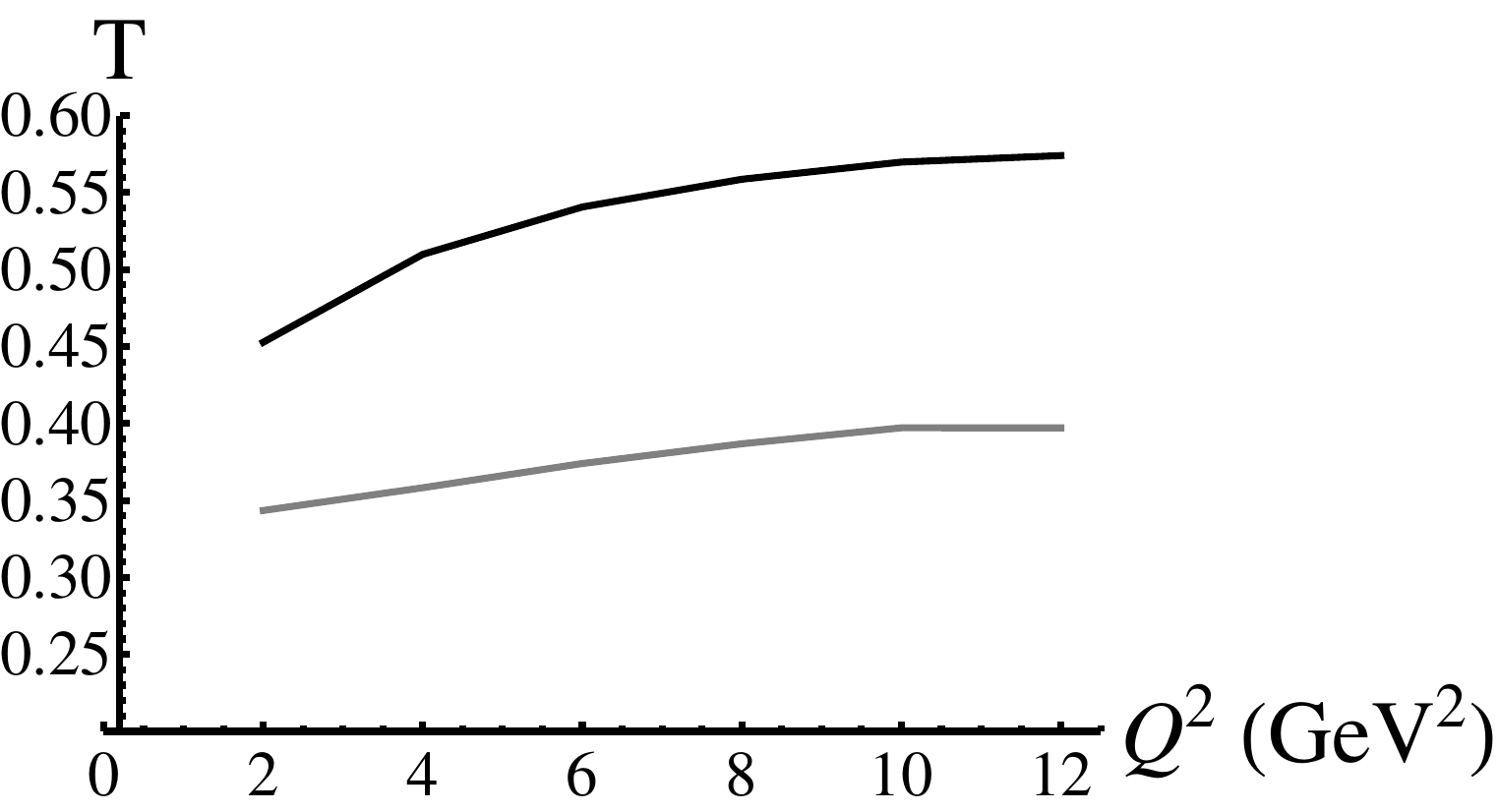}
        }\\ 
        \subfigure[ $b=8$ GeV$^{-2}$]{%
            \label{fig:A=12rho3}
            \includegraphics[width=0.4\textwidth,height=1.5in]{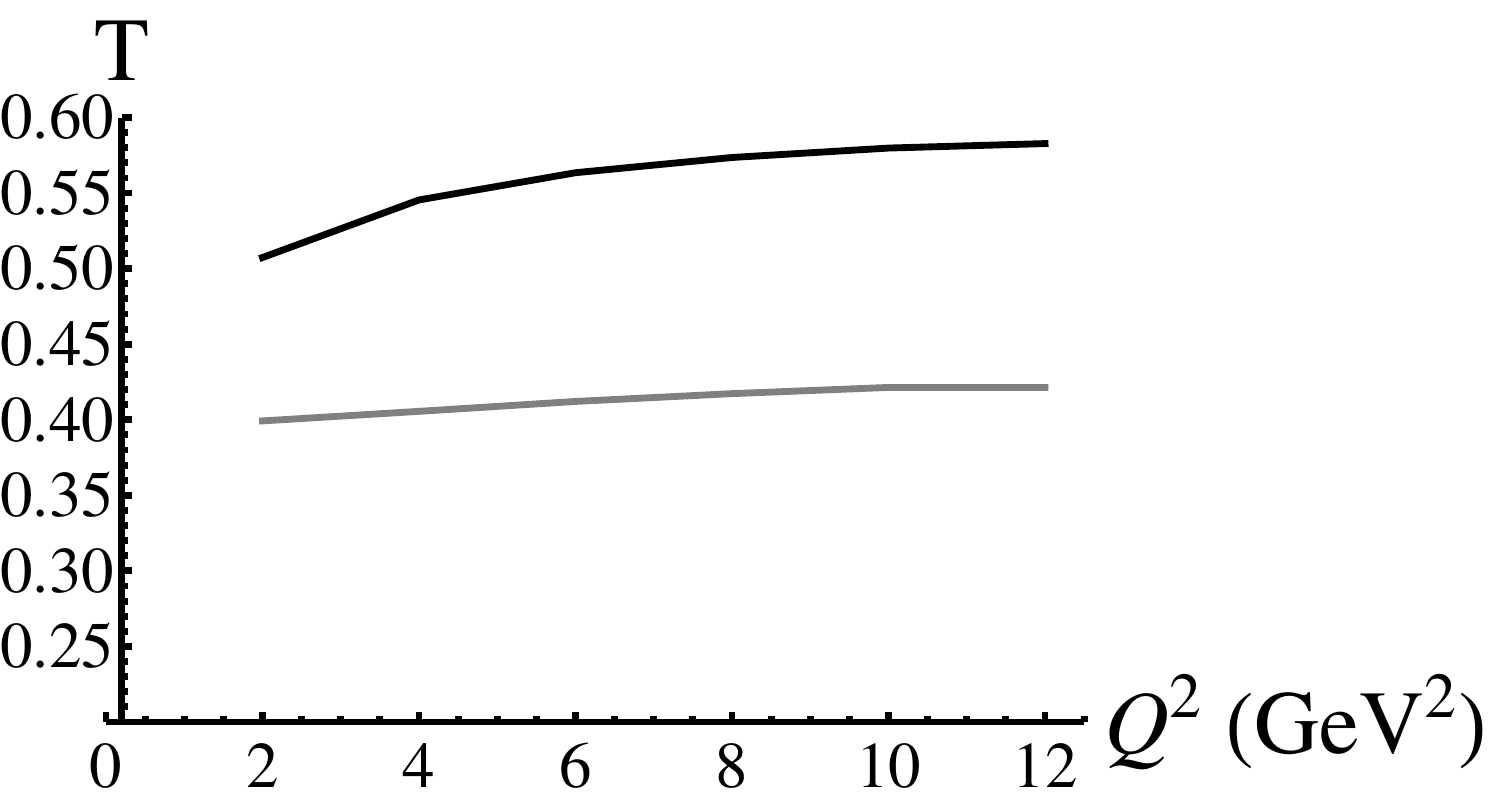}
        }%
    \end{center}
    \caption{%
        Integrated transparency $T$ for  $A=12$, $t=-2$ GeV$^2$, $l_c=5$ fm.  
        The bottom curves (gray) are the Glauber result; the top curves (black) are the CT result.  The value of the elastic $\rho$-nucleon $t$-slope parameter $b$ used in the calculation is indicated for each graph; VMD corresponds to $b_{\gamma V}=b$.
     }%
   \label{fig:intTA=12b}
\end{figure}

\begin{figure}[tbp]
     \begin{center}

        \subfigure[ VMD]{%
            \label{fig:A=56rho1}
            \includegraphics[width=0.4\textwidth,height=1.5in]{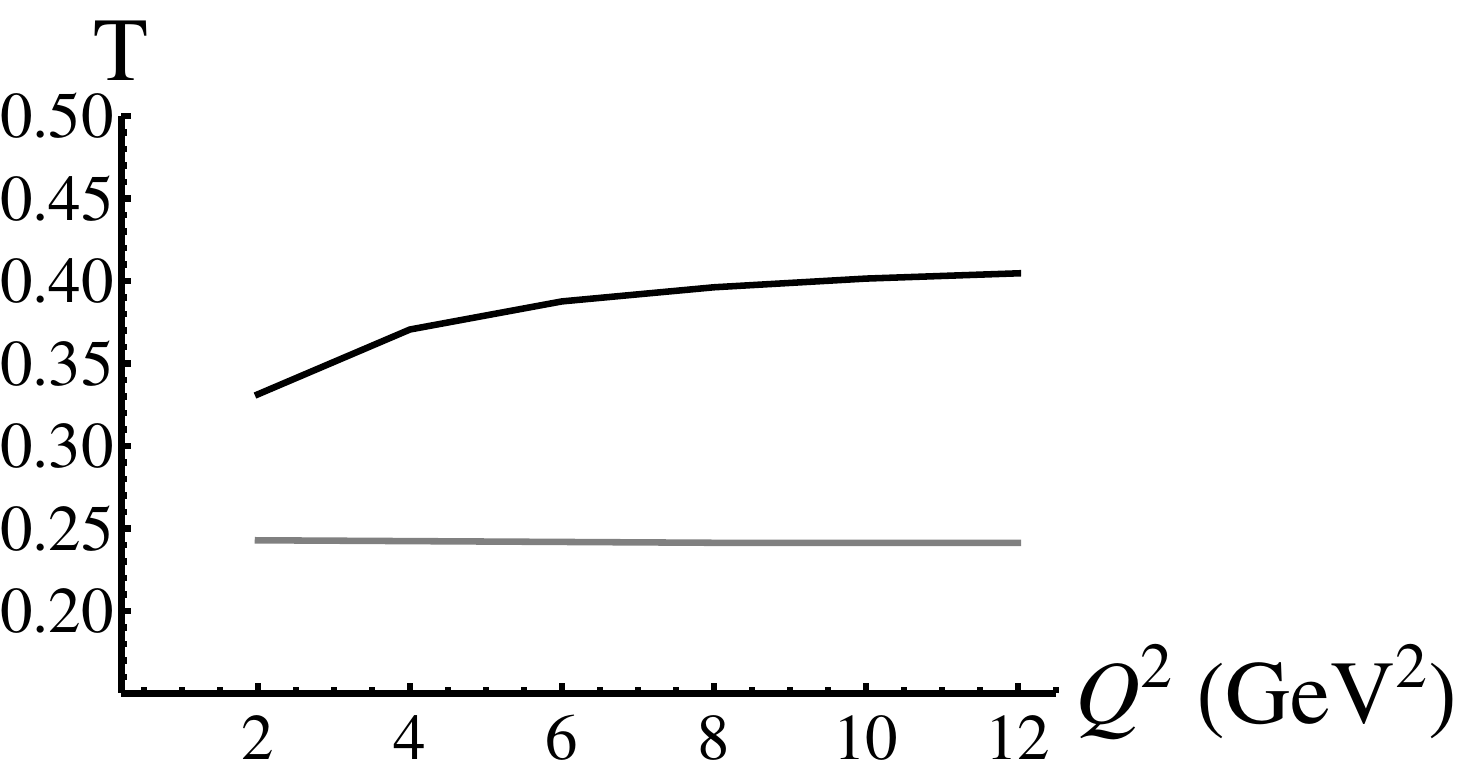}
        }%
    \hspace{0.5in}
        \subfigure[  $b=7$ GeV$^{-2}$]{%
           \label{fig:A=56rho2}
           \includegraphics[width=0.4\textwidth,height=1.5in]{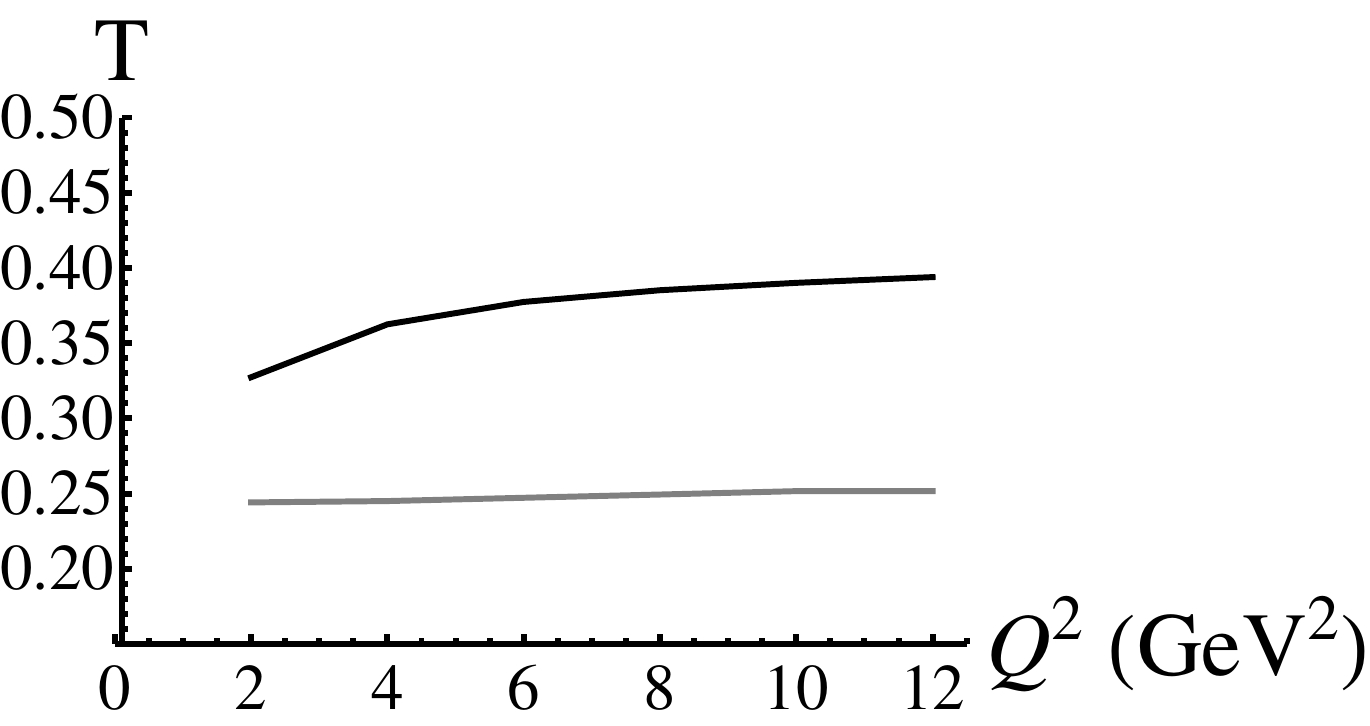}
        }\\ 
        \subfigure[ $b=8$ GeV$^{-2}$]{%
            \label{fig:A=12rho3}
            \includegraphics[width=0.4\textwidth,height=1.5in]{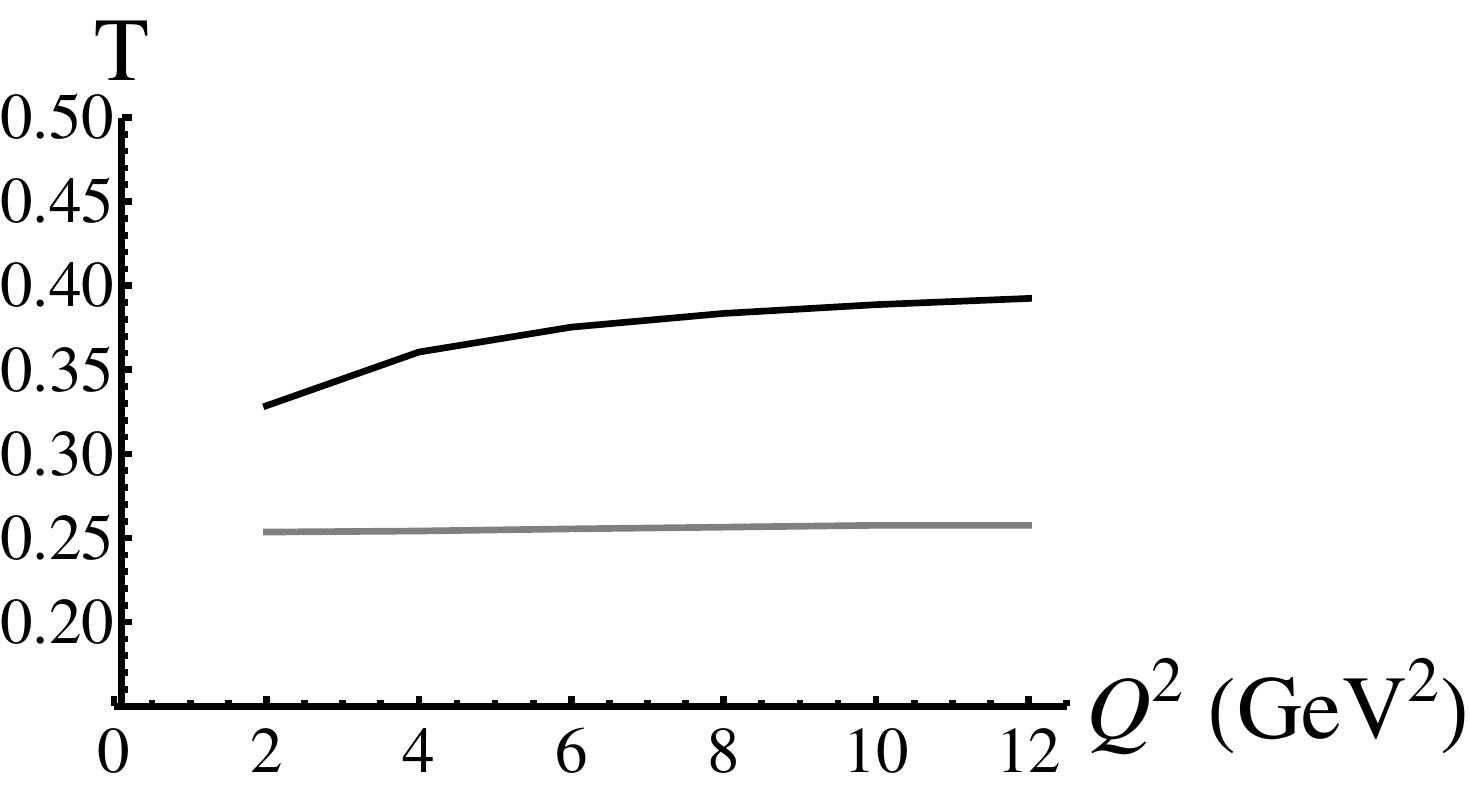}
        }%
    \end{center}
    \caption{%
        Integrated transparency $T$ for  $A=40$, $t=-2$ GeV$^2$, $l_c=2$ fm.  
        The bottom curves (gray) are the Glauber result; the top curves (black) are the CT result.  The value of the elastic $\rho$-nucleon $t$-slope parameter $b$ used in the calculation is indicated for each graph; VMD corresponds to $b_{\gamma V}=b$.
     }%
   \label{fig:intTA=40}
\end{figure}

\begin{figure}[tbp]
     \begin{center}

        \subfigure[ VMD]{%
            \label{fig:A=56rho1}
            \includegraphics[width=0.4\textwidth,height=1.5in]{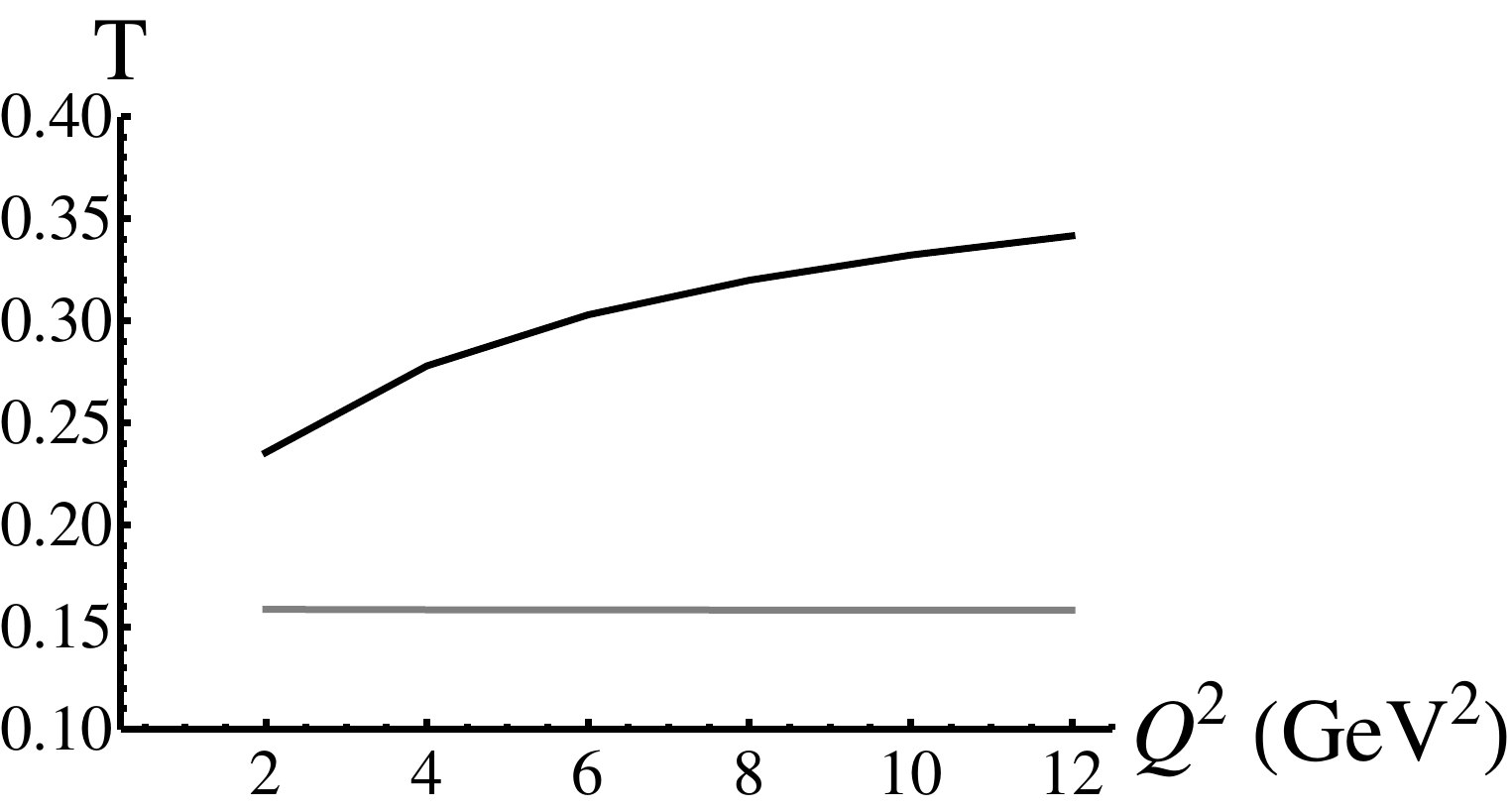}
        }%
    \hspace{0.5in}
        \subfigure[  $b=7$ GeV$^{-2}$]{%
           \label{fig:A=56rho2}
           \includegraphics[width=0.4\textwidth,height=1.5in]{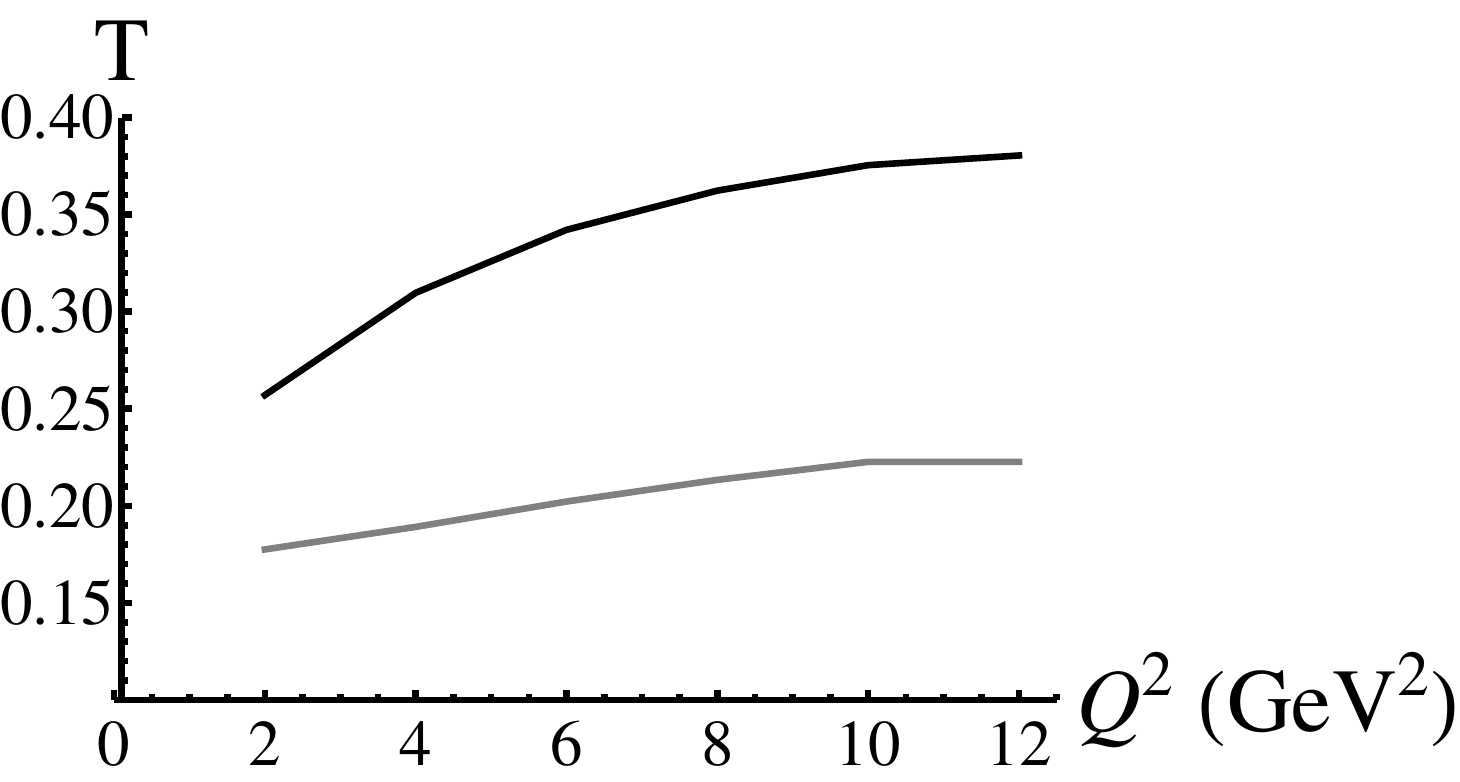}
        }\\ 
        \subfigure[ $b=8$ GeV$^{-2}$]{%
            \label{fig:A=12rho3}
            \includegraphics[width=0.4\textwidth,height=1.5in]{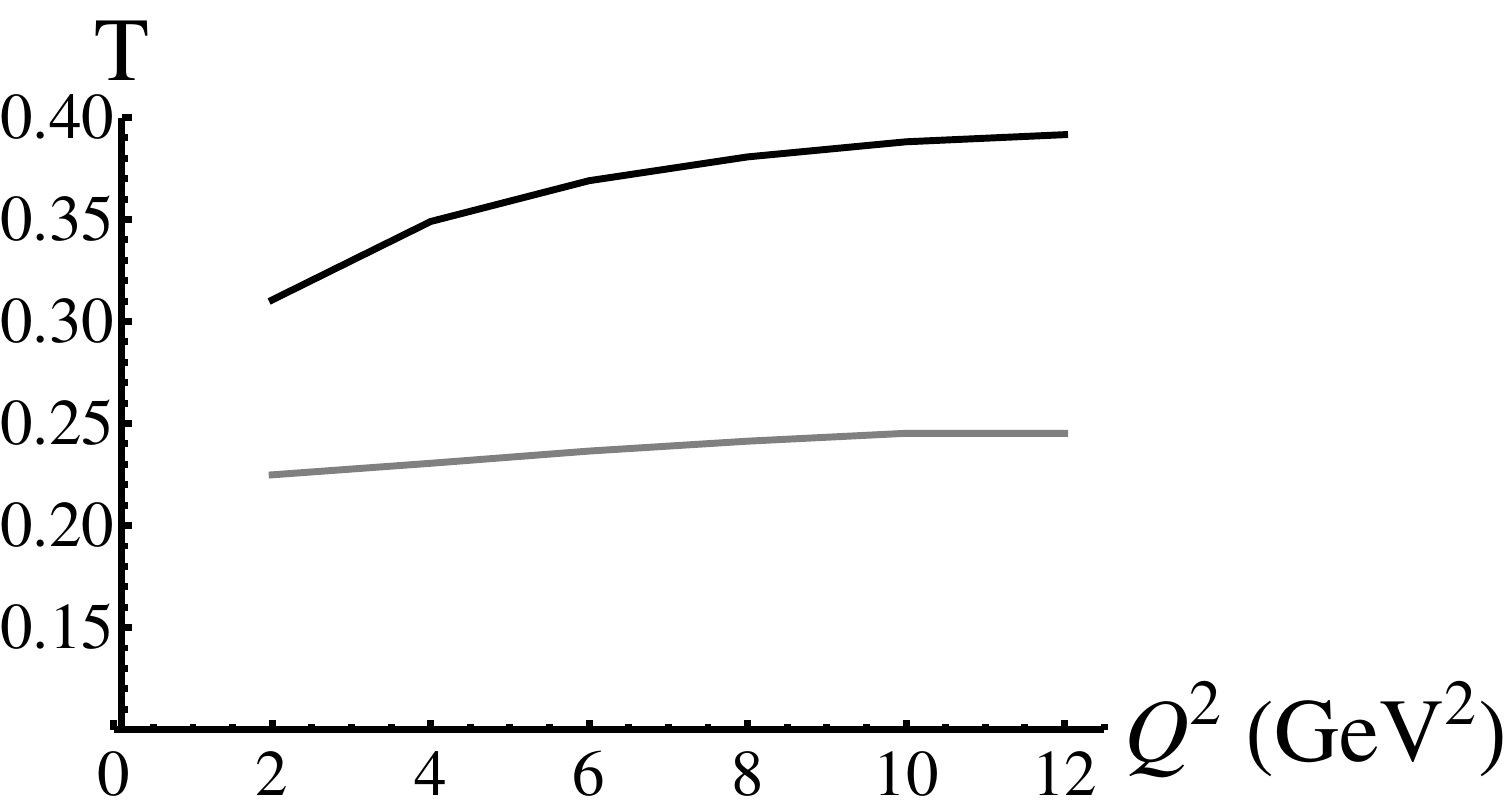}
        }%
    \end{center}
    \caption{%
        Integrated transparency $T$ for  $A=40$, $t=-2$ GeV$^2$, $l_c=5$ fm.  
        The bottom curves (gray) are the Glauber result; the top curves (black) are the CT result.  The value of the elastic $\rho$-nucleon $t$-slope parameter $b$ used in the calculation is indicated for each graph; VMD corresponds to $b_{\gamma V}=b$.
     }%
   \label{fig:intTA=40b}
\end{figure}

\begin{figure}[tbp]
     \begin{center}
        \subfigure[ $A=12$, $Q^2=0.5$ GeV$^2$, $l_c=5$ fm]{%
            \label{fig:A=56rho1}
            \includegraphics[width=0.7\textwidth,height=2.2in]{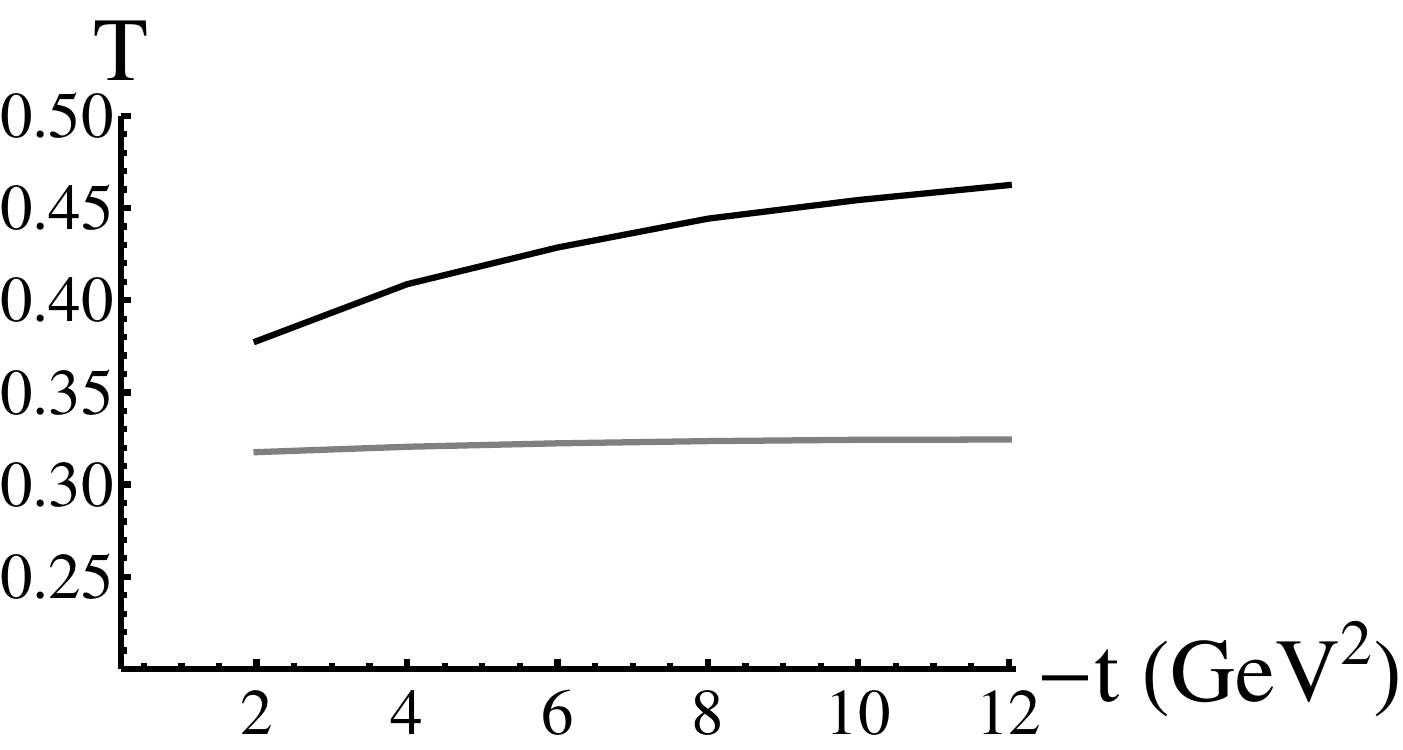}
        }%
 \hspace{0.5in}
        \subfigure[  $A=40$, $Q^2=0.5$ GeV$^2$, $l_c=5$ fm]{%
           \label{fig:A=56rho2}
           \includegraphics[width=0.7\textwidth,height=2.2in]{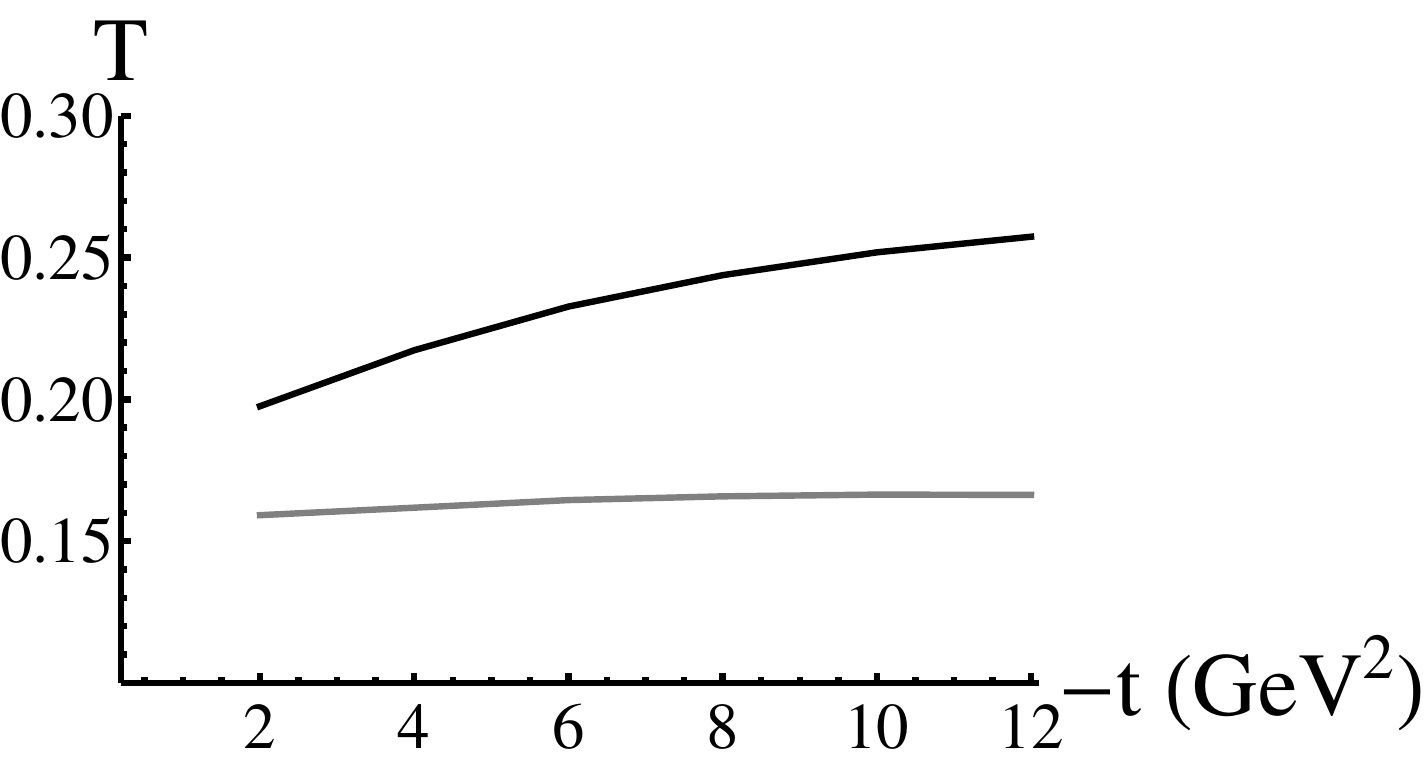}
        }\\ 
%

    \end{center}
    \caption{%
        Integrated transparency $T$ for fixed $Q^2$ and $l_c$ and varying $t$.  The bottom curves (gray) are the Glauber result; the top curves (black) are the CT result.
     }%
   \label{fig:intTfixedQ}
\end{figure}

The integrated transparency was calculated for $A=12$ and $A=40$, for a range of values of $t$ and $Q^2$.  In Figs. \ref{fig:intTA=12a} - \ref{fig:intTA=40b}, the transparency is shown  for fixed $t$ as a function of $Q^2$, for two different values of the coherence length.  The same values of $b$ and $b_{\gamma V}$ were used as for the $T(\mathbf{p}_m=0)$ calculation; VMD corresponds to $b=b_{\gamma V}$.

The same overall features of the graphs are present as were seen for the $\mathbf{p}_m=0$ transparency.  In addition, here one can see that for a given $A$ and $Q^2$, the transparency increases as the coherence length $l_c$ decreases, which agrees with expectations.  For the whole range of $Q^2$ from $2$ to $12$ GeV$^2$, the difference between the CT transparency and the Glauber transparency is significant.  For the higher values of $Q^2$, the CT value is of the order of $1.5$ times as large as the Glauber transparency, for $A=12$, and $2$ times as large as the Glauber transparency for $A=40$. The integrated transparency is significantly smaller than the values for ${\bf p}_m=0$. This is a relevant feature for experimentalists to note.

In Fig. \ref{fig:intTfixedQ}, the transparency is shown for fixed $Q^2$ as a function of $t$.  In that figure, $Q^2=0.5$ GeV$^2$, which is small enough that for the rescattering terms (Eqs. \ref{eq:h2} and \ref{eq:h3}) the produced $q\bar{q}$ (at either $z_2$ or $z_3$) is a normal $\rho$-meson.  Thus no Color Transparency effects occur as it propagates from the point where it was produced to the point where it undergoes the hard scatter of momentum transfer $\mathbf{q}$ which knocks out the nucleon.  But the large-momentum transfer scattering  at $z_1$ causes the outgoing $\rho$-like configuration to be in a small-sized configuration.  Hence the outgoing $\rho$ experiences reduced interactions on its way out of the nucleus (the knocked-out proton also experiences reduced interactions).  This is a manifestation of Color Transparency effects for small $Q^2$ (but large $t$).  The difference between the CT result and the Glauber result is not as significant, however, as  in the case of large $Q^2$. The range of $t$ shown is such that production angle of the outgoing $\rho$ is small, which is necessary for the validity of the Glauber model.  For the same $Q^2=0.5$ GeV$^2$ but for $l_c=2$ fm, the maximum allowable $\vert t\vert$ such that the $\rho$ production angle is small is only around $\vert t\vert=2$ GeV$^2$; hence plots for this value of $l_c$ are not shown since the range of $t$ would be small.

\section{Conclusion}
\label{sec:conclusion}

We have calculated the transparency for $\gamma^*+A\to \rho+p+(A-1)^*$, both without inclusion of CT effects (Glauber case) and with inclusion of CT effects, for several different combinations of $A$ and $l_c$.  The transparencies clearly exhibit the coherence length effect, i.e. the decrease of the transparency as $l_c$ is increased, which is not due to Color Transparency.  Thus to observe the effects of CT it is necessary to keep $l_c$ fixed while varying $\nu$ and $Q^2$.  The quantity of experimental interest, namely the integrated transparency, is smaller in general than the transparency evaluated at missing momentum $\mathbf{p}_m=0$.  However, the difference between the Glauber transparency and the CT transparency is marked, particularly as $Q^2$ is increased while $t$ is fixed.  However, it should still be possible to observe the effects of CT when $Q^2$ is small, if $t$ is large enough.  This represents an as yet unexplored kinematic region in the search for CT effects in electroproduction of vector mesons, namely small $Q^2$ but large $t$.  The difference between the CT prediction and the Glauber prediction for the transparency in this case is not as large as it is in the case of large $Q^2$.




 
\chapter{Low- and intermediate-energy $J/\psi$ electroproduction on the deuteron at JLAB}

\section{Introduction}

With the impending $12\;GeV$ upgrade at JLab, electroproduction of the $J/\psi$ will be possible.  With the mass of the $J/\psi$ being $3.097\;GeV$, the threshold photon energy for photoproduction on a single nucleon is $8.2\;GeV$, and is thus accessible with a $12\;GeV$ electron beam.  Most of the existing data on $J/\psi$ photo- and electroproduction is at much higher energy.  The $12\;GeV$ upgrade provides the opportunity to measure $J/\psi$ production near threshold~\cite{jlab12}.  In addition, measuring electroproduction on the deuteron provides the opportunity to measure the $J/\psi$-nucleon elastic scattering amplitude at lower energies than it has previously been measured at, if the rescattering of the produced $J/\psi$ on the spectator nucleon in the deuteron is significant.  

The motivation for the work in the first part of this chapter (Secs. 4.2 - 4.4) was a proposal at JLab~\cite{jlab10} to measure the $J/\psi $-nucleon scattering length by the reaction $\gamma^*+d\to J/\psi +p+n$, where the $J/\psi$ is produced on one nucleon in the deuteron and then re-scatters from the other nucleon.  The reason the $J/\psi $-nucleon scattering length is of interest is that several authors have argued that a nuclear bound state of the $J/\psi$ may exist~\cite{savage92,brodsky97}.  They propose that the force between a $J/\psi$ and a nucleon is purely gluonic in nature, and therefore is the analogue in QCD of the van der Waals force in electrodynamics, since the hadrons are of course color neutral objects.  There is very little experimental data on elastic $J/\psi$-nucleon scattering.  There has only been one experimental measurement of it, at SLAC in 1977, where the $J/\psi$-nucleon total cross-section was extracted by measuring production of $J/\psi$'s on nuclei and using an optical model for the re-scattering of the $J/\psi$ on the spectator nucleons~\cite{psidata77}.  

Measurement of the scattering length provides information on the bound states of the two particles involved in the scattering.  In particular, for an attractive potential, if the scattering length is positive then there exists a bound state.  Since the scattering length is the (negative of) the zero-energy scattering amplitude, in order to measure this it is necessary for the two particles to scatter with small relative-momentum.  In the case of $\gamma^*+d\to J/\psi +p+n$ at the energies which are kinematically allowed in the proposed JLab experiment, it isn't possible to have an on-mass-shell nucleon and $J/\psi$ scatter at small relative momentum.  For an incident virtual photon of energy $\nu=9\;GeV$, and an outgoing $J/\psi$-neutron pair with zero relative momentum, the minimum possible momentum of the neutron in the LAB frame (deuteron at rest) is $\simeq 0.85\;GeV$; for $\nu=6.5\;GeV$ and zero relative momentum of the $J/\psi$-neutron pair, the minimum LAB momentum of the neutron is $\simeq 1\;GeV$ (see Fig. \ref{fig:kin}).  For zero relative momentum of the outgoing pair, the initial LAB momentum of the neutron (before the collision with the $J/\psi$) must equal the final LAB momentum of the neutron.  Therefore, the momentum of the neutron inside the deuteron would have to be $0.85\;GeV$ (for $\nu=9\;GeV$). However,  the deuteron wavefunction at that momentum is very small (essentially zero).  

So although the proposed experiment~\cite{jlab10} may not be able to measure the $J/\psi$-nucleon scattering length, it might still be possible to measure the on-mass-shell $J/\psi$-nucleon scattering amplitude, but at higher relative energies.  The relative energy of the $J/\psi$-neutron pair would still be significantly smaller than in the only existing data (from the 1977 experiment at SLAC).  Under certain kinematic conditions, the dominant contributions to the amplitude will come from p-n rescattering and/or $J/\psi-n$ rescattering after the $J/\psi$ is produced.  If we fix the magnitude of the outgoing neutron's momentum at a moderately large value (here taken to be 0.5 GeV) the contribution of the impulse diagram (where the $J/\psi$ is produced on the proton and the neutron recoils freely) will be negligible, since the impulse diagram is proportional to the value of the deuteron wavefunction at that momentum  (see Fig. \ref{fig:diagrams2} for the impulse and rescattering diagrams).  This higher-energy rescattering is the subject of the second part of this chapter (Sec. 4.5). 

Note that when the relative energy of the produced particle and nucleon is small, the Glauber theory that was used in the previous two chapters is not applicable.  In the Glauber theory, the projectile or produced particle always moves at high speed relative to the nucleons.  To determine the scattering length, we have the opposite situation:   the produced $J/\psi$ needs to be moving slowly relative to the nucleons.
Also, in Glauber theory no account is made for the Fermi motion of the nucleons.  But for the case of near-threshold production, the Fermi motion has a large effect on the amplitude and must be taken into account.  Therefore a different method must be used to calculate the scattering amplitude in this case.  For the calculations in this chapter, a covariant Feynman diagram method is used.   

This chapter is organized as follows.  In Sec. \ref{sec:diag} the diagrammatic approach is discussed in a heuristic manner, as well as its reduction to the Glauber theory under certain kinematic conditions.  In Sec. \ref{sec:electroproduction}, electroproduction of a particle from a nucleus is discussed.   In Sec. \ref{sec:kinematics} the kinematics for the case of zero and small relative momentum of the outgoing $J/\psi$-neutron pair is discussed.  In Sec. \ref{sec:invaramps} the calculation of the invariant amplitudes for $\gamma^*+D\to J/\psi +p+n$ are presented, including the one-loop diagrams corresponding to the $p-n$ and $J/\psi$-nucleon rescattering processes.  In order to calculate the amplitude corresponding to the low-energy $J/\psi$-neutron scattering, which involves the scattering length, model $J/\psi$-neutron scattering wavefunctions and potentials are used, and it is shown that the resulting amplitude is insensitive to the model used.  In addition, it is shown that for the kinematic conditions of the JLab experiment, the dominant amplitude is the impulse diagram, corresponding to $J/\psi$ production on the neutron with the proton recoiling freely, with no rescattering of any particles.  This demonstrates that the measurement of the $J/\psi$-nucleon scattering length is not feasible for the JLab experiment.  Finally, in Sec. \ref{sec:intermedenergy} calculations of the amplitude for $\gamma^*+D\to J/\psi +p+n$ are presented under different kinematic conditions (not restricting the outgoing $J/\psi$-neutron pair to small relative momentum).  There it is shown that if the $J/\psi$-neutron elastic scattering amplitude is somewhat larger than the value measured at SLAC at higher energy, it may be possible to extract this amplitude from the Jlab experiment.

\section{Diagrammatic approach:  Heuristic discussion}
\label{sec:diag}

Many years ago it was shown that the Glauber approximation can be derived using a diagrammatic approach~\cite{harrington64,abers66,bertocchi67,bassel68,bertocchi72} where each Feynman diagram represents a Lorentz invariant amplitude, under the kinematic conditions for which the Glauber theory is valid, i.e. neglect of longitudinal momentum transfers, neglect of energy transfer from the projectile, neglect of the Fermi motion of the nucleons in the nucleus, and neglect of off-mass-shell effects.  In this section, I'll heuristically outline this derivation.  Here I omit various factors from expressions, in order to just show the dependence of quantities on the amplitudes $f$ and wavefunction $\Psi$.

For example, for the case of elastic hadron-deuteron scattering, i.e. $h+d\to h+d$, which is the case that is treated in~\cite{harrington64,abers66,bertocchi67,bassel68}, the  Glauber result for the scattering amplitude consists of two types of terms, a single-scattering term and a double scattering term.  The single-scattering term is just the product of the hadron-nucleon scattering amplitude and the deuteron form factor:
\be
\label{glaubdiag1}
F_1(\mathbf{q})\sim f(\mathbf{q}) G\Bigl(\frac{\mathbf{q}^2}{4}\Bigr),
\ee
where $\mathbf{q}$ is the 3-momentum transfer from the incident hadron $h$, $f$ is the 2-body hadron-nucleon scattering amplitude, and $G$ is the deuteron form factor:
\be
 G\Bigl(\frac{\mathbf{q}^2}{4}\Bigr)=\int d^3k \Psi^*(\mathbf{k}+\mathbf{q}/4)\Psi(\mathbf{k}-\mathbf{q}/4),
\ee
where $\Psi$ is the deuteron momentum-space wavefunction. 
The second term in the Glauber approximation is the double scattering term, where the projectile first scatters elastically from the proton and then scatters elastically from the neutron (or vice versa), and has the form
\be
\label{glaubdiag2}
F_2\sim\int d^2p \;G(p^2)f(\mathbf{q}/2 +\mathbf{p})f(\mathbf{q}/2 -\mathbf{p}).
\ee
The two amplitudes $F_1$ and $F_2$ can be represented diagrammatically by the diagrams in Fig. \ref{fig:deutsingledouble}.  If we interpret these diagrams as covariant Feynman diagrams, with the solid gray dot representing the deuteron vertex function $\Gamma(n)$ (the invariant amplitude for the virtual dissocation $D\to p+n$) and the circle-with-cross representing the invariant amplitude for hadron-nucleon elastic scattering ${\cal M}$, then the first diagram would have the following Lorentz invariant expression (leaving out various constants):
\be
F_a\sim \int d^4n \frac{\Gamma(p)\Gamma(p_2){\cal M}(p,p_3,p_4)}{(n^2-m^2+i\epsilon)(p^2-m^2+i\epsilon)(p_2^2-m^2+i\epsilon)}
\ee
where the factors in the denominators are due to the propagators of the internal lines, and $n$, $p$, etc. are 4-momenta.  The second diagram would have the Lorentz invariant form
\be
\begin{split}
F_b\sim \int d^4n\;d^4p_2& \frac{\Gamma(n)\Gamma(p_2){\cal M}(p,p_3,k){\cal M}(n,k,p_4)}{(n^2-m^2+i\epsilon)(p_2^2-m^2+i\epsilon)}\\
&\times\frac{1}{(p^2-m^2+i\epsilon)(n_2^2-m^2+i\epsilon)(k^2-m_h^2+i\epsilon)}
\end{split}
\ee

\begin{figure}[tbp]
     \begin{center}
        \subfigure[single-scattering term]{%
            \label{fig:lab}
            \includegraphics[width=0.4\textwidth]{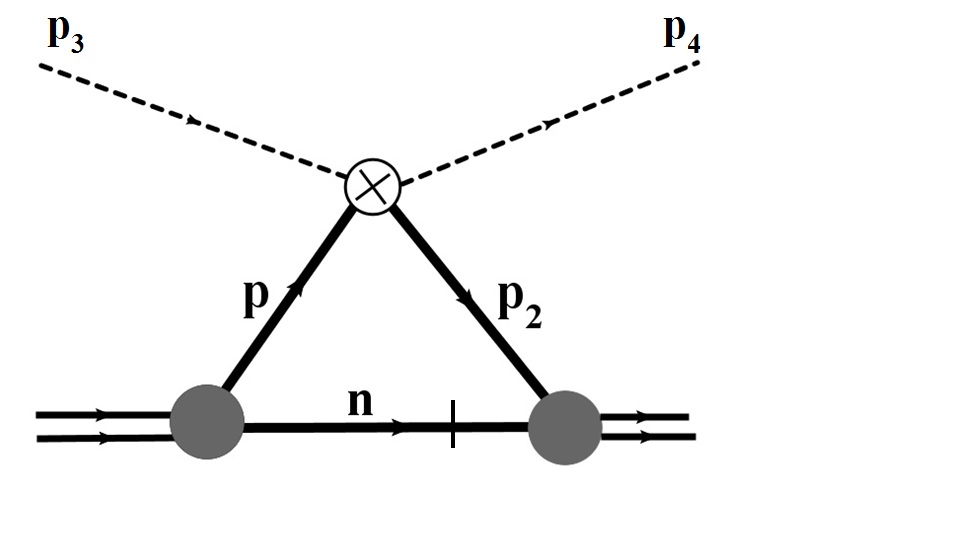}
        }%
        \subfigure[double-scattering term]{%
           \label{fig:cm}
           \includegraphics[width=0.4\textwidth]{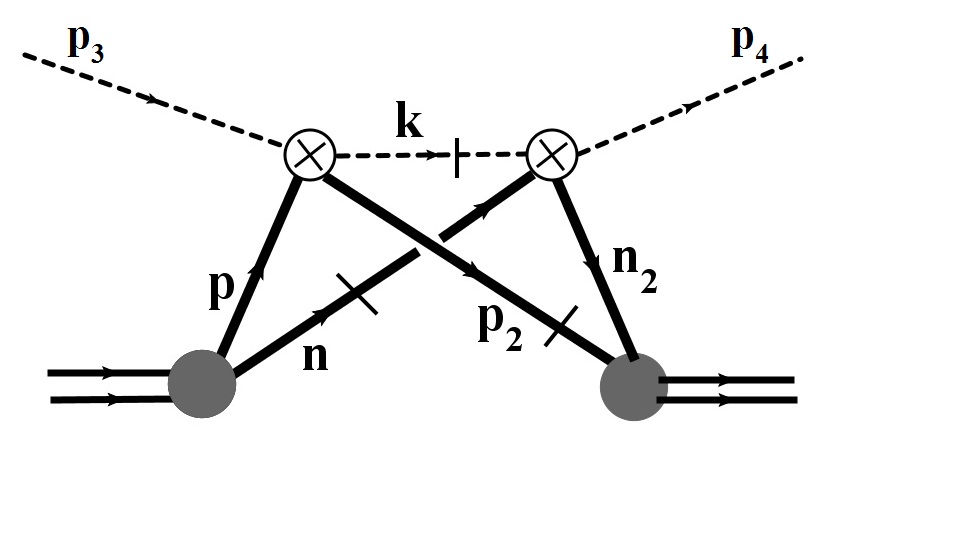}
        }\\ 
%
%
    \end{center}
    \caption{%
        Single- and double-scattering diagrams for hadron-deuteron elastic scattering.  The internal lines with bars on them are on-mass-shell in the Glauber approximation.
     }%
   \label{fig:deutsingledouble}
\end{figure}

The two amplitudes $F_a$ and $F_b$ can be shown to be equal to the Glauber amplitudes $F_1$ and $F_2$, under the kinematic conditions where the Glauber approximation is valid:  high-energy of the incident hadron, neglect of energy loss by the incident hadron, neglect of longitudinal momentum transfer at each of the $h$-nucleon scatterings, and the non-relativistic limit $\vert \mathbf{n}\vert,\;\vert \mathbf{p}_2\vert\ll m$.  In the non-relativistic limit we have (leaving out various constants): 
\be
 \frac{\Gamma(p)}{p^2-m^2+i\epsilon}\sim \Psi(p) 
\ee
and
\be
\frac{\Gamma(p_2)}{p_2^2-m^2+i\epsilon}\sim \Psi(p_2)
\ee
The integrals over 4-momenta are reduced to integrals over 3-momenta by integrating over $n^0$ and $p_2^0$ using the residue theorem and keeping only the contributions due to the positive-energy poles at $n^0=\sqrt{m^2+\mathbf{n}^2}-i\epsilon$ and $p_2^0=\sqrt{m^2+\mathbf{p}_2^2}-i\epsilon$ coming from the propagator denominators $n^2-m^2+i\epsilon$ and $p_2^2-m^2+i\epsilon$, respectively.  Contributions from all other poles in $n^0$ and $p_2^0$ are neglected.  In $F_a$, if the dependence of ${\cal M}$ on the Fermi momentum $n$ is neglected, one obtains
\be
F_a\sim{\cal M}\int d^3n \Psi(\mathbf{n})\Psi(\mathbf{n}-\mathbf{q}).
\ee
which is equal to $F_1$ when the various constants are included.  In $F_b$, in addition one more momentum component is integrated over by using
\be
\frac{1}{k^2-m_h^2+i\epsilon}=-i\pi\delta(k^2-m_h^2)+{\cal P}\frac{1}{k^2-m_h^2}
\ee
( with ${\cal P}$ indicating the principal value) and only keeping the contribution from the delta function.  This leaves $F_b$ in the schematic form
\be
F_b\sim\int d^3n\;d^2q' \Psi^*\Psi{\cal M}{\cal M}
\ee
and it is shown in~\cite{harrington64,abers66,bertocchi67,bassel68} that in the Glauber limit, $F_b=F_2$.

The Glauber approximation is thus equivalent to the sum of the covariant Feynman diagrams, but keeping only the positive-energy poles from some of the propagators; contributions from all other poles are neglected.  Note that the integrations that were done by taking poles from the indicated propagators enforced that those particles are on-mass-shell in the final result.  In  Fig. \ref{fig:deutsingledouble}, those particles are indicated with a bar on the line.  Since at each deuteron vertex one of the nucleon lines coming out of it is on-mass-shell, that means that the other line coming out of the deuteron vertex must necessarily be off-mass-shell.  However, under the conditions of validity of the Glauber approximation, they are not very far off-shell, and the amplitudes ${\cal M}$ can be approximated by their on-shell expressions.  In addition, by keeping only the indicated poles, the covariant diagrams become time-ordered diagrams, with a definite time-ordering to the sequence of hadron-nucleon scattering events.

In the work of ~\cite{harrington64,abers66,bertocchi67,bassel68} the covariant expressions are further used to calculate the scattering amplitude under general conditions, when the Glauber conditions on momentum- and energy-transfer are not satisfied.  This generalizes the Glauber approximation to arbitrary momentum and energy transfers.  In particular it is valid for the case where the projectile or produced particle is moving slowly relative to the nucleons, which is the case we are interested in for determination of the $J/\psi$-nucleon scattering length.  One benefit of the diagrammatic approach with invariant amplitudes is that it allows one to account for the Fermi motion of the nucleons in the deuteron, by including the exact dependence of the scattering amplitudes ${\cal M}$ on the 4-momenta of the particles; this dependence on Fermi momentum is completely neglected in the Glauber approximation (it assumes that the nucleons are at rest, and uses scattering amplitudes $f(\mathbf{q})$ assuming at-rest nucleons).  This becomes important when the scattering or particle production is near threshold, where the 2-body scattering amplitude ${\cal M}$ has strong dependence on the momentum of the struck nucleon.

In ~\cite{bertocchi72} and ~\cite{gribov70}, the diagrammatic method is shown to reduce to the Glauber result for nuclei with arbitrary nucleon number $A$, under the Glauber conditions.  In addition, it is shown that the dominant diagrams that contribute to the total amplitude, under the Glauber conditions, are only those diagrams for which the nucleons do not interact with each other while the projectile is traversing the nucleus; diagrams where, e.g. 2 nucleons scatter with each other in between 2 projectile-nucleon scatterings, are suppressed in the Glauber limit, compared to the other diagrams.  This limits the number of projectile-nucleon scatterings in any diagram to at most $A$, and so there are a finite set of diagrams to evaluate.
\begin{figure}[tbp]
     \begin{center}
        \subfigure[]{%
            \label{fig:inglaub}
            \includegraphics[width=0.4\textwidth]{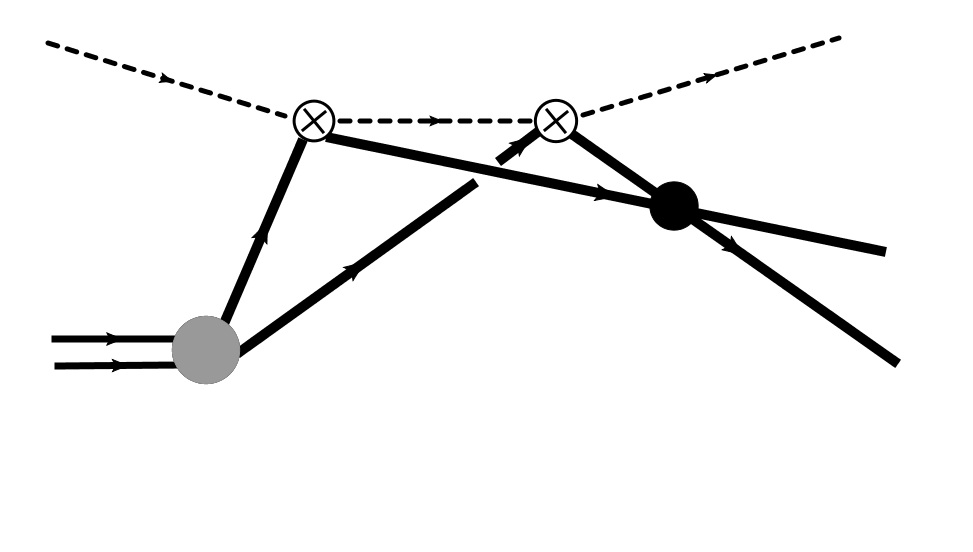}
        }%
        \subfigure[]{%
           \label{fig:notinglaub}
           \includegraphics[width=0.4\textwidth]{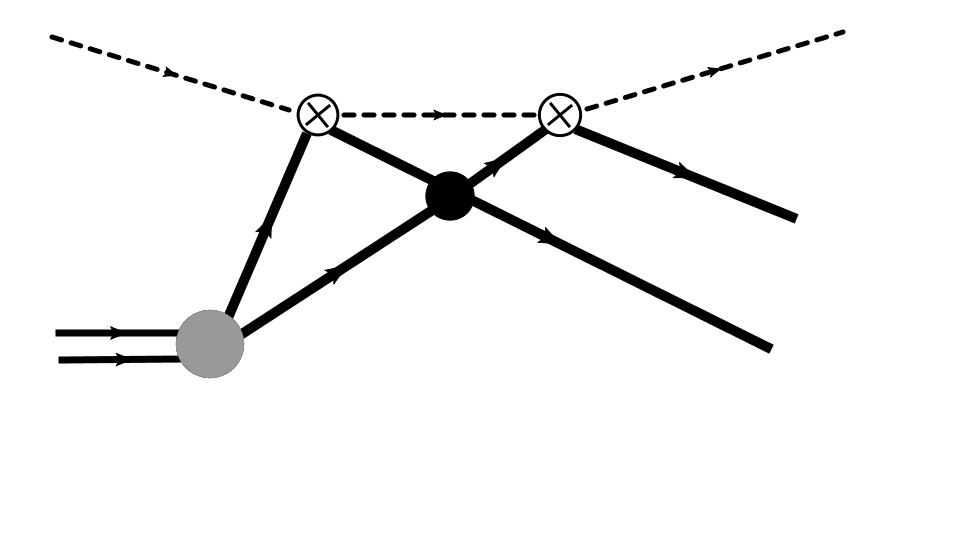}
        }\\ 

%
    \end{center}
    \caption{%
        a)  Diagram included in Glauber approximation;  b)  Diagram not included in Glauber approximation.  The solid black circle represents the proton-neutron scattering amplitude; the solid gray circle is the deuteron vertex function. 
     }%
   \label{fig:deuttriple}
\end{figure}

For example, for the case of incoherent scattering of a hadron by the deuteron, wherein the deuteron breaks up into an unbound proton and neutron in the final state,  Fig. \ref{fig:inglaub} shows a diagram which contributes in the Glauber limit (since the proton and neutron interact after the last projectile-nucleon interaction), while Fig. \ref{fig:notinglaub} shows a diagram which would be negligible in the Glauber limit.  If the kinematic conditions do not satisfy the Glauber conditions, then in general both of these diagrams must be evaluated.

An improvement on the Glauber approximation, called the Generalized Eikonal Approximation (GEA), was derived by Frankfurt, Strikman, Sargsian, et al ~\cite{SS97,SS97b,sarg2001,SS2005,SS2006,sarg2010} by evaluating Feynman diagrams in the high-energy limit but retaining the dependence on the longitudinal momentum transfers (which is neglected in the Glauber approximation) and the dependence of the 2-body amplitudes on the internal momenta.  The advantage of the GEA over the Glauber approximation is that it accounts for the exact momentum transfers, and therefore extends the range of validity beyond the Glauber limit.  In addition, by virtue of its diagrammatic derivation the scattering amplitudes that appear at vertices have the correct dependence on the Fermi momentum of the nucleons.  Finally, because it still assumes high-energy scattering, the number of Feynman diagrams that have to be evaluated is limited to those for which the nucleons do not interact with each other while the projectile is traversing the nucleus; any other diagrams are suppressed, in the high-energy limit.  Again, this yields a finite set of diagrams to evaluate.

The GEA would not be valid, however, for the determination of the $J/\psi$-nucleon scattering length, since the $J/\psi$-nucleon rescattering takes place at low energy, not high energy.  Therefore in principle there would be a large number of diagrams to evaluate.  Here we will content ourselves with only evaluating the first few diagrams.  Previous calculations using the diagrammatic method, and keeping a finite number of diagrams, have been performed for $\gamma D\to NN\pi$~\cite{laget81}, $\pi^+ D\to \pi^+ pn$~\cite{gyles86}, $\gamma D\to\pi^0 D$~\cite{garcilazo95},  $\pi^- D\to \eta nn$~\cite{garcilazo99}, $np\to\eta d$~\cite{garcilazo2005}, $e+D\to e p n$, $e+^3He\to e p D$, $e+^3He\to e p (pn)$~\cite{kaptari05}, $e+^3He\to e p p n$~\cite{SS2005}, $\gamma D\to \phi D$~\cite{SS2006}, and $e+D\to e p n$~\cite{sarg2010}; the last four references calculate within the GEA.

For the calculations in this chapter, it will be assumed that the diagrammatic expansion gives the scattering amplitude accurately, upon retaining only the first few diagrams (up through one-loop order, which represent the processes where the $J/\psi$ is produced on one nucleon and then the other (spectator) nucleon scatters afterward, either from the first nucleon or from the $J/\psi$).

\section{Electroproduction from a Nucleus}
\label{sec:electroproduction}

We consider here electron scattering from the deuteron with production of a vector meson, with the final state of the proton-neutron system being a continuum state.  The formalism for the cross-section for electroproduction from a nucleus can be found in~\cite{Raskin:1988kc}.  Here we summarize the relevant facts.   We consider here the completely unpolarized electron (initial and final) cross-section.  Then the cross-section can be written in terms of the amplitudes for $\gamma^*+d\to p+n+V$, i.e. vector meson production from virtual photons.  In the LAB frame, with $\epsilon^\prime$ the final electron energy , $\Omega^\prime$ the final electron solid angle, $\Omega_V$ the vector meson solid angle , and $\mathbf{p}_{pn}^*$ the final proton-neutron relative momentum in the $p-n$ center-of-mass frame,    the 8-fold differential cross-section has the form
\be
\frac{d^8\sigma}{d\epsilon^\prime d\Omega^\prime d\Omega_V d^3p_{pn}^*}=(kinematic\;factors)\times (v_T R^T_{fi}+v_{TT} R^{TT}_{fi}+v_L R^L_{fi}+v_{TL} R^{TL}_{fi}),
\ee
where the factors
\be
 R^T_{fi}=\vert \langle f\vert J_{+1}(\mathbf{q})\vert i \rangle\vert^2+\vert \langle f\vert J_{-1}(\mathbf{q})\vert i \rangle\vert^2
\ee 
\be
 R^{TT}_{fi}=2 \operatorname{Re}\langle f\vert J^*_{+1}(\mathbf{q})\vert i \rangle \langle f\vert J_{-1}(\mathbf{q})\vert i \rangle
\ee 
\be
 R^{TL}_{fi}=-2 \operatorname{Re}\langle f\vert \rho^*(\mathbf{q})\vert i \rangle (\langle f\vert J_{+1}(\mathbf{q})\vert i \rangle - \langle f\vert J_{-1}(\mathbf{q})\vert i \rangle)
\ee 
\be
 R^{L}_{fi}=\vert\langle f\vert \rho(\mathbf{q})\vert i \rangle\vert^2, 
\ee 
are in terms of the matrix elements of the spherical vector components of the electromagnetic current operator $J$ between the initial deuteron state $\vert i\rangle$ and final hadron ($p+n+J/\psi$) state $\vert f\rangle$.  $v_T$, $v_{TT}$, etc., are kinematic factors that only depend on the electron momenta.  $ R^T_{fi}$ is the sum of the squares of the amplitudes for $\gamma^*+d\to p+n+V$ for transversely polarized virtual photons, whereas $R^{TT}_{fi}$ is an interference term between these two amplitudes.  $R^{L}_{fi}$ is the square of the amplitude for a longitudinally polarized virtual photon, while $ R^{TL}_{fi}$ is an interference term between the amplitudes for production from transverse and longitudinally polarized photons. The matrix element $R^{TL}_{fi}$ is proportional to $\cos{\phi}$, while $R^{TT}_{fi}$ is proportional to $\cos{2\phi}$, where $\phi$ is the angle between the plane including the initial and final electron momenta, and the plane including the 3-momentum transfer $\mathbf{q}$ and the $J/\psi$ momentum $\mathbf{p}_V$.  Thus if we integrate the cross-section over $\phi$, the terms $R^{TT}$ and $R^{TL}$ drop out.  Or, if we assume helicity conservation (i.e. the helicity of the outgoing $J/\psi$ is equal to the helicity of the photon) then $R^{TT}=R^{TL}=0$.  Moreover, several theoretical models~\cite{Dosch:2003dh,Fiore:2009xk} indicate that for small $Q^2$, the amplitude for $J/\psi$ electroproduction from transverse virtual photons is much larger than the amplitude for production from longitudinally polarized virtual photons; for $Q^2=0$ (photoproduction) the production amplitude for longitudinal photon polarization is of course exactly zero.  Therefore in what follows we will neglect  $R^{L}_{fi}$, and so the differential cross-section is simply given by $ R^T_{fi}$ multiplied by  kinematic factors.  Thus our task is to calculate
\be
 R^T_{fi}=\vert \langle f\vert J_{+1}(\mathbf{q})\vert i \rangle\vert^2+\vert \langle f\vert J_{-1}(\mathbf{q})\vert i \rangle\vert^2\equiv \vert F_+\vert^2 +\vert F_-\vert^2
\ee
where $F_\pm$ are the amplitudes for $J/\psi$ production from positive and negative helicity virtual photons.  In the following we will calculate the amplitude for $\gamma^*+d\to p+n+V$ by evaluating Feynman diagrams corresponding to the various processes contributing to it.

\section{Kinematics for small relative energy of the $J/\psi$-neutron pair}
\label{sec:kinematics}

Since the scattering length is the zero-energy limit of the scattering amplitude, in order to measure it the relative momentum of the $J/\psi-n$ system must be small.  An estimate of how small can be obtained by requiring only S-wave scattering, meaning the contribution of higher partial waves should be negligible.  The classical relation between impact parameter and angular momentum yields an estimate for the maximum $l$ that contributes.  If the relative momentum is $p^*$ and the impact parameter is $b$, then the orbital angular momentum is $L=p^*b=b\sqrt{2\mu T^*}$ where $T^*$ is the total kinetic energy of the $J/\psi$-n pair in their c.m. frame, and $\mu$ is the reduced mass.  The largest angular momentum wave which is scattered is obtained by setting $b$ equal to the range of the potential.  With $L^2=l(l+1)$ (we take $\hbar=1$), the condition for only S-wave scattering is that $l\ll1$, which implies $L^2=b^2 2\mu T^*\ll 1$.   Taking the range of the interaction to be $\simeq 1\;fm$ yields $T^*\ll\;30 MeV$.

Experimentally, perhaps the simplest quantity to measure is the total production cross-section, integrated over all available phase space, for a given incident photon energy.  However, if we restrict the photon energy such that the maximum $J/\psi$-neutron c.m. kinetic energy $T^*_{max}$ is small in of all the available phase space, then that means that the maximum proton-neutron relative energy will also be small everywhere in the available phase space: for a given value of the Mandlestam variable $s$ for a system consisting of 3 particles, the total kinetic energy of any two of the particles (say 1 and 2) in their c.m. frame satisfies
\be
T^*_{12}\le \sqrt{s} - m_1-m_2-m_3.
\ee 
The low-energy $J/\psi-n$ scattering amplitude is expected to be much smaller than the low-energy $p-n$ scattering amplitude, and therefore  the $p-n$ rescattering would dominate over the $J/\psi-n$ rescattering, as contributions to the total production cross-section.  Thus we need to restrict our considerations to a kinematic range where the $p-n$ rescattering is at relatively high energy, while the $J/\psi-n$ rescattering is at very low energy, in order to have the possibility that the $J/\psi-n$ rescattering makes a noticeable contribution to the differential cross-section.

The ideal situation would be to have the final $J/\psi$ and neutron sitting at rest in the LAB, with the proton moving off at high velocity.  Such a final state is kinematically allowed for other reactions, e.g. $\pi^+ d\to\eta pp$, $\gamma^* d\to\eta p n$, but it is not possible for the reaction $\gamma^* d\to J/\psi\; p n$, for any real or virtual photon 4-momentum.

\subsection{The $T^*=0$ case}

The kinematics for the case of zero relative momentum of the $J/\psi-n$ pair is the simplest to analyze, since for this case the two particles are each moving with the same velocity (in any reference frame), and so kinematically they are identical to a single particle of mass $M=m+m_{V}$.  If their common velocity in a given frame is $\beta$, then the energy in that frame of the $J/\psi$ (neutron) is $\beta m_V$ ($\beta m$), and the momentum of the $J/\psi$ (neutron) in that frame is $\gamma\beta m_V$ ($\gamma\beta m$).  Thus the energies and momenta of the two particles are in the ratio $E_V/E_n = p_V/p_n=m_V/m$.  Defining $E_{Vn}\equiv E_V+E_n$ and $\mathbf{p}_{Vn}\equiv \mathbf{p}_V+\mathbf{p}_n$ (I use the subscript capital $V$ to stand for the $J/\psi$ throughout this chapter) we have:
\begin{equation}
\begin{array}{rcl}

E_V & = & \frac{m_V}{M}E_{Vn} \\
E_n&=&\frac{m}{M}E_{Vn}\\
\mathbf{p}_V&=&\frac{m_V}{M}\mathbf{p}_{Vn}\\
\mathbf{p}_n&=&\frac{m}{M}\mathbf{p}_{Vn}\\

 \end{array}  \end{equation}

and also 
\begin{equation}
E_{Vn}^2-\mathbf{p}_{Vn}^2=M^2\;,
\end{equation}
and conservation of energy and momentum is

\begin{equation}\begin{array}{rcl}
\nu+E_D&=&E_p+E_{Vn}\\

\mathbf{q}+\mathbf{p}_D&=&\mathbf{p}_p +\mathbf{p}_{Vn}
\end{array} \end{equation}
where the photon's 4-momentum is $q=(\nu,\mathbf{q})$ and the deuteron's is $p_D=(E_D,\mathbf{p}_D)$.

\begin{figure}[bth]
     \begin{center}
        \subfigure[ LAB frame]{%
            \label{fig:lab}
            \includegraphics[width=0.4\textwidth]{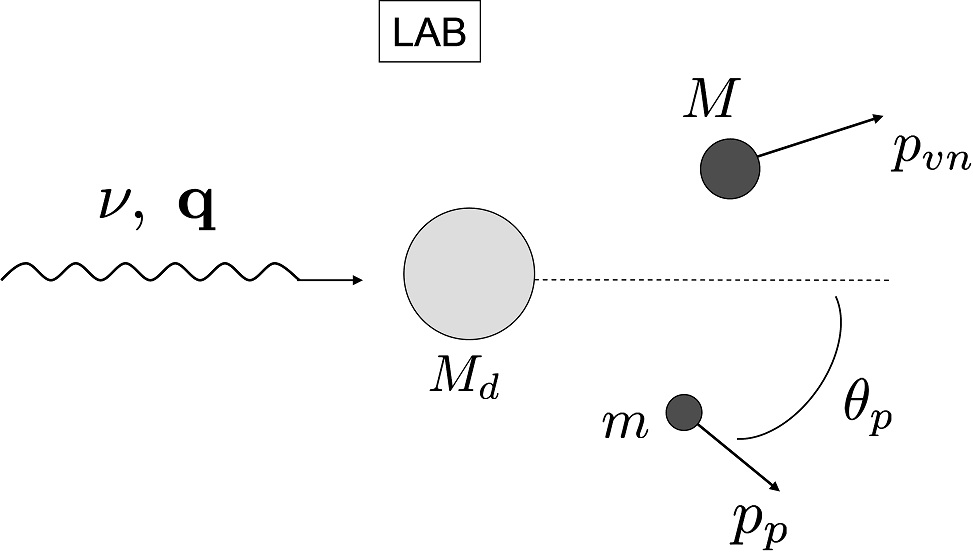}
        }%
        \subfigure[ overall center-of-mass frame]{%
           \label{fig:cm}
           \includegraphics[width=0.4\textwidth]{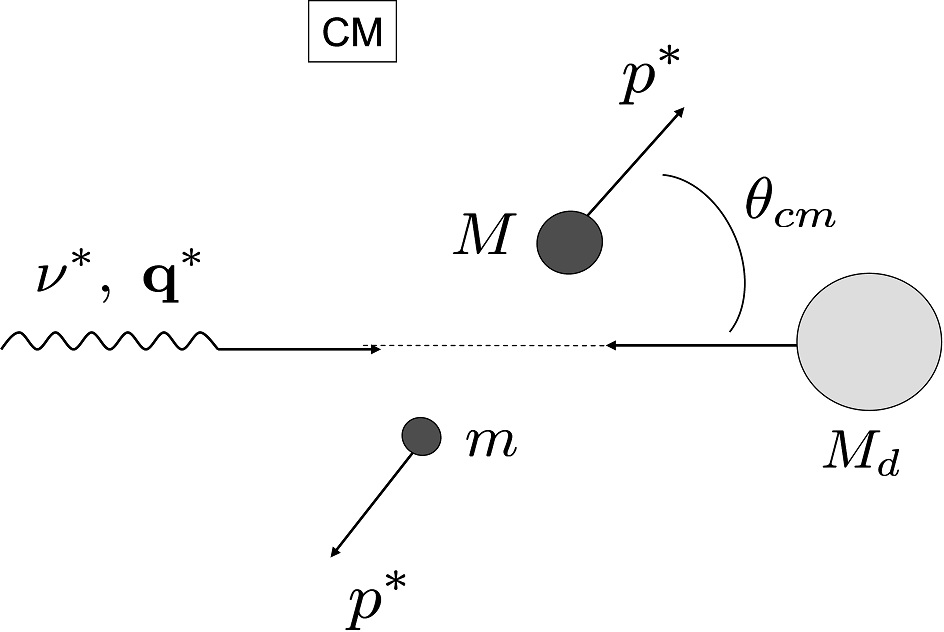}
        }\\ 

%
    \end{center}
    \caption{%
        The reaction for $T^*=0$ in the LAB and CM frames
     }%
   \label{fig:labandcm}
\end{figure}

Thus, just as in $2-2$ scattering, there is only one independent variable in the final state, which can be taken as $\theta_{cm}$ (see Figure \ref{fig:labandcm}).  Then we can plot the momenta of the proton, neutron, and $J/\psi$ in the LAB frame as a function of $\theta_{cm}$.  Figure \ref{fig:kin} shows these plots for two different photon energies, one for $\nu=9$ GeV (which is above the threshold energy for $J/\psi$ production on a single nucleon) and one for $\nu=6.5$ GeV (which is below threshold for production on a single nucleon). 
\begin{figure}[h]
     \begin{center}
        \subfigure[ proton (solid) and neutron (dashed) momenta for $\nu=9$ GeV]{%
            \label{fig:mom9}
            \includegraphics[width=0.4\textwidth]{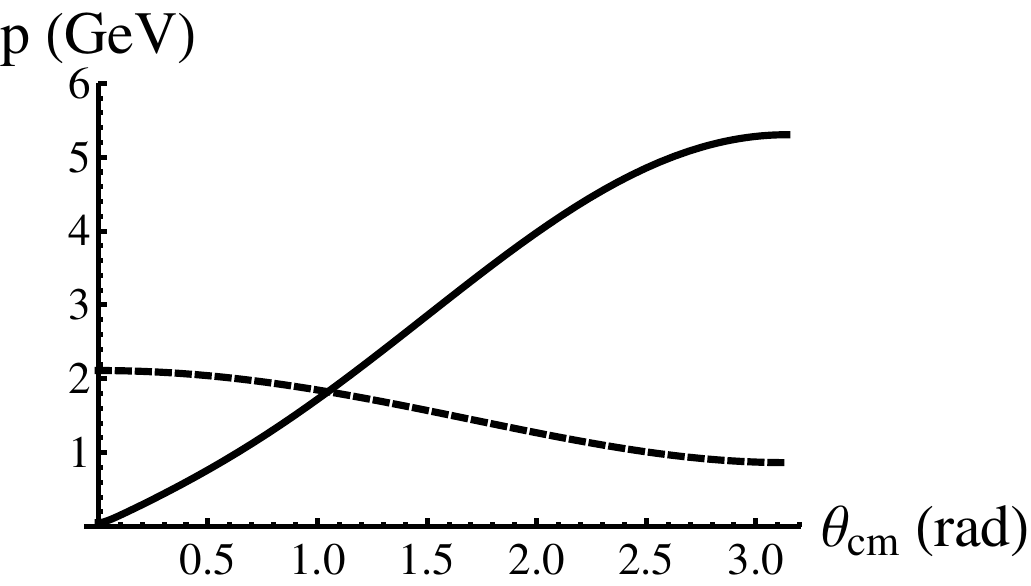}
        }%
        \hspace{0.5in}
         \subfigure[ proton (solid) and neutron (dashed) momenta for $\nu=6.5$ GeV]{%
            \label{fig:momentum65}
            \includegraphics[width=0.4\textwidth]{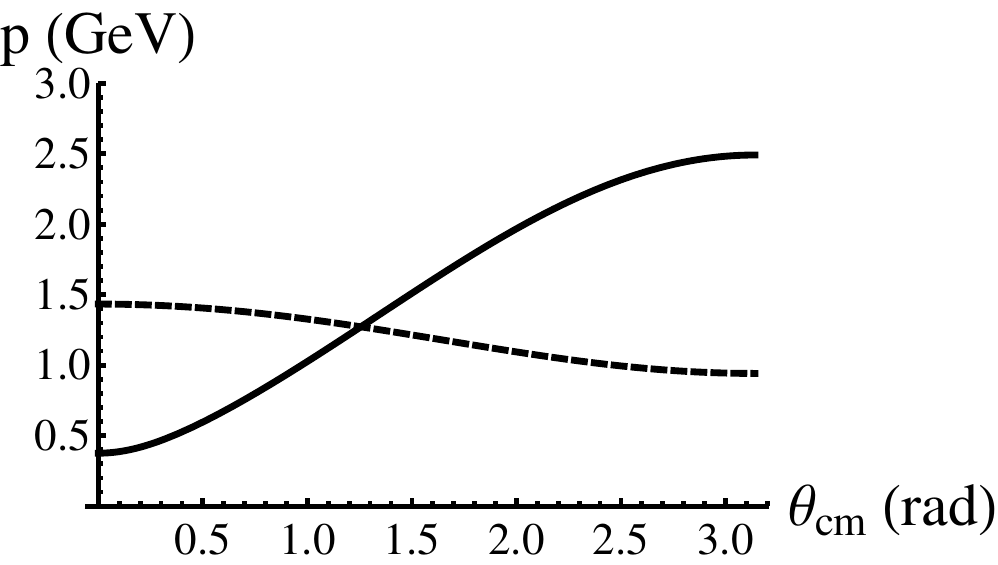}
        }\\ 


    \end{center}
    \caption{%
        Kinematics for $T^*=0$
     }%
   \label{fig:kin}
\end{figure}

%

\subsection{The $T^*\neq0$ case}

Fig. \ref{fig:nandpmomentum} shows the minimum possible outgoing neutron LAB momentum and the corresponding proton LAB momentum vs. $\theta_{cm}$ for fixed $T^*_{Vn}=30$ MeV (see Fig. \ref{fig:cmkin} for the definition of $\theta_{cm}$ for the case $T^*_{Vn}\neq 0$); graphs are shown for photon LAB energy $\nu=9$ GeV and for $\nu=6.5$ GeV.   One can see that the neutron's momentum is always greater than at least $0.6$ GeV.  Since the maximum nucleon momentum in the deuteron is around $0.3$ GeV (the deuteron momentum-space wavefunction is negligible for momenta larger than that) that means that in order for these final-state kinematics to occur,  the neutron must have acquired its large momentum through a scattering event.  In fact it will turn out that the dominant process corresponds to the impulse approximation wherein the $J/\psi$ is produced on the neutron itself, and the proton simply recoils freely.  For the kinematics of interest here, rescattering processes (e.g. $J/\psi$-neutron rescattering, $J/\psi$-proton rescattering, proton-neutron rescattering)  make  very small contributions to the total amplitude. 

We use the following notation throughout this chapter:  $q=(\nu,\mathbf{q})$ is the virtual photon 4-momentum in the LAB, with $q^2=-Q^2<0$; $\mathbf{p}_p$ is the outgoing proton LAB 3-momentum; $\mathbf{p}_n$ is the outgoing neutron LAB 3-momentum; $\mathbf{p}_V$ is the $J/\psi$ LAB 3-momentum; and the same variables with $cm$ superscripts denote their values in the overall (3-body) center-of-mass frame.  $\theta_p$, $\theta_n$, and $\theta_V$ denote the angle that the outgoing proton, neutron, and $J/\psi$ momenta, respectively, make with $\mathbf{q}$, in the LAB frame.

\begin{figure}[bth]
     \begin{center}
        \subfigure[  $\nu=9$ GeV]{%
            \label{fig:nandpmoma}
            \includegraphics[width=0.4\textwidth]{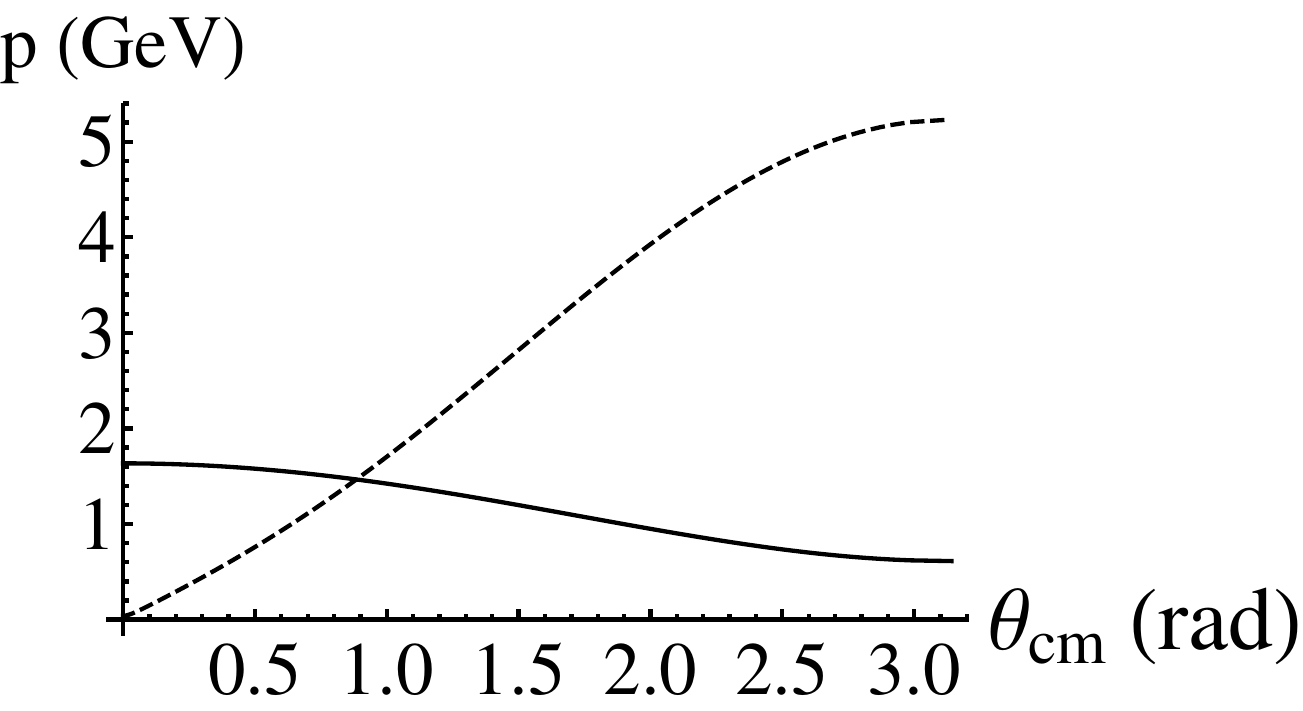}
        }%
        \subfigure[ $\nu=6.5$ GeV]{%
           \label{fig:nandpmomb}
           \includegraphics[width=0.4\textwidth]{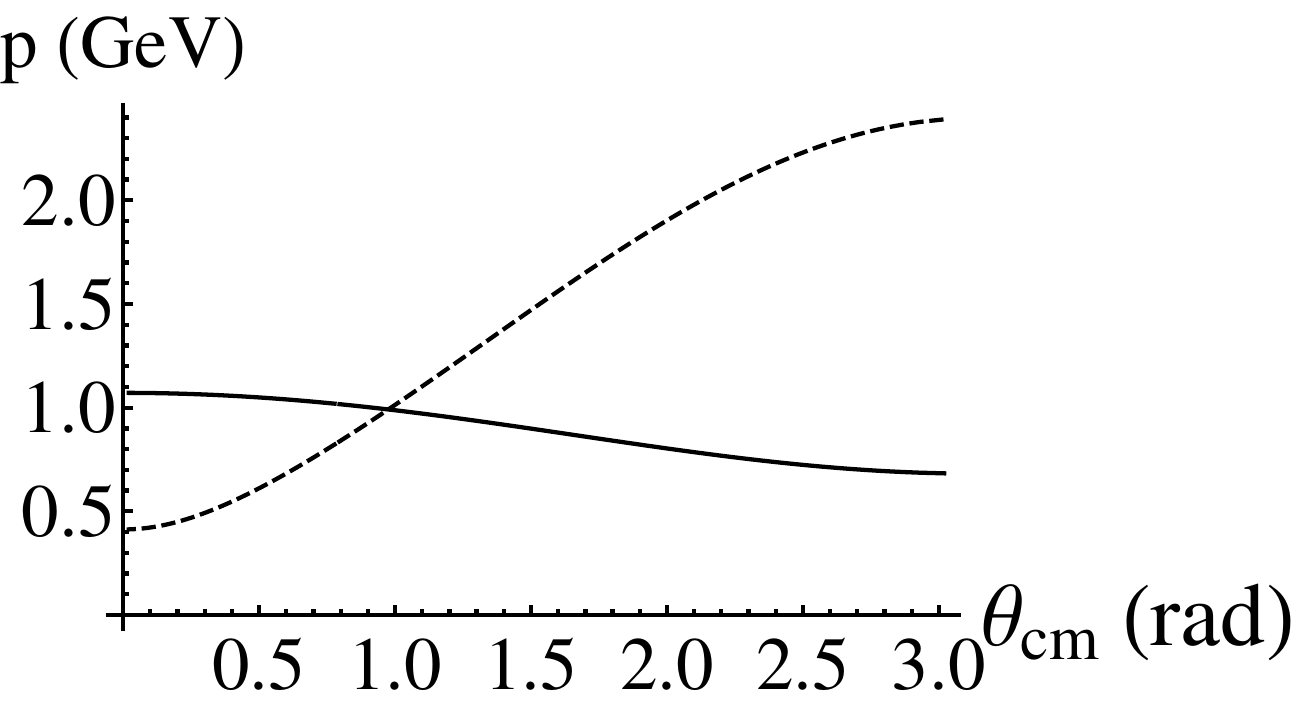}
        }\\ 

%
    \end{center}
    \caption{%
       Minimum possible neutron LAB momentum (solid curve), and the corresponding proton LAB momentum (dashed curve), vs. $\theta_{cm}$, for $T^*_{Vn}=30$ MeV and two values of photon LAB energy $\nu$.
     }%
   \label{fig:nandpmomentum}
\end{figure}

\begin{figure}[tbp]
     \begin{center}

            \includegraphics[width=4.5in,height=2.5in]{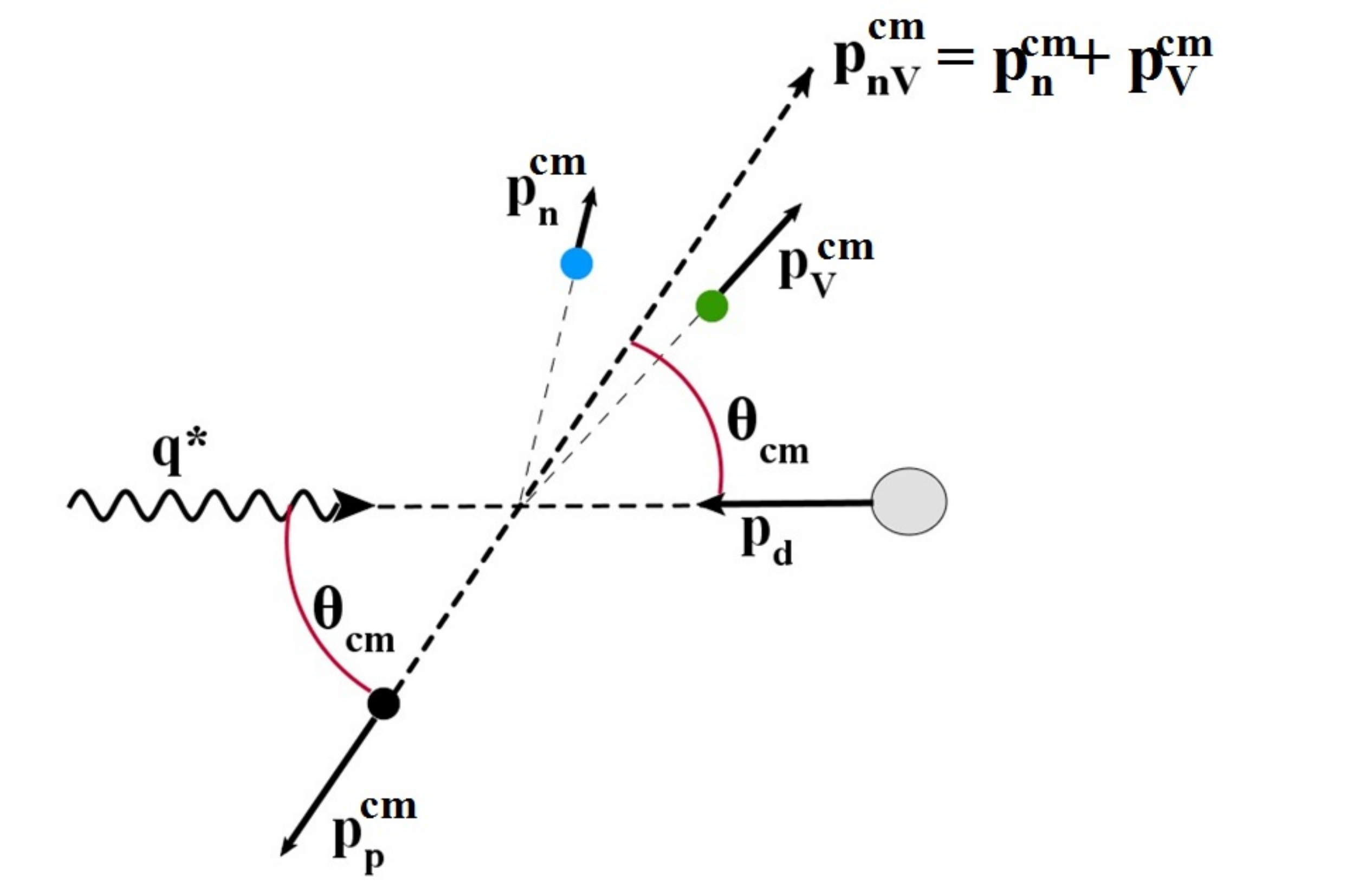}

    \end{center}
    \caption{%
        Momenta and angles in the overall c.m. frame, for coplanar kinematics.  
     }%
   \label{fig:cmkin}
\end{figure}

\section{Invariant Scattering Amplitudes}
\label{sec:invaramps}

\begin{figure}[tbp]
     \begin{center}
        \subfigure[$F_{1a}$:  Impulse diagram]{%
            \label{fig:F1a}
            \includegraphics[width=0.4\textwidth]{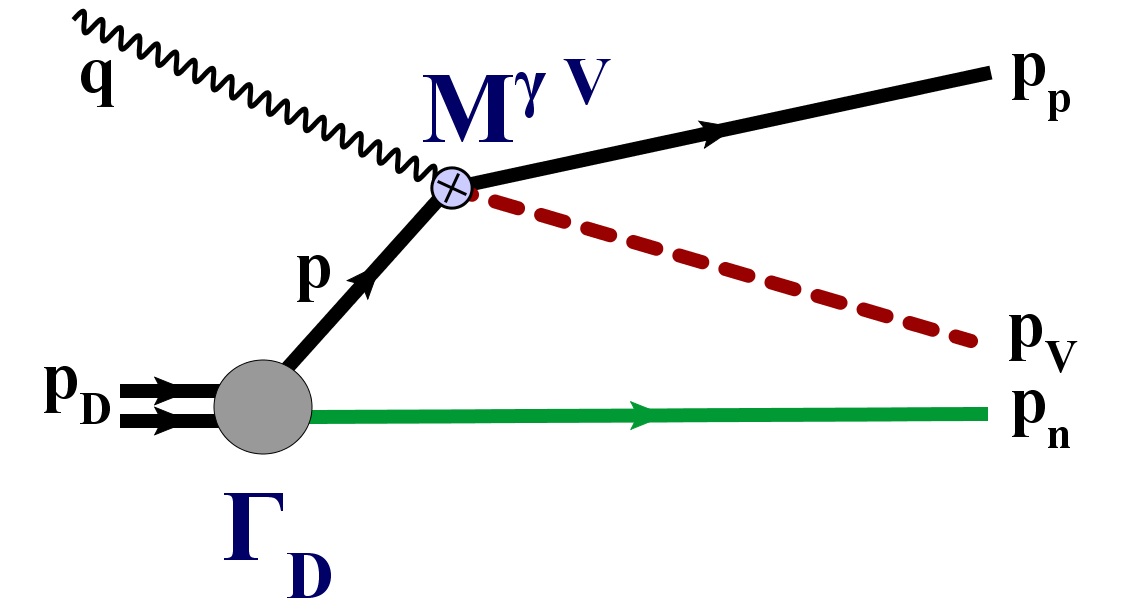}
        }%
        \hspace{0.5in}
         \subfigure[$F_{2a}$:  p-n rescattering diagram]{%
           \label{fig:F2a}
           \includegraphics[width=0.4\textwidth]{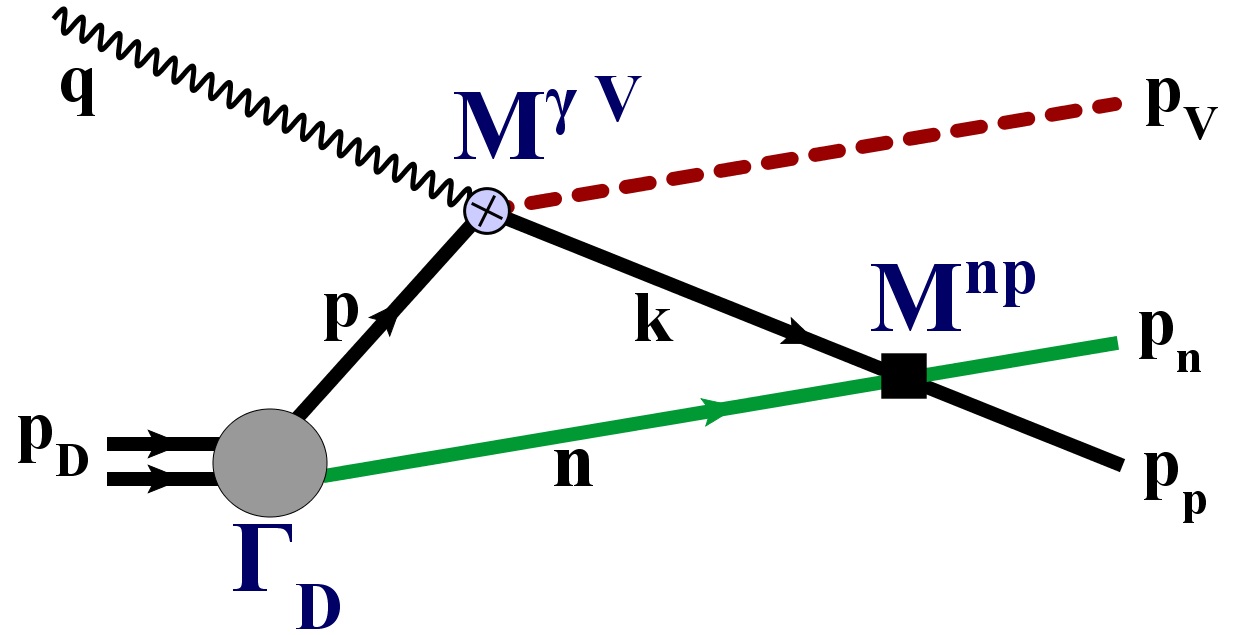}
        }\\ 

        \subfigure[ $F_{3a}$:  $J/\psi$-n rescattering diagram]{%
            \label{fig:F3a}
            \includegraphics[width=0.4\textwidth]{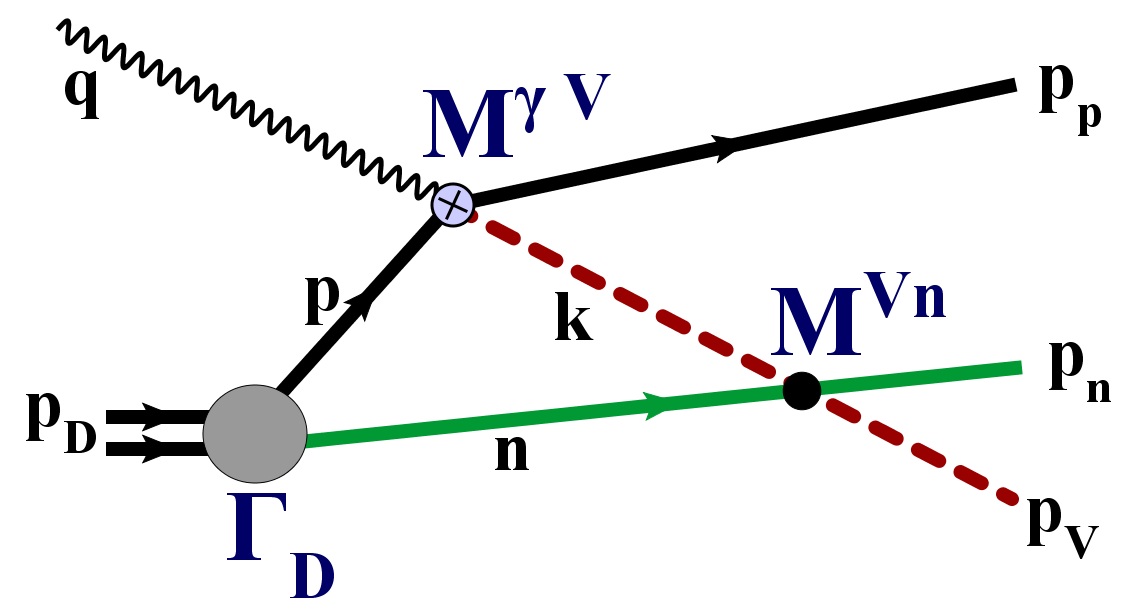}
        }%

    \end{center}
    \caption{%
       Feynman diagrams for $\gamma^*+d\to J/\psi + p+n$, for production on the proton.
     }%
   \label{fig:diagrams}
\end{figure}

\begin{figure}[tbp]
     \begin{center}
        \subfigure[$F_{1b}$:  Impulse diagram]{%
            \label{fig:F1b}
            \includegraphics[width=0.4\textwidth]{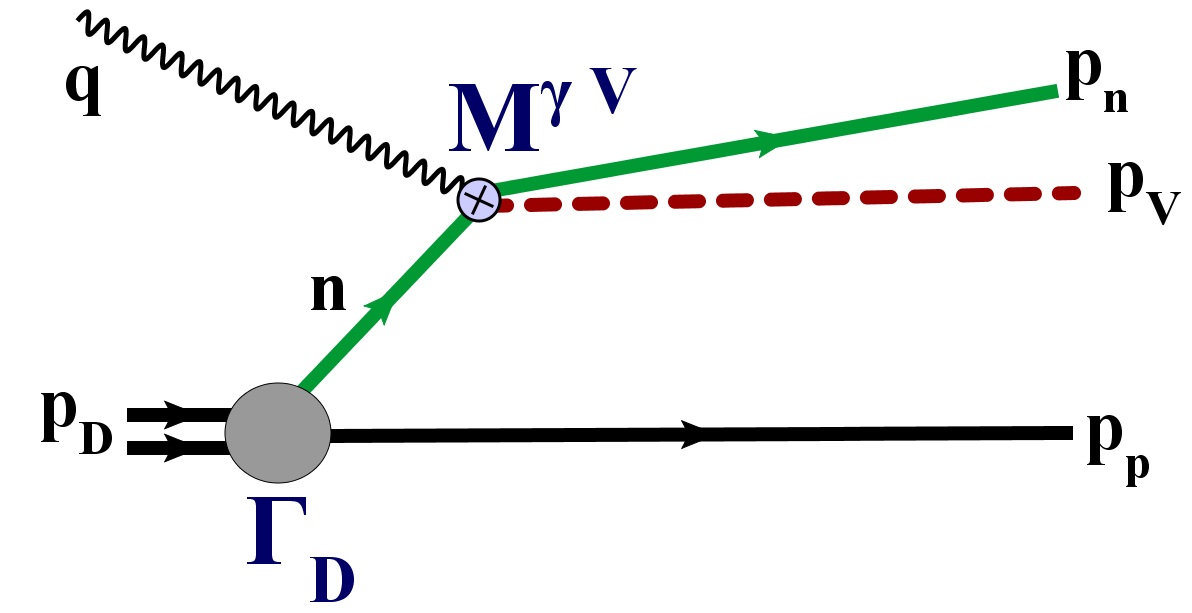}
        }%
        \hspace{0.5in}
         \subfigure[$F_{2b}$:  p-n rescattering diagram]{%
           \label{fig:F2b}
           \includegraphics[width=0.4\textwidth]{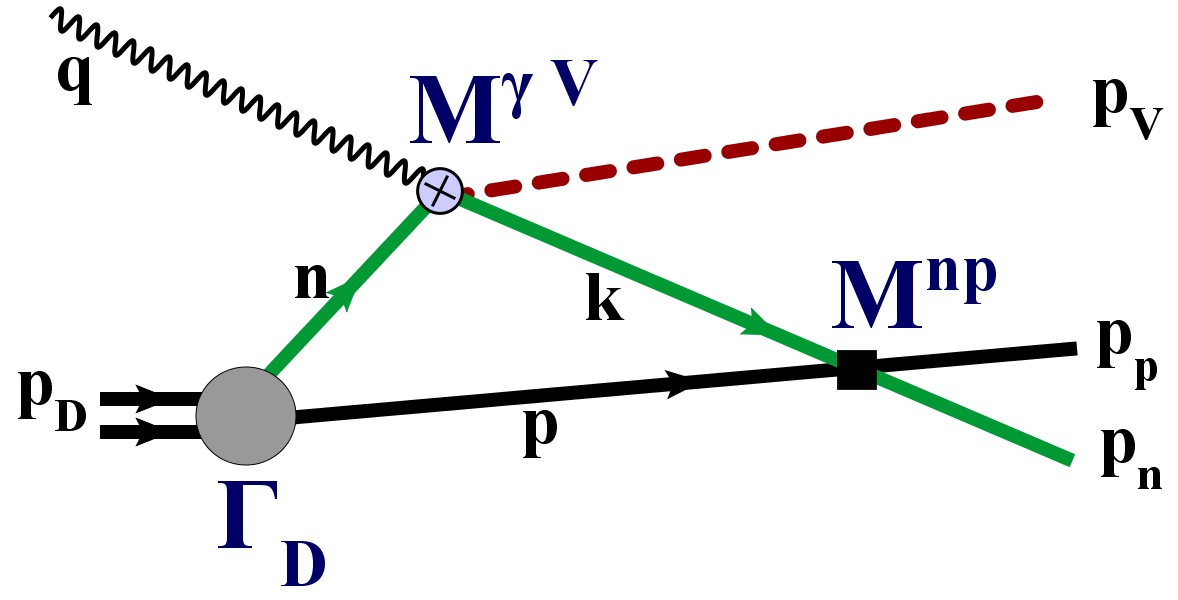}
        }\\ 

        \subfigure[ $F_{3b}$:  $J/\psi$-p rescattering diagram]{%
            \label{fig:F3b}
            \includegraphics[width=0.4\textwidth]{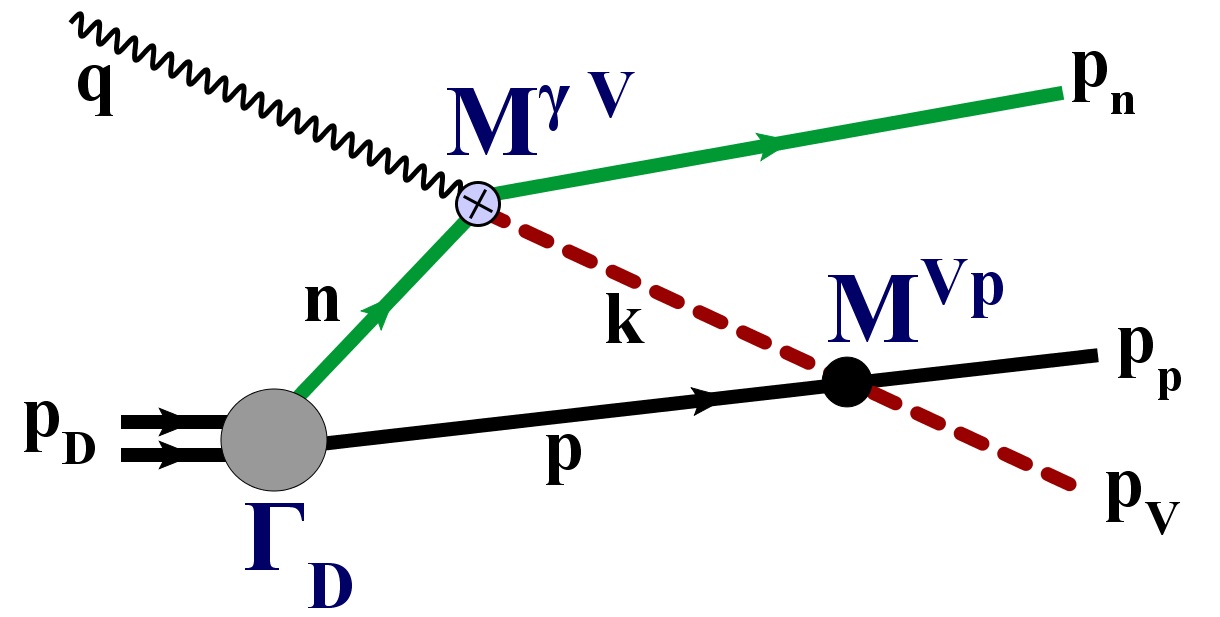}
        }%

    \end{center}
    \caption{%
       Feynman diagrams for $\gamma^*+d\to J/\psi + p+n$, for production on the neutron.
     }%
   \label{fig:diagrams2}
\end{figure}

The diagrams considered here are shown in Figs. \ref{fig:diagrams} and \ref{fig:diagrams2}.  There are 3 diagrams for production on the proton, and 3 similar diagrams where the $J/\psi$ is produced on the neutron.  In all cases we are interested in kinematics where the $J/\psi$ and neutron have small relative momentum.  The diagrams are covariant, and hence give Lorentz invariant amplitudes.  In the diagrams,   ${\cal M}^{\gamma V}$ is the Lorentz invariant amplitude for the quasi-2-body process $\gamma^*+N\to V+N$ (where $N$ is a nucleon, and $V$ stands for the $J/\psi$), while ${\cal M}^{V n}$ is the  Lorentz invariant amplitude for the elastic scattering process $V+n\to V+n$ (with $n$ meaning neutron), and ${\cal M}^{V p}$ and ${\cal M}^{np}$ are the same for elastic $J/\psi$-proton scattering and neutron-proton scattering, respectively.

\subsection{Impulse Diagrams}

Amplitudes $F_{1a}$ and $F_{1b}$ are the impulse diagrams, where the $J/\psi$ is produced on one of the nucleons and the other nucleon (the ``spectator") recoils freely without interacting with the other particles.  In $F_{1a}$ the vector meson is produced on the proton and the neutron is the spectator, while in $F_{1b}$ the production occurs on the neutron and the proton is the spectator.  The invariant amplitudes in this case are
\begin{equation}
F_{1a}={\cal M}^{\gamma V}(s_{1a},t_{1a})\;\frac{\Gamma_D(p)}{D(p)}\end{equation} 
\begin{equation}
F_{1b}={\cal M}^{\gamma V}(s_{1b},t_{1b})\;\frac{\Gamma_D(n)}{D(n)}.\end{equation} 
Here ${\cal M}^{\gamma V}$ is the Lorentz invariant amplitude for the quasi-2-body process $\gamma^*+N\to V+N$ (where $N$ is a nucleon), $\Gamma_D$ is the covariant vertex function for the virtual dissociation $D\to p+n$, and $D(p)$ is the propagator denominator for the intermediate-state nucleon, $D(p)\equiv-p^2+m^2-i\epsilon$.  Evaluated in the LAB frame, and neglecting any contributions to the deuteron vertex from antinucleons, the deuteron vertex function is related to the nonrelativistic deuteron wavefunction by~\cite{gross80}
\begin{equation}
\psi_D(\mathbf{k}_{rel})=\frac{-\Gamma_D(p)}{\sqrt{2p^0 (2\pi)^3}\;D(p)}\end{equation}
where in the LAB frame, $\mathbf{k}_{rel}=\mathbf{p}=-\mathbf{p}_n$ for $F_{1a}$, and $\mathbf{k}_{rel}=\mathbf{p}_p=-\mathbf{n}$ for $F_{1b}$ ($\mathbf{k}_{rel}$ is the proton's momentum inside the deuteron, in the LAB frame, for both).

\subsubsection{Normalization of Amplitudes}
All 2-body amplitudes ${\cal M}^{\gamma V}$, ${\cal M}$ are related to the corresponding 2-body differential cross-section by 
\begin{equation}
\label{crossandM}
\frac{d\sigma}{dt}=\frac{1}{16\pi\lambda(s,m_1^2,m_2^2)}\vert{\cal M}\vert^2,
 \end{equation}
where the flux factor $\lambda$ is given in terms of the incident particle masses $m_1$ and $m_2$ by
\begin{equation}
\lambda(s,m_1^2,m_2^2)=(s-m_1^2-m_2^2)^2-4m_1^2m_2^2,\end{equation}
and $s$ and $t$ are the Mandelstam variables for the 2-body process. For $\gamma^*+N\to V+N$, we have $m_1^2=q^2=-Q^2$ and $m_2^2=p^2$ (for $F_{1a}$) or $m_2^2=n^2$ (for $F_{1b}$).  Note that the internal nucleon lines are not on-mass-shell.  For example,  $p^2=(p_D-p_n)^2=M_D^2+m^2-2M_D E_n=5m^2-4m E_n$, and so only for $\mathbf{p}_n=0$ do we have $p^2=m^2$.  

The Mandelstam invariants that the elementary amplitudes ${\cal M}^{\gamma V}$ depend on are given by:
\begin{equation}
\begin{array}{rcl}

s_{1a}&=&(q+p)^2=(p_p+p_V)^2\\
t_{1a}&=&(q-p_V)^2\\
s_{1b}&=&(q+n)^2=(p_n+p_V)^2\\
t_{1b}&=&(q-p_V)^2\\

 \end{array}  \end{equation}

Note that for the particular case of zero $J/\psi$-neutron relative momentum ($T^*=0$) we have $s_{1b}=(p_n+p_V)^2=(m+m_V)^2= M^2$ where $M\equiv m+m_V$, so that the $J/\psi$ production amplitude ${\cal M}^{\gamma V}$ in $F_{1b}$ is always at threshold.

In terms of the deuteron wavefunction, the amplitudes are thus:
\begin{equation}
F_{1a}=-{\cal M}^{\gamma V}(s_{1a},t_{1a})\;\psi_D(-\mathbf{p}_{n})\sqrt{2m (2\pi)^3}\end{equation}
\begin{equation}
F_{1b}=-{\cal M}^{\gamma V}(s_{1b},t_{1b})\;\psi_D(\mathbf{p}_{p})\sqrt{2m (2\pi)^3}.\end{equation}
The amplitudes ${\cal M}^{\gamma V}$ used in calculations are to be taken from experimental data on $J/\psi$ production on a single nucleon.

In the above expressions for the amplitudes $F_{1a}$ and $F_{1b}$, spin labels have been suppressed.  The initial virtual photon and the deuteron are in specific spin states, the final hadrons are in specific spin states, and there is a sum over the spin states of the intermediate-state virtual nucleon.  For example, the amplitude $F_{1b}$, including spin state specification, is explicitly:
\begin{equation}
F_{1b}=-\sqrt{2m (2\pi)^3}    \sum_{m_1}{\cal M}^{\gamma V}(m_1,\lambda,m_n,\lambda_V)\;\psi_D^M(\mathbf{p}_{p},m_1,m_p),
\end{equation}
where $m_1$ is the spin state of the intermediate-state neutron, i.e. the line with momentum $n$ in the Feynman diagram, $m_n$ and $m_p$ are the spin states of the final neutron and proton, $\lambda$ is the photon polarization, $\lambda_V$ is the $J/\psi$ polarization, and $M$ is the deuteron spin state.  In what follows we will assume that the elementary amplitudes for spin-flip are negligible compared to the non-spin-flip amplitudes, and so the amplitudes are diagonal in the nucleon spin, and also in the photon and $J/\psi$ spin.    In that case we are able to calculate the spin-averaged squares of the various amplitudes $F_{1a}$, $F_{2a}$, etc., and the spin-averaged square of the total amplitude.  We have included the contribution from the $D$-state in the deuteron wavefunction. For $\nu=9$ GeV the $D$-state was found to not make a significant contribution to the amplitudes, but for $\nu=6.5$ GeV the $D$-state did contribute significantly, especially for the impulse diagram $F_{1b}$.  The deuteron wavefunction used was the Argonne $v18$ wavefunction.

\subsubsection{Parameterization of the  amplitudes ${\cal M}^{\gamma V}$}

If the cross-section for $J/\psi$ production on a single nucleon is parametrized as
\begin{equation}
\frac{d\sigma}{dt}=A_1e^{B_1t},  \end{equation}
with the parameters $A_1$, $B_1$ dependent on energy (in principle), then the elementary production amplitude ${\cal M}^{\gamma V}$ is given by 
\begin{equation}
{\cal M}^{\gamma V}=-i\;\sqrt{16\pi A_1 \lambda(s,-Q^2,m^2)}e^{\frac{1}{2}B_1t} \end{equation}
where $s$ and $t$ are either $s_{1a}$, $t_{1a}$ or $s_{1b}$, $t_{1b}$.

The parameters $A_{1}$ and $B_{1}$ that are needed for the elementary $J/\psi$ production amplitude ${\cal M}^{\gamma V}$ were estimated from the (scant) existing data on exclusive $J/\psi$ production on a nucleon.  The only available data for the incident photon energy $\nu\simeq 10\;GeV$ is from a photoproduction experiment at Cornell in 1975~\cite{psidata75}.  For $\nu$ in the range 9.3 to 10.4 GeV, they determined $A_{1}=1.1\pm0.17\;nb/GeV^2=(2.8\pm0.43)\times 10^{-6}\;GeV^{-4}$ and $B_{1}=1.31\pm0.19\;GeV^{-2}$.  Those are the values used in this analysis.

\subsubsection{$F_{1a}$ and $F_{1b}$}

Since $F_{1a}$ is proportional to $\Psi(\mathbf{p}_n)$, and as seen in Fig. \ref{fig:nandpmomentum}  the outgoing neutron's momentum is always greater than $\simeq 0.6\;GeV$, the amplitude $F_{1a}$ will therefore be very small, since the deuteron wavefunction is negligible for those values of momentum.  The amplitude $F_{1b}$, on the other hand, is proportional to $\Psi(\mathbf{p}_p)$; thus as seen in Fig. \ref{fig:nandpmomentum} for $\theta_{cm}<0.3$ rad $F_{1b}$ should be non-negligible for $\nu=9\;GeV$ since the proton momentum is less than $0.4$ GeV over that range of $\theta_{cm}$.

\subsection{One-loop diagrams}
\begin{figure}[tbp]
     \begin{center}

            \includegraphics[width=4.5in,height=2.5in]{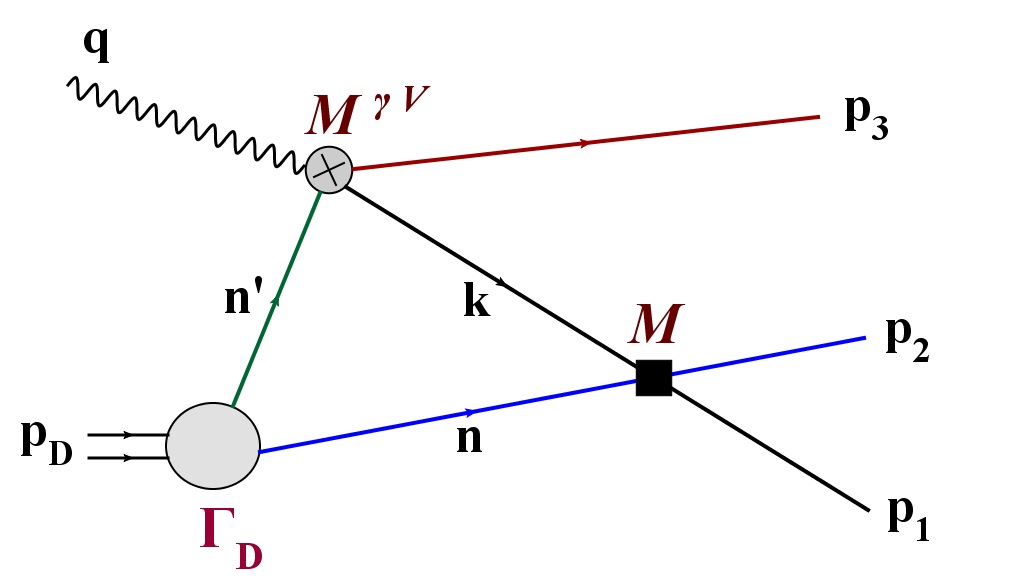}

    \end{center}
    \caption{%
        General one-loop diagram. $n$ and $p_2$ are the same particle (either neutron or proton).
     }%
   \label{fig:oneloopdiag}
\end{figure}

The covariant expression for a general one-loop diagram (see Fig. \ref{fig:oneloopdiag} ) is
\be
\label{eq:generaloneloop}
F=-\int \frac{d^4n}{i(2\pi)^4} \frac{\Gamma(p)}{D(p)}\frac{{\cal M}^{\gamma V}{\cal M}}{D(n)D(k)}
\ee
where $p,\;n,\;k$ are the internal momenta indicated in the figure, ${\cal M}$ stands for either ${\cal M}^{pn}$,  ${\cal M}^{Vn}$, or ${\cal M}^{Vp}$ (elastic scattering amplitude for proton-neutron, V-neutron, or V-proton scattering, respectively), and $D(p)=p^2-m_p^2+i\epsilon$, etc., are propagator denominators.  Spin labels have been suppressed in Eq. \ref{eq:generaloneloop}; in particular, there is an implicit sum over the spin states of the intermediate state particles (the lines labelled $p$, $k$, and $n$).  There are 4 diagrams total for a given set of outgoing proton, neutron and $J/\psi$ momenta.  Taking $\mathbf{p}_2=\mathbf{p}_n$, $\mathbf{p}_1=\mathbf{p}_p$ (so that the internal line $n$ is the neutron, and $p$ and $k$ are the proton) gives one diagram (p-n rescattering diagram).  
The other 3 are:   $\mathbf{p}_2=\mathbf{p}_n$, $\mathbf{p}_1=\mathbf{p}_V$, where the internal line $n$ is the neutron, $k$ is the $J/\psi$, and $p$ and $p_3$ are the proton (V-n rescattering diagram);  $\mathbf{p}_2=\mathbf{p}_p$, $\mathbf{p}_1=\mathbf{p}_n$, where the internal line $n$ is the proton, and $p$ and $k$ are the neutron (another p-n rescattering diagram); and  $\mathbf{p}_2=\mathbf{p}_p$, $\mathbf{p}_1=\mathbf{p}_V$, where the internal line $n$ is the proton, $k$ is the $J/\psi$, and $p$ and $p_3$ are the neutron (V-p rescattering diagram).

All of the one-loop diagrams can be evaluated in the same manner; see e.g.~\cite{laget81}, \cite{SS2006}, \cite{garcilazo99}, \cite{garcilazo2005}, \cite{sarg2010}.   We first integrate over $n^0$ by identifying the poles in the integrand and using the residue theorem.  The contribution of a given pole corresponds to a particular time-ordered diagram.  We make the approximation of neglecting the anti-nucleon contribution to the deuteron wavefunction; this corresponds to only keeping the positive energy pole at $n^0=\omega_n-i\epsilon$, where $\omega_n=\sqrt{m^2+\bf{n}^2}$, coming from the zero in the propagator denominator $D(n)=n^2-m^2+i\epsilon$.  After doing this, and using the relation $ \frac{\Gamma(p)}{D(p)}=\sqrt{2m(2\pi)^3}\Psi(\vert\mathbf{p}\vert)= \frac{\Gamma(p)}{D(p)}=\sqrt{2m(2\pi)^3}\Psi(\vert\mathbf{n}\vert)$ where $\Psi$ is the momentum-space deuteron wavefunction (and we work in the rest frame of the deuteron), we obtain
\be
F=\sqrt{\frac{2m}{(2\pi)^3}}\int \frac{d^3n}{2\omega_n}\Psi(n)\frac{{\cal M}^{\gamma V}{\cal M}}{D(k)}
\ee
\be
\simeq \frac{1}{\sqrt{2m(2\pi)^3}}\int d^3n\Psi(n)\frac{{\cal M}^{\gamma V}{\cal M}}{D(k)} 
\ee
where we've approximated $\omega_n\simeq m$, which is valid since the presence of the deuteron wavefunction implies that only small internal momenta $n$ contribute to the integral.  Note that in this expression, the internal nucleon line $n$ is now on-mass-shell, since $n^0=\omega_n$.  Thus in the time-ordered diagram, only the lines $p$ and $k$ can be off-shell; all the rest are on-shell.

\begin{figure}[tbp]
     \begin{center}

            \includegraphics[width=3.5in,height=2in]{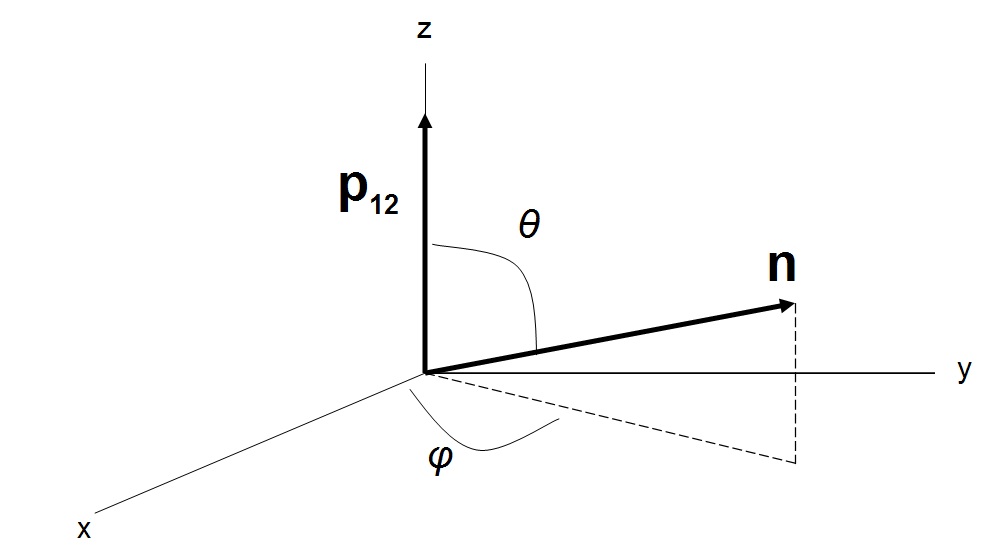}

    \end{center}
    \caption{%
        Coordinate system used.  For the on-shell amplitude, $\theta$ is fixed for a given $n\equiv\vert\mathbf{n}\vert$ and $\mathbf{p}_1$, $\mathbf{p}_2$.
     }%
   \label{fig:coordsystem}
\end{figure}

The above expression for the amplitude $F$ can be separated into two terms, one term in which the line $k$ is on-mass-shell and one in which $k$ is off-mass-shell; we follow closely the method of~\cite{laget81}.  This separation is useful since the elementary amplitudes ${\cal M}^{\gamma V}$, ${\cal M}$ can be determined (at least in principle) directly from experimental data on the relevant 2-body scattering processes only when all 4 particles involved (2 initial and 2 final) are on-mass-shell.  This is not the case if one of the particles involved is off-mass-shell.  But if for reasonable choices of the off-mass-shell amplitudes the off-mass-shell part is small compared to the on-shell part, then the off-mass-shell part will not play an important role.   Taking the $n_z$ axis along the direction of the vector $\mathbf{p}_{12}\equiv\mathbf{p}_1+\mathbf{p}_2$ gives
\be
D(k)=k^2-m_k^2+i\epsilon=(p_1+p_2-n)^2-m_k^2+i\epsilon=s_{12}+m^2-m_k^2-2E_{12}\omega_n+2\vert\mathbf{p}_{12}\vert n\cos{\theta}+i\epsilon
\ee
where $s_{12}\equiv(p_1+p_2)^2$ is the Mandelstam $s$-variable for the elastic scattering of particles 1 and 2, $m_k$ is the mass of the real particle which the line $k$ represents, $E_{12}\equiv E_1+E_2$, and $\theta$ is as shown in Fig. \ref{fig:coordsystem} .  This allows us to write $F$ as
\be
 \label{onandoff}
F=\frac{1}{\sqrt{2m(2\pi)^3}}\frac{1}{2\vert\mathbf{p}_{12}\vert}\int_0^{2\pi}d\phi\int_0^{\infty}dn\; n\Psi(n)\int d\cos\theta \frac{  {\cal M}^{\gamma V} {\cal M} }{f_{12}(n)+\cos{\theta}+i\epsilon}
\ee
where 
\be
f_{12}(n)\equiv\frac{        s_{12}+m^2-m_k^2-2E_{12}\omega_n}{    2\vert\mathbf{p}_{12}\vert n}
\ee
 is independent of $\theta$ and $\phi$.  Now using the relation
\be
\frac{1}{x+i\epsilon}=-i\pi\delta(x)+{\cal P }\frac{1}{x}
\ee
with ${\cal P}$ representing the principal value, Eq. \ref{onandoff} can be written in terms of its on-mass-shell and off-mass-shell parts:  the delta function gives the part where $k$ is on-mass-shell, since it's only non-zero (with $x\equiv\cos\theta  +f_{12}(n)$) for $\cos\theta = -f_{12}(n)$, which implies $k^2=m_k^2$.  Thus we have
\be
F=F^{on}+F^{off}
\ee
where
\be
\label{Fon}
F^{on}=-i\pi \frac{1}{\sqrt{2m(2\pi)^3}}\frac{1}{2\vert\mathbf{p}_{12}\vert}\int_0^{2\pi}d\phi\int_{\vert n_-\vert}^{n_+}dn\; n\Psi(n) {\cal M}^{\gamma V} {\cal M}
\ee
and
\be
\label{Foff}
F^{off}=\frac{1}{\sqrt{2m(2\pi)^3}}\frac{1}{2\vert\mathbf{p}_{12}\vert}\int_0^{2\pi}d\phi\int_0^{\infty}dn\; n\Psi(n)\;\;{\cal P}\int d\cos\theta \frac{  {\cal M}^{\gamma V} {\cal M} }{f_{12}(n)+\cos{\theta}}.
\ee
In $F^{on}$, the intermediate particle line $k$ is now on-mass-shell, so $k^2=m_k^2$ with $k^0=E_{12}-\omega_n$.  The limits of integration $\vert n_-\vert$ and $n_+$ are the solutions of
\be
f_{12}(n_{\pm})^2=1,
\ee
which are
\be
\label{nplusminus}
n_{\pm}=\frac{E_2^*}{\sqrt{s_{12}}}\vert\mathbf{p}_{12}\vert \pm \frac{p_2^*}{\sqrt{s_{12}}}E_{12}
\ee
where $p_2^*$, $E_2^*$ are the momentum and energy of outgoing particle 2 in the c.m. frame of particles 1 and 2, and particle 2 is the \underline{same} particle as the internal line with momentum $\mathbf{n}$.  Thus we have the relation 
\be
s_{12}=E_{12}^2-\mathbf{p}_{12}^2=\Bigl(\sqrt{m^2+p^{*2}}+\sqrt{m_k^2+p^{*2}}\Bigr)^2.
\ee
 
The range of $n$ given by  $\vert n_-\vert\le n \le n_+$ is the range of $n$ for which it is kinematically possible for the line $k$ to be on-mass-shell (given that $n$, $p_1$, and $p_2$ are on-shell), and in $F^{on}$ the value of $\cos\theta$ is fixed at
\be
\cos\theta=-f_{12}(n)=-\frac{        s_{12}+m^2-m_k^2-2E_{12}\omega_n}{    2\vert\mathbf{p}_{12}\vert n}.
\ee
The amplitudes $ {\cal M}^{\gamma V},\; {\cal M}$ are evaluated, for a given $n$ and $\phi$, at that value of $\cos\theta$.  The amplitude ${\cal M}$ is now fully on-shell, i.e. all 4 particle lines $n$, $k$, $p_1$, and $p_2$ are on-mass-shell.  The amplitude ${\cal M}^{\gamma V} $ has only one particle off-shell ($p$), but given that the magnitude of $\mathbf{n}$ is small (due to the deuteron wavefunction), $p$ is almost on-shell:  $p^0=M_d-\omega_n\simeq M_d-m\simeq m$, and so $p^2=m^2+{\cal O}(\frac{n^2}{m^2})$.

In the amplitude $F^{off}$, $k$ is never on-mass-shell; the principal value imposes this, since for $k$ to be on-mass-shell, $\cos\theta$ must equal $-f_{12}(n)$, which never occurs in the principal value.  Thus the amplitudes  $ {\cal M}^{\gamma V},\; {\cal M}$ that enter into $F^{off}$ have either one particle off-mass-shell (for ${\cal M}$) or two particles off-shell (for  $ {\cal M}^{\gamma V}$).  It will be shown below that the amplitude $F^{off}$ is much smaller than $F^{on}$, for the kinematics of interest here, and for reasonable expressions for the off-shell amplitudes $ {\cal M}^{\gamma V},\; {\cal M}$.

\subsection{General features of the one-loop diagrams}

The on-shell part of a given one-loop amplitude is dictated largely by the behavior of $\vert n_-\vert$ as a function of $\theta_{cm}$, where for a given diagram $n_-$ is given by Eq. \ref{nplusminus}.
A necessary condition for the on-shell part of a given one-loop diagram to be non-negligible is that the corresponding $\vert n_-\vert$ must be small enough so that the range of integration in Eq. \ref{Fon} includes the momenta where the deuteron wavefunction is significant (see Fig. \ref{fig:psiandppsi}).  In fact, since it is $n\Psi(n)$ which enters into the integral in Eq. \ref{Fon}, and this quantity is fairly sharply peaked at $n\simeq 0.05\; GeV$ (see Fig. \ref{fig:psiandppsi2}), it is necessary to have $n_-\lesssim 0.05\;GeV$ in order for the on-shell part of the amplitude to be non-negligible.
\begin{figure}[tbp]
     \begin{center}
        \subfigure[]{%
            \label{fig:psiandppsi1}
            \includegraphics[width=0.4\textwidth]{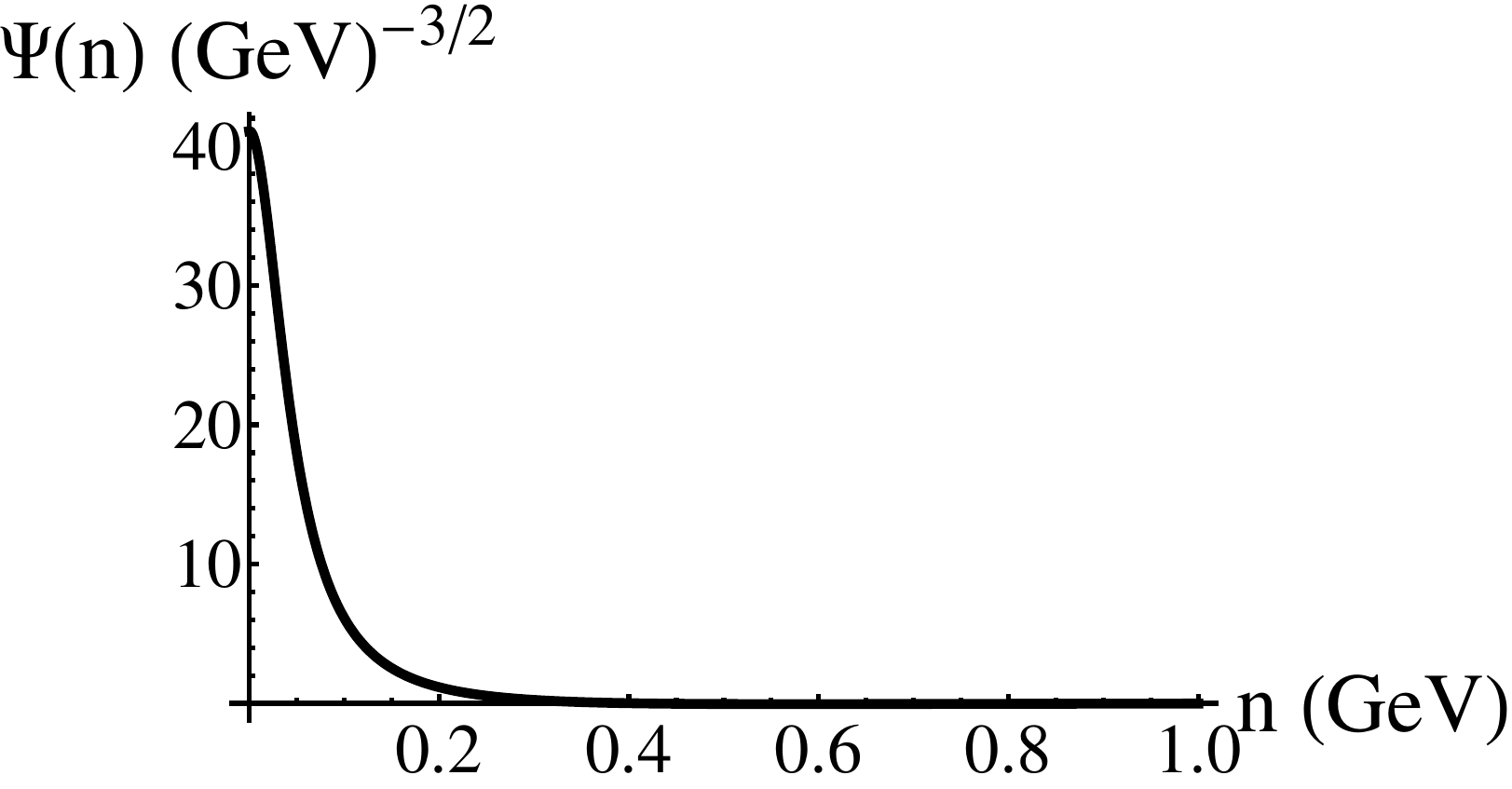}
        }%
        \hspace{0.5in}
         \subfigure[]{%
           \label{fig:psiandppsi2}
           \includegraphics[width=0.4\textwidth]{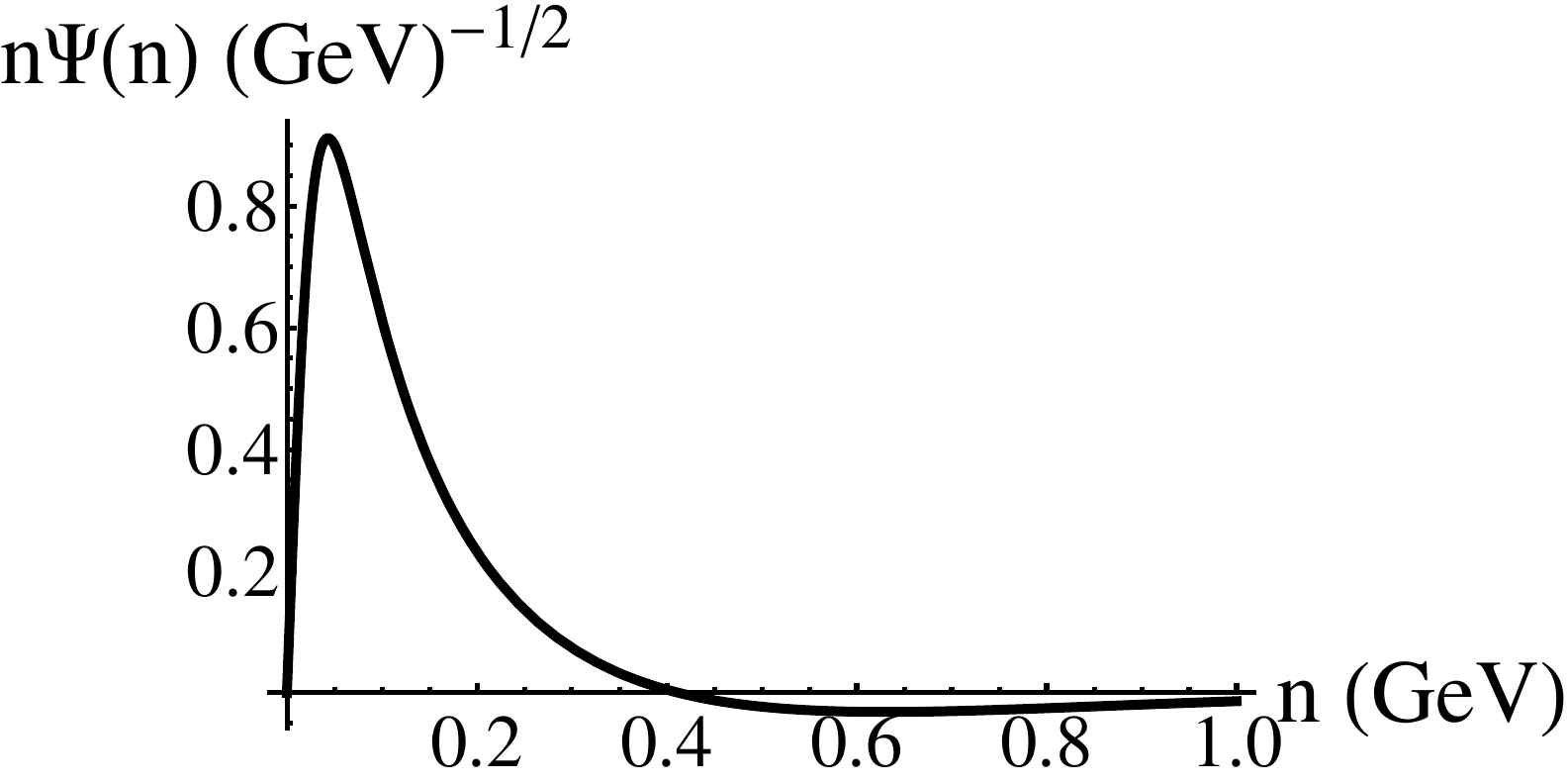}
        }\\ 


    \end{center}
    \caption{%
       (a)  Deuteron momentum-space wavefunction (S-state only);  (b) $n\Psi(n)$ vs. $n$
     }%
   \label{fig:psiandppsi}
\end{figure}

For the diagram of most interest, $F_{3a}$, particles $1$ and $2$ are the $J/\psi$ and neutron, respectively.    Fig. \ref{fig:nminusplus3a1a} shows $n_{\pm}$ vs. $\theta_{cm}$ for $T^*_{Vn}=30\;MeV$, for $\nu=9\;GeV$, and Fig. \ref{fig:nminusplus3a1b} shows the same for $\nu=6.5\;GeV$, for the diagram $F_{3a}$.   At both of these photon energies, $n_-$ is greater than $\sim0.6\;GeV$ for all $\theta_{cm}$, and so the on-shell amplitude will be negligible since the deuteron wavefunction is negligible for that momentum.   Note that  for the $T^*_{Vn}=0$ case (zero relative momentum of the $J/\psi-n$ pair), we have $p_2^*=0$ in Eq. \ref{nplusminus}, and so $n_-=n_+$; thus  the on-shell amplitude, given by Eq. \ref{Fon}, is exactly zero for  $T^*_{Vn}=0$. 

\begin{figure}[tbp]
     \begin{center}
        \subfigure[$\nu=9$ GeV]{%
            \label{fig:nminusplus3a1a}
            \includegraphics[width=0.4\textwidth]{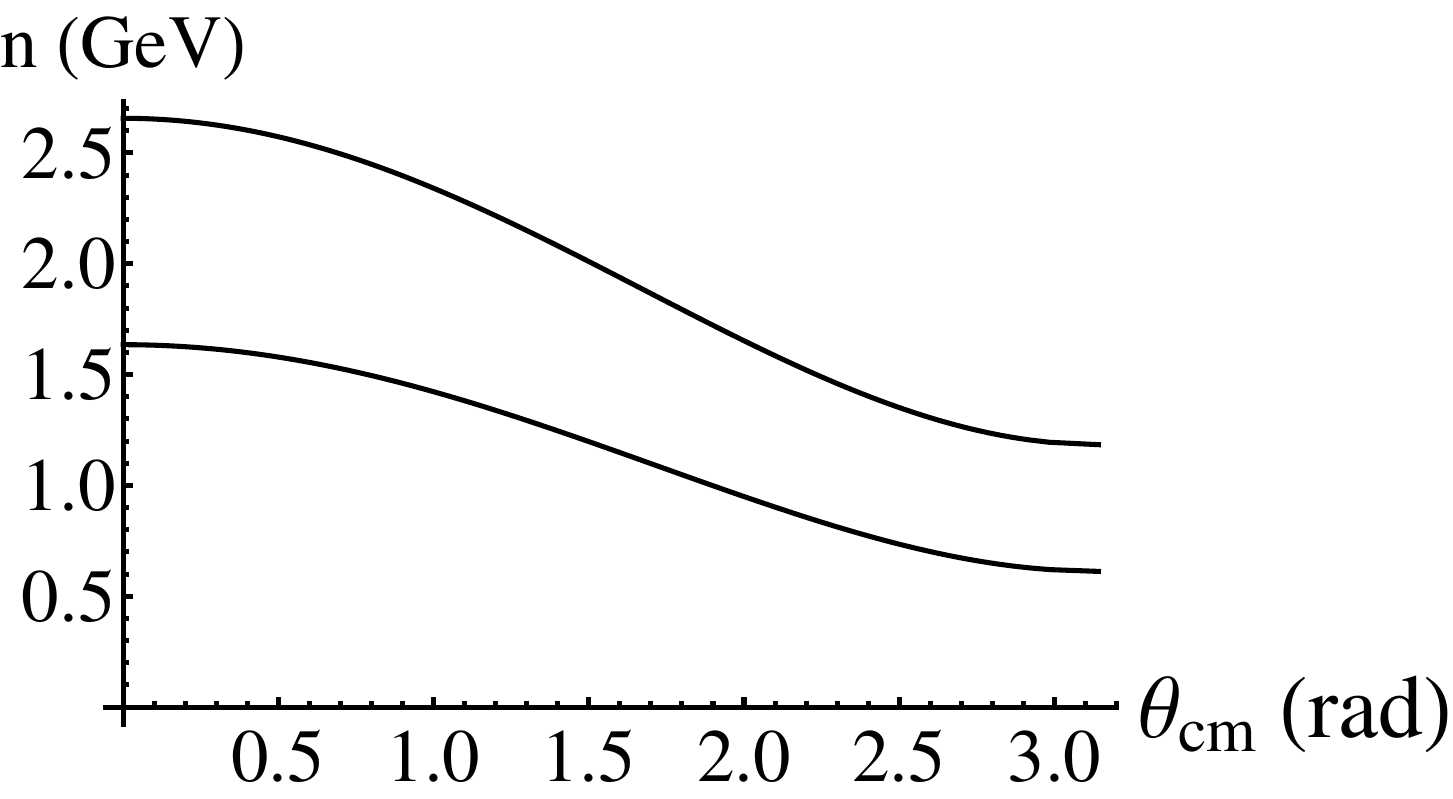}
        }%
        \hspace{0.5in}
         \subfigure[$\nu=6.5$ GeV]{%
           \label{fig:nminusplus3a1b}
           \includegraphics[width=0.4\textwidth]{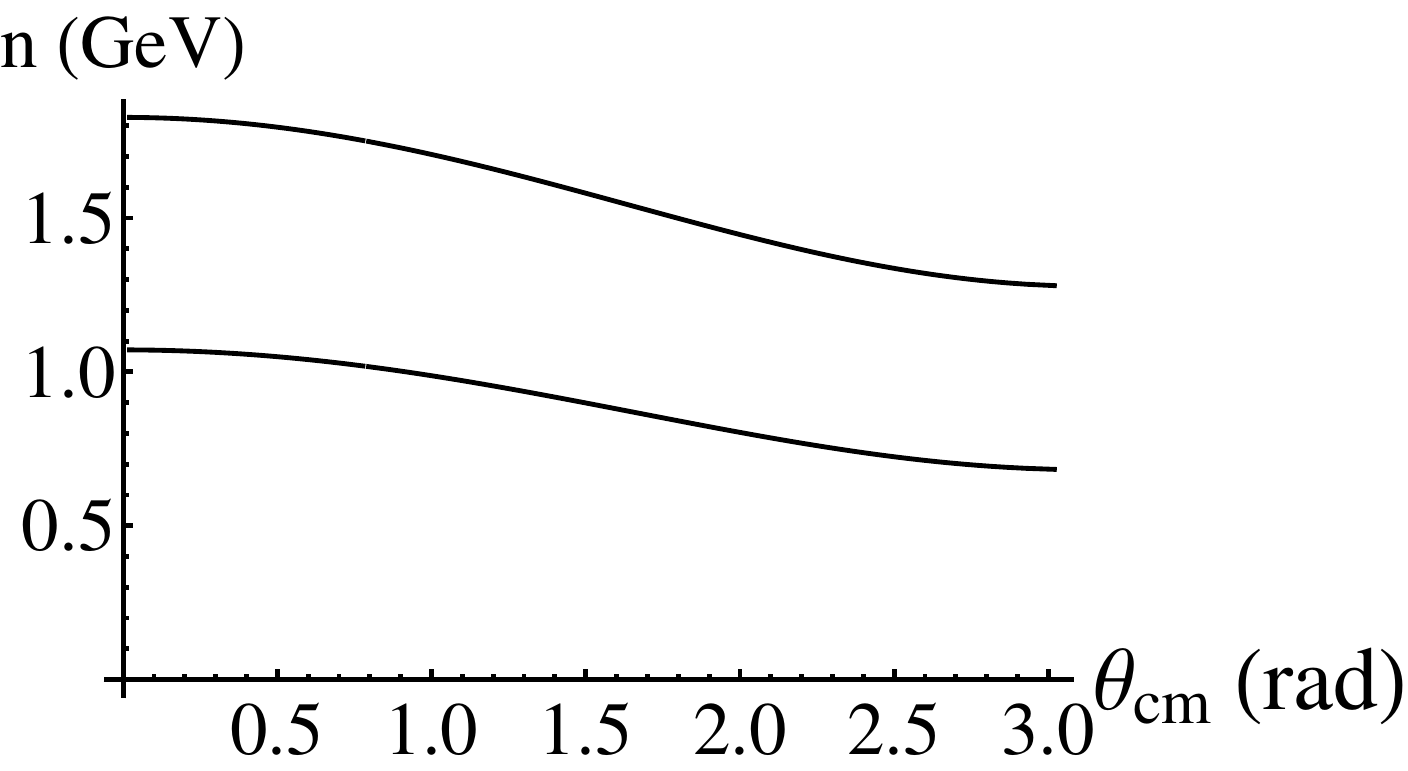}
        }\\ 


    \end{center}
    \caption{%
         $n_{\pm}$ vs. $\theta_{cm}$ for the amplitude $F_{3a}$ ($J/\psi$-neutron rescattering), for $T^*_{Vn}=30\;MeV$, and $\nu=9.0\;GeV$ and $\nu=6.5\;GeV$.  Upper curve is $n_+$, lower curve is $n_-$.
     }%
   \label{fig:nminusplus3a1}
\end{figure}

\begin{figure}[tbp]
     \begin{center}

            \includegraphics[width=3.5in,height=2in]{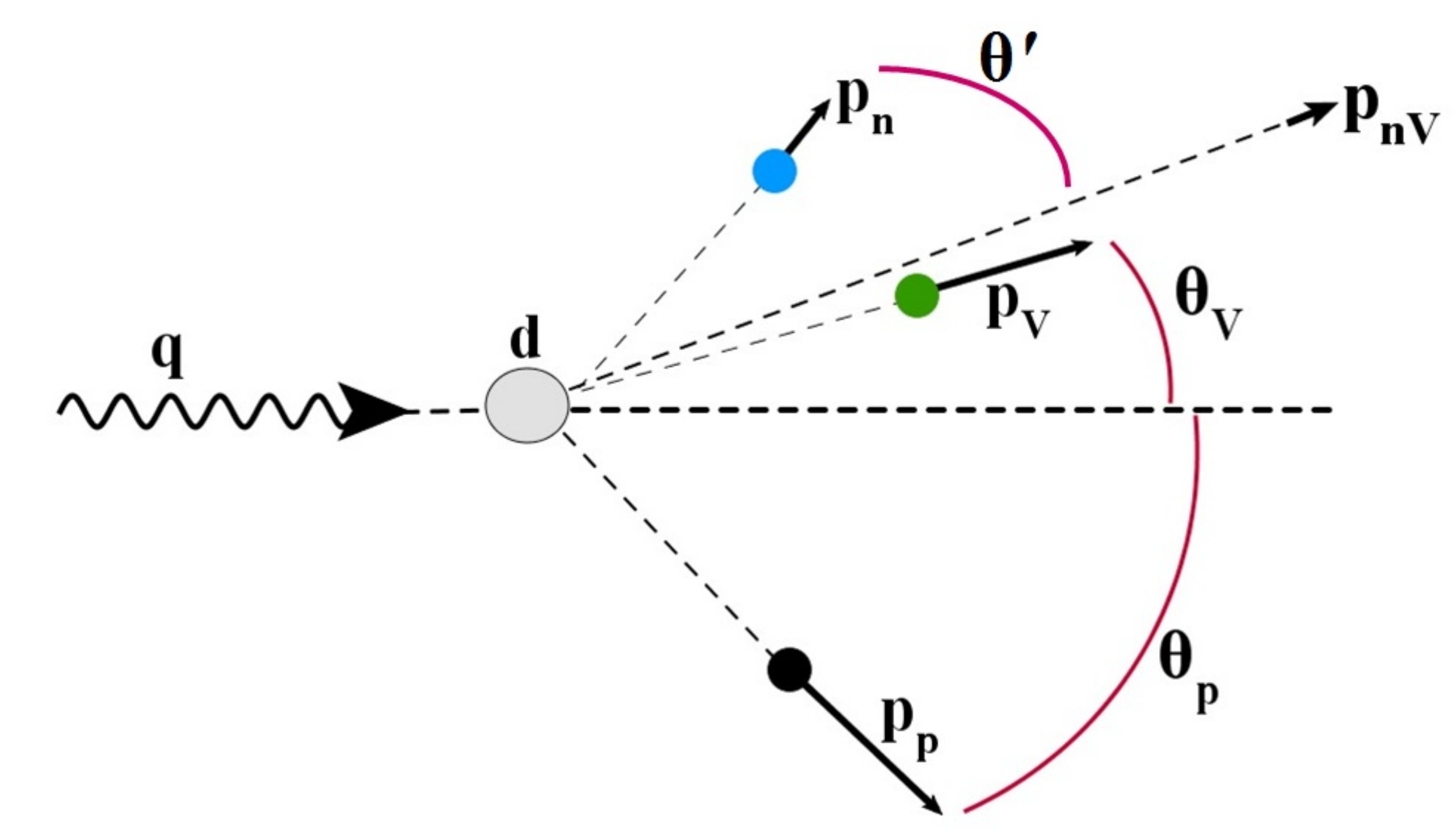}

    \end{center}
    \caption{%
        LAB frame momenta.  For a given $\theta_{cm}$ and $T^*_{Vn}$, the neutron LAB momentum can range between a minimum and maximum value, both of which correspond to $\theta^{\prime}=0$ in the figure.  $\mathbf{p}_{nV}=\mathbf{p}_V+\mathbf{p}_n$.
     }%
   \label{fig:labkinematics}
\end{figure}

For the amplitudes $F_{2a}$, $F_{2b}$ and $F_{3b}$, the corresponding $n_-$ graphs are shown in Fig. \ref{fig:nminus2b3b} for $\nu=9\;GeV$ and $\nu=6.5\;GeV$, for $T^*_{Vn}=30$ MeV.  Note that at a given value of $\theta_{cm}$ and $T^*_{Vn}$, the neutron LAB momentum can range from a minimum to a maximum allowed value (with these two values both corresponding to $\mathbf{p}_n$ and $\mathbf{p}_V$ pointing in the same direction in the LAB, with $\theta^{\prime}=0$ in Fig. \ref{fig:labkinematics}), and the values of $n_-$ for $F_{2a}$, $F_{2b}$ and $F_{3b}$ depend on the neutron LAB momentum in addition to  $\theta_{cm}$ and $T^*_{Vn}$.   For our calculations, we have fixed the neutron LAB momentum for a given $\theta_{cm}$ (and $T^*_{Vn}=30$ MeV) at its minimum value; we denote this value by $p_{n,min}$.   One can see from these graphs that for $\nu=9\;GeV$, there are intervals of the variable $\theta_{cm}$ for which the on-shell parts of $F_{2a}$, $F_{2b}$, and $F_{3b}$ should be non-negligible, since $n_- <0.05\;GeV$ there  (note that for the diagram $F_{3b}$, the $J/\psi$-proton rescattering occurs at relatively high energy, if the $J/\psi$-neutron relative energy is small; so $F_{3b}$ is not directly related to the $J/\psi$-nucleon scattering length).  For $\nu=6.5\;GeV$, $n_-$ is larger than $\simeq 0.4$ GeV, and so these on-shell amplitudes should be small.  This is born out by the exact calculations, where the one-loop on-shell  amplitudes for $\nu=6.5\;GeV$ are in general much smaller than those for $\nu=9\;GeV$.

\begin{figure}[tbp]
     \begin{center}
        \subfigure [$\nu=9\;GeV$ ]{%
            \label{fig:nminus2bnu9}
            \includegraphics[width=0.4\textwidth]{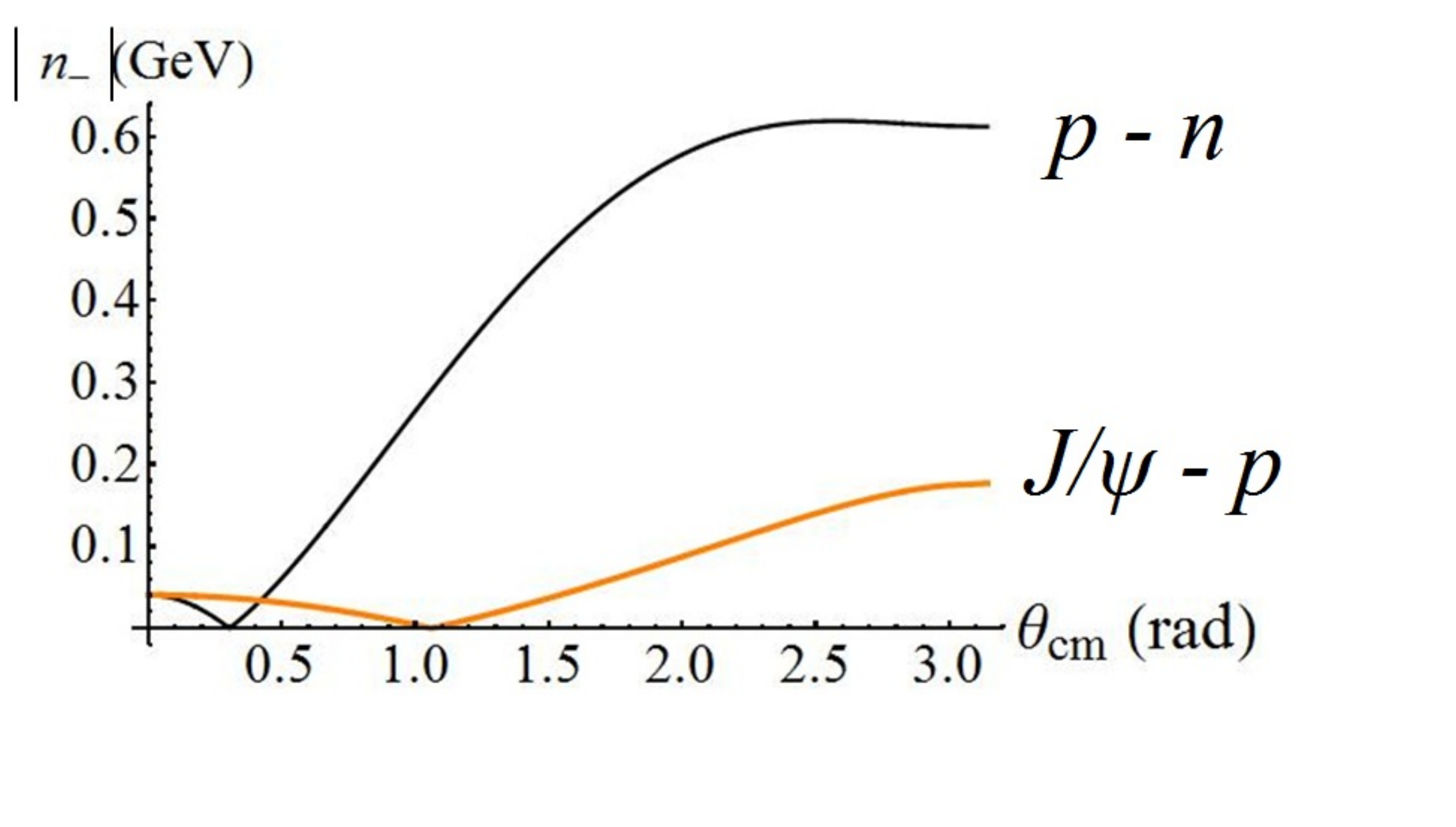}
        }%
        \hspace{0.5in}
         \subfigure[ $\nu=6.5\;GeV$]{%
           \label{fig:nminus3bnu9}
           \includegraphics[width=0.4\textwidth]{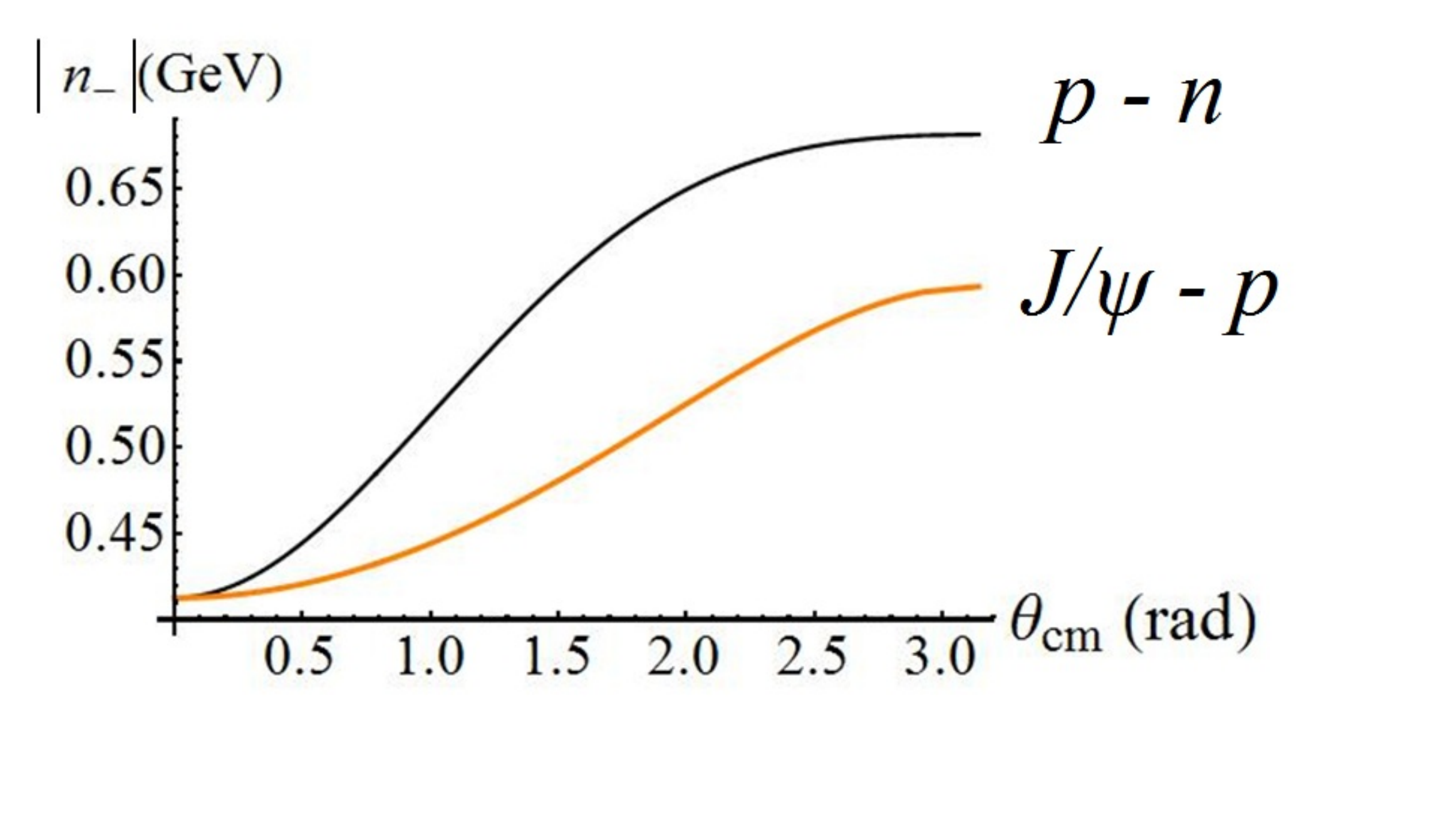}
        }\\ 


    \end{center}
    \caption{%
       $n_-$ vs. $\theta_{cm}$ for proton-neutron rescattering and $J/\psi$-proton rescattering amplitudes, for $T^*_{Vn}=30$ MeV, $p_n=p_{n,min}$.
     }%
   \label{fig:nminus2b3b}
\end{figure}

\subsection{Parameters used in the elementary amplitudes}
\label{sec:parameterssubsec}

For the calculation of the amplitudes, using Eq. \ref{Fon} and Eq. \ref{Foff}, the elementary amplitudes ${\cal M}^{\gamma V}$ and ${\cal M}$ are phenomenological amplitudes obtained from existing experimental data.   In the diffractive regime, the individual 2-body differential cross-sections for elastic scattering are of the form:
\begin{equation}
\frac{d\sigma^{pn}}{dt}=A_{pn}e^{B_{pn}t}  
\end{equation}
and
\begin{equation}
\frac{d\sigma^{Vp}}{dt}=A_{Vp}e^{B_{Vp}t}  
\end{equation}
where the $A$'s and $B$'s can depend on energy.  These are related to the elementary amplitudes ${\cal M}$ by  
\begin{equation}
\frac{d\sigma}{dt}=\frac{1}{16\pi\lambda(s,m_1^2,m_2^2)}\vert {\cal M}\vert^2 
\end{equation}
where $\lambda(s,m_1^2,m_2^2)=(s-m_1^2-m_2^2)^2-4m_1^2m_2^2$ is the flux factor.  Thus we have
\be
{\cal M}=-i\sqrt{16\pi\lambda(s,m_1^2,m_2^2) A}\;e^{\frac{1}{2}Bt}
\ee
relating ${\cal M}$ to $A$ and $B$.

The values of $A$ and $B$ depend on the relative momentum (or energy) of the rescattering pair.  Table \ref{table:parameters} lists the values of the momentum $p$ of the neutron in the proton's rest frame (for the $p-n$ subsystem) and the momentum $p$ of the $J/\psi$ in the proton's rest frame (for the $V-p$ subsystem), for the case of $T^*_{Vn}=0$.  Note that these quantities are independent of $\theta_{cm}$ (easily shown in the overall c.m. frame).  Also given in \ref{table:parameters} is the total kinetic energy of the pair in the c.m. frame of that pair, $T^*_{12}$.

\begin{table}[ht]
\caption{Parameters used in elementary scattering amplitudes} 
\centering 
\begin{tabular}{c c c c c c c c} 
\hline\hline 
$\nu$ (GeV) & Subsystem  & $p$ (GeV) & $T^*_{12}$ (GeV) & $B$ (GeV$^{-2}$) & $\sigma_{tot}$ (mb) & $A$ (GeV$^{-4}$) \\ [0.5ex] 
\hline 
9 & $n-p$ & 2.25 & 0.64 & 5.7 - 6.2 & 43 - 46 & 260  \\ 
9 & $V-p$ & 7.4 & 1.02 & 1.31 & 3.5 & 1.61  \\
6.5 & $n-p$ & 0.86 & 0.16 & 6.9 & 35 & 160  \\
6.5 & $V-p$ & 2.84 & 0.247 & 1.31 & 3.5 & 1.61  \\ [1ex] 
\hline 
\end{tabular}
\label{table:parameters} 
\end{table}

Given the values of the momentum $p$, we can determine the parameters that enter into the elementary amplitudes ${\cal M}^{pn}$, ${\cal M}^{Vp}$. 

\subsubsection{p-n rescattering}

For $\nu=9\;GeV$, the existing data~\cite{Perl69} at this momentum $p$ give $B_{pn}=5.7\;to\;6.2\;GeV^{-2}$.  The value of $A_{pn}$ can be obtained from the total $p-n$ cross-section by using the optical theorem and neglecting the real part of the scattering amplitude:
\begin{equation}
\label{Aandsigma}
\frac{d\sigma}{dt}\vert_{t=0}=\frac{1}{16\pi}\sigma_{tot}^2=A_{pn}  
\end{equation}
The measured value of $\sigma_{tot}$ given in the table is then used to calculate $A_{pn}$.

For $\nu=6.5$ GeV, the existing data give $B_{pn}=6.9\;GeV^{-2}$~\cite{Perl69} and $A_{pn}=160\;GeV^{-4}$~\cite{bugg66}.

\subsubsection{$J/\psi$-nucleon rescattering}

There is very little data on elastic $J/\psi$-proton scattering from which to determine the parameters $A_{Vp}$ and $B_{Vp}$ that are needed for the $J/\psi$-proton rescattering amplitude $M^{Vp}$.  For the present analysis, I have assumed the validity of Vector Meson Dominance~\cite{feyn72} for which the $t$-slope for elastic $J/\psi$-nucleon scattering is equal to the $t$-slope for the process $\gamma^*+N\to J/\psi+N$, and so I've taken $B_{Vp}=B_{\gamma V}=1.31\pm0.19\;GeV^{-2}$.  We can obtain $A_{Vp}$ from the total $J/\psi$-nucleon cross-section, using the optical theorem; however, there has only been one measurement of $\sigma_{tot}^{J/\psi\; N}$~\cite{brambilla2011}.  In an experiment in 1977 at SLAC~\cite{psidata77} $J/\psi$ photoproduction was measured on beryllium and tantalum targets, and the total $J/\psi$-nucleon cross-section was extracted by using an optical model for the rescattering of the produced $J/\psi$ from the other nucleons in the nucleus.  The value they obtained was $\sigma_{tot}^{J/\psi\; N}=3.5\pm0.8\;mb$, which gives via the optical theorem $A_{Vp}=1.61\pm0.4\;GeV^{-4}$.  In that paper, however, they also note that the measured $J/\psi$-photoproduction cross-section together with vector meson dominance arguments would give a $J/\psi$-nucleon total cross-section of $\simeq 1\;mb$.   So we can assume the value of the $J/\psi$-nucleon total cross-section to be not very well known.  In addition, the photon energy in the SLAC experiment was $20\;GeV$, and so assuming forward production of the $J/\psi$ then the energy of the $J/\psi$ in the LAB frame would also be $\simeq 20\;GeV$, giving a kinetic energy in the LAB of $\simeq 17\;GeV$.  This is significantly larger than the kinetic energy of the $J/\psi$ in the proton rest frame considered here, where for $\nu=9\;GeV$ it is $4.94\;GeV$ and for $\nu=6.5\;GeV$ it is $1.1\;GeV$.  This introduces more uncertainty in the value of $A_{Vp}$ to be used.  In~\cite{brodsky97}, a theoretical calculation of the $J/\psi$-nucleon scattering length yields a value for the total $J/\psi$-nucleon cross-section at threshold of $7\;mb$, and it is argued that the total cross-section should decrease as the energy is increased from threshold.  Thus at the energy of the $J/\psi$-proton rescattering here, the value of $A_{Vp}$ may be larger than the value measured in the experiment at SLAC.  For the purpose of calculating the amplitude $F_{3b}$, however, we will use the value measured at SLAC.  (In Sec. \ref{sec:intermedenergy} on intermediate-energy $J/\psi$ production, the uncertainty in the value of $A_{Vp}$ will be put to good use, as $A_{Vp}$ will be considered a free parameter, which might in fact be measured by the proposed experiment at JLab).

\subsection{Subthreshold $J/\psi$ production}

The threshold photon energy for production on a single nucleon at rest is 
\be
\nu_{thresh}=m_V+\frac{m_V^2+Q^2}{2m}
\ee
while for production on the deuteron it is
\be
\nu_{thresh}=m_V+\frac{m_V^2+Q^2}{2M_d}
\ee
For $Q^2=0.5 \;GeV$, these are $8.47 \;GeV$ and $5.78 \;GeV$, respectively.

For $\nu=6.5\;GeV$, which is below threshold for $J/\psi$ production on a single nucleon at rest, we assume that the production mechanism is the same as for production on a free nucleon.  The Fermi motion of the nucleon in the deuteron is what allows the production to occur, e.g. if the nucleon is moving towards the photon with a large enough momentum then the value of $s_1=(q+p)^2$, where $p$ is the 4-momentum of the nucleon in the deuteron, will be above the threshold value.  In the calculation of the amplitudes for $\nu=6.5\; GeV$ this condition was imposed on the internal nucleon momentum in the integrals involved.

\subsection{Calculated On-shell and Off-shell amplitudes}

Using the parameters in Table \ref{table:parameters}, the on-shell and off-shell parts of the amplitudes were calculated.  The squares of the individual amplitudes $F_{2a}$, $F_{2b}$, and $F_{3b}$ are shown in Figs. \ref{fig:ampstotandon} and \ref{fig:ampstotandon65}; shown in the graphs is a curve which includes only the (square of the) on-shell part of the amplitude, and also a curve which is the square of the total amplitude including both the on-shell and off-shell parts.  For the off-shell parts, the same parametrizations of the elementary amplitudes ${\cal M}^{\gamma V}$ and ${\cal M}$ were used as for the on-shell parts.  As stated previously, the off-shell parts are very small compared to the on-shell parts, which means that knowledge of the exact forms of the off-shell elementary amplitudes ${\cal M}^{\gamma V}$ and ${\cal M}$ are not needed.

\begin{figure}[hbp]
     \begin{center}
        \subfigure[]{%
            \label{fig:31}
            \includegraphics[width=0.6\textwidth]{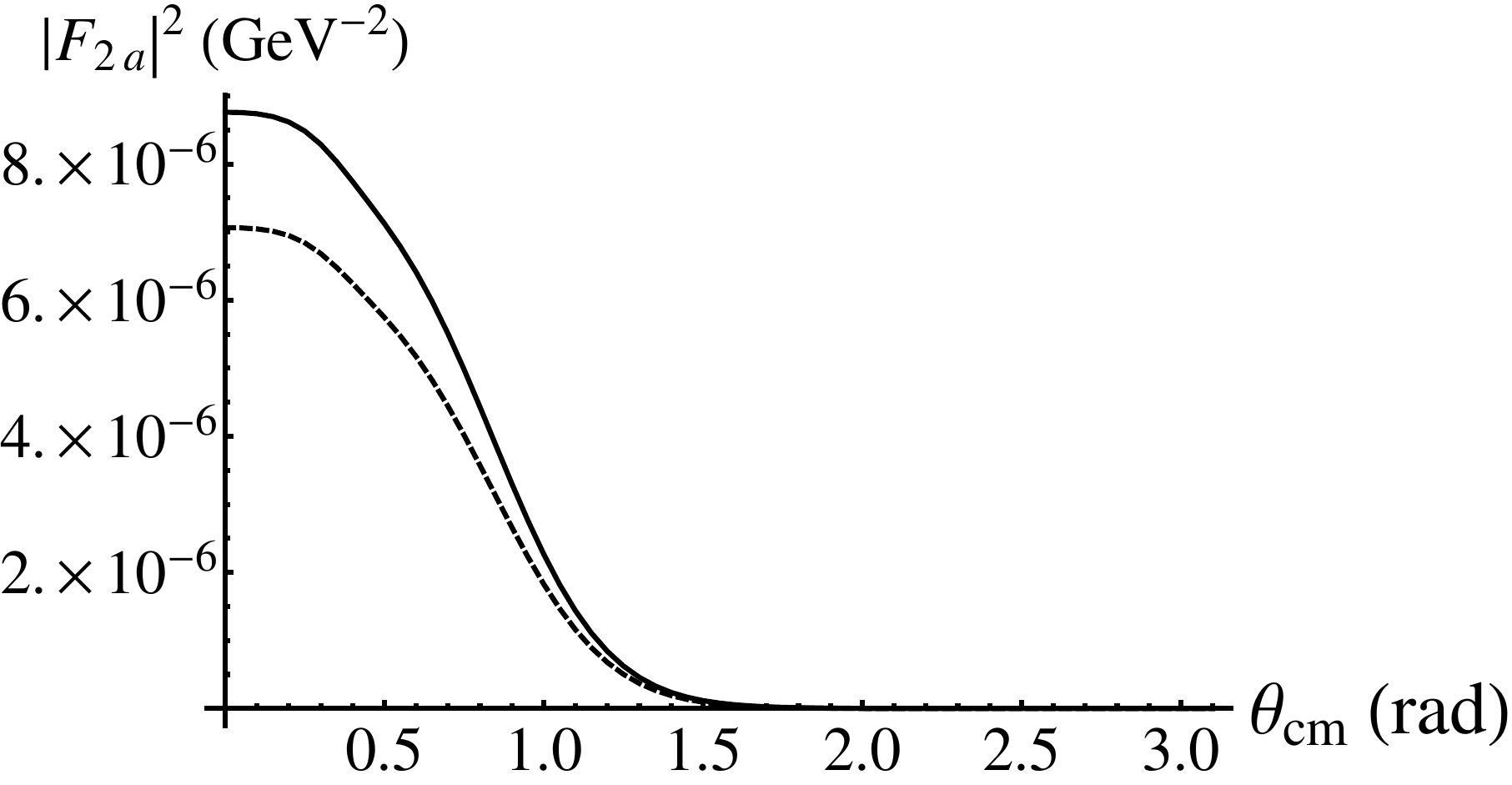}
        }%
        \hspace{0.5in}
         \subfigure[]{%
           \label{fig:32}
           \includegraphics[width=0.6\textwidth]{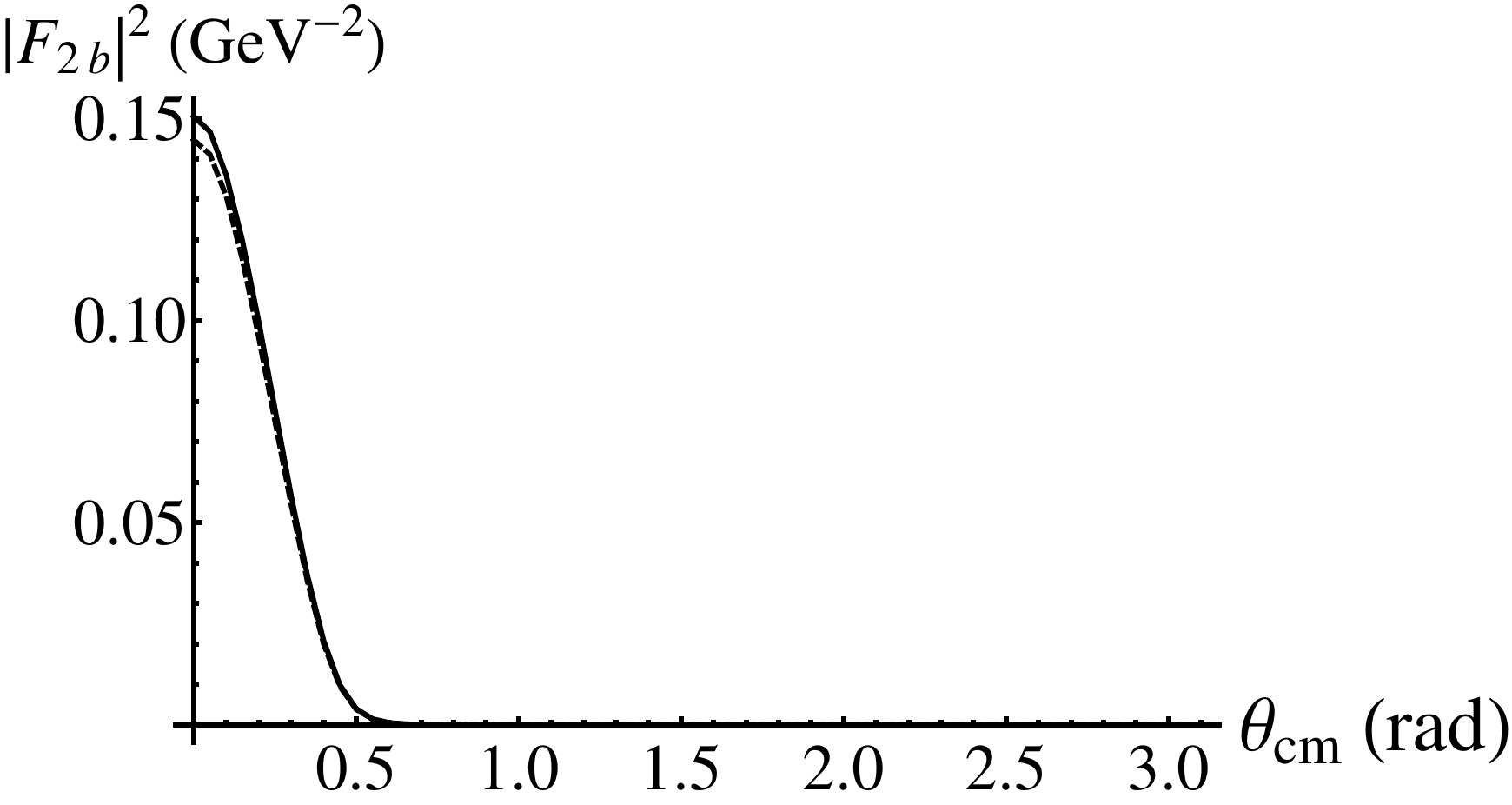}
        }\\ 

        \subfigure[]{%
            \label{fig:33}
            \includegraphics[width=0.6\textwidth]{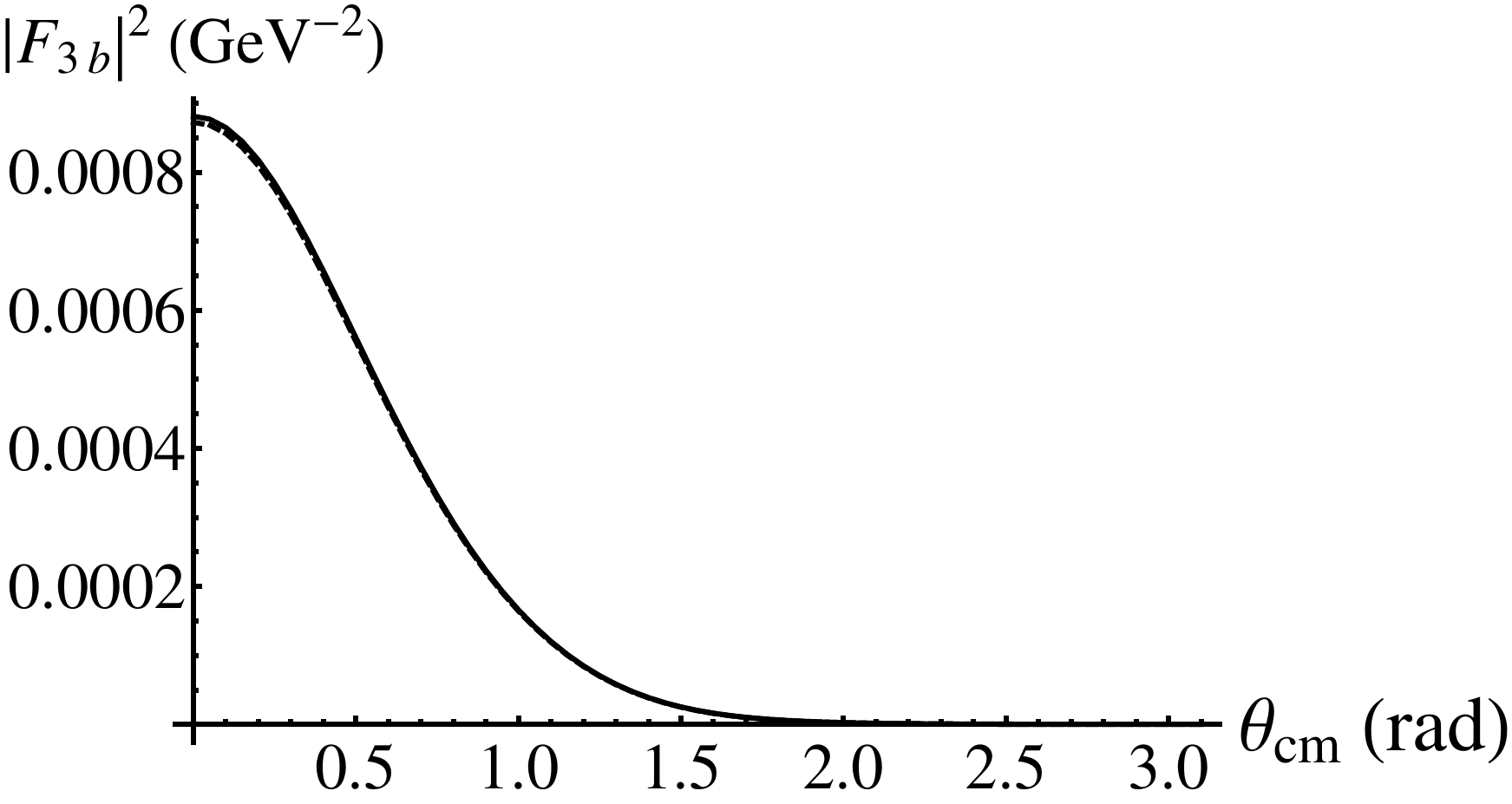}
        }%

    \end{center}
    \caption{%
       Squares of amplitudes for $\nu=9.0\;GeV$, $T^*_{Vn}=30$ MeV, $p_n=p_{n,min}$.  The solid curve is the total (on-shell plus off-shell parts), while the dashed curve is only including the on-shell part of the amplitude.
     }%
   \label{fig:ampstotandon}
\end{figure}

\begin{figure}[!hbp]
     \begin{center}
        \subfigure[]{%
            \label{fig:34}
            \includegraphics[width=0.6\textwidth]{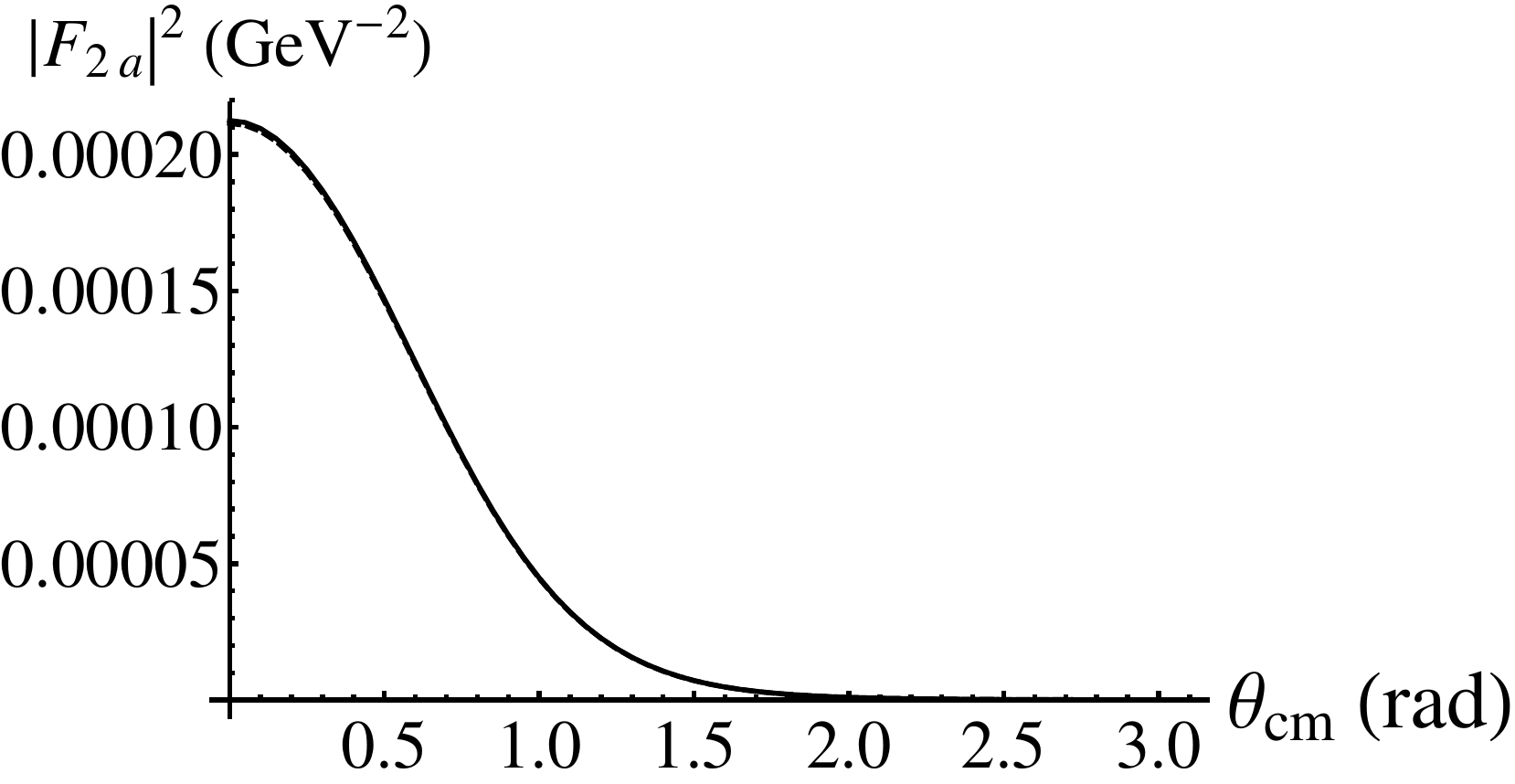}
        }%
        \hspace{0.5in}
         \subfigure[]{%
           \label{fig:35}
           \includegraphics[width=0.6\textwidth]{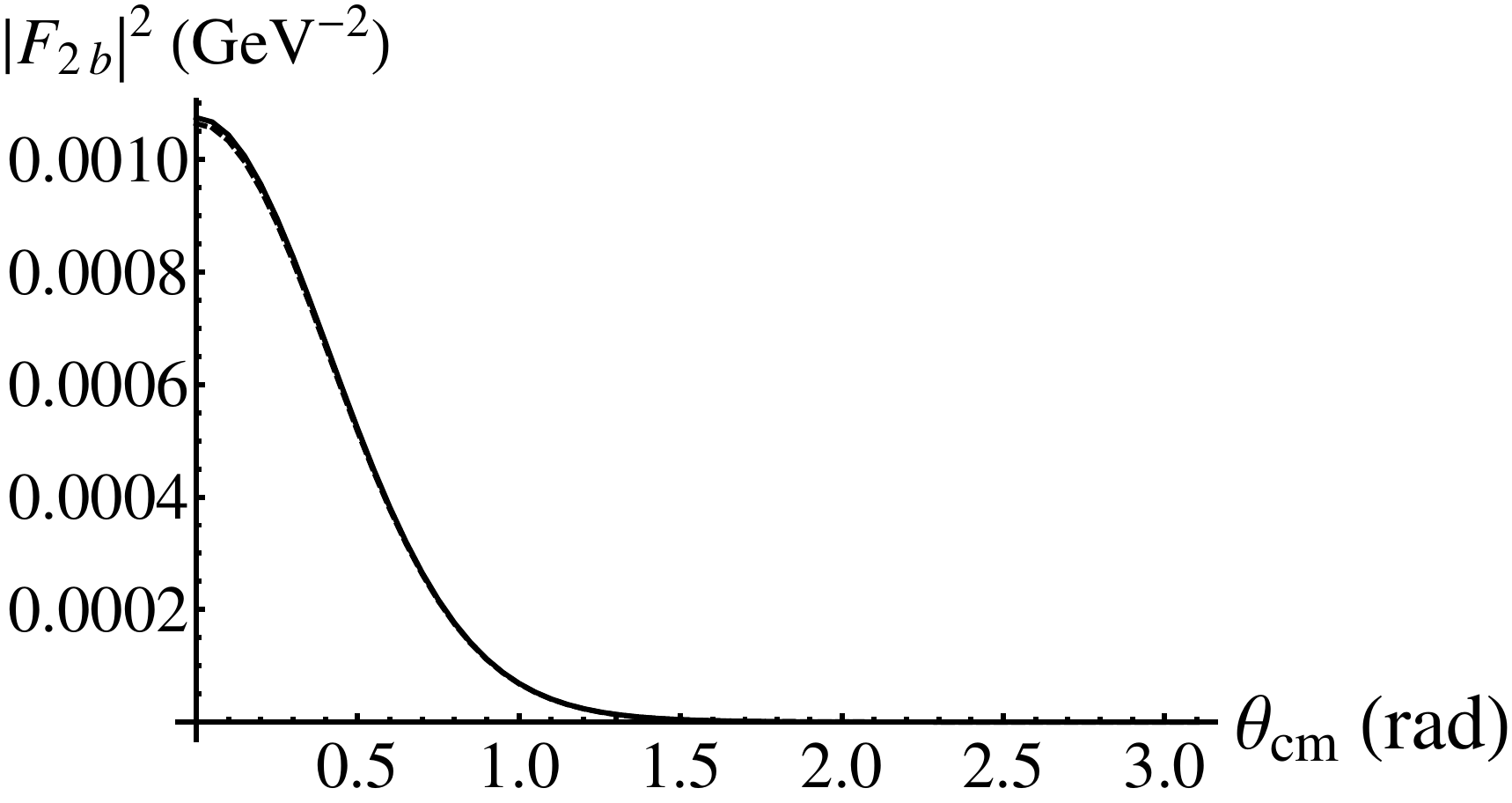}
        }\\ 

        \subfigure[]{%
            \label{fig:36}
            \includegraphics[width=0.6\textwidth]{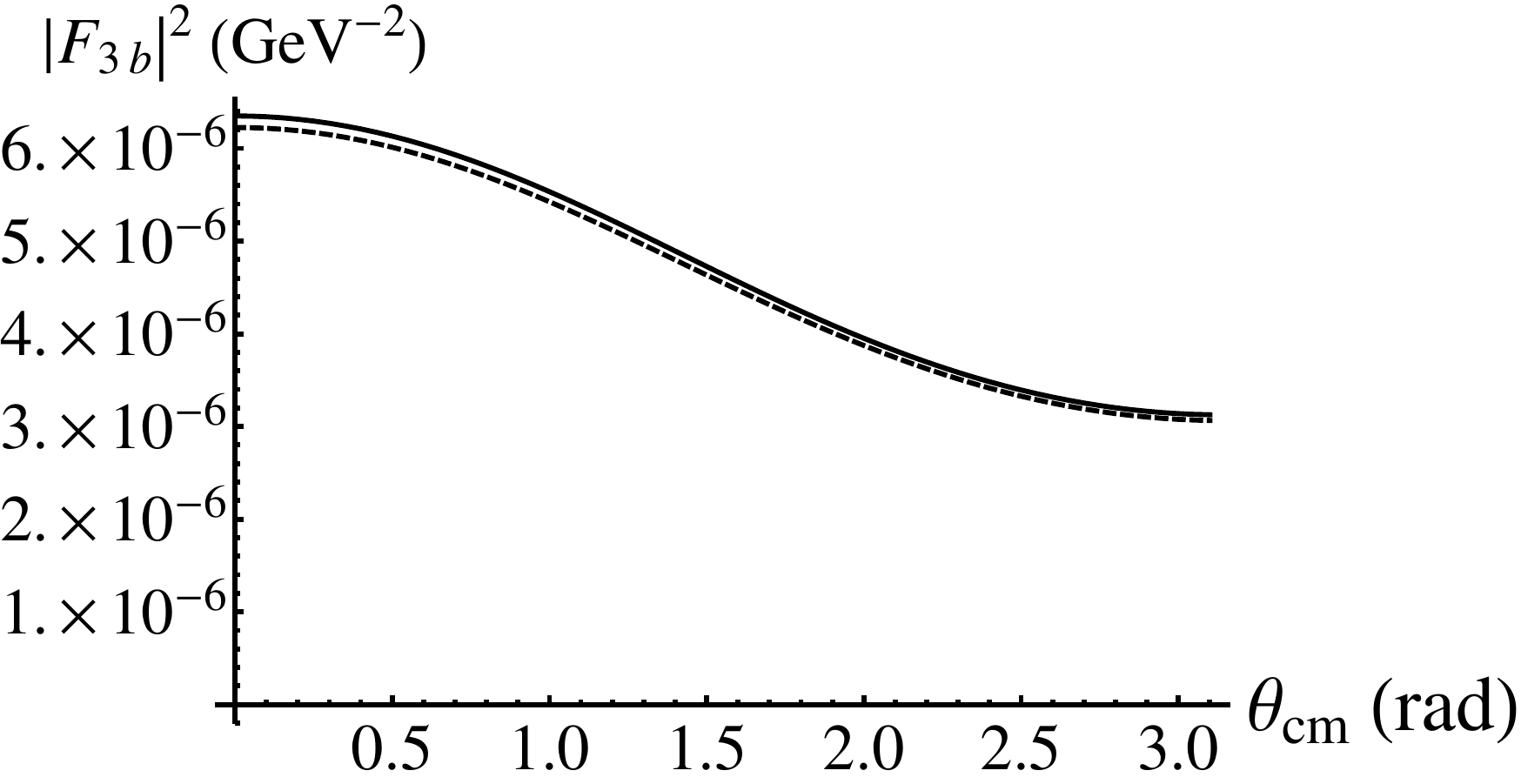}
        }%

    \end{center}
    \caption{%
       Squares of amplitudes for $\nu=6.5\;GeV$, $T^*_{Vn}=30$ MeV, $p_n=p_{n,min}$.  The solid curve is the total (on-shell plus off-shell parts), while the dashed curve is only including the on-shell part of the amplitude.
     }%
   \label{fig:ampstotandon65}
\end{figure}

Since the $J/\psi$-nucleon scattering length is expected to be small (much smaller than e.g. the proton-neutron scattering length), the $J/\psi$-neutron rescattering diagram $F_{3a}$ should be a small contribution to the total amplitude.  This is born out in the next subsection, where $F_{3a}$ is calculated using a model potential and wavefunction, for a value of the scattering length of the order of that predicted by theoretical models.

\subsection{$J/\psi$-neutron Rescattering Diagrams and the Scattering Length}

Diagram $F_{3a}$ (Fig. \ref{fig:scattlengthdiag}) is the $J/\psi$-neutron rescattering diagram.  This is the diagram where the $J/\psi$ and neutron scatter from each other with small relative momentum; hence this amplitude will involve the scattering length for the $J/\psi$-neutron interaction. 

\subsubsection{The amplitude $F_{3a}$}

\begin{figure}[tbp]
     \begin{center}
            \includegraphics[width=0.4\textwidth]{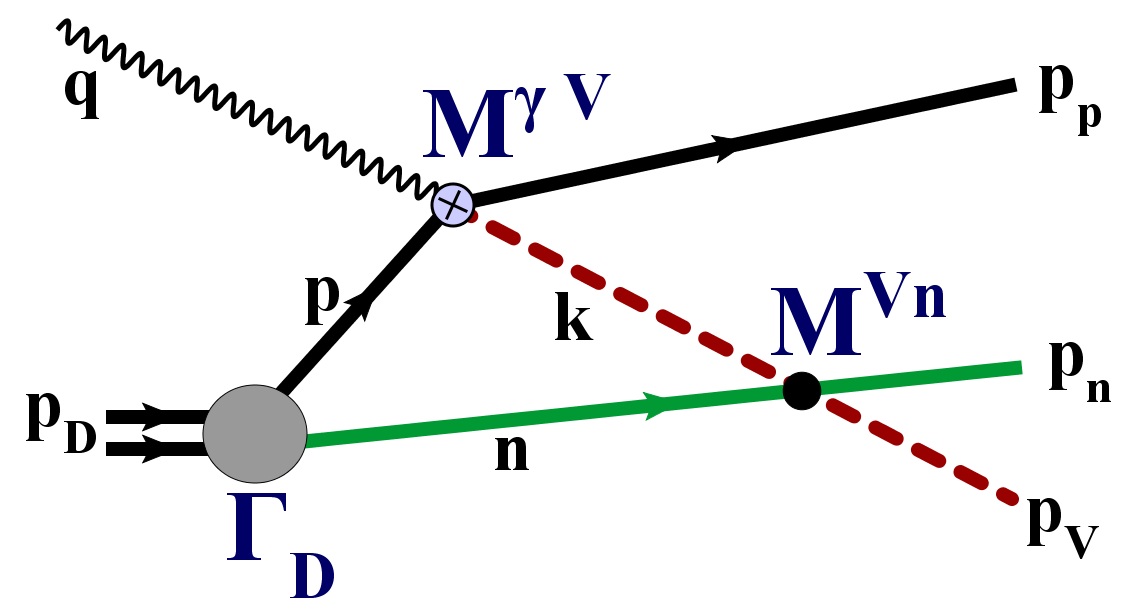}

    \end{center}
    \caption{%
        $F_{3a}$
     }%
   \label{fig:scattlengthdiag}
\end{figure}

The on-shell and off-shell parts of $F_{3a}$ are
\be
\label{F3aon}
F_{3a}^{on}=-i\pi \frac{1}{\sqrt{2m(2\pi)^3}}\frac{1}{2\vert\mathbf{p}_{Vn}\vert}\int_0^{2\pi}d\phi\int_{\vert n_-\vert}^{n_+}dn\; n\Psi(n) {\cal M}^{\gamma V} {\cal M}^{Vn}
\ee
and
\be
\label{F3aoff}
F_{3a}^{off}=\frac{1}{\sqrt{2m(2\pi)^3}}\frac{1}{2\vert\mathbf{p}_{Vn}\vert}\int_0^{2\pi}d\phi\int_0^{\infty}dn\; n\Psi(n)\;\;{\cal P}\int d\cos\theta \frac{  {\cal M}^{\gamma V} {\cal M}^{Vn} }{f_{12}(n)+\cos{\theta}}.
\ee
where $\mathbf{p}_{Vn}\equiv \mathbf{p}_V +\mathbf{p}_n$, and $\theta$ is the angle between $\mathbf{p}_{Vn}$ and $\mathbf{n}$ (see Fig. \ref{fig:coordsystem}).  For $T^*_{Vn}=0$ we have $n_+=\vert n_-\vert$ (see Eq. \ref{nplusminus}), and so $F_{3a}^{on}=0$.  For $T^*_{Vn}=30$ MeV, $F_{3a}^{on}$ will be small because $n_-\gtrsim 0.6\;GeV$ for the possible JLab kinematics .  Thus the main contribution to $F_{3a}$ is from $F_{3a}^{off}$, for which the intermediate-state $J/\psi$ is always off-mass-shell.  However, because of the propagator denominator $f_{12}(n)+\cos{\theta}$, where 
\be
f_{12}(n)+\cos{\theta}=\frac{s_{Vn}+m^2-m_V^2-2E_{Vn}\omega_n}{2\vert\mathbf{p}_{Vn}\vert n}+\cos{\theta}=\frac{1}{2\vert\mathbf{p}_{Vn}\vert n}\Bigl(k^2-m_V^2\Bigr),
\ee
contributions to $F_{3a}^{off}$ from values of $\mathbf{n}$ for which $k$ is far off-mass-shell will be small.  To obtain estimates of $F_{3a}^{off}$, we will therefore evaluate it using on-mass-shell values of $ {\cal M}^{\gamma V}$ and ${\cal M}^{Vn}$.

For small relative momentum of the $J/\psi$-neutron pair, ${\cal M}^{Vn}$ is related to the  $J/\psi$-neutron scattering length.  The relation between the invariant amplitude ${\cal M}^{Vn}$ and the scattering amplitude $f(k,\theta)$ is~\cite{pdg}
\begin{equation}
{\cal M}=-8\pi\sqrt{s_{Vn}}\; f(k,\theta)  \end{equation}
for the on-energy-shell amplitudes.   Note that in $F_{3a}^{off}$ the amplitude ${\cal M}^{Vn}$ is off-energy-shell if it's on-mass-shell, since the magnitude of the final $J/\psi$-neutron relative momentum is not equal to the magnitude of their initial relative momentum, since $\mathbf{n}$ is being integrated over.

\subsubsection{Scattering Length}

The definition of the scattering length for a given 2-body interaction is that it is the (negative of) the zero-energy limit of the scattering amplitude in the center of mass frame.  For low-energy scattering, only the $S$-wave will contribute, and so the scattering amplitude is just a constant (independent of scattering angle), i.e.
\begin{equation}
\lim_{k\to 0} f(k,\theta)=-a  \end{equation}
which defines the scattering length $a$~\cite{joachain75}.  The scattering amplitude $f$ here is the on-energy-shell amplitude, so the initial and final relative momenta are $\mathbf{k}_i$ and $\mathbf{k}_f$ with $\vert \mathbf{k}_i\vert=\vert \mathbf{k}_f\vert=k$.  We require the off-energy-shell amplitude, which is given by
\begin{equation}
\label{offshellamp}
f^{VN}(\mathbf{k}_1,\mathbf{k}_2)=-(2\pi)^2\mu \;\langle\mathbf{k}_2\vert V \vert \Psi_{\mathbf{k}_1}^{(+)} \rangle  = -(2\pi)^2\mu \;\langle\Psi_{\mathbf{k}_2}^{(-)}\vert V \vert {\mathbf{k}_1} \rangle = -(2\pi)^2\mu \;\langle\mathbf{k}_1\vert V \vert \Psi_{\mathbf{k}_2}^{(-)} \rangle^*  \end{equation}
where $\mu$ is the reduced mass; $\mathbf{k}_1$ is the initial relative momentum (in terms of $\mathbf{n}$, $\mathbf{k}$ in Fig. \ref{fig:scattlengthdiag}); $\mathbf{k}_2$ is the final relative momentum (in terms of $\mathbf{p}_V$, $\mathbf{p}_n$ in Fig. \ref{fig:scattlengthdiag}); and $\Psi_{\mathbf{k}_2}$ is the exact scattering wavefunction for asymptotic relative momentum $\mathbf{k}_2$.  
Since this off-energy-shell scattering amplitude depends on the scattering wavefunction $\Psi_{\mathbf{k}_2}^{(-)}$ and the potential $V$, both of which are unknown for $J/\psi$-nucleon elastic scattering, we will resort to models in order to estimate the amplitude.

We normalize our $S$-wave wavefunction $\Psi$, and define the radial wavefunction $u(r)$, by:
\begin{equation}
\label{wfdef}
\Psi(r)=\frac{1}{\sqrt{(2\pi)^3}}\;e^{i\delta(k)}\frac{u(r)}{r} 
 \end{equation}
where $\delta(k)$ is the $S$-wave phase shift.  
In order to calculate the matrix element, we can either specify a model potential and solve the Schrodinger equation for the wavefunction $u(r)$, or instead specify a model zero-energy wavefunction $u^0(r)$ which determines the potential $V(r)$ via the Schrodinger equation, and use that potential to solve for the wavefunction for $k\ne 0$.  We will do the second procedure, choosing a model zero-energy wavefunction which satisfies the minimal constraints imposed by the Schrodinger equation.  

We assume the $J/\psi$-nucleon potential is of finite range, and so is zero for $r$ larger than some distance $R$.   
The phase-shift $\delta(k)$ satisfies the following well-known properties~\cite{joachain75} as $k\to 0$ :  
\begin{enumerate}
\item for a repulsive potential, or an attractive potential that doesn't admit a bound state:  $\delta\to -ak$ as $k\to 0$;
\item for an attractive potential which admits a single bound state:  $\delta\to \pi-ak$ as $k\to 0$
\end{enumerate}
The zero-energy wavefunction $\Psi(r)$ outside the range of the potential is then
\begin{equation}
\label{psiout}
\Psi_{out}^0(r)=\frac{1}{\sqrt{(2\pi)^3}}\;e^{i\delta(0)}\frac{u_{out}^0(r)}{r}=\frac{1}{\sqrt{(2\pi)^3}}\;\frac{r-a}{r}  
\end{equation}
for both cases, while the zero-energy radial wavefunction $u_{out}^0$ differs by a minus sign for the two cases; this is purely due to including the factor $e^{i\delta(k)}$ in the definition of $\Psi$ in \eq{wfdef}  (see Appendix \ref{wavefuncappend} for more details).  The superscript $0$ indicates $k=0$.  Furthermore, my normalization conventions give $a>0$ for either a repulsive potential or an attractive potential with a bound state, and $a<0$ for an attractive potential that doesn't admit a bound state.  In all cases $a$ is the intercept on the $r$-axis of $u_{out}^0(r)$ (Fig. \ref{fig:aForPotentials}).  

Theoretical calculations~\cite{brodsky97,kawanai2010} give values of $\vert a \vert\simeq 0.3\;fm$, with effective range $r_e\simeq 2.0\;fm$.  It is thought that the potential is attractive, but too weak to support a bound state.  Below, calculations of $F_{3a}$ are made for both cases of an attractive potential:  $a>0$ (bound state) and $a<0$ (no bound state).

\begin{figure}[tbp]
     \begin{center}
        \subfigure[Attractive potential that possesses a bound state.  $a> 0$ in this case.]{%
            \label{fig:boundstate}
            \includegraphics[width=0.6\textwidth]{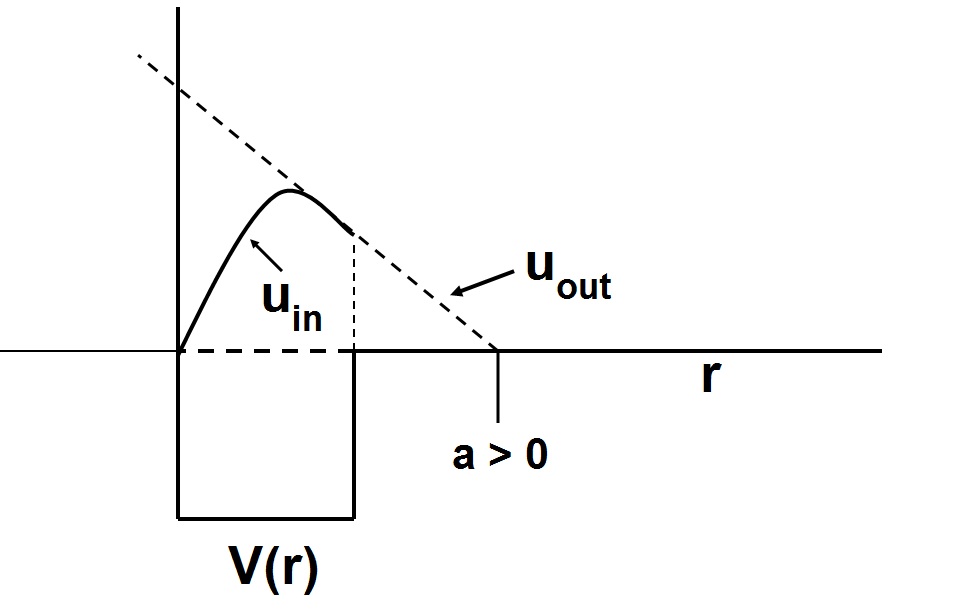}
        }%
        \hspace{0.5in}
         \subfigure[Attractive potential with no bound states.  $a<0$ in this case.]{%
           \label{fig:noboundstate}
           \includegraphics[width=0.6\textwidth]{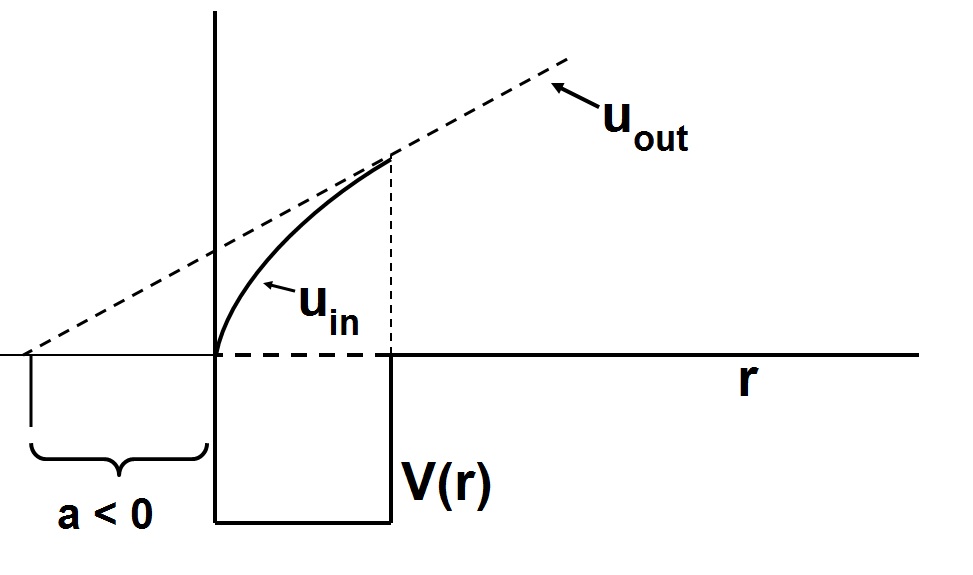}
        }\\ 

        \subfigure[Repulsive potential.  $a> 0$ in this case.]{%
            \label{fig:repulsive}
            \includegraphics[width=0.6\textwidth]{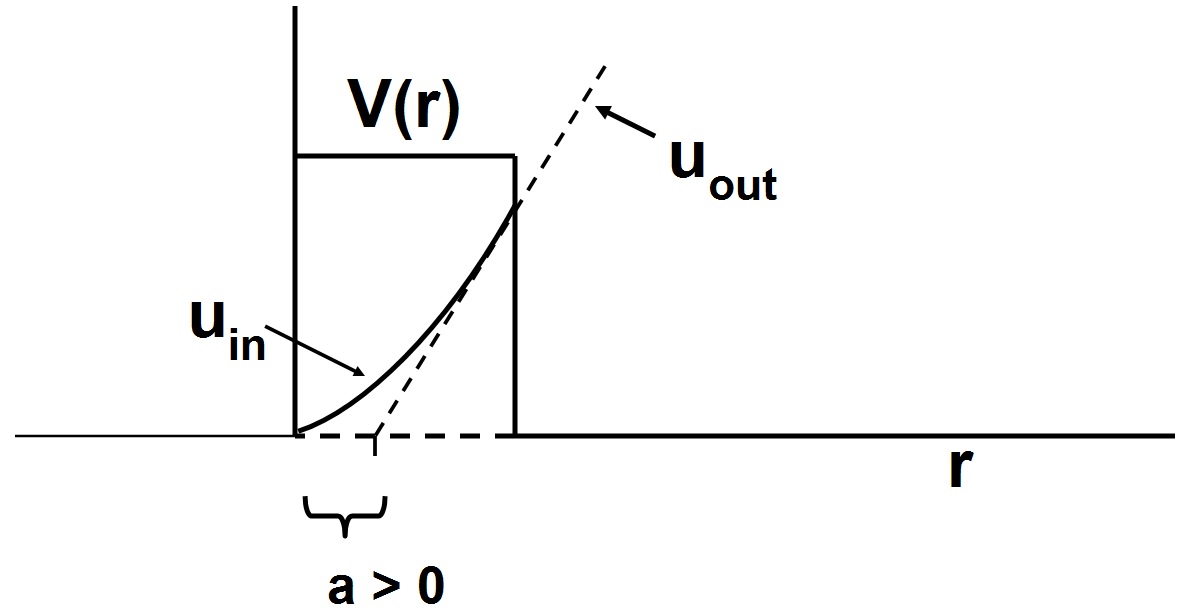}
        }%

    \end{center}
    \caption{%
       Scattering length, $u_{out}(r)$ (dashed line), and $u_{in}(r)$ (solid curve) for different types of potentials.
     }%
   \label{fig:aForPotentials}
\end{figure}

\subsection{Positive scattering length $a$}

For the case of a positive scattering length and attractive potential (which possesses a bound state), we have $u_{out}^0(r)=-(r-a)$ .  The properties imposed on the zero-energy wavefunction $u$ for $r<R$, required by the Schrodinger equation, are then:
\begin{equation}
\label{wfprop1}
u_{in}(0)=0  
\end{equation}
\begin{equation}
\label{wfprop2}
u_{in}(R)=u_{out}(R)=-(R-a)  
\end{equation}
\begin{equation}
\label{wfprop3}
u_{in}'(R)=u_{out}'(R)=-1  
\end{equation}
\begin{equation}
\label{wfprop4}
u_{in}''(R)=0  
\end{equation}
The first 3 are the standard requirements imposed by the Schrodinger equation, while the last one IS the Schrodinger equation at $r=R$ because $V(R)=0$.  We will assume the simplest form for $u$ that's consistent with these requirements, which is that $u$ is a cubic function of $r$.

\subsubsection{Model wavefunction for $k=0$}

The wavefunction $u_{in}$ that satisfies Eqs. \ref{wfprop1} - \ref{wfprop4} is
\begin{equation}
u_{in}^0(r)=\Bigl(-1+\frac{3a}{R}\Bigr) r + \Bigl(-\frac{3a}{R^2}\Bigr) r^2 +\Bigl (\frac{a}{R^3} \Bigr) r^3.
 \end{equation}
One further requirement on $u_{in}^0$ is that $u_{in}^0$ have no zeros on the interval $[0,R]$ (besides at $r=0$).  This ensures that the corresponding potential $V(r)$ is non-singular, since for $k=0$, $V(r)=\frac{1}{2\mu}\frac{u''}{u}$.  The zeros of $u_{in}^0$ are at $r=0$ and
\begin{equation}
\frac{r}{R}=\frac{3}{2}\pm\frac{1}{2}\sqrt{4\frac{R}{a}-3}  \end{equation}
and the right-hand-side must lie outside the range $[0,1]$.  This requires either
\begin{equation}
R<a \end{equation}
or
\begin{equation}
R>3a  \end{equation}
The potential is
\begin{equation}
\label{modelpot}
V(r)=\frac{1}{2\mu}\frac{1}{r}\frac{r-R}{R^2(3-\frac{R}{a})-3R r +r^2}  \end{equation}
and one can see that for $R>3a$ the potential is repulsive.  Therefore we require $R<a$.

\subsubsection{Model wavefunction for $k\ne 0$}

Given this model potential we can proceed to calculate the off-shell scattering amplitude \eq{offshellamp} and the amplitude $F_{3a}^{off}$, \eq{F3aoff}, once we calculate the wavefunction $u$ for non-zero $k$ for a given $a$ and $R$.  The potential does not admit analytic solutions for non-zero $k$, so we calculate them numerically.  However, we only require them for small $k$ (less than 100 MeV or so).  Taking $a=0.3\;fm$, $R=0.1\;fm$, Fig. \ref{fig:wfandoff1} and Fig. \ref{fig:wfandoff2} show $u_{in}$ for $k=0,\;50\;MeV$, and $100\;MeV$; there's virtually no difference between them.  Figs. \ref{fig:wfandoff3} and \ref{fig:wfandoff4} show the off-energy-shell amplitude $f^{Vn}(p,k)$ for $k=0,\;50\;MeV$, and $100\;MeV$; again there's virtually no variation of $f^{Vn}(p,k)$ for $k$ up to $100\;MeV$.
\begin{figure}[tbp]
     \begin{center}
        \subfigure[$u_{in}(r)$ for $k=0$, $50$, and $100$ MeV.  Solid black curve is for $k=0$, solid gray curve is for $k=50$ MeV; dashed curve is for $k=100$ MeV.]{%
            \label{fig:wfandoff1}
            \includegraphics[width=0.4\textwidth]{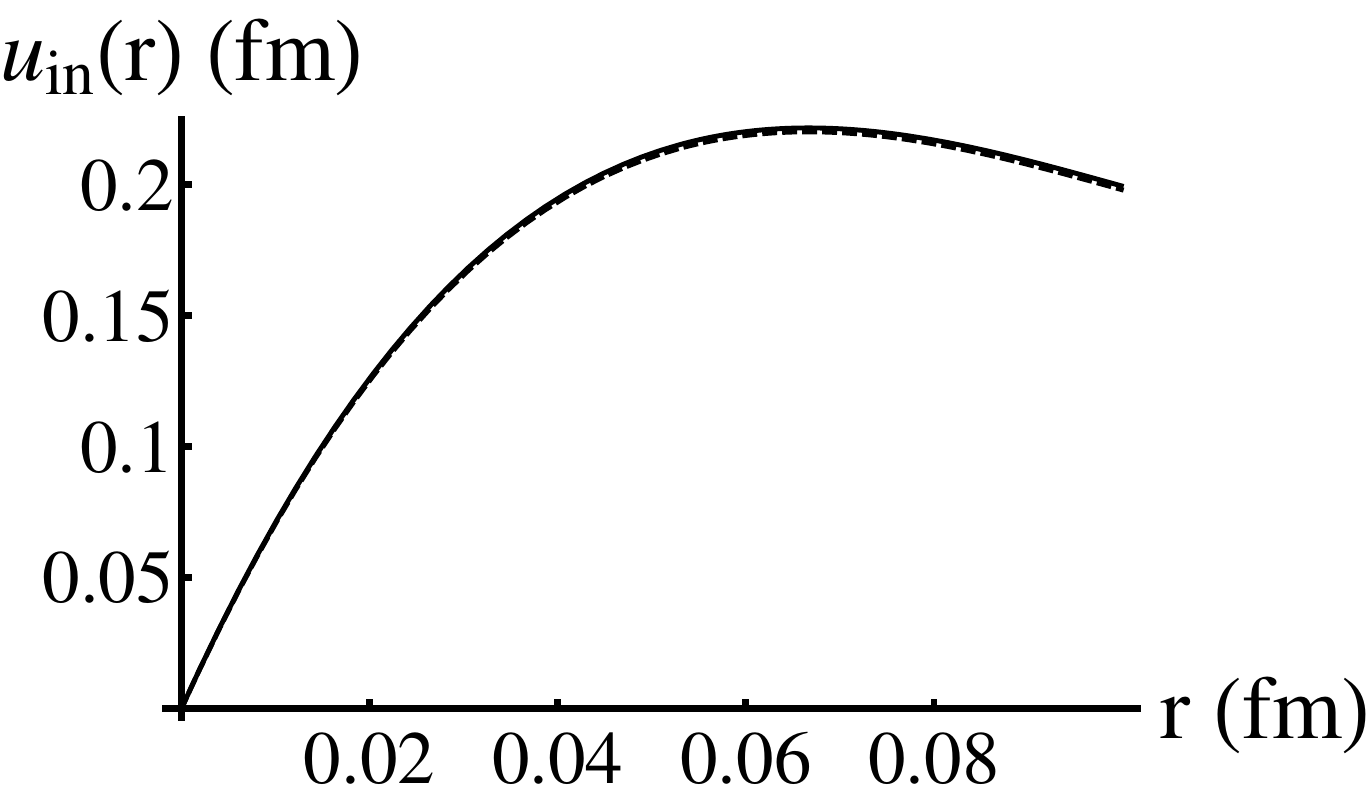}
        }%
        \hspace{0.5in}
         \subfigure[same as (a)]{%
           \label{fig:wfandoff2}
           \includegraphics[width=0.4\textwidth]{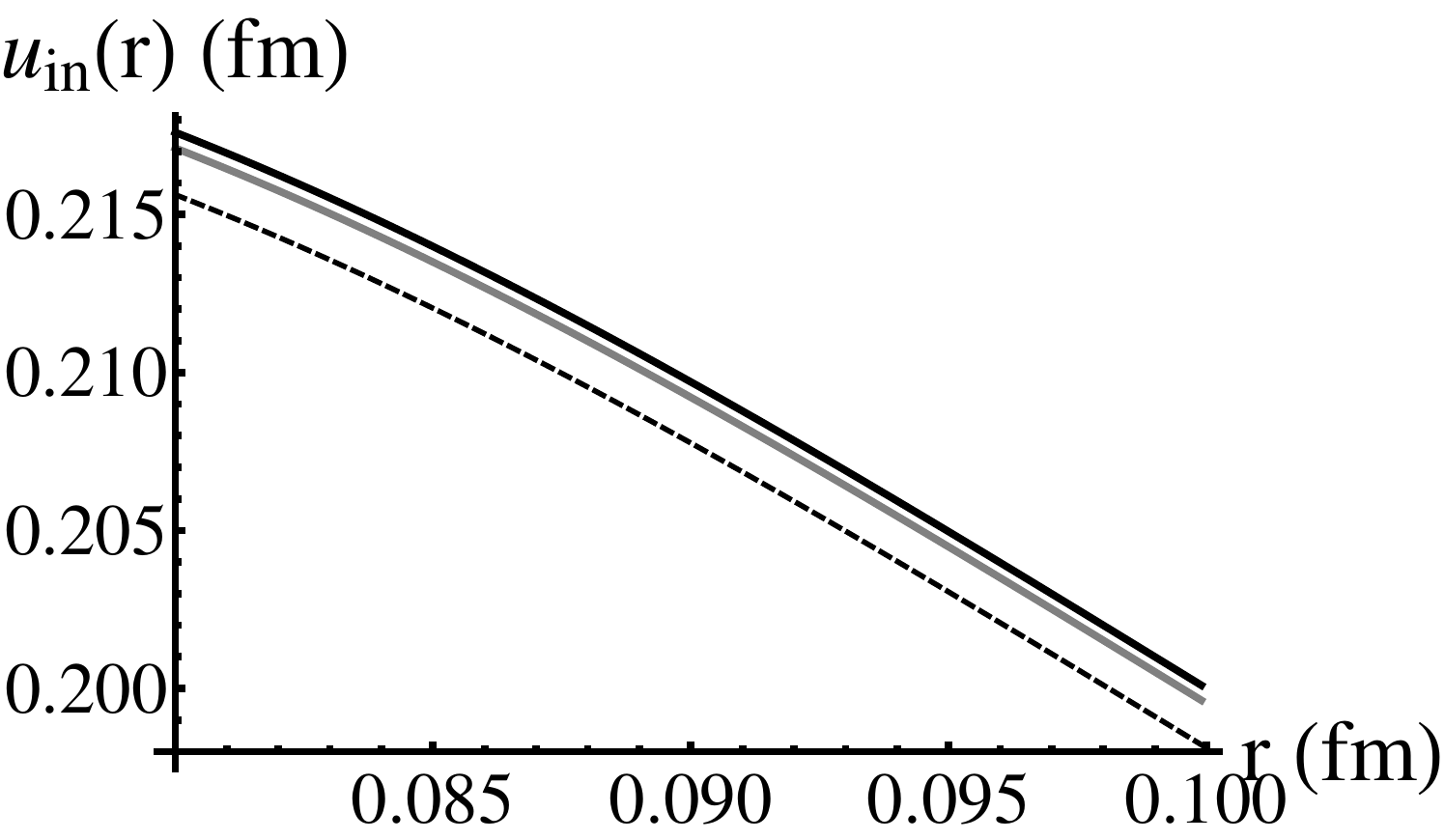}
        }\\ 

        \subfigure[Off-shell amplitude for $k=0$ (solid black), $50$ (black dashed), and $100$ MeV (gray dashed)]{%
            \label{fig:wfandoff3}
            \includegraphics[width=0.4\textwidth]{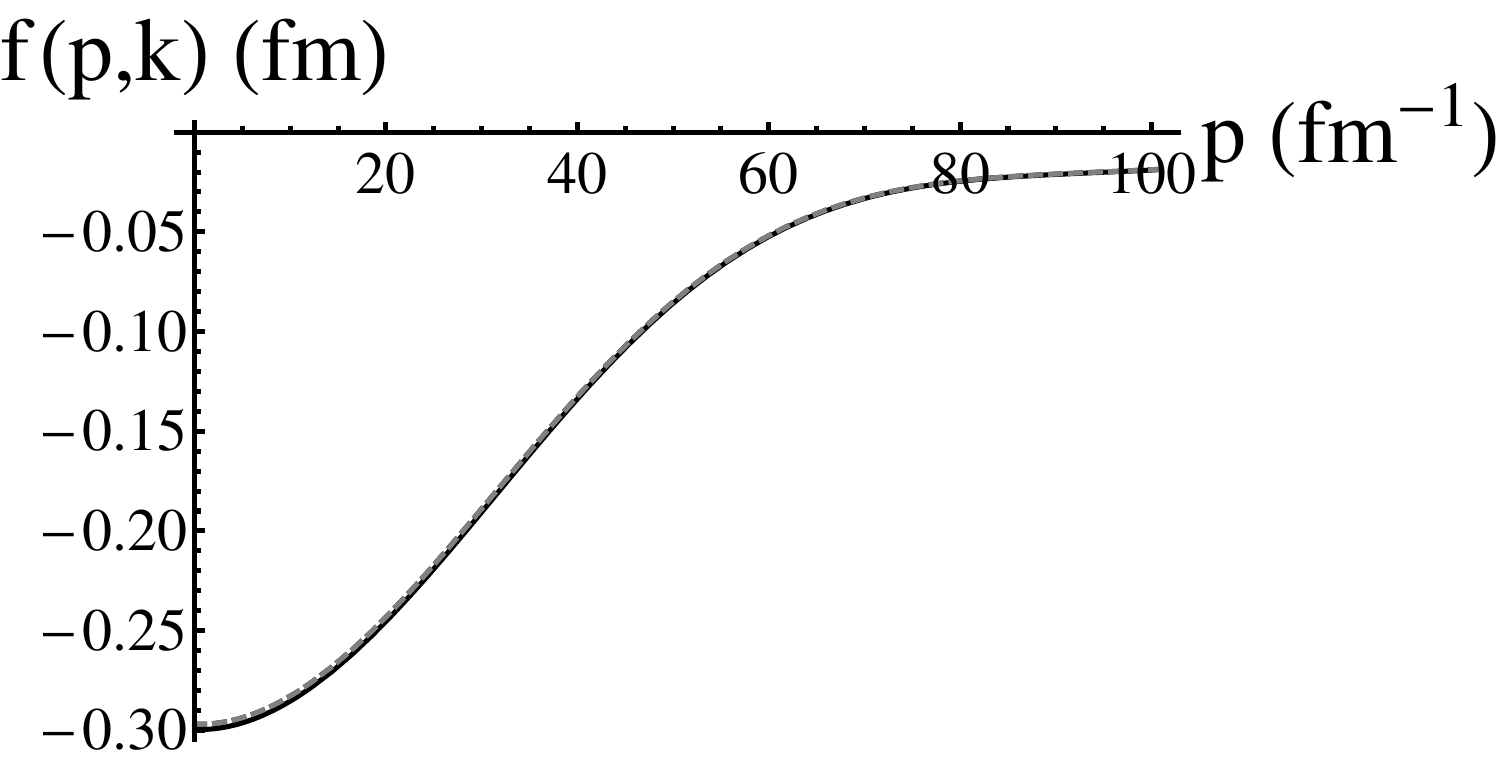}
        }%
         \hspace{0.5in} 
        \subfigure[same as (c)]{%
            \label{fig:wfandoff4}
            \includegraphics[width=0.4\textwidth]{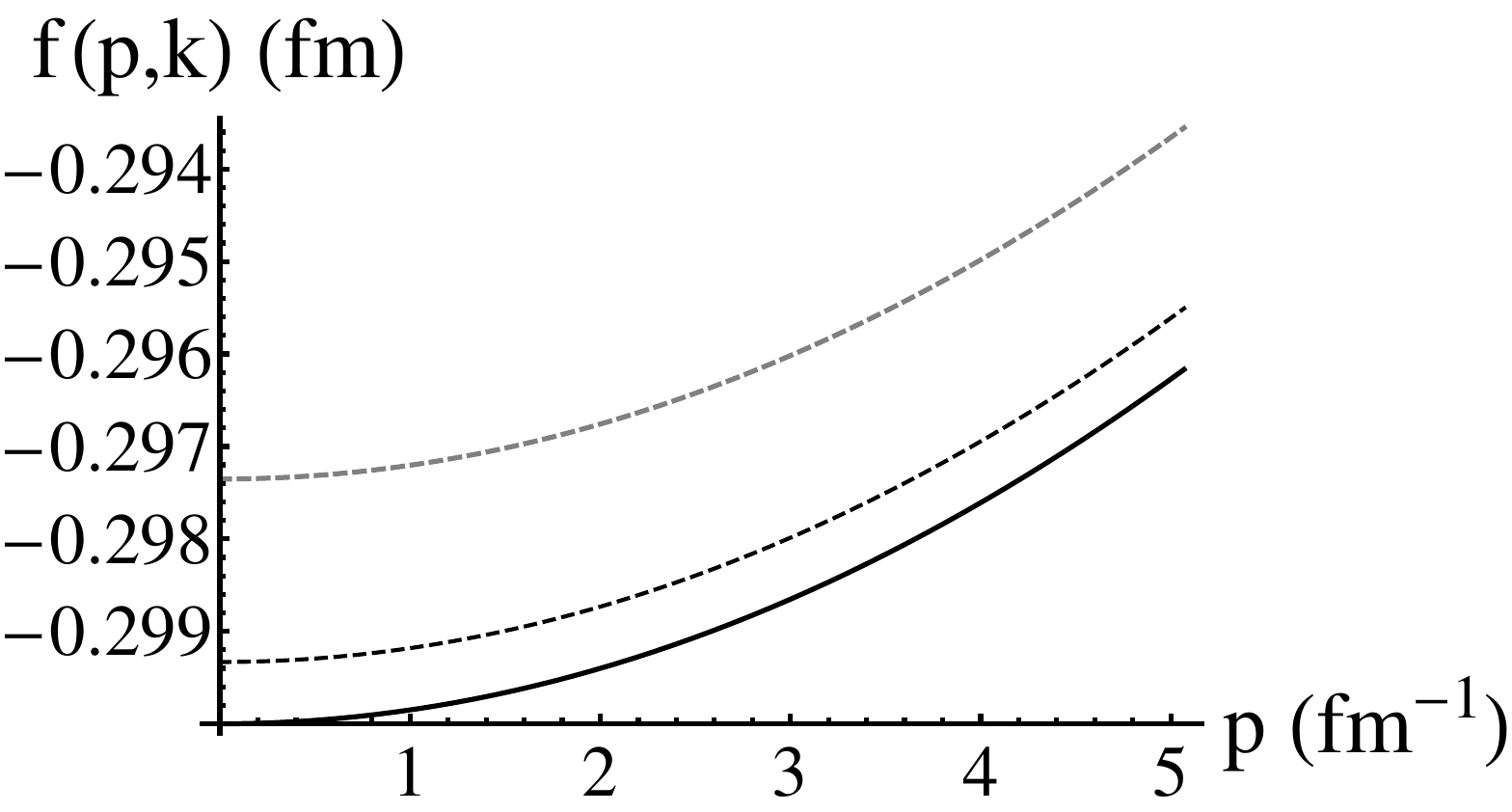}
        }%

    \end{center}
    \caption{%
       Wavefunction and off-shell amplitudes for $a=0.3\;fm$, $R=0.1\;fm$, using the model potential of Eq. \ref{modelpot}.
     }%
   \label{fig:wfandoff}
\end{figure}

Since we are only interested in the $J/\psi$-neutron relative momentum up to around 100 MeV, it is legitimate to approximate the off-energy-shell amplitude $f^{Vn}(p,k)\simeq f^{Vn}(p,0)$ for the range of $k$ we are interested in.  For our model wavefunction $u^0_{in}$ we can evaluate $ f^{Vn}(p,0)$ analytically:
\be
\begin{split}
f^{Vn}(\mathbf{p},0)&= -(2\pi)^2\mu \;\langle\mathbf{p}\vert V \vert \Psi_{0}^{(-)} \rangle^*=-2\pi^2  \;\langle\mathbf{p}\vert U \vert \Psi_{0}^{(-)} \rangle^*\\
&=\frac{-2\pi^2}{\sqrt{(2\pi)^3}}\int d^3r e^{-i\mathbf{p}\cdot\mathbf{r}}U(r)\Psi(r)=\frac{1}{p}\int_0^R dr\;\sin{pr}\;U(r)\;u_{in}^0(r) \\
f^{Vn}(\mathbf{p},0)&=\frac{6a}{p^2 R^2}\Bigl(-1+\frac{\sin{pR}}{pR}\Bigr).\\ 
\end{split} 
\ee
The momentum $p$ appearing in $f(p,0)$ is the relative momentum of the $J/\psi$-neutron pair in their center-of-mass frame, before they scatter in diagram $F_{3a}$; it is thus the magnitude of $\mathbf{n}$ (or -$\mathbf{k}$) in the outgoing $V-n$ center-of-mass frame, and so we must boost $n$ to that frame.

\subsection{Negative scattering length $a$}

If the potential is attractive but too weak to support a bound state, then $a<0$ and we have $u_{out}^0(r)=r-a$ .  The properties imposed on the zero-energy wavefunction $u$ for $r<R$, required by the Schrodinger equation, are then:

\begin{equation}
\label{wfprop1b}
u_{in}(0)=0  
\end{equation}
\begin{equation}
\label{wfprop2b}
u_{in}(R)=u_{out}(R)=R-a  
\end{equation}
\begin{equation}
\label{wfprop3b}
u_{in}'(R)=u_{out}'(R)=1  
\end{equation}
\begin{equation}
\label{wfprop4b}
u_{in}''(R)=0  
\end{equation}
In this case the zero-energy wavefunction is
\begin{equation}
\label{negau}
u_{in}^0(r)=\Bigl(1-\frac{3a}{R}\Bigr) r + \Bigl(\frac{3a}{R^2}\Bigr) r^2 -\Bigl (\frac{a}{R^3} \Bigr) r^3.  
\end{equation}
The requirement that $u$ have no zeros on $[0,R]$ imposes no restriction on $a$ and $R$ in this case.  Theoretical calculations give $a$ around $-0.3\;fm$, and effective range $r_e\simeq 2.0\;fm$~\cite{brodsky97,kawanai2010}.  Using these values with our model wavefunction and potential implies $R=1.3\;fm$.  The wavefunction and off-energy-shell scattering amplitude for this case are shown in Fig. \ref{fig:wfandoffnega}.  Again there's very little difference between the curves for $k$ from $0$ to $0.1\;GeV$, so to calculate $F_{3a}$ in this case we will again approximate $f^{Vn}(\mathbf{p},\mathbf{k})\simeq f^{Vn}(\mathbf{p},0)$.  We have for $a<0$:
\be
f^{Vn}(\mathbf{p},0)=\frac{6a}{p^2 R^2}\Bigl(1-\frac{\sin{pR}}{pR}\Bigr) .
\ee

\begin{figure}[tbp]
     \begin{center}
        \subfigure[$u_{in}(r)$ for $k=0$, $50$, and $100$ MeV.  Solid black curve is for $k=0$, solid gray curve is for $k=50$ MeV; dashed curve is for $k=100$ MeV.]{%
            \label{fig:wfandoffNega1}
            \includegraphics[width=0.4\textwidth]{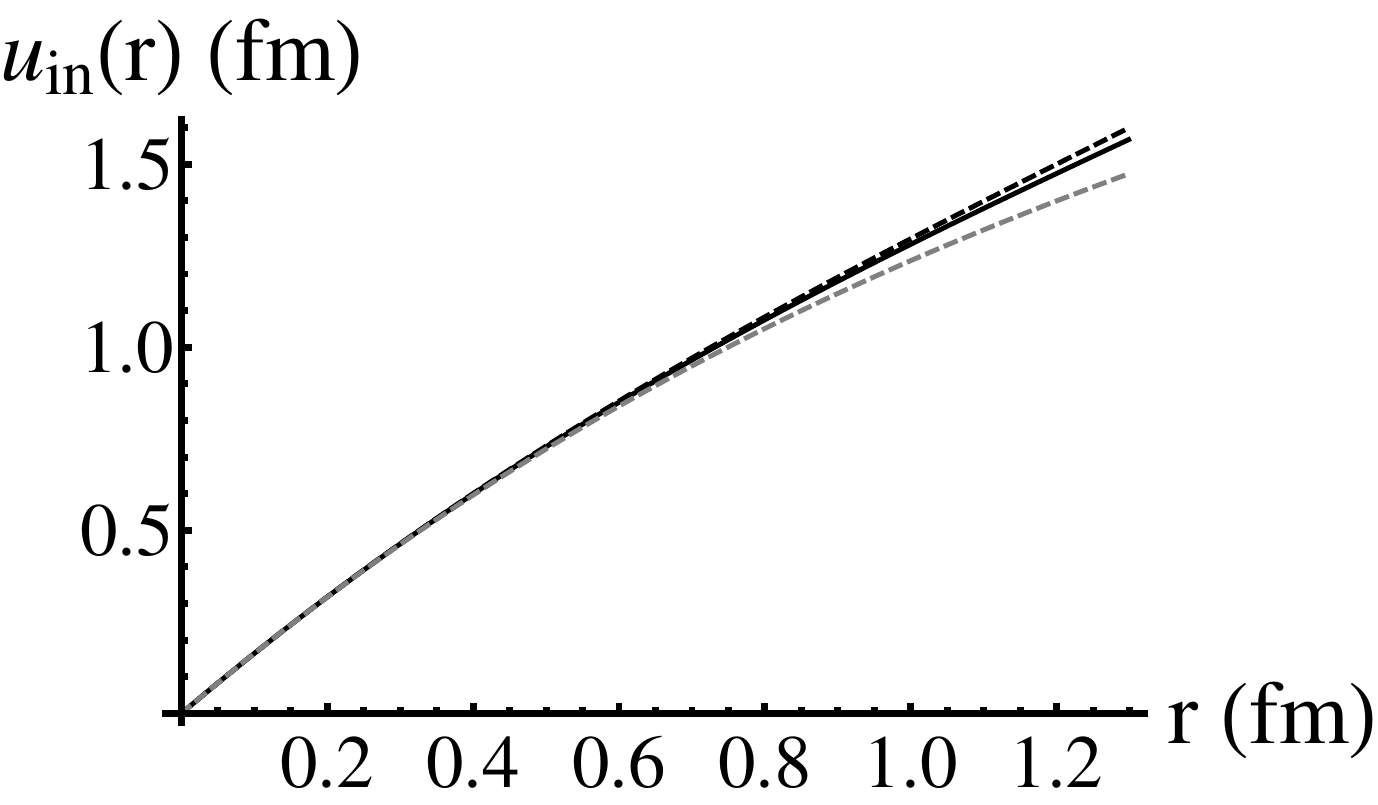}
        }%
        \hspace{0.5in}
         \subfigure[Off-shell amplitude for $k=0$ (solid black), $50$ (black dashed), and $100$ MeV (gray dashed)]{%
           \label{fig:wfandoffNega2}
           \includegraphics[width=0.4\textwidth]{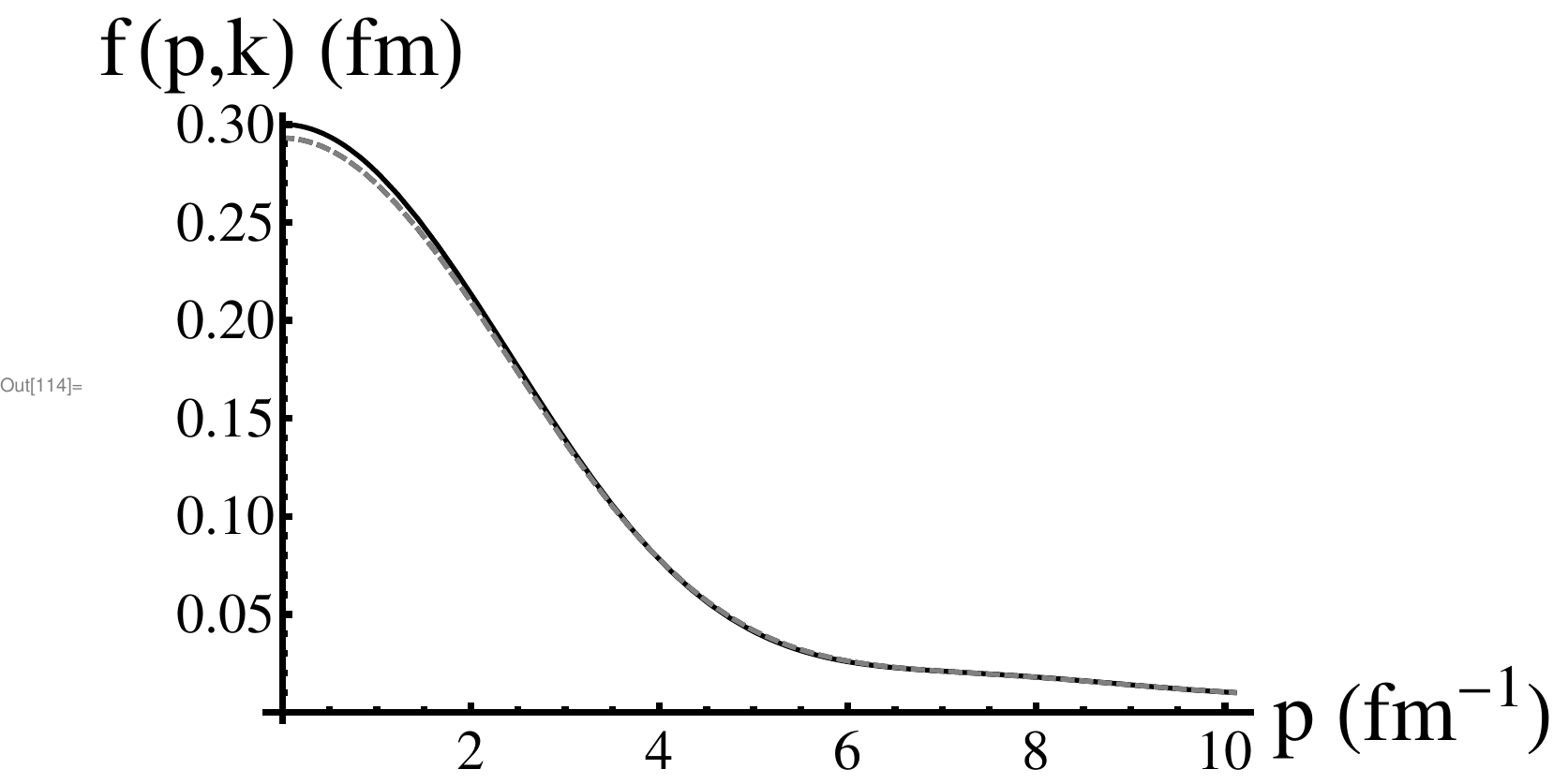}
        }\\ 
%

    \end{center}
    \caption{%
       Wavefunction and off-shell amplitudes for $a=-0.3\;fm$ and $r_e=2.0\;fm$ for the model potential Eq. \ref{modelpot}.
     }%
   \label{fig:wfandoffnega}
\end{figure}

\subsection{Results}

\begin{figure}[tbp]
     \begin{center}
        \subfigure[Squares of individual amplitudes.   Dashed curve is the square of the total amplitude.  Not shown are $F_{1a}$, $F_{2a}$, which are negligible.]{%
            \label{fig:squares9a}
            \includegraphics[width=0.8\textwidth]{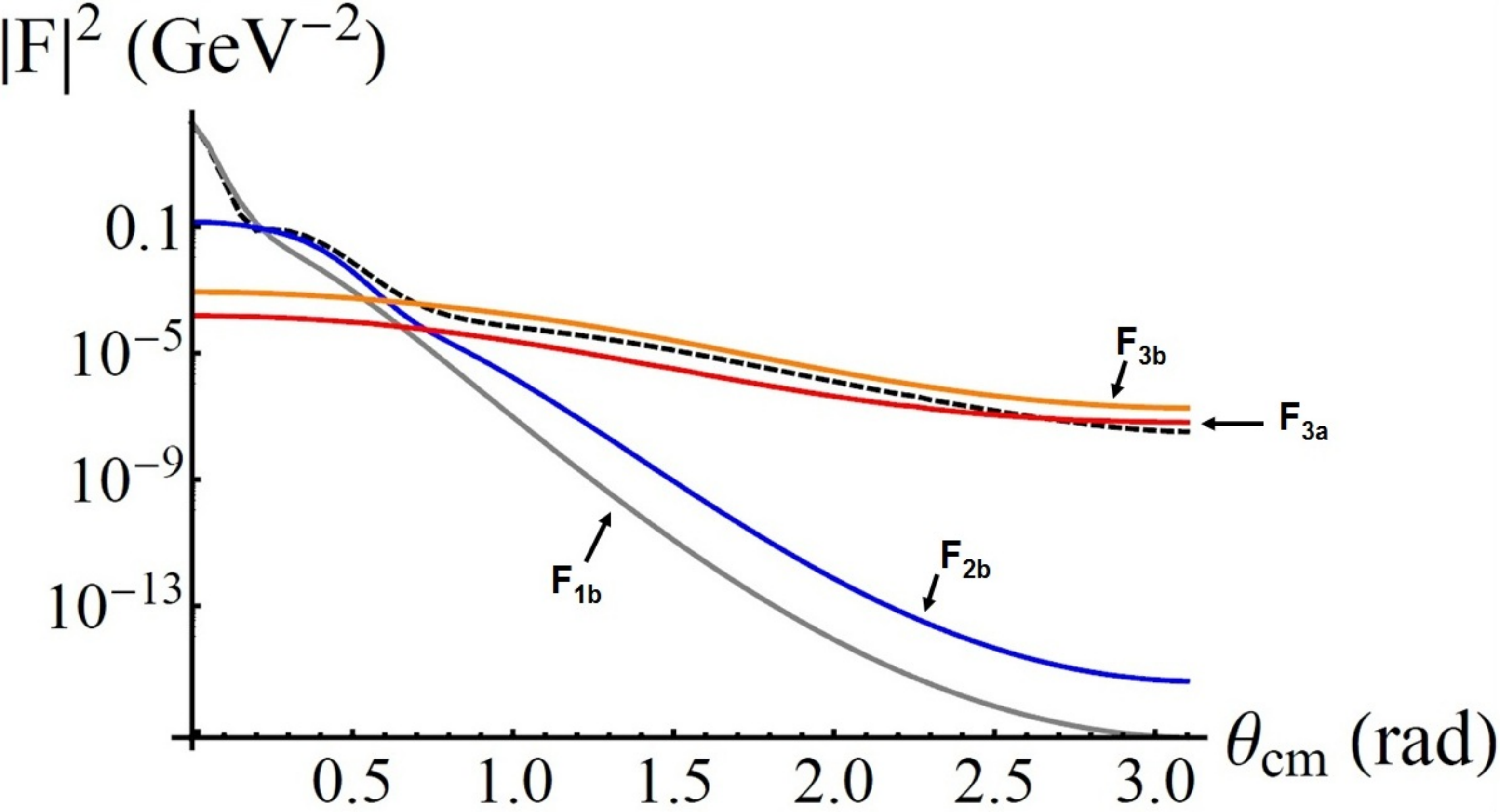}
        }%
        \hspace{0.5in}
         \subfigure[Square of total amplitude.  Solid: includes all amplitudes.  Dashed: omitting $F_{3a}$.]{%
           \label{fig:squaretot9b}
           \includegraphics[width=0.8\textwidth]{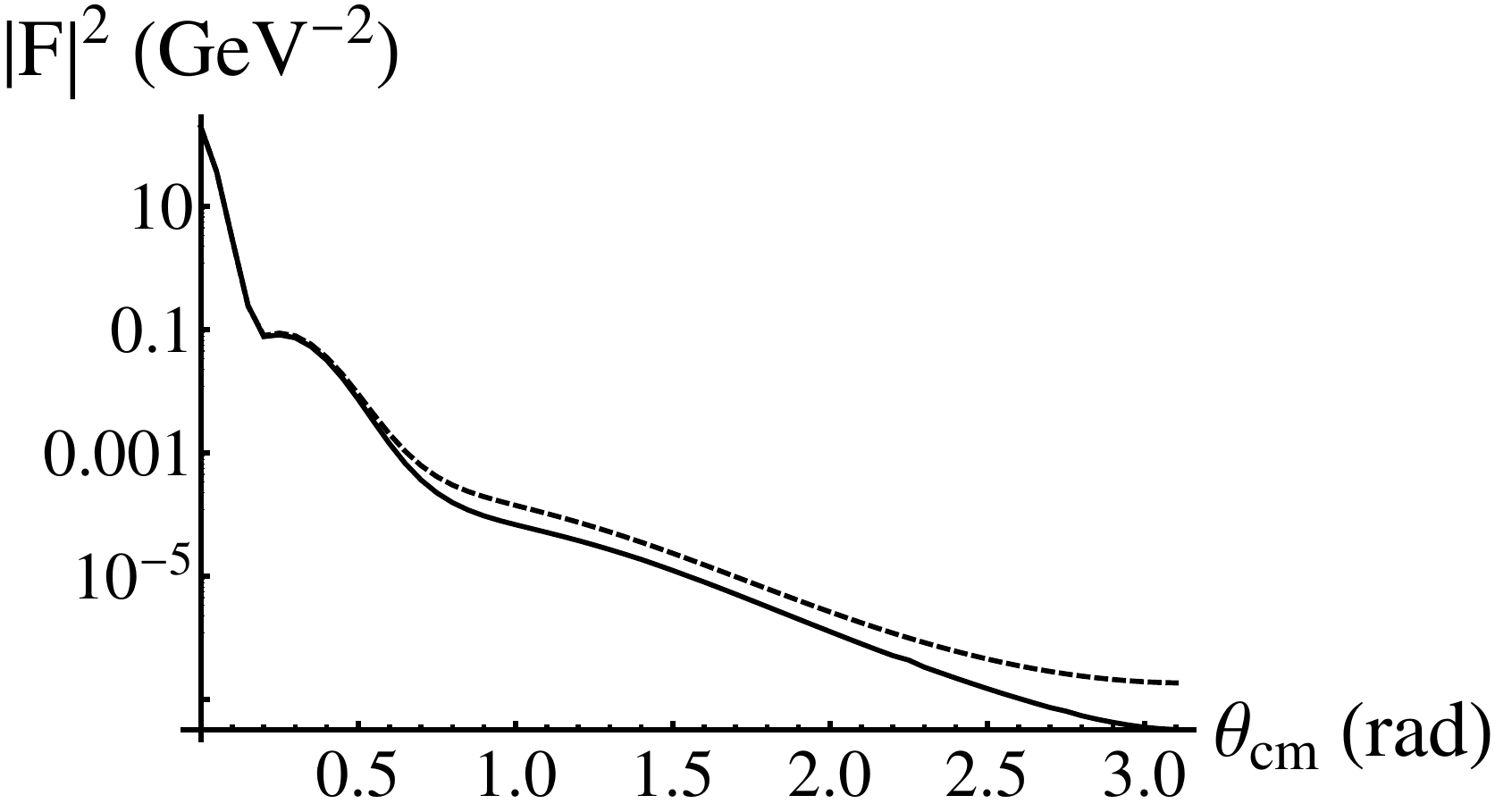}
        }\\ 
%

    \end{center}
    \caption{%
     Squares of amplitudes for $\nu=9\;GeV$, $T^*=0$, and $a=0.3\;fm$, $R=0.1\;fm$.
     }%
   \label{fig:ampsSquared9}
\end{figure}

\begin{figure}[tbp]
     \begin{center}
        \subfigure[Squares of individual amplitudes.  Dashed curve is the square of the total amplitude.]{%
            \label{fig:squares65a}
            \includegraphics[width=0.8\textwidth]{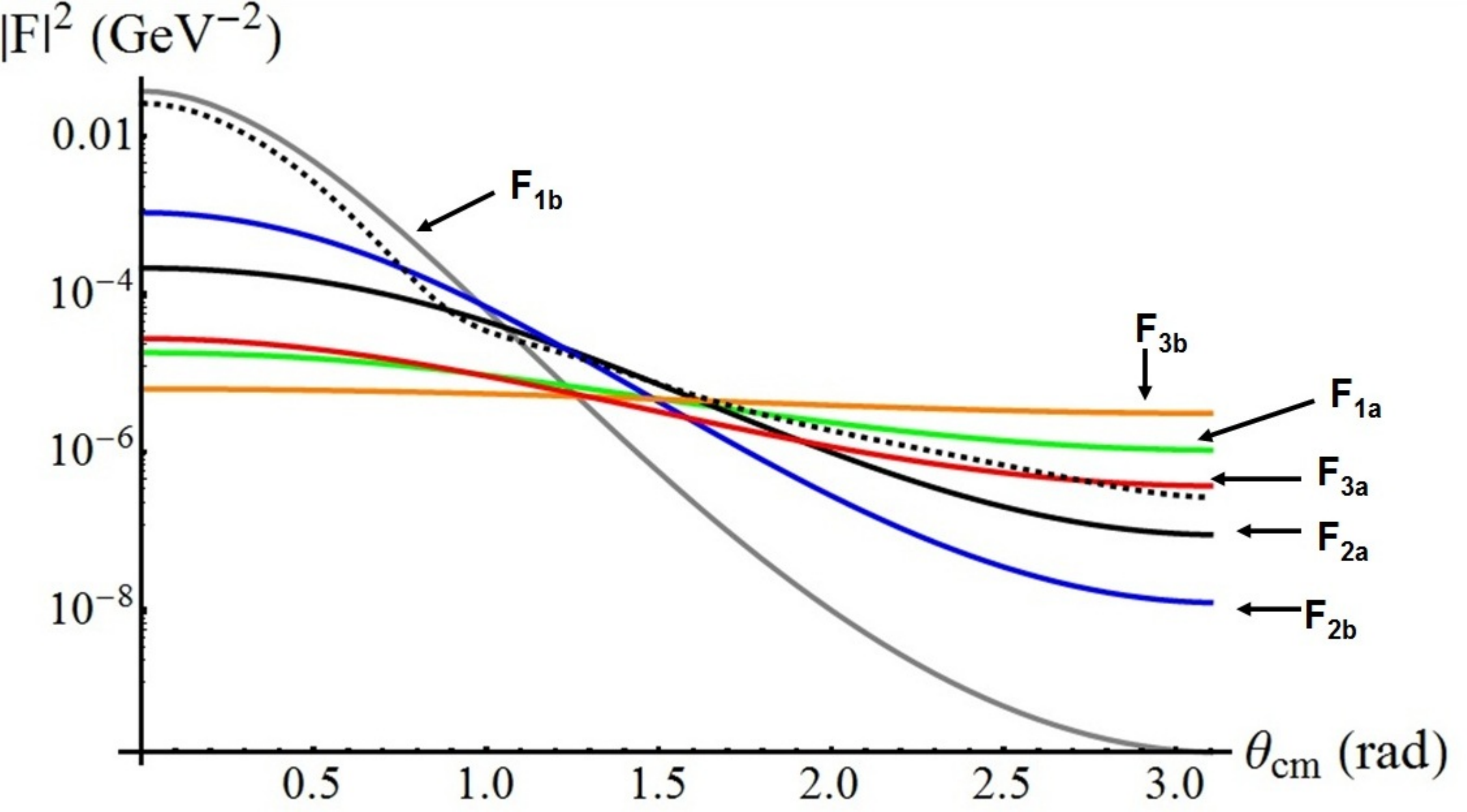}
        }%
        \hspace{0.5in}
         \subfigure[Square of total amplitude.  Solid: includes all amplitudes.  Dashed: omitting $F_{3a}$.]{%
           \label{fig:squaretot65b}
           \includegraphics[width=0.8\textwidth]{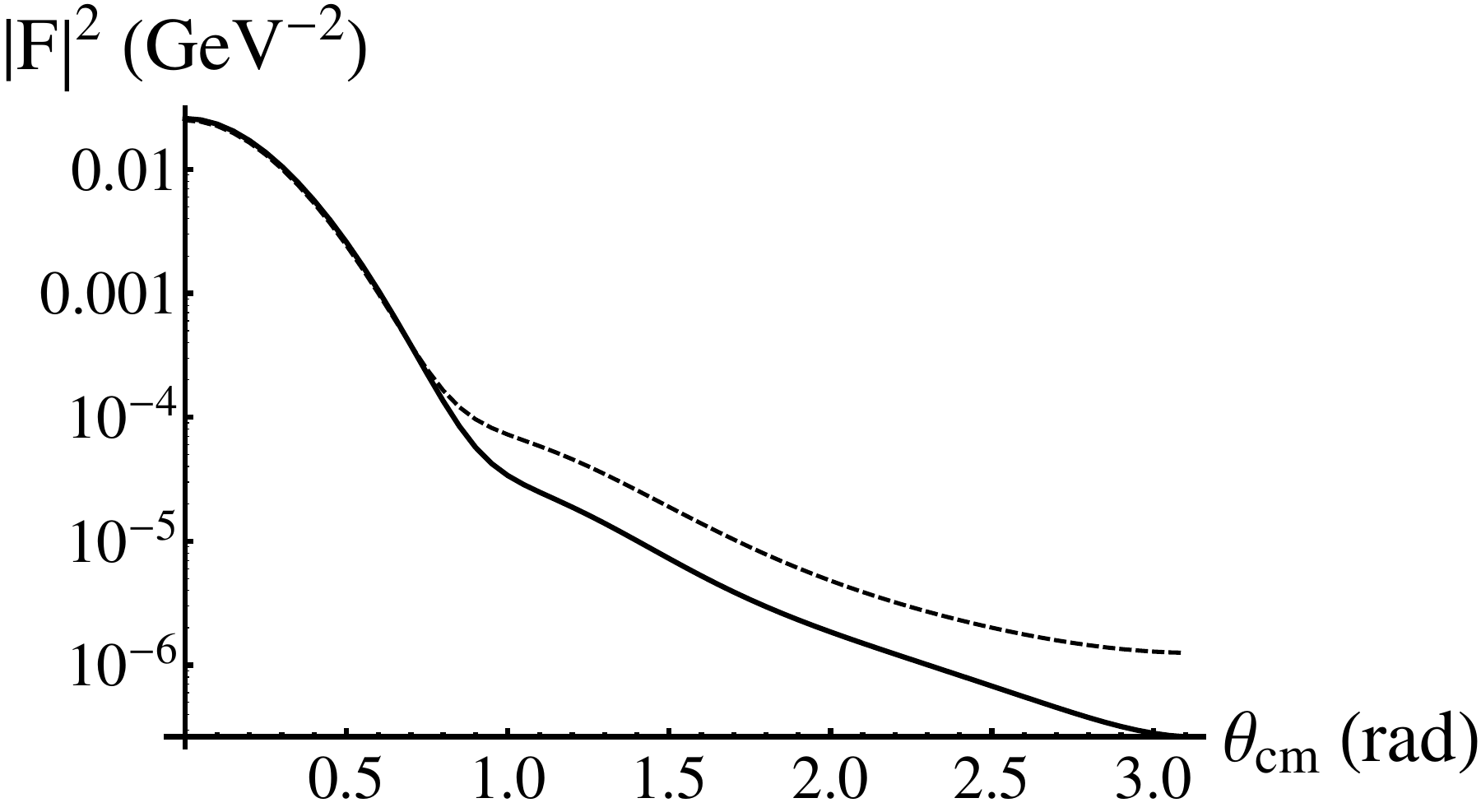}
        }\\ 
%

    \end{center}
    \caption{%
      Squares of amplitudes for $\nu=6.5\;GeV$, $T^*=0$, and $a=0.3\;fm$, $R=0.1\;fm$.
     }%
   \label{fig:ampsSquared65}
\end{figure}

\begin{figure}[tbp]
     \begin{center}
        \subfigure[Squares of individual amplitudes.   Dashed curve is the square of the total amplitude.  Not shown are $F_{1a}$, $F_{2a}$, which are negligible.]{%
            \label{fig:squares9c}
            \includegraphics[width=0.8\textwidth]{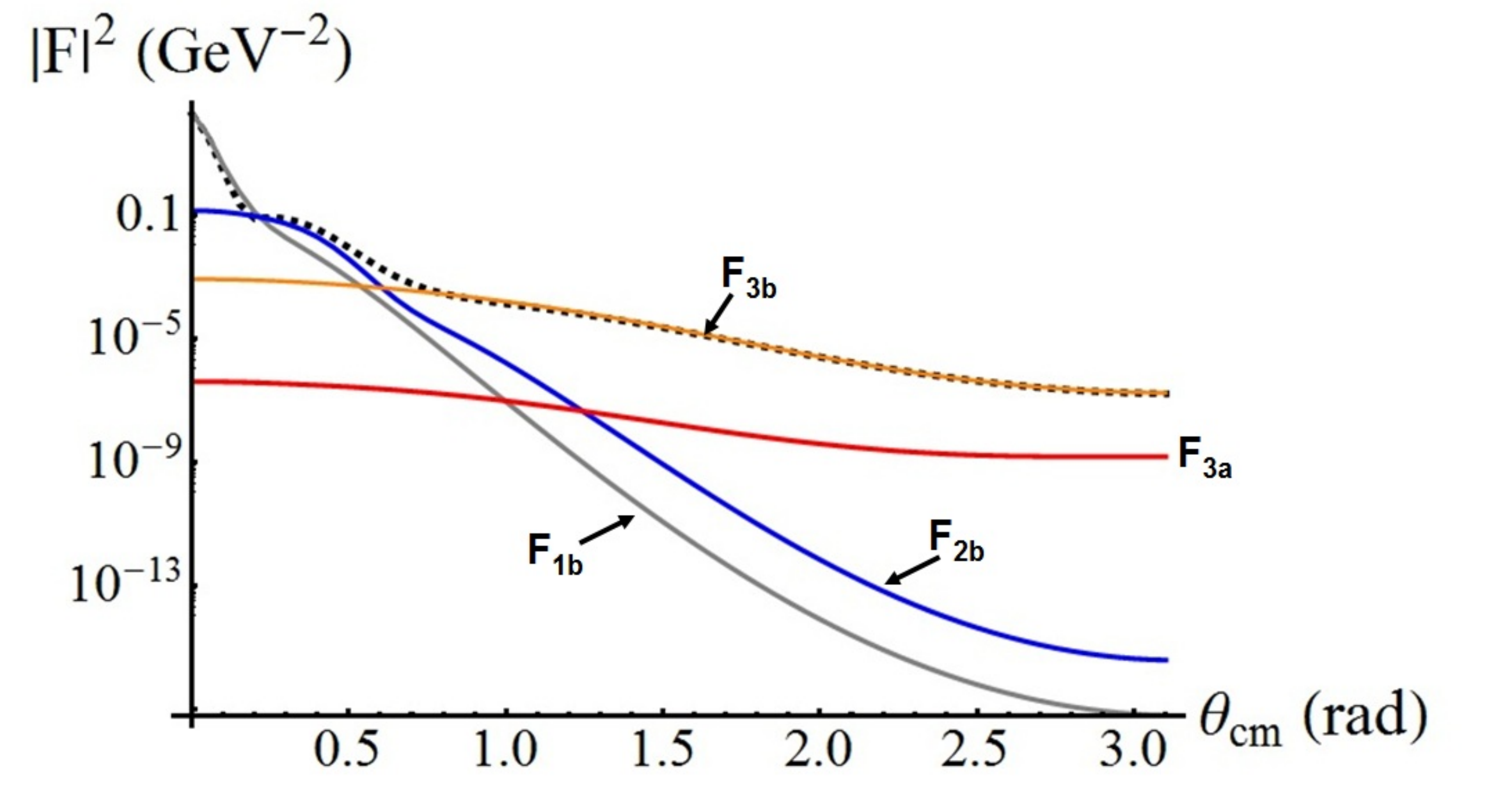}
        }%
        \hspace{0.5in}
         \subfigure[Square of total amplitude.  Solid: includes all amplitudes.  Dashed: omitting $F_{3a}$ (not distinguishable from the solid line at this scale).]{%
           \label{fig:squaretot9d}
           \includegraphics[width=0.8\textwidth]{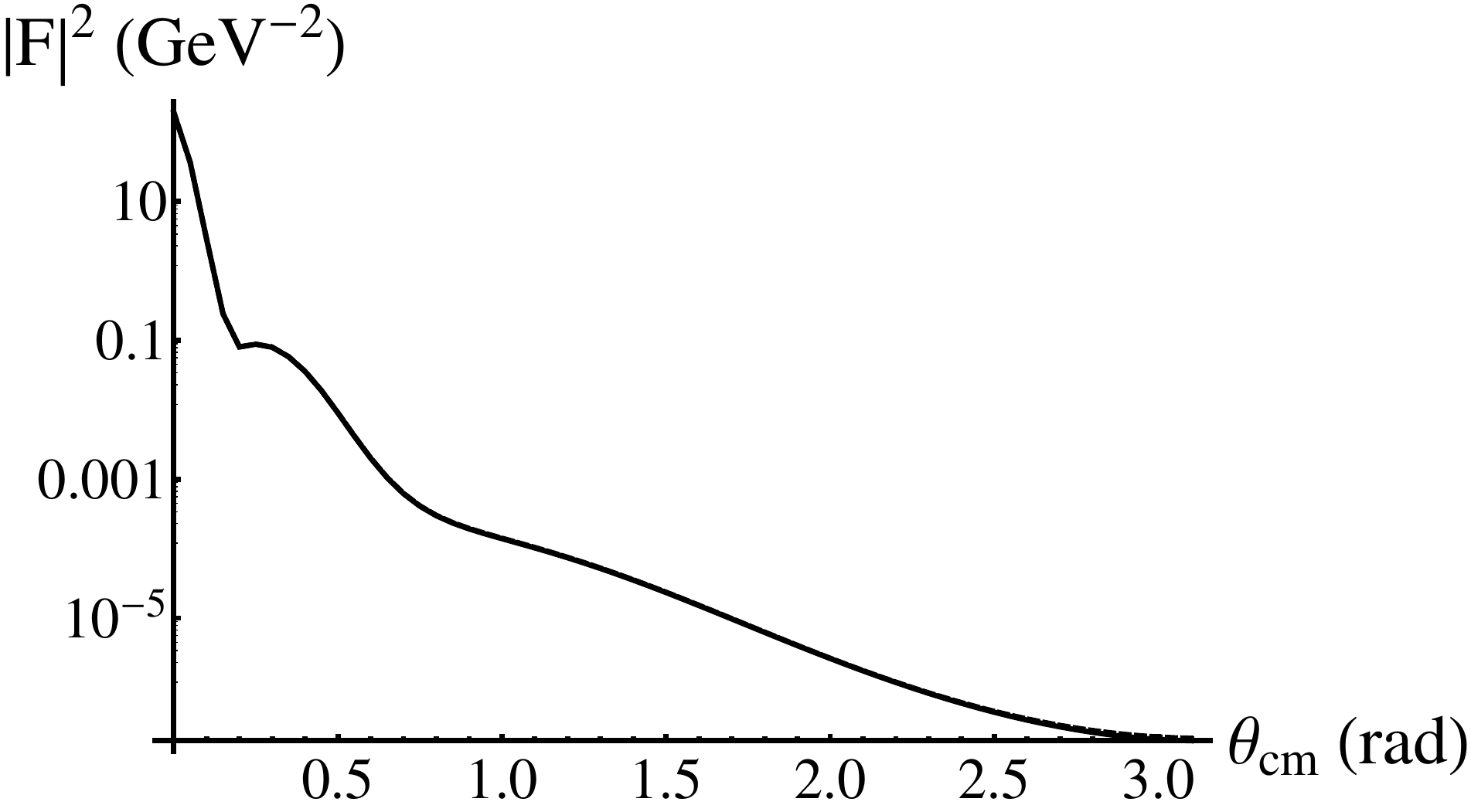}
        }\\ 
%

    \end{center}
    \caption{%
     Squares of amplitudes for $\nu=9\;GeV$, $T^*=0$, for $a=-0.3\;fm$, $R=1.3\;fm$, $r_e=2.0\;fm$, for the model wavefunction of \eq{negau}.  
     }%
   \label{fig:ampsSquared9nega}
\end{figure}

\begin{figure}[tbp]
     \begin{center}
%
           \label{fig:squaretot65d}
           \includegraphics[width=0.8\textwidth]{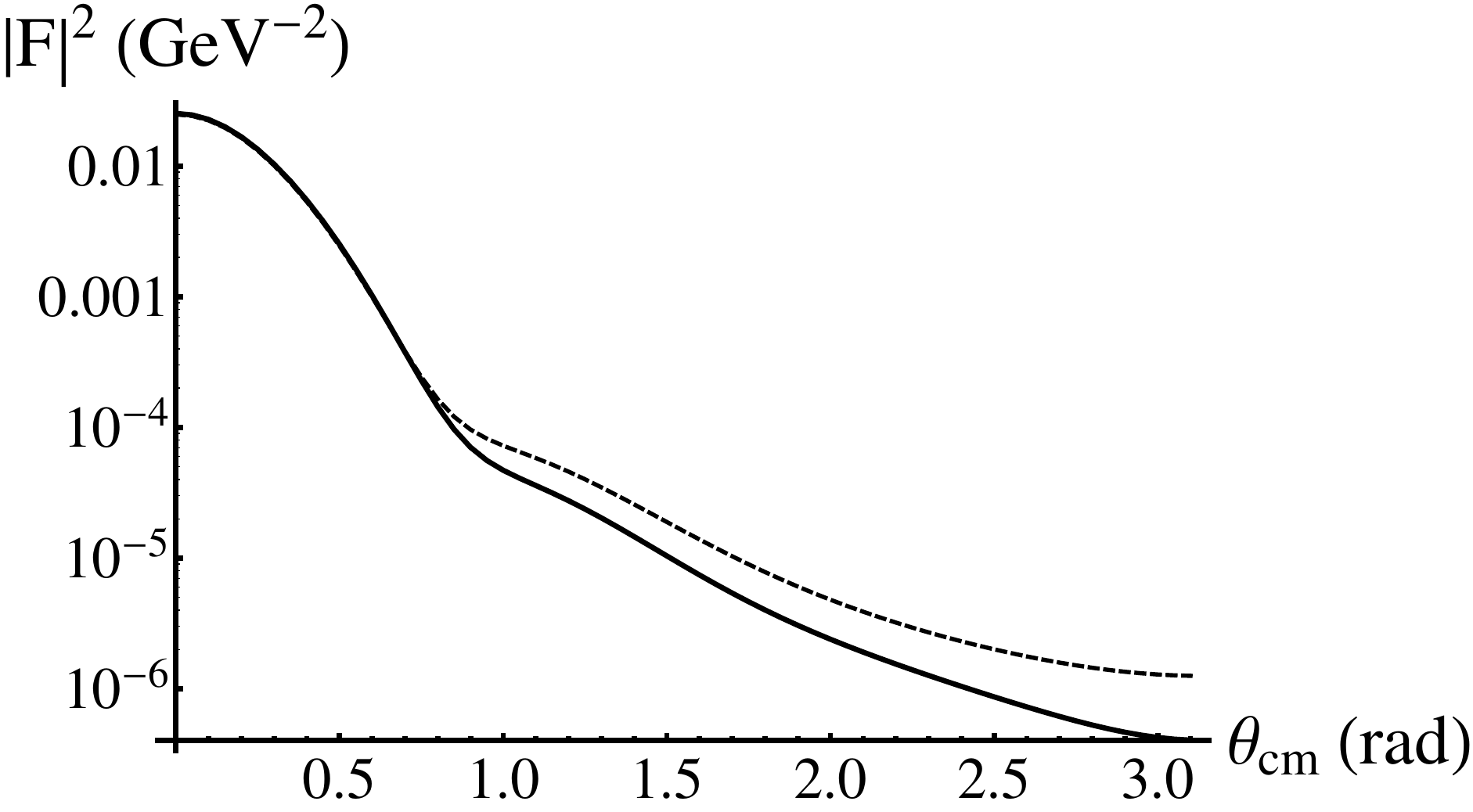}
%

    \end{center}
    \caption{%
     Square of total amplitude for $\nu=6.5\;GeV$, $T^*=0$, for $a=-0.3\;fm$, $R=1.3\;fm$, $r_e=2.0\;fm$, for the model wavefunction of \eq{negau}.  Solid: includes all amplitudes.  Dashed: omitting $F_{3a}$.  
     }%
   \label{fig:ampsSquared65nega}
\end{figure}

\begin{figure}[tbp]
     \begin{center}
        \subfigure[Off-shell amplitude for $k=0$ for a square-well potential, for $\nu=9.0\;GeV$, with $a=0.3\;fm$, $R=0.1\;fm$ (the depth is $U_0=13.13 \;GeV^{-2}=337.3\;fm^2$).]{%
            \label{fig:offshellsqwell}
            \includegraphics[width=0.4\textwidth]{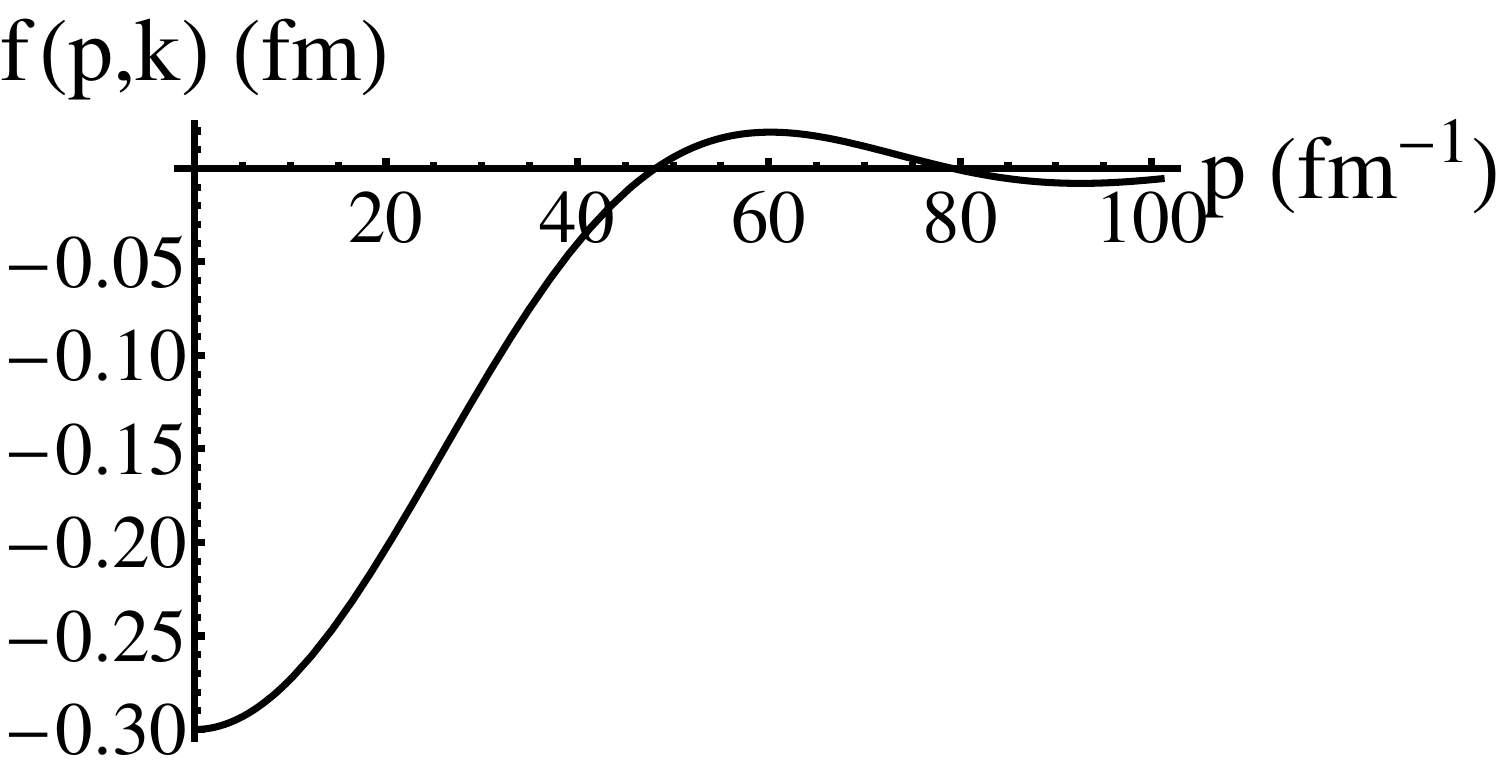}
        }%
        \hspace{0.5in}
         \subfigure[$\vert F_{3a}\vert^2$ for 4 different potentials ($\nu=9.0\;GeV$).   From top to bottom the curves are:  cubic wavefunction with $a=0.3$ fm, $R=0.1$ fm; square-well potential with $a=0.3$ fm, $R=0.1$ fm; cubic wavefunction with $a=0.3$ fm, $R=0.29$ fm; square-well potential with $a=0.3$ fm, $R=0.29$ fm. ]{%
           \label{fig:sqwell9}
           \includegraphics[width=0.4\textwidth]{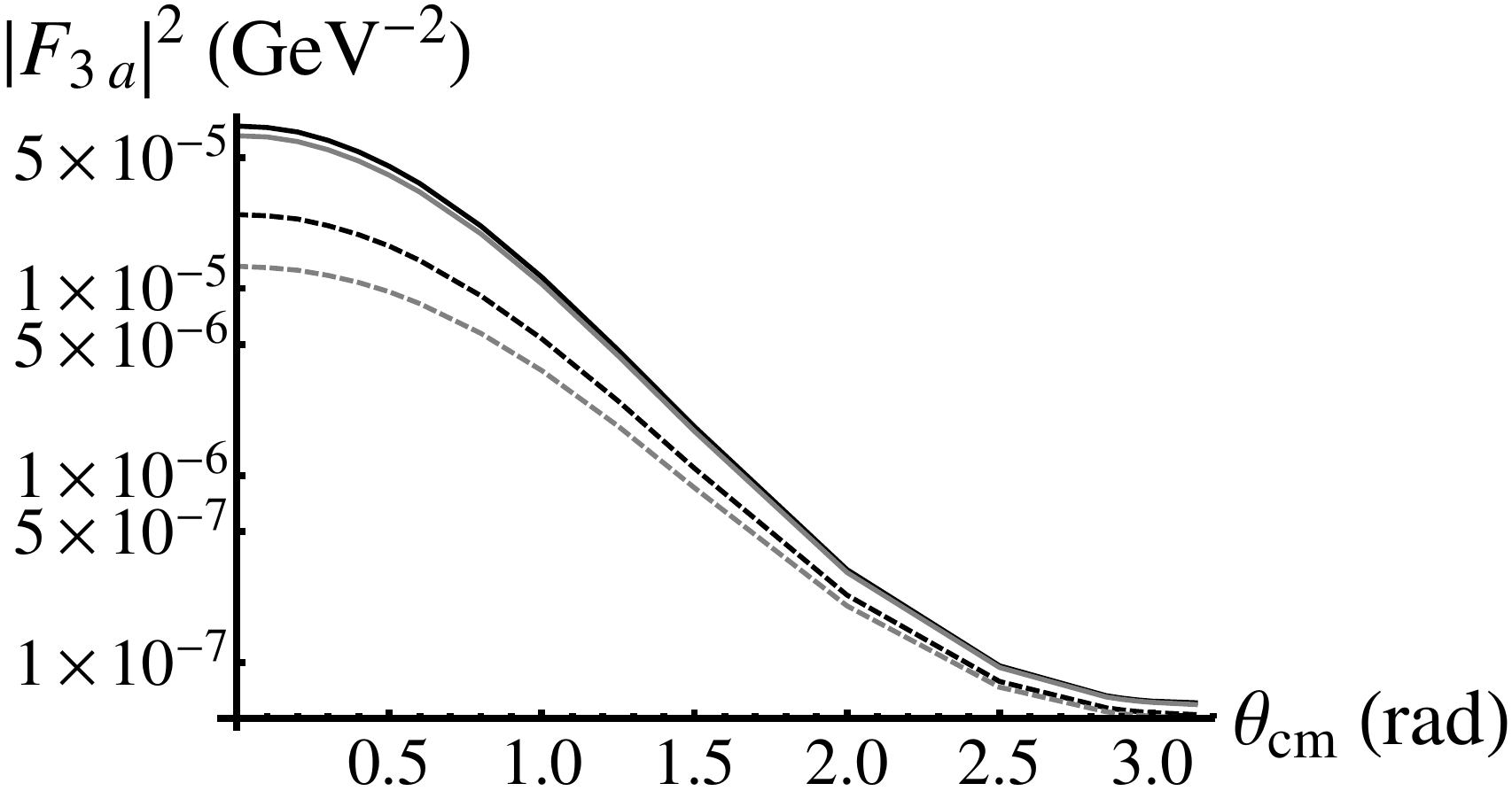}
        }\\ 

        \subfigure[$\vert F_{3a}\vert^2$ for 4 different potentials, for $\nu=6.5\;GeV$.  Top curve is for both the cubic wavefunction and the square-well for $a=0.3$ fm, $R=0.1$ fm. Bottom curve is for both the cubic wavefunction and the square-well for  $a=0.3$ fm, $R=0.29$ fm. ]{%
            \label{fig:sqwell65}
            \includegraphics[width=0.4\textwidth]{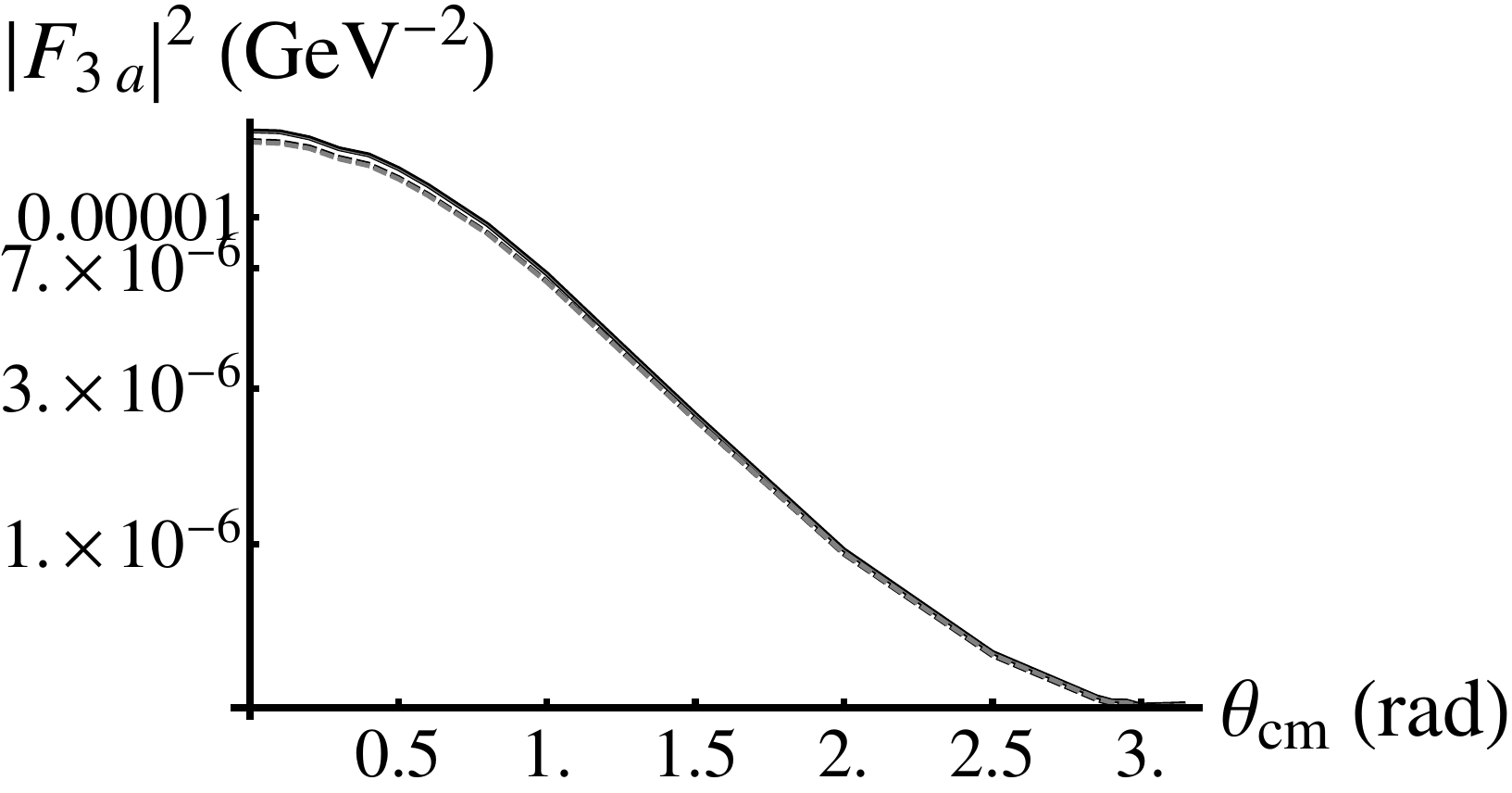}
        }%
         \hspace{0.5in} 
        \subfigure[same as (c)]{%
            \label{fig:protangle65}
            \includegraphics[width=0.4\textwidth]{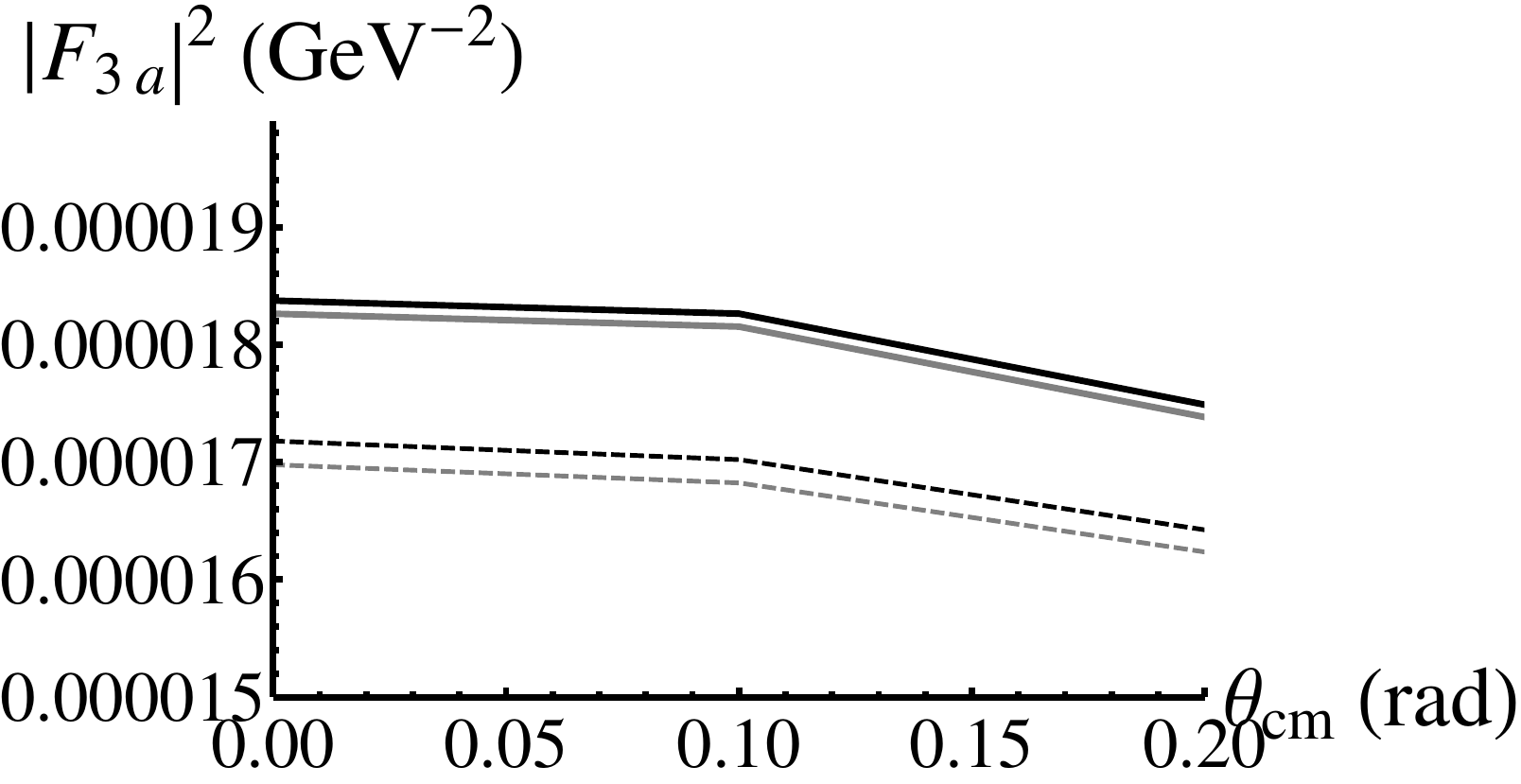}
        }%

    \end{center}
    \caption{%
      Off-shell amplitudes for various potentials, for $T^*=0$.
     }%
   \label{fig:variouspot}
\end{figure}

The results for all amplitudes are shown in Figs. \ref{fig:ampsSquared9} and \ref{fig:ampsSquared65} for positive $a$, and Figs. \ref{fig:ampsSquared9nega} and \ref{fig:ampsSquared65nega} for negative $a$, for the case of $T^*=0$..   As can be seen in the graphs, the amplitude $F_{3a}$ is smaller than $F_{3b}$, but in the range of $\theta_{cm}$ where these two diagrams (the $J/\psi$-nucleon rescattering diagrams) dominate, the square of the total amplitude is much smaller than the square of the total amplitude at its maximum, by a factor of $10^{5}$ to $10^{3}$.  The total amplitude is due almost exclusively to the impulse diagram $F_{1b}$, in the region where the total amplitude is non-negligible.

$F_{3a}$ was also calculated using 3 other potentials, for $T^*=0$:  a square-well potential yielding $a=0.3\;fm$ and $R=0.1\;fm$, and also the potential of Eq. \ref{modelpot} but with $a=0.3\;fm$, $R=0.29\;fm$, and a square-well potential yielding $a=0.3\;fm$ but with $R=0.29\;fm$.  The results are shown in Fig. \ref{fig:variouspot}.  It can be seen that there's not much difference in the value of $F_{3a}$ for different potentials with the same scattering length.  For $\nu=6.5\;GeV$, there's virtually no difference at all.

\subsection{Non-zero $T^*$ and Differential Cross-sections}

The electroproduction differential cross-sections were calculated for the case of $T^*=30\;MeV$.  Coplanar kinematics were used (i.e. all 3 final momenta $\mathbf{p}_p$, $\mathbf{p}_n$, and $\mathbf{p}_V$ lie in the same plane).  At a given value of the proton momentum $\mathbf{p}_p$, for a fixed value of $T^*_{Vn}$, the magnitude of the neutron's momentum in the LAB satisfies
\be
\vert p_{n,min}\vert \le p_n \le p_{n,max}
\ee
where
\be
\label{pnmin}
p_{n,min}=\frac{E_n^*}{\sqrt{s_{Vn}}}\vert\mathbf{p}_{Vn}\vert - \frac{p_n^*}{\sqrt{s_{Vn}}}E_{Vn}
\ee
and
\be
\label{pnmax}
p_{n,min}=\frac{E_n^*}{\sqrt{s_{Vn}}}\vert\mathbf{p}_{Vn}\vert + \frac{p_n^*}{\sqrt{s_{Vn}}}E_{Vn}
\ee
and where $E_n^*$ is the neutron energy in the $J/\psi$-neutron c.m. frame, $p_n^*$ is the neutron momentum in  the $J/\psi$-neutron c.m. frame, $E_{Vn}$ is the total energy of the  $J/\psi$-neutron pair in the LAB frame, $\mathbf{p}_{Vn}$ is the total momentum of the  $J/\psi$-neutron pair in the LAB frame, and $s_{Vn}=(p_V+p_n)^2$.  Note that $p_{n,min}$ and $p_{n,max}$ depend on $\mathbf{p}_p$, for a given value of $T^*_{Vn}$.

For the differential cross-section calculated here, at a given value of $\mathbf{p}_p$ the value of the magnitude of the neutron's momentum was set at the $p_{n,min}$ corresponding to that $\mathbf{p}_p$.  This corresponds to the neutron and $J/\psi$ momenta both pointing in the same direction in the LAB frame (collinear momenta); in the $J/\psi$-neutron c.m. frame, the neutron momentum points opposite the direction that the total neutron plus $J/\psi$ momentum points in the LAB, while the $J/\psi$ momentum in the $J/\psi$-neutron c.m. frame points in the same direction that the total neutron plus $J/\psi$ momentum points in the LAB.  

The electroproduction differential cross-section in the LAB frame is given by
\be
\label{electrocross}
\frac{d^8\sigma}{dE^{\prime} d\Omega^{\prime} dp_p d\Omega_p d\Omega_n}=\frac{ v_0 V_T\;E^{\prime}}{8 (2\pi)^3 M_d\; E}\;\times\frac{1}{8(2\pi)^5}\frac{p_p^2}{E_p}\frac{p_n^3}{ \vert E_V\;p_n^2 - E_n \mathbf{p}_n \cdot \mathbf{p}_V \vert}\;\vert F \vert^2
\ee 
where $F=F_{1a}+F_{1b}+F_{2a}+F_{2b}+F_{3a}+F_{3b}$ is the total amplitude for $J/\psi$ production from a virtual photon, and
\be
v_0=\sqrt{16 E^2 E^{\prime 2} - Q^4}=4 E E^{\prime} \cos^2(\theta^{\prime}/2),
\ee
\be
V_T = \frac{1}{2}\frac{Q^2}{\mathbf{q}^2}+\frac{Q^2}{v_0},
\ee
and
\be
E^{\prime}=E-\nu=12\;GeV - \nu.
\ee
In the above, $E$ is the initial electron energy (taken to be $12$ GeV), $E^{\prime}$ is the final electron energy, and $\theta^{\prime}$ is the scattering angle of the electron relative to the initial electron momentum (all quantities in the LAB frame).

The results are shown in Figs. \ref{fig:ampsSquared9b}, \ref{fig:ampsSquared65b} and \ref{fig:electrocross65} for the case of $T^*_{Vn}=30$ MeV.  For $\nu=9$ GeV, the squares of the individual amplitudes are shown in Fig. \ref{fig:squares9b}, along with the square of the total amplitude including all diagrams.   As can be seen in that graph, the amplitude $F_{3a}$ makes a negligible contribution to the total amplitude.  There are intervals of $\theta_{cm}$ where the amplitudes  $F_{1b}$, $F_{2b}$, and $F_{3b}$ individually dominate the  total amplitude.  However, by comparing Fig. \ref{fig:squares9b} with Fig. \ref{fig:diff9b}, which shows the electroproduction differential cross-section on a linear scale, over the range of $\theta_{cm}$ for which the cross-section is non-negligible (for $0<\theta_{cm}<0.2$ rad) the cross-section is due exclusively to the impulse diagram $F_{1b}$.  (The very small ``bump" visible in Fig. \ref{fig:diff9b} at $\theta_{cm}\simeq 0.3$ rad is due to the proton-neutron rescattering amplitude $F_{2b}$.) 

For $\nu=6.5$ GeV, the difference between the cross-section including $F_{3a}$ and omitting $F_{3a}$ is visible in the logarithmic-scale graphs (Figs. \ref{fig:diff65a} and \ref{fig:diff65b}), but not in the linear-scale graph, Fig. \ref{fig:diff65c}.  However, by comparing Fig. \ref{fig:ampsSquared65b} with Fig. \ref{fig:diff65c}, one can see that  over the range of $\theta_{cm}$ for which the cross-section is non-negligible (for $0<\theta_{cm}<0.6$ rad) the cross-section is due exclusively to the impulse diagram $F_{1b}$.  Thus the rescattering effects are negligible, for these kinematics, for both $\nu=9$ GeV and $\nu=6.5$ GeV.

\begin{figure}[tbp]
     \begin{center}
        \subfigure[Squares of individual amplitudes, for positive $a$.   Dashed curve: square of total amplitude.    Impulse:  $F_{1b}.\;\;$  $p-n$: $F_{2b}.\;\;$  $V-p$:  $F_{3b}.\;\;$  $V-n$: $F_{3a}$.  Not shown are $F_{1a}$, $F_{2a}$, which are negligible ]{%
            \label{fig:squares9b}
            \includegraphics[width=0.6\textwidth]{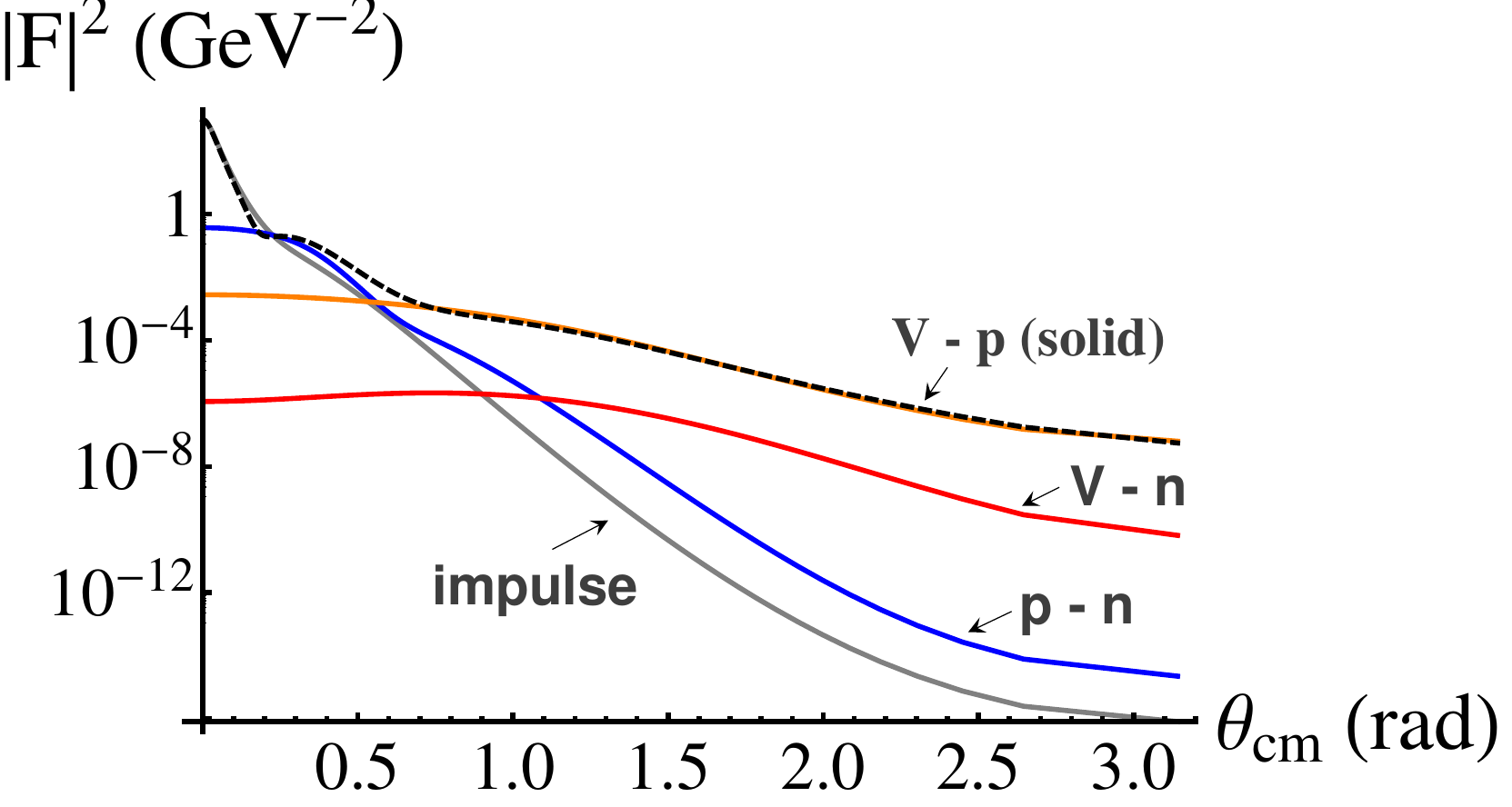}
        }%
        \hspace{0.5in}
         \subfigure[Electroproduction differential cross-section (logarithmic scale). ]{%
           \label{fig:diff9a}
           \includegraphics[width=0.4\textwidth]{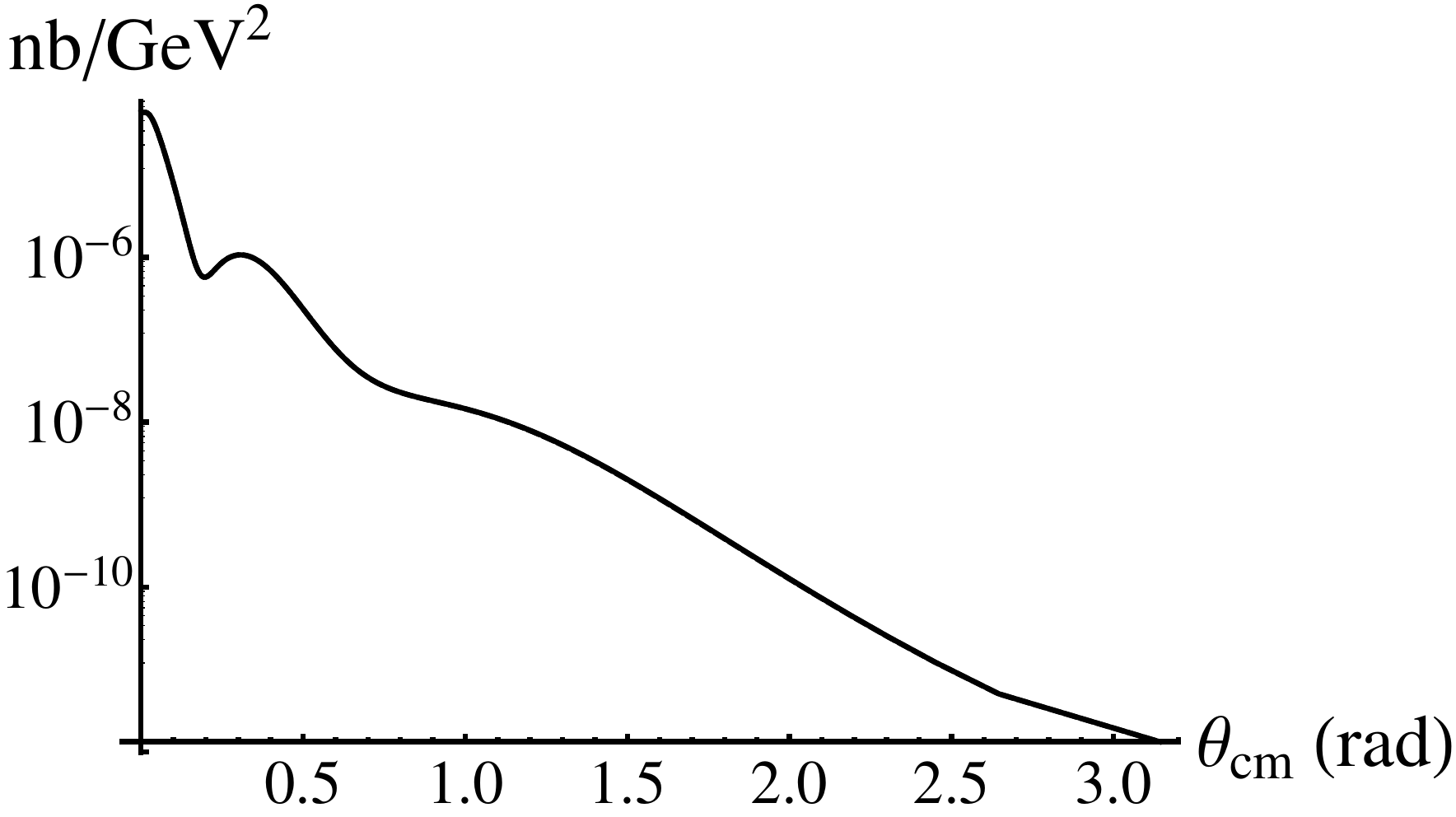}
        }\\ 

        \subfigure[Electroproduction differential cross-section (linear scale). ]{%
           \label{fig:diff9b}
           \includegraphics[width=0.4\textwidth]{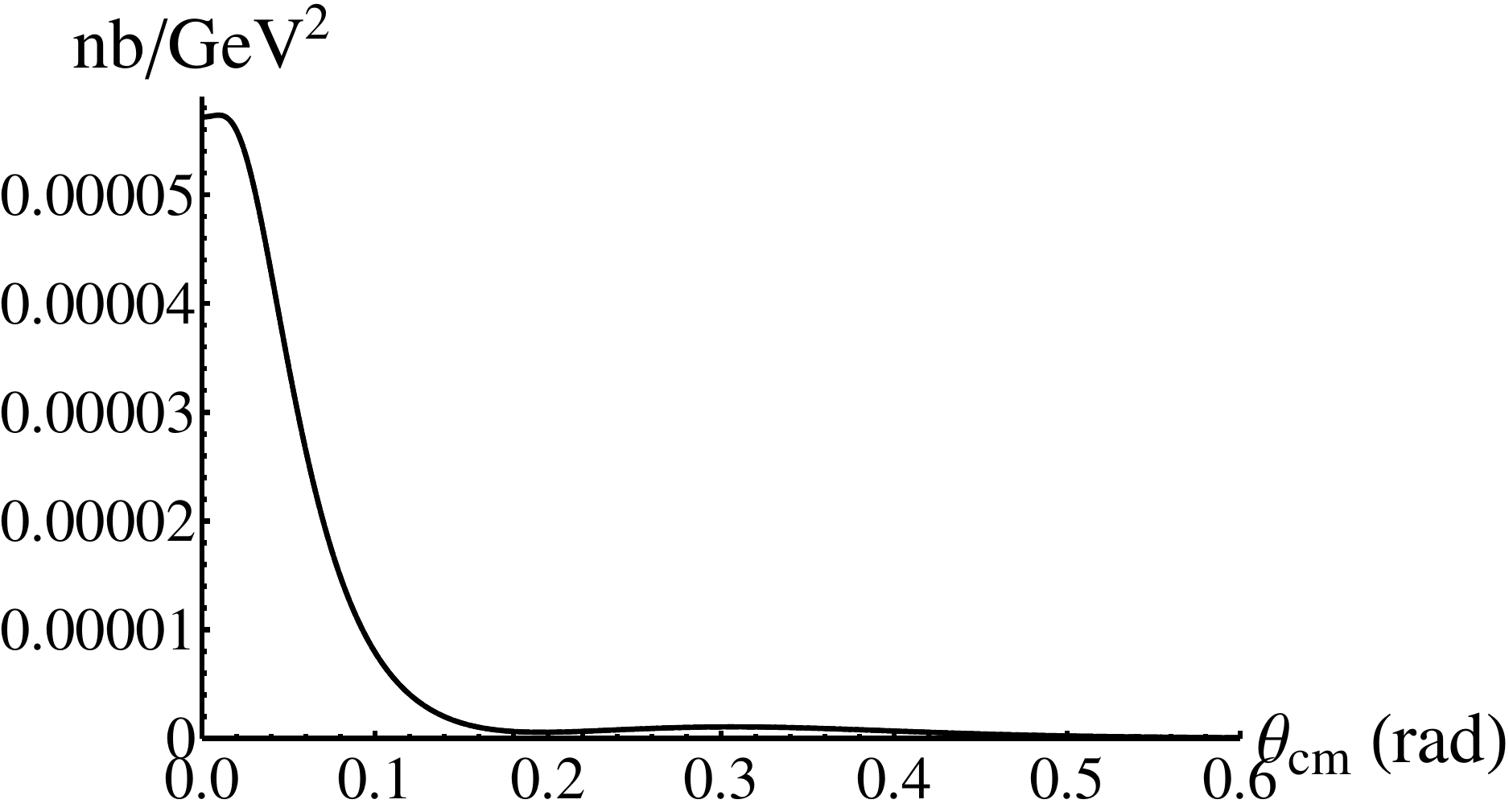}
        }%

    \end{center}
    \caption{%
    Squares of amplitudes, and electroproduction differential cross-section \eq{electrocross}, for $\nu=9\;GeV$,  $T^*_{Vn}=30$ MeV, $p_n=p_{n,min}$.  Solid curves in (b) and (c) includes all amplitudes,  dashed curves (not distinguishable from the solid curve) omit $F_{3a}$,  for the model potential of \eq{modelpot}.
     }%
   \label{fig:ampsSquared9b}
\end{figure}

\begin{figure}[tbp]
     \begin{center}
            \includegraphics[width=0.8\textwidth]{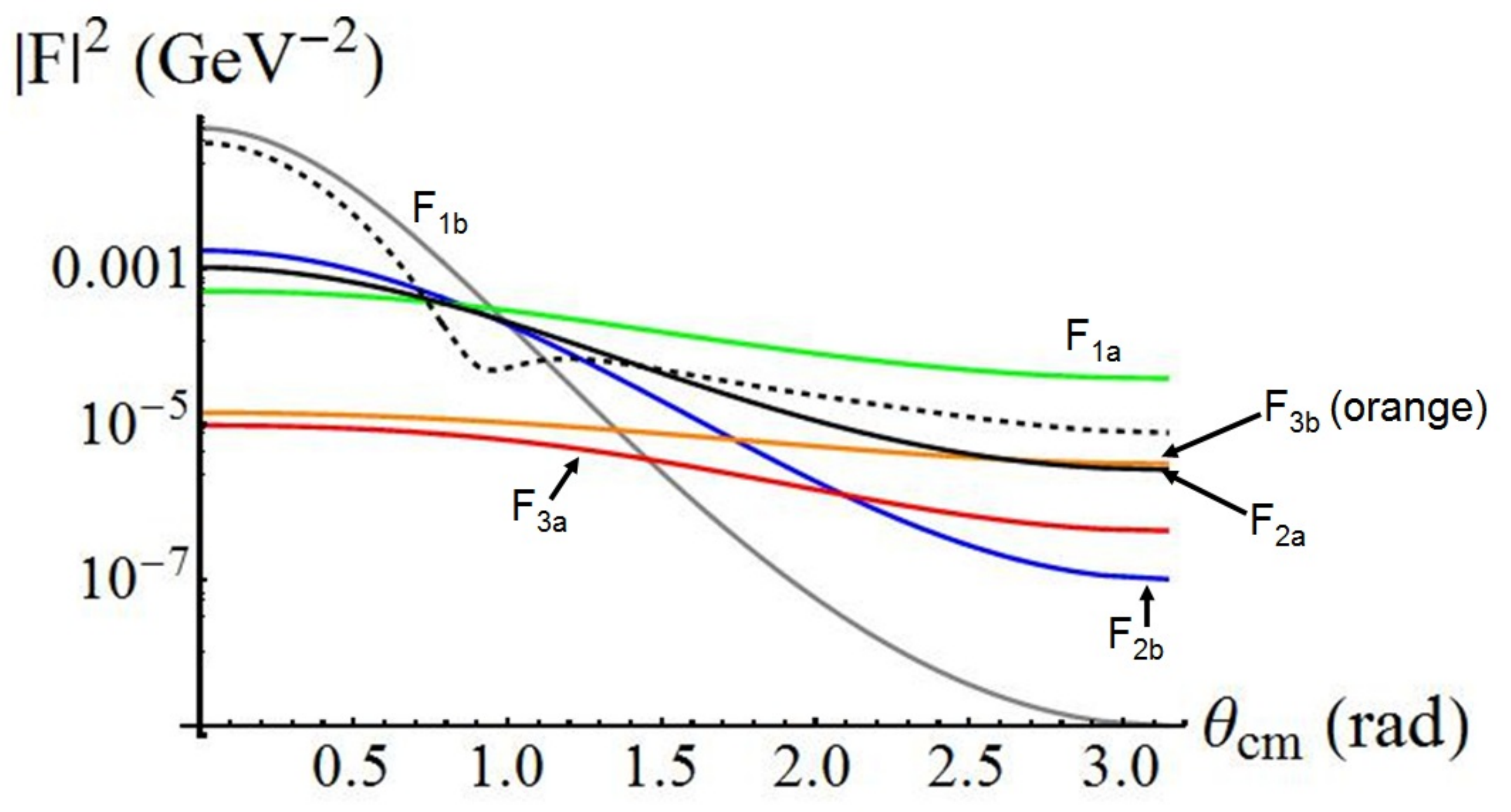}

    \end{center}
    \caption{%
       Squares of individual amplitudes  for $\nu=6.5\;GeV$, $T^*_{Vn}=30$ MeV, $p_n=p_{n,min}$.   Dashed curve is the square of the total amplitude.  $a=0.3\;fm$.
     }%
   \label{fig:ampsSquared65b}
\end{figure}

\begin{figure}[tbp]
     \begin{center}
        \subfigure[Electroproduction cross-section.  Negative scattering length $a$.  Solid curve includes all amplitudes, dashed is omitting $F_{3a}$. ]{%
           \label{fig:diff65a}
           \includegraphics[width=0.4\textwidth]{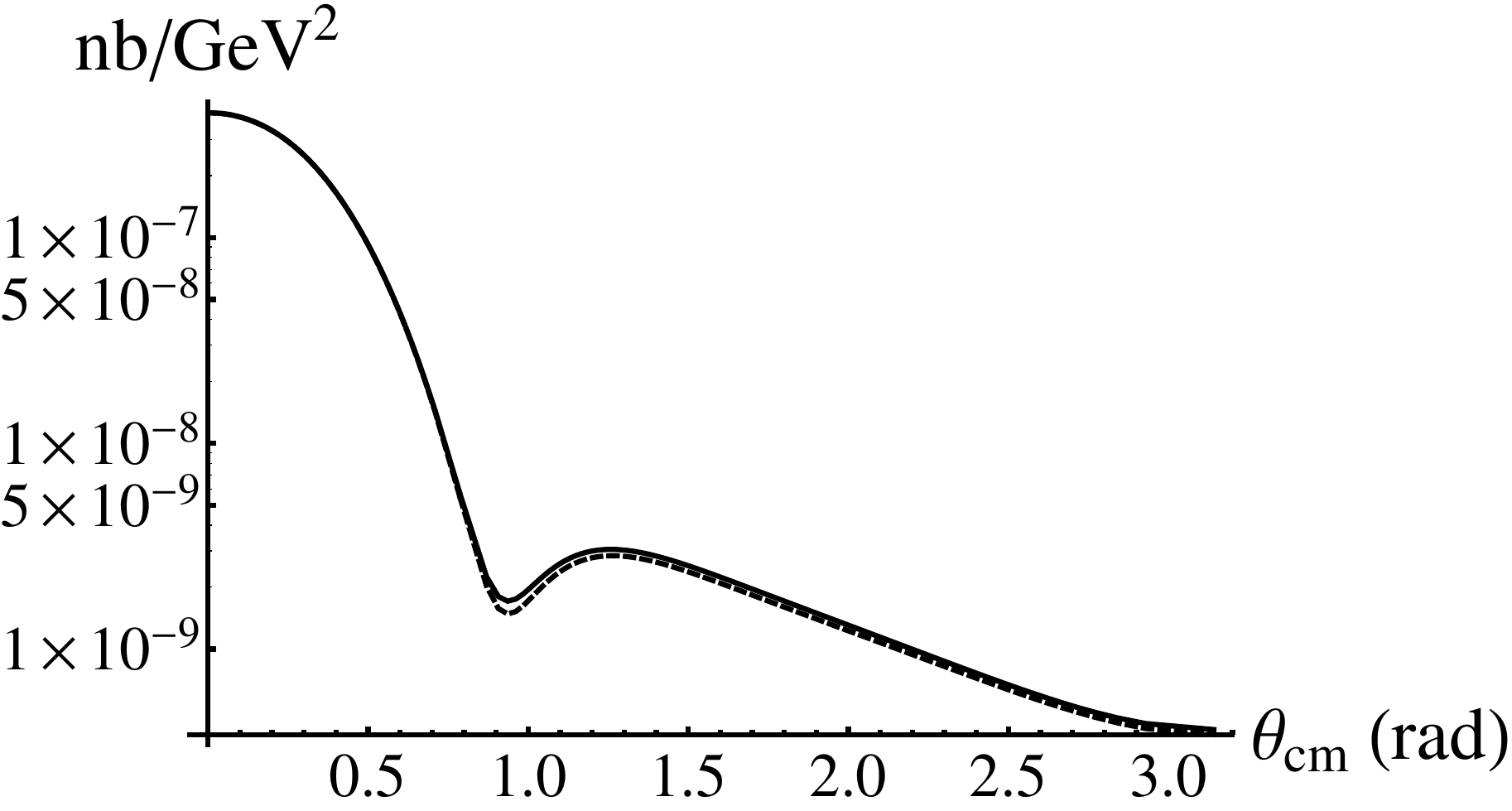}
        }%
        \hspace{0.5in}
        \subfigure[Electroproduction cross-section. Positive scattering length $a$.  Solid curve includes all amplitudes, dashed is omitting $F_{3a}$.]{%
           \label{fig:diff65b}
           \includegraphics[width=0.4\textwidth]{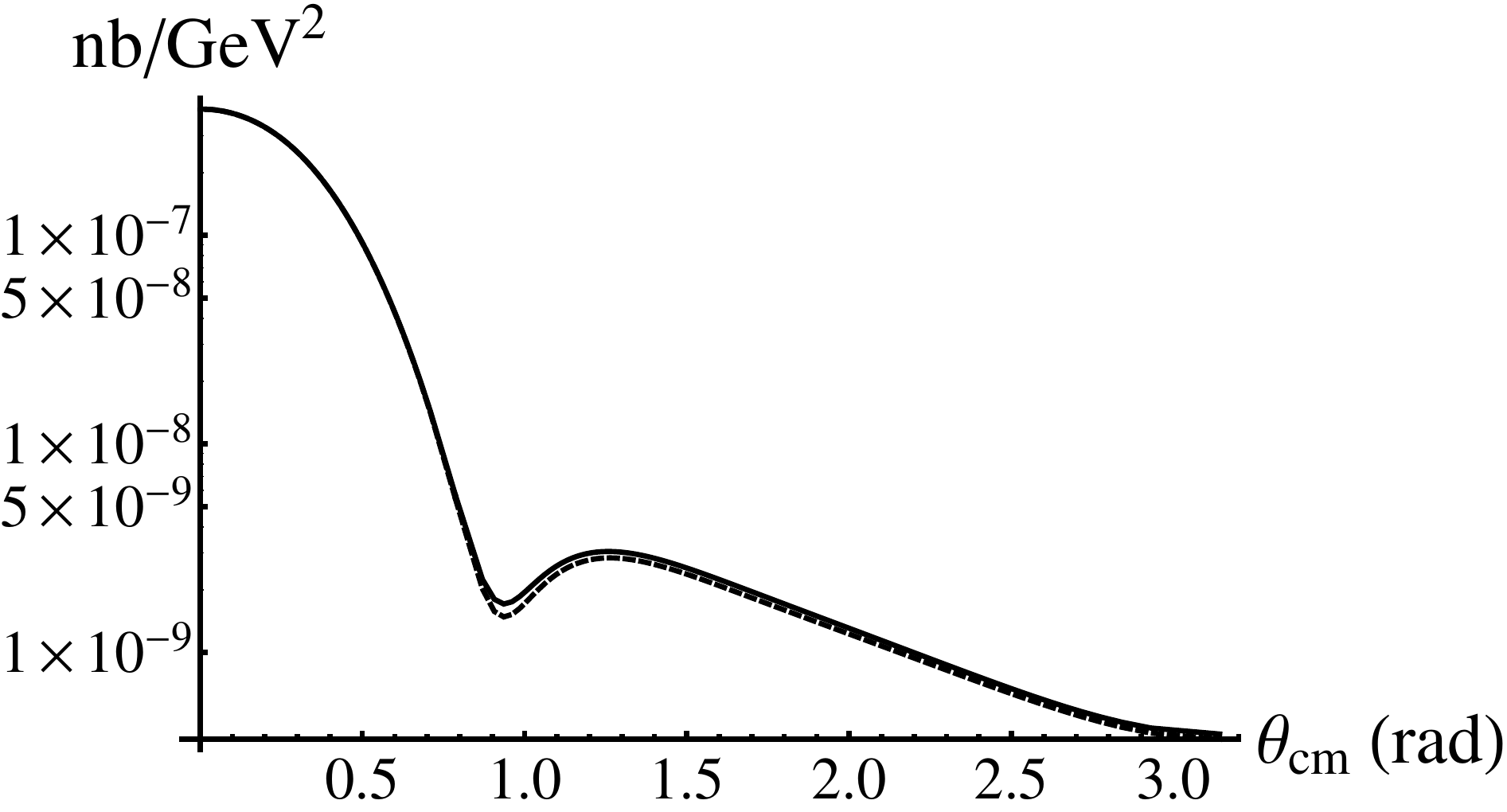}
        }\\ 

       \subfigure[Electroproduction cross-section.  Linear scale]{%
            \label{fig:diff65c}
            \includegraphics[width=0.6\textwidth]{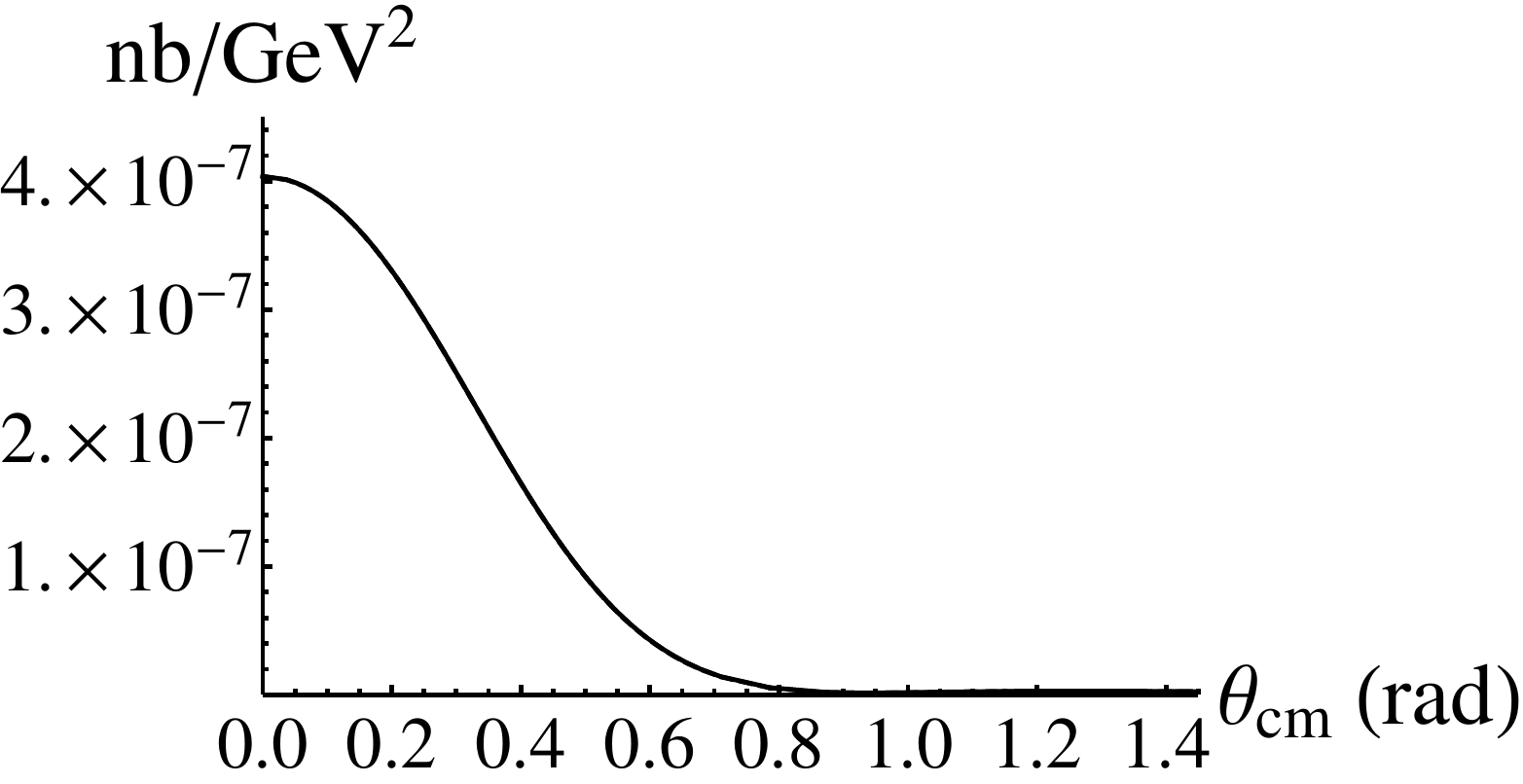}
        }%

    \end{center}
    \caption{%
      Electroproduction differential cross-section \eq{electrocross} for $\nu=6.5\;GeV$, $T^*_{Vn}=30$ MeV, $p_n=p_{n,min}$, and $a=\pm0.3\;fm$.  In (c), the dashed curves (not distinguishable from solid curve) are omitting $F_{3a}$ for both positive and negative $a$,  for the model potential of \eq{modelpot}.
     }%
   \label{fig:electrocross65}
\end{figure}

\subsection{Conclusion}

It does not appear to be possible to measure the $J/\psi$-nucleon scattering length via production on the deuteron, under the kinematic conditions available at JLab.  For small values of the relative momentum of the outgoing $J/\psi$-neutron pair, the initial momentum of the neutron inside the deuteron that is required for on-mass-shell rescattering of the $J/\psi$-neutron pair is larger than $\sim 0.6\;GeV$ (see Fig. \ref{fig:nminusplus3a1}), where the deuteron wavefunction is negligible.  The off-mass-shell part of the rescattering amplitude was calculated using model $J/\psi$-nucleon potentials and was found to make a negligible contribution to the total amplitude.   The vast majority of $J/\psi$ production events, for $T_{Vn}^*\le 0.03\;GeV$, will be at small values of $\theta_{cm}$, where the impulse diagram $F_{1b}$ dominates, and therefore information on $J/\psi$-nucleon elastic scattering at small relative energy cannot be obtained.

\section{Intermediate energy $J/\psi$ production on the deuteron}
\label{sec:intermedenergy}

It may be possible to extract the $J/\psi$-nucleon elastic scattering amplitude from the $\gamma^*+D\to J/\psi +p+n$ experiment, at higher relative energy of the $J/\psi$-nucleon pair, under different kinematic conditions for the final-state particles than was considered in the previous sections of this chapter.
Under certain kinematic conditions, the dominant contributions to the amplitude will come from rescattering diagrams (p-n rescattering and $J/\psi-n$ rescattering).  If we fix the magnitude of the outgoing neutron's momentum at a moderately large value (here taken to be 0.5 GeV) the contribution of the impulse diagram will be negligible, since the impulse diagram is proportional to the value of the deuteron wavefunction at that momentum (see Fig. \ref{fig:diagrams2} for the impulse and rescattering diagrams). For the analysis presented here, we:
 \begin{itemize}
   \item use coplanar kinematics
   \item fix the magnitude of the outgoing neutron momentum at $p_n=0.5\;GeV$
   \item fix the 4-momentum-transfer-squared $t=(q-p_V)^2$ at a particular value
   \item plot amplitudes or differential cross-sections vs. $\theta_n$ (the angle that the outgoing neutron momentum $\mathbf{p}_n$ makes with direction of the incoming photon momentum) for fixed $p_n$ and $t$	
\end{itemize}
(see Fig. \ref{fig:kindiag}).  
For some range of $\nu$ and $t$, these graphs will display peaks due to $p-n$ and $J/\psi-n$ on-mass-shell rescattering.  For $\nu=10\;GeV$, the peak due to $J/\psi-n$ rescattering is evident (see Fig. \ref{fig:2a3atot}), but for $\nu=9\;GeV$ it is not evident (see Fig.  \ref{fig:nu9}).  This analysis is similar to what has been done in~\cite{laget81} for the reaction $\gamma+D\to \pi+N+N$.

The kinematics here are very different than in the previous sections.  There it was the relative energy of the $J/\psi$-neutron system which was kept fixed, at a small value, while the parameter which was varied was the angle of the proton momentum in the overall center-of-mass system.

\begin{figure}[tbp]
     \begin{center}

            \includegraphics[width=4.5in,height=2.5in]{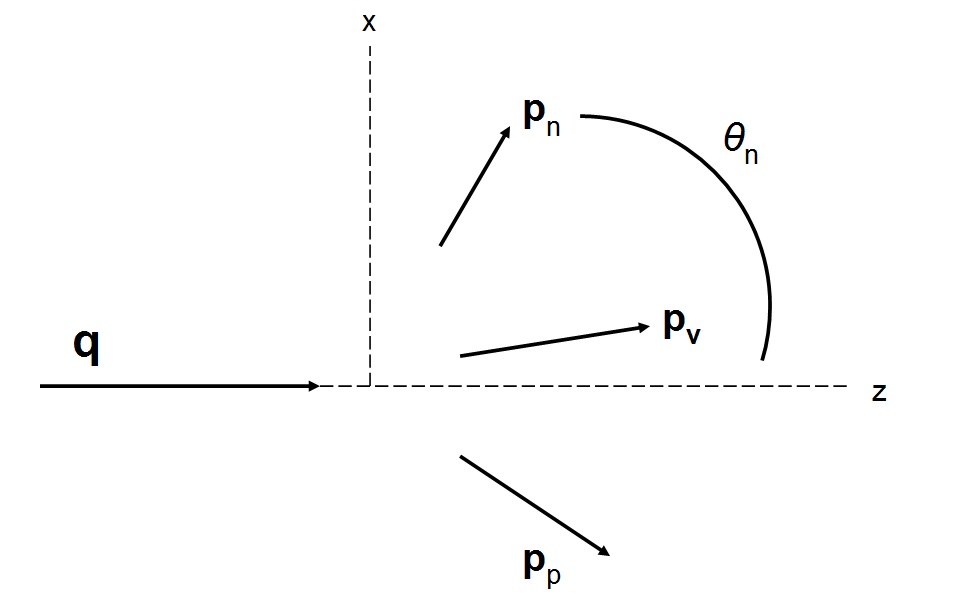}

    \end{center}
    \caption{%
        General orientation of momentum vectors, for coplanar kinematics.  $\mathbf{q}$ is the photon momentum vector.  
     }%
   \label{fig:kindiag}
\end{figure}

\subsection{Intermediate-energy $J/\psi$ production}
\label{higherJpsi}

We are interested here in kinematics available at JLab after the 12 GeV upgrade.  The maximum (virtual) photon energy is then around 11 GeV.  Here we evaluate the amplitude for virtual photon 4-momentum $q=(\nu,\mathbf{q})$ with $\nu=10\;GeV$, $Q^2=-q^2=0.5\;GeV$, keeping $p_n=0.5\;GeV$ and $t$ fixed.  The results presented here are for $t=-2\;GeV^2$; calculations were done for larger values of $\vert t\vert$, with similar results (although the total amplitude decreases with increasing $\vert t\vert$).  We consider the same set of diagrams as before, shown in Figs. \ref{fig:diagrams} and \ref{fig:diagrams2}.  The impulse diagrams, $F_{1a}$ and $F_{1b}$, are negligible for these kinematics. 

\subsubsection{One-loop diagrams - General features}
\begin{figure}[tbp]
     \begin{center}

            \includegraphics[width=4.5in,height=2.5in]{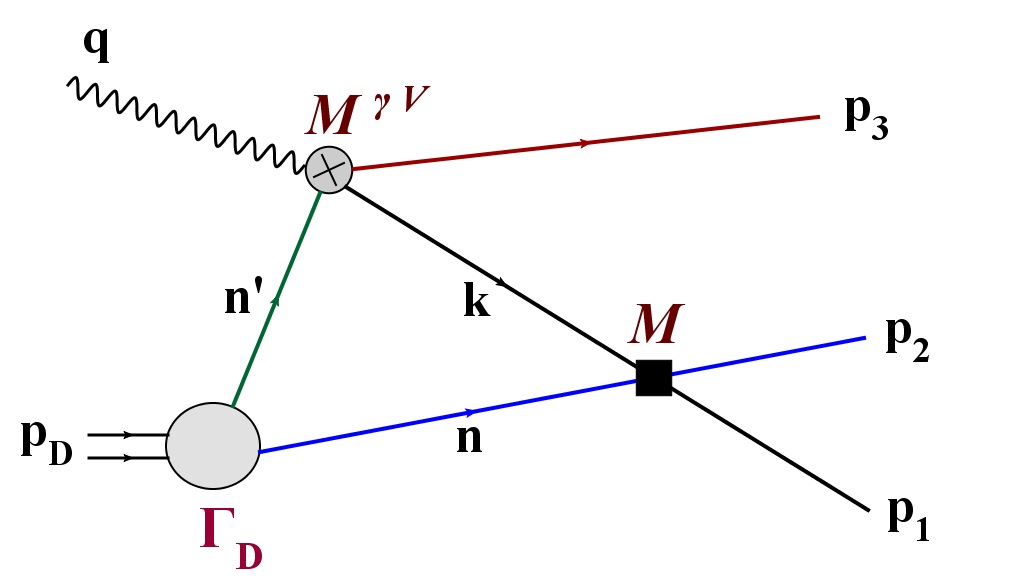}

    \end{center}
    \caption{%
        General one-loop diagram. $n$ and $p_2$ are the same particle (either neutron or proton).
     }%
   \label{fig:oneloopdiag}
\end{figure}
Following~\cite{laget81}, the main features of the on-shell part of the one-loop amplitudes in this case can be seen by first approximating the elementary amplitudes $ {\cal M}^{\gamma V}$ and ${\cal M}$ to be constants, evaluated at a typical value of the intermediate-state momentum $\mathbf{n}$ (Fig. \ref{fig:oneloopdiag}).  In that case we obtain for the on-shell amplitude:
\be
 \label{constMonshell}
F^{on}=-i\pi\frac{1}{\sqrt{2m(2\pi)^3}}\frac{2\pi}{2\vert\mathbf{p}_{12}\vert} {\cal M}^{\gamma V} {\cal M}\int_{\vert n_-\vert}^{n_+}dn\; n\Psi(n) 
\ee
For a given value of $\vert\mathbf{p}_n\vert$ and $t$, most of the dependence on $\theta_n$ in the above expression comes from the dependence of $n_-$ on $\theta_n$:  if $\vert n_-\vert$ is larger than around 0.4 GeV, the amplitude is essentially zero since the deuteron wavefunction is essentially zero for $n>0.4\;GeV$.  The maximum of $F^{on}$ as a function of $\theta_n$ occurs at the value of $\theta_n$ for which $n_-=0$.  The momentum $\vert\mathbf{p}_{12}\vert$ also varies with $\theta_n$, but it is much more nearly constant than $n_-$ is (Figs. \ref{fig:p12} and \ref{fig:pvp}).

\begin{figure}[tbp]
     \begin{center}
        \subfigure[$p_{pn}$]{%
            \label{fig:ppnplus}
            \includegraphics[width=0.4\textwidth]{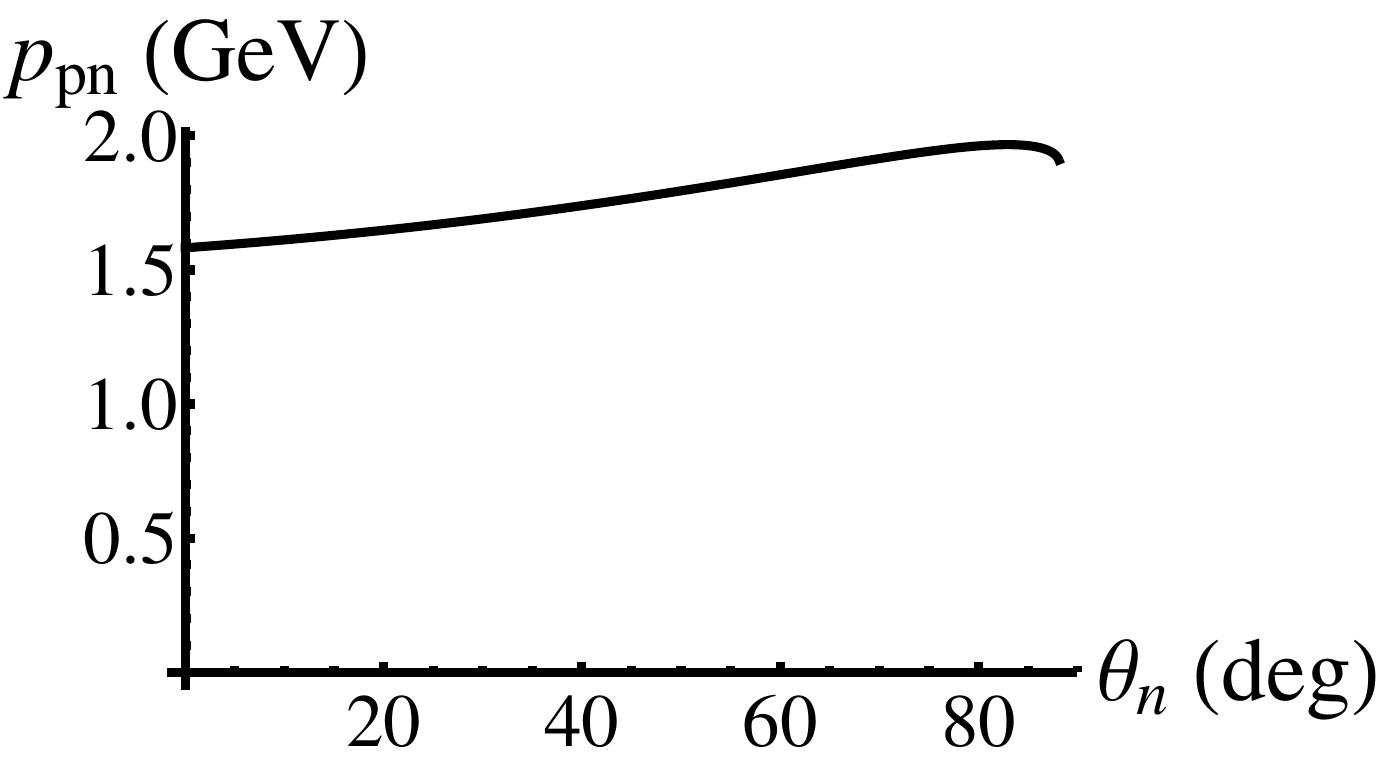}
        }%
        \hspace{0.5in}
         \subfigure[$p_{Vn}$]{%
           \label{fig:pvnplus}
           \includegraphics[width=0.4\textwidth]{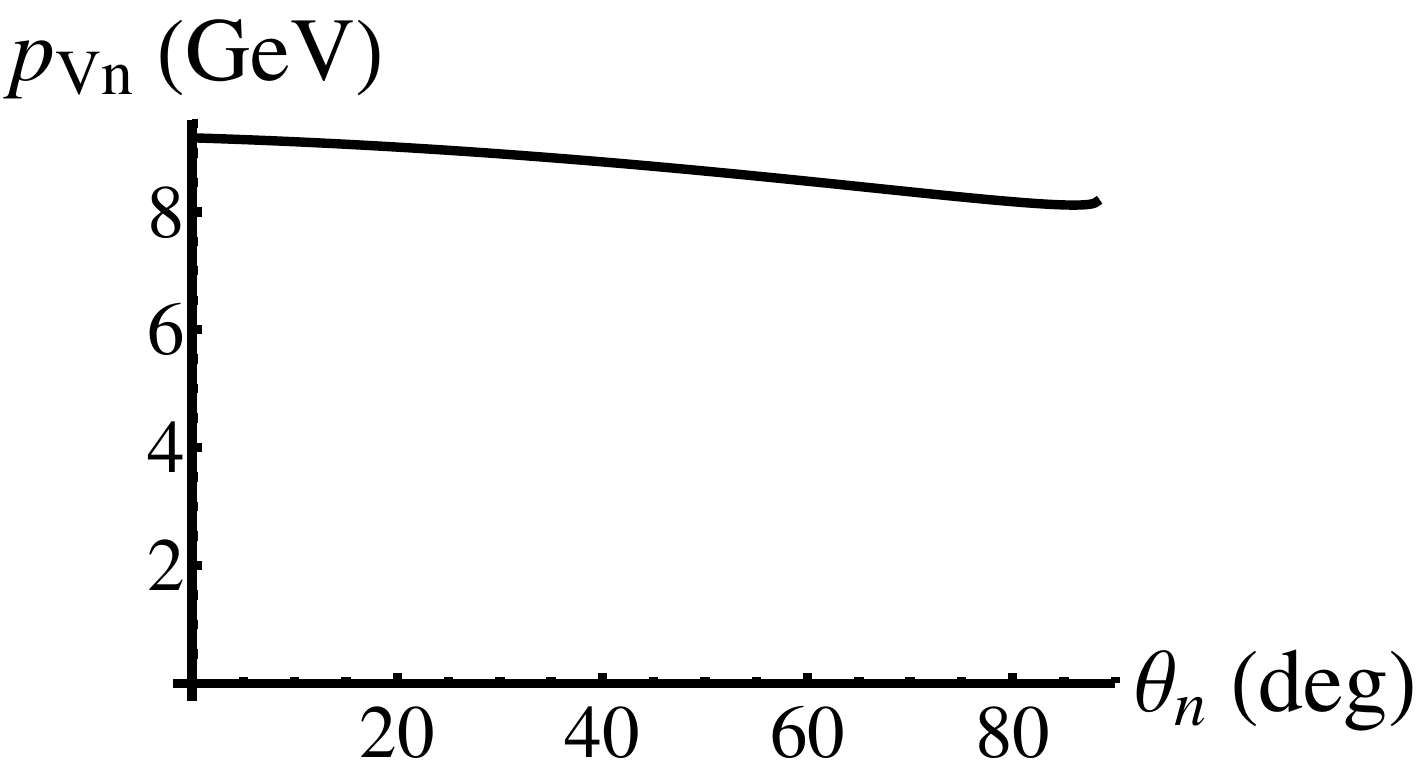}
        }\\ 

        \subfigure[$p_{pn}$]{%
            \label{fig:ppnmin}
            \includegraphics[width=0.4\textwidth]{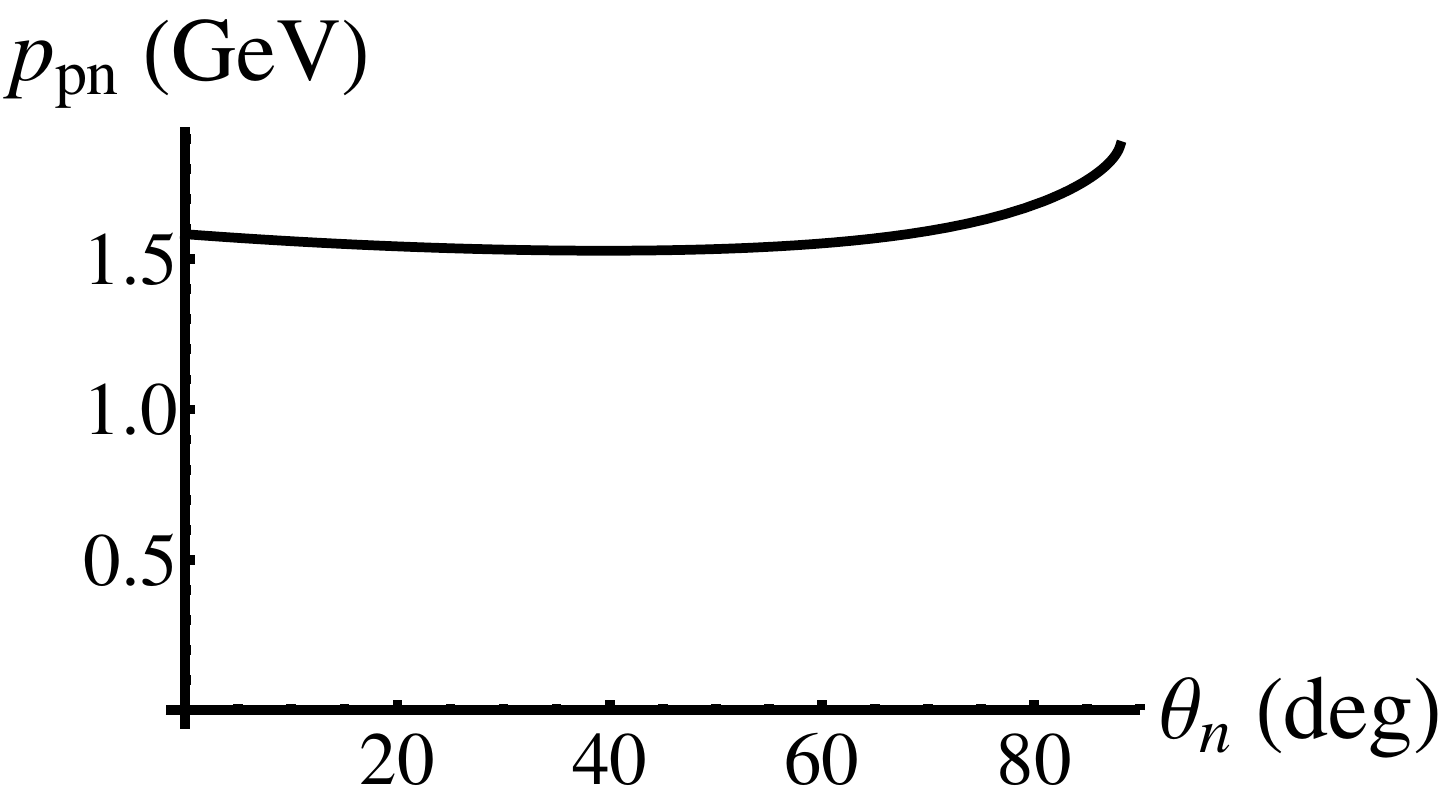}
        }%
         \hspace{0.5in} 
        \subfigure[$p_{Vn}$]{%
            \label{fig:pvnmin}
            \includegraphics[width=0.4\textwidth]{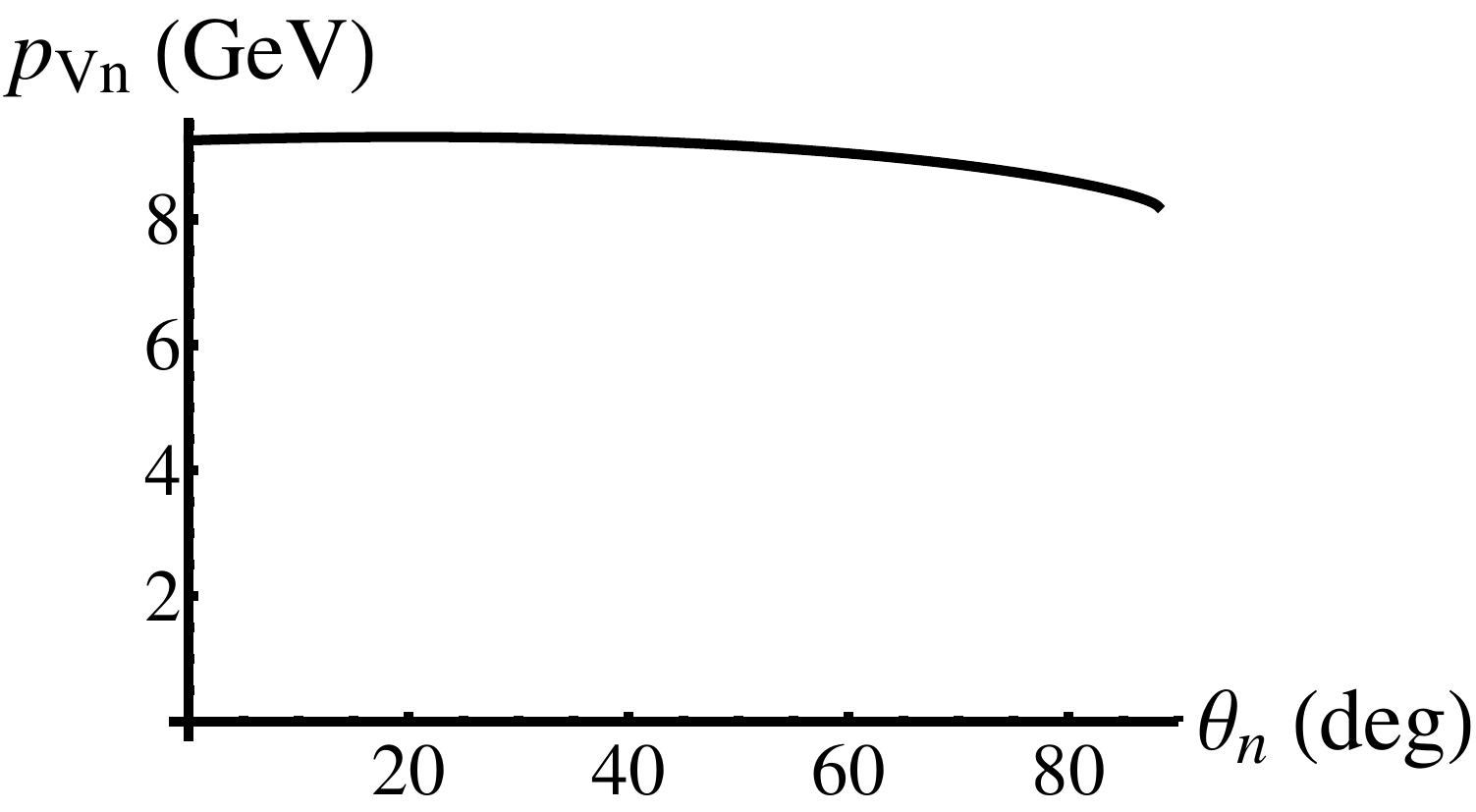}
        }%

    \end{center}
    \caption{%
       $p_{12}$ vs. $\theta_n$, for photon energy $\nu=10\;GeV$, and $t=-2\;GeV^2$.  (a) and (b) are for the ``plus" kinematics, (c) and (d) are for the ``minus" kinematics.
     }%
   \label{fig:p12}
\end{figure}

\begin{figure}[tbp]
     \begin{center}
        \subfigure[$p_{Vp}$]{%
            \label{fig:pvpplus}
            \includegraphics[width=0.4\textwidth]{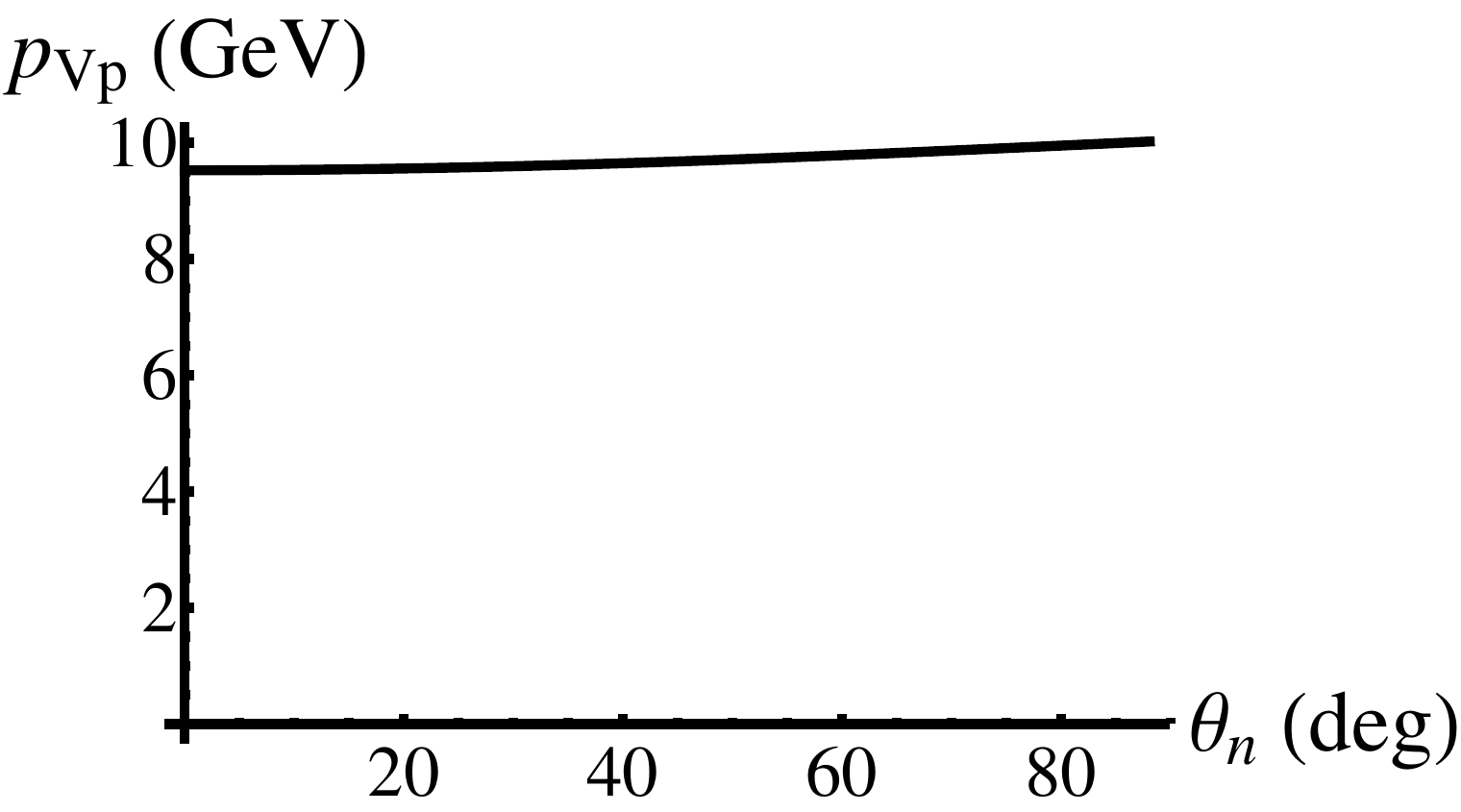}
        }%
        \hspace{0.5in}
         \subfigure[$p_{Vp}$]{%
           \label{fig:pvpmin}
           \includegraphics[width=0.4\textwidth]{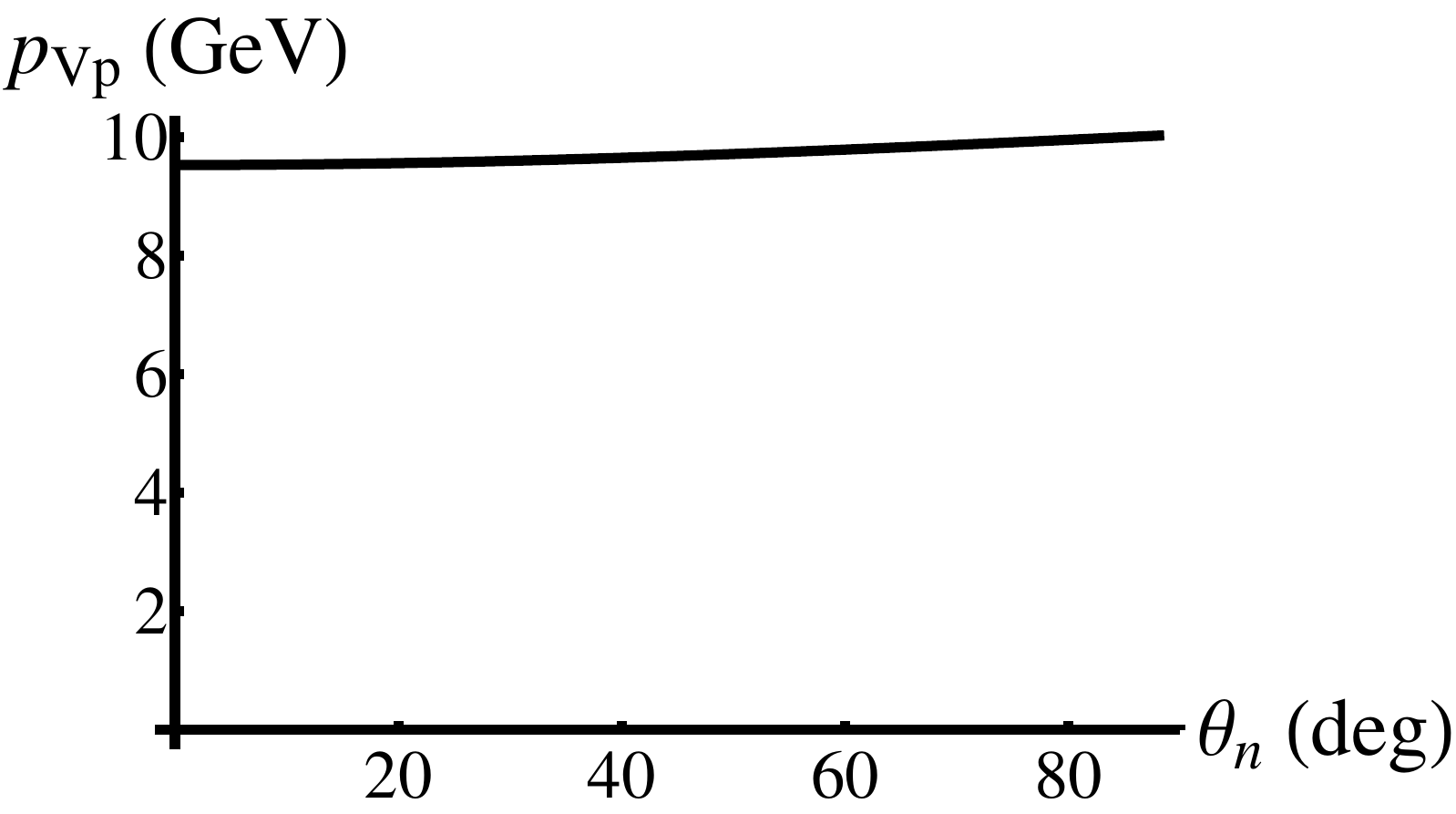}
        }\\ 


    \end{center}
    \caption{%
       $p_{Vp}$ vs. $\theta_n$, for photon energy $\nu=10\;GeV$, and $t=-2\;GeV^2$.  (a) is for the ``plus" kinematics, (c) is for the ``minus" kinematics.
     }%
   \label{fig:pvp}
\end{figure}

Note that for a given $t$, $p_n$, and $\theta_n$, there are 2 sets of allowed values of the proton and $J/\psi$ momentum $\lbrace \mathbf{p}_p,\mathbf{p}_V \rbrace$; I've called the two sets the ``plus" set and the ``minus" set.  If we define $x$ and $z$ axes as in Fig. \ref{fig:plusminpic}, with the $x$-component of the neutron momentum always positive, then the ``plus" kinematics is as shown in Fig. \ref{fig:pluskin} and the ``minus" kinematics is as shown in Fig. \ref{fig:minuskin}. As seen from Fig. \ref{fig:plusandminuskin}, for the ``plus" kinematics, $p_{px}$ is negative for all $\theta_n$ (while $p_{Vx}$ takes both positive and negative values over the range of $\theta_n$), while for the ``minus" kinematics, $p_{Vx}$ is negative for all $\theta_n$ (while $p_{px}$ takes both positive and negative values over the range of $\theta_n$).
\begin{figure}[tbp]
     \begin{center}
        \subfigure[]{%
            \label{fig:pluskin}
            \includegraphics[width=0.7\textwidth]{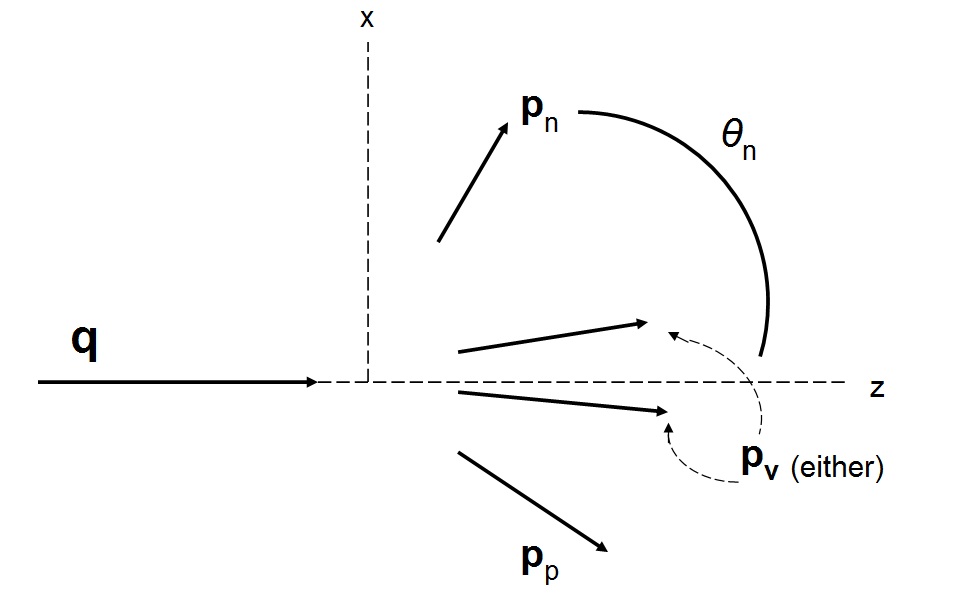}
        }%
        \hspace{0.5in}
         \subfigure[]{%
           \label{fig:minuskin}
           \includegraphics[width=0.7\textwidth]{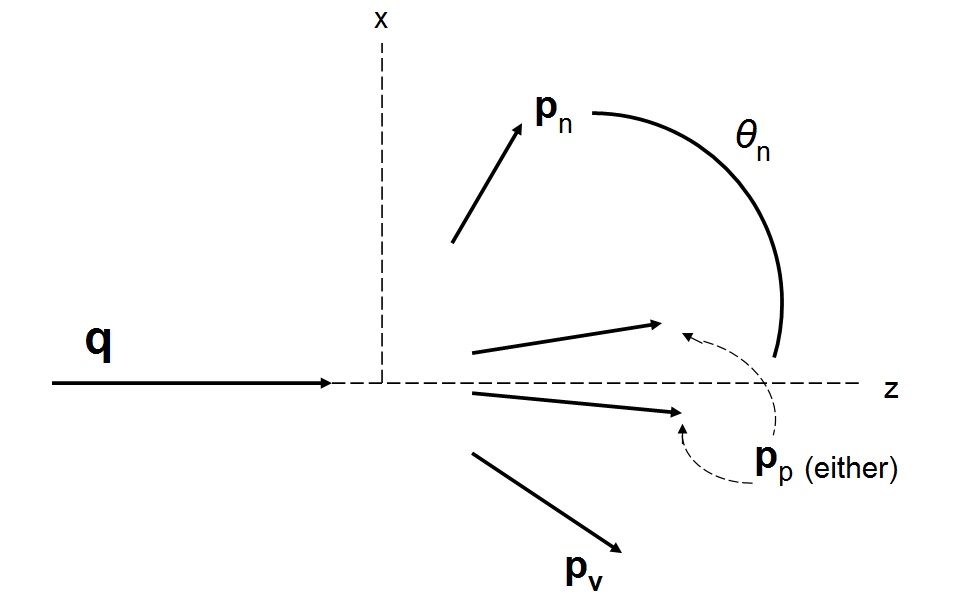}
        }\\ 


    \end{center}
    \caption{%
       (a) ``plus" kinematics and (b) ``minus" kinematics.  For ``plus", $\mathbf{p}_p$ is always on the opposite side of the photon momentum $\mathbf{q}$ direction as the neutron momentum.  For ``minus", $\mathbf{p}_V$ is always on the opposite side of the photon momentum $\mathbf{q}$ direction as the neutron momentum.
     }%
   \label{fig:plusminpic}
\end{figure}

\begin{figure}[tbp]
     \begin{center}
        \subfigure[]{%
            \label{fig:ppxgraph}
            \includegraphics[width=0.7\textwidth]{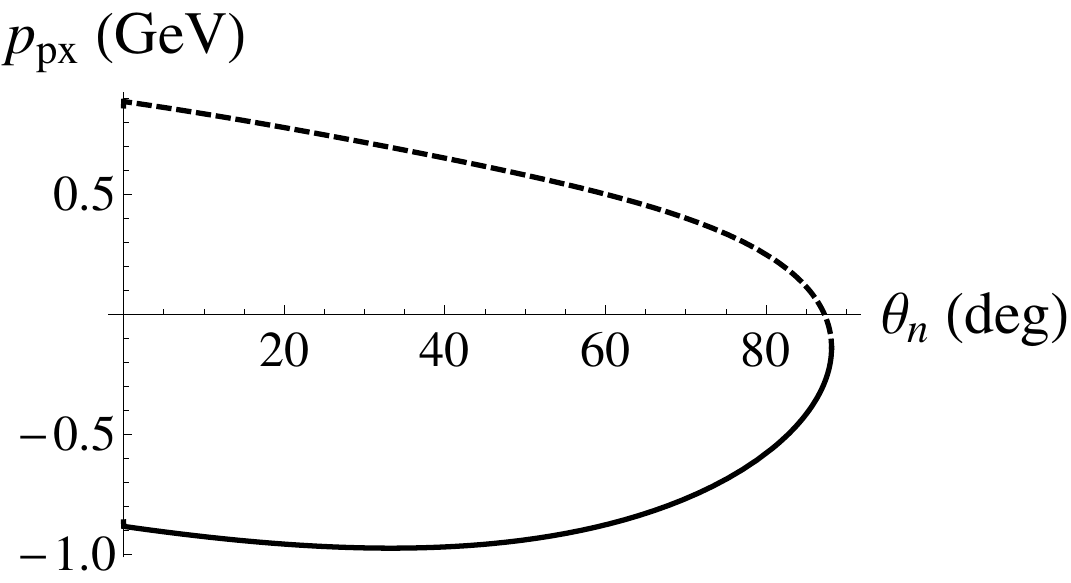}
        }%
        \hspace{0.5in}
         \subfigure[]{%
           \label{fig:pvxgraph}
           \includegraphics[width=0.7\textwidth]{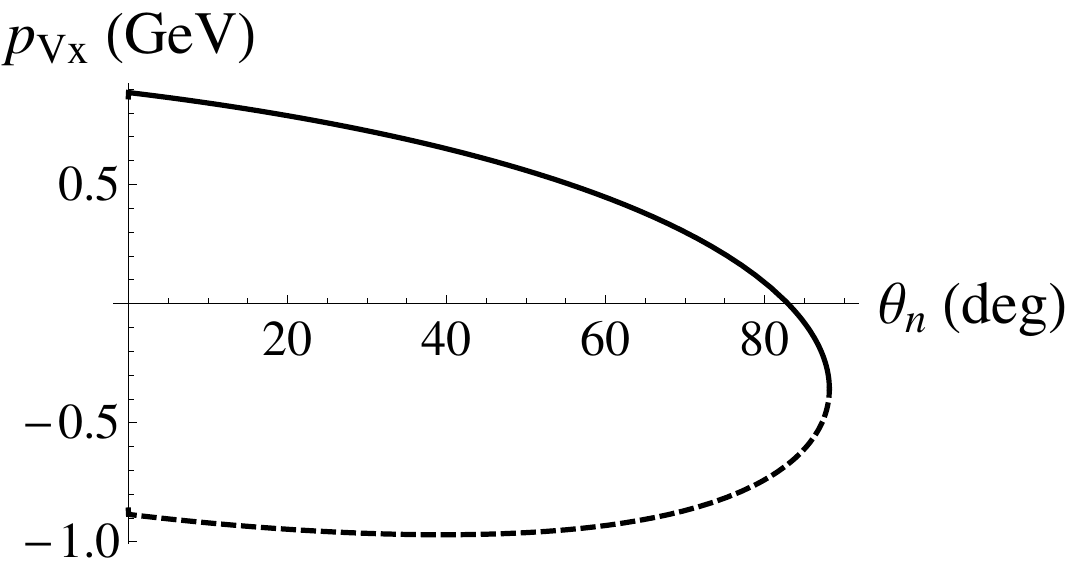}
        }\\ 


    \end{center}
    \caption{%
       (a) $p_{px}$ vs. $\theta_n$ and (b) $p_{Vx}$ vs. $\theta_n$, for photon energy $\nu=10\;GeV$, and $t=-2\;GeV^2$.  The solid curves are the ``plus" kinematics, and the dashed curves are the ``minus" kinematics. 
     }%
   \label{fig:plusandminuskin}
\end{figure}

\begin{figure}[tbp]
     \begin{center}
        \subfigure[p-n rescattering]{%
            \label{fig:nminpnmin}
            \includegraphics[width=0.4\textwidth]{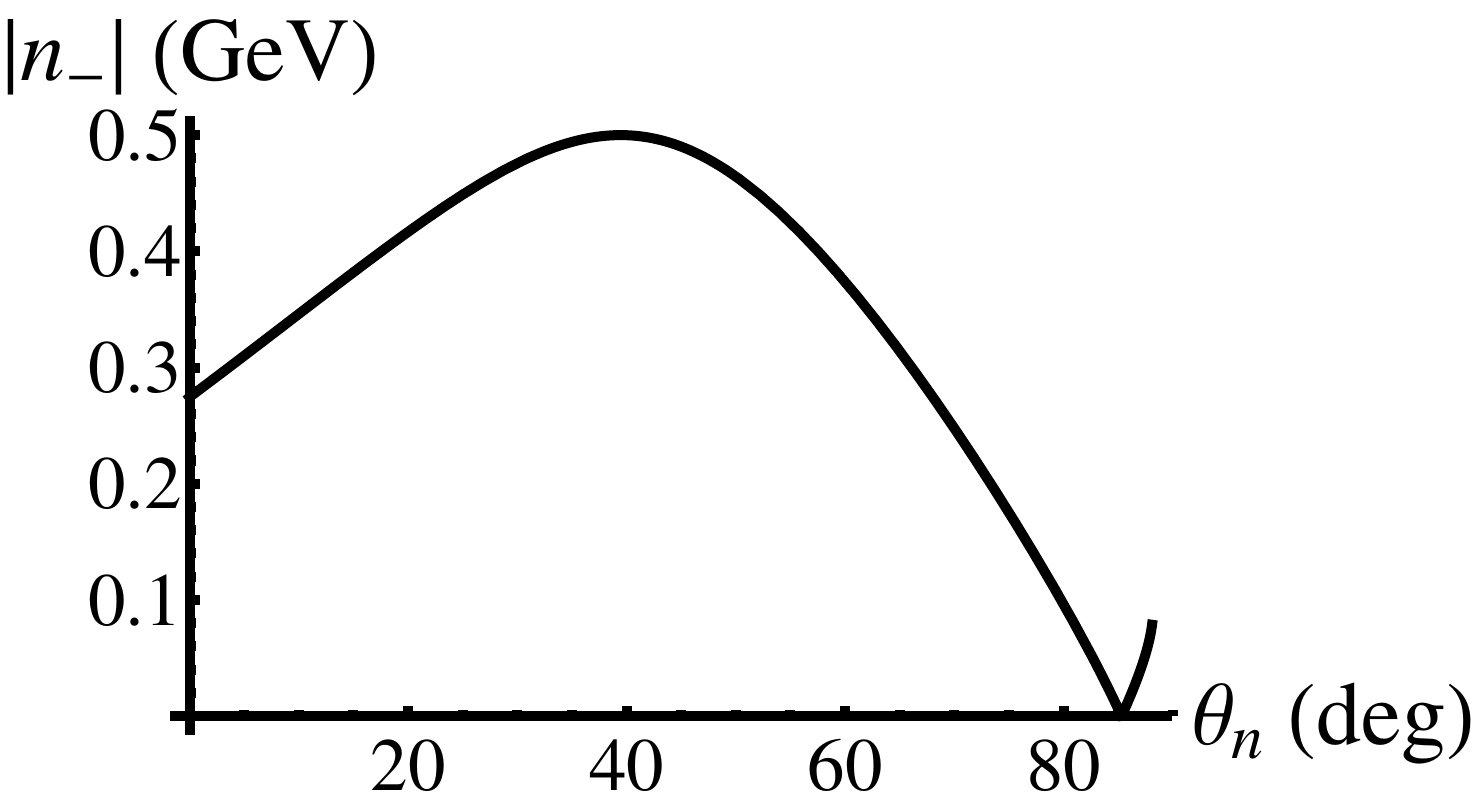}
        }%
        \hspace{0.5in}
         \subfigure[$J/\psi$-n rescattering]{%
           \label{fig:nminvnmin}
           \includegraphics[width=0.4\textwidth]{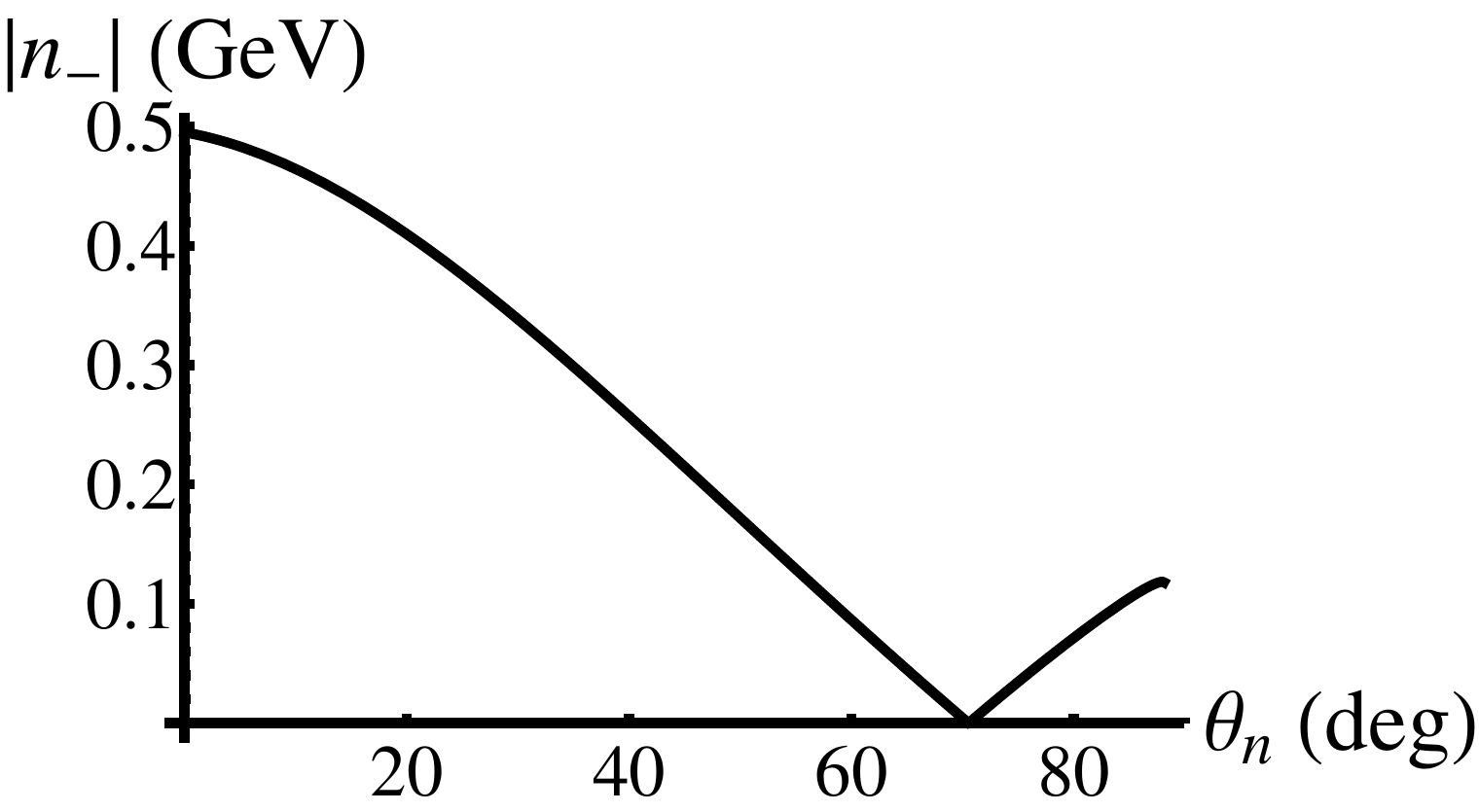}
        }\\ 

        \subfigure[p-n rescattering]{%
            \label{fig:nminpnpl}
            \includegraphics[width=0.4\textwidth]{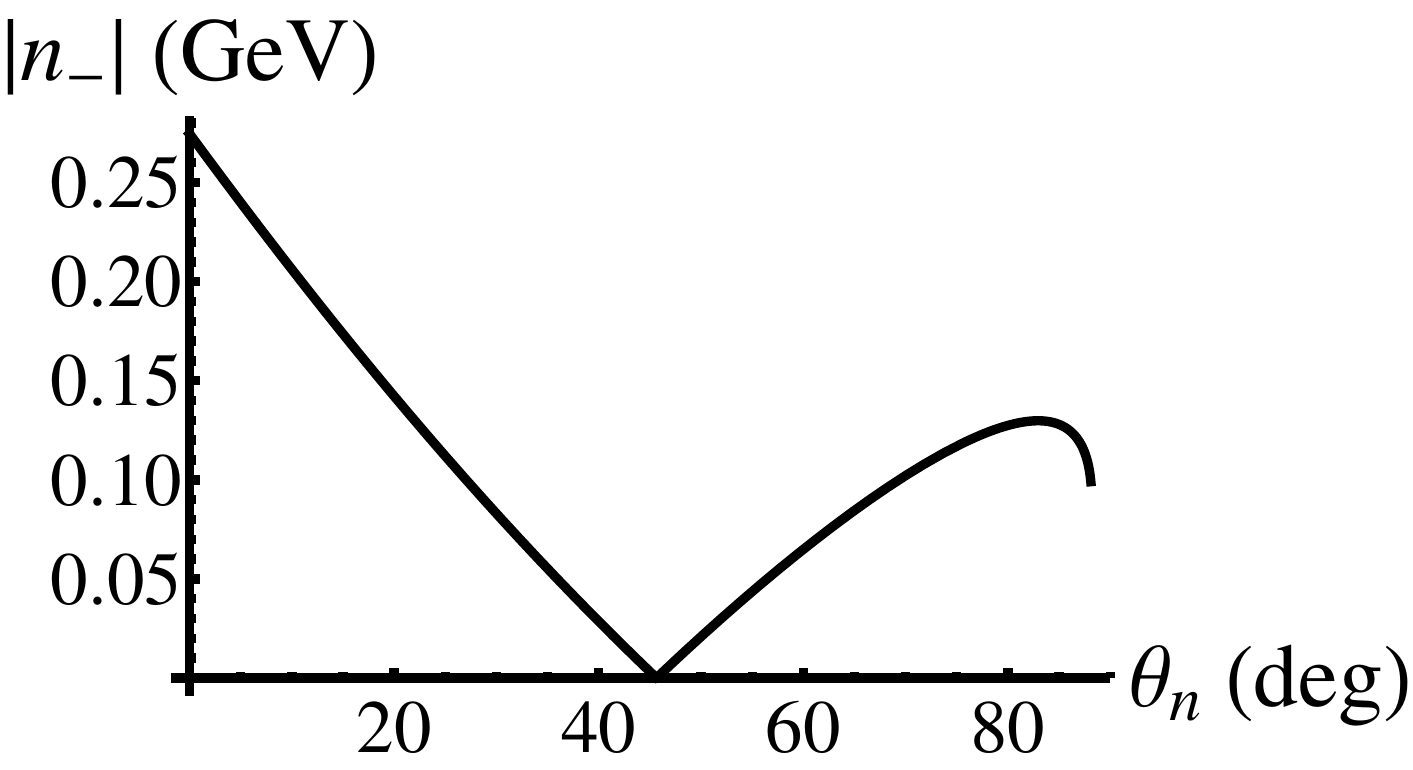}
        }%
         \hspace{0.5in} 
        \subfigure[$J/\psi$-n rescattering]{%
            \label{fig:nminvnpl}
            \includegraphics[width=0.4\textwidth]{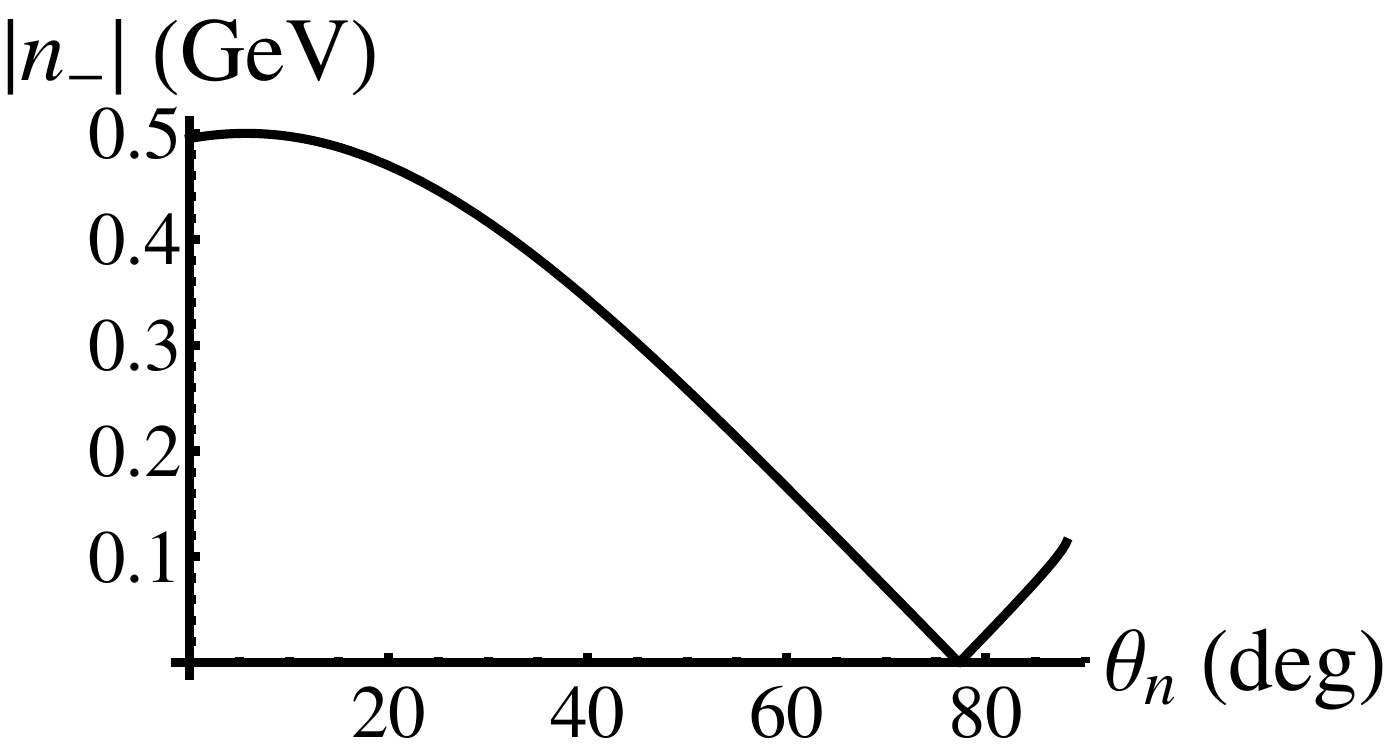}
        }%

    \end{center}
    \caption{%
       $\vert n_-\vert$ vs. $\theta_n$, for photon energy $\nu=10\;GeV$, and $t=-2\;GeV^2$.  (a) and (b) are for the ``minus" kinematics, (c) and (d) are for the ``plus" kinematics.
     }%
   \label{fig:nmin}
\end{figure}

\begin{figure}[tbp]
     \begin{center}
        \subfigure[$J/\psi$-p rescattering]{%
            \label{fig:nminpnmin}
            \includegraphics[width=0.4\textwidth]{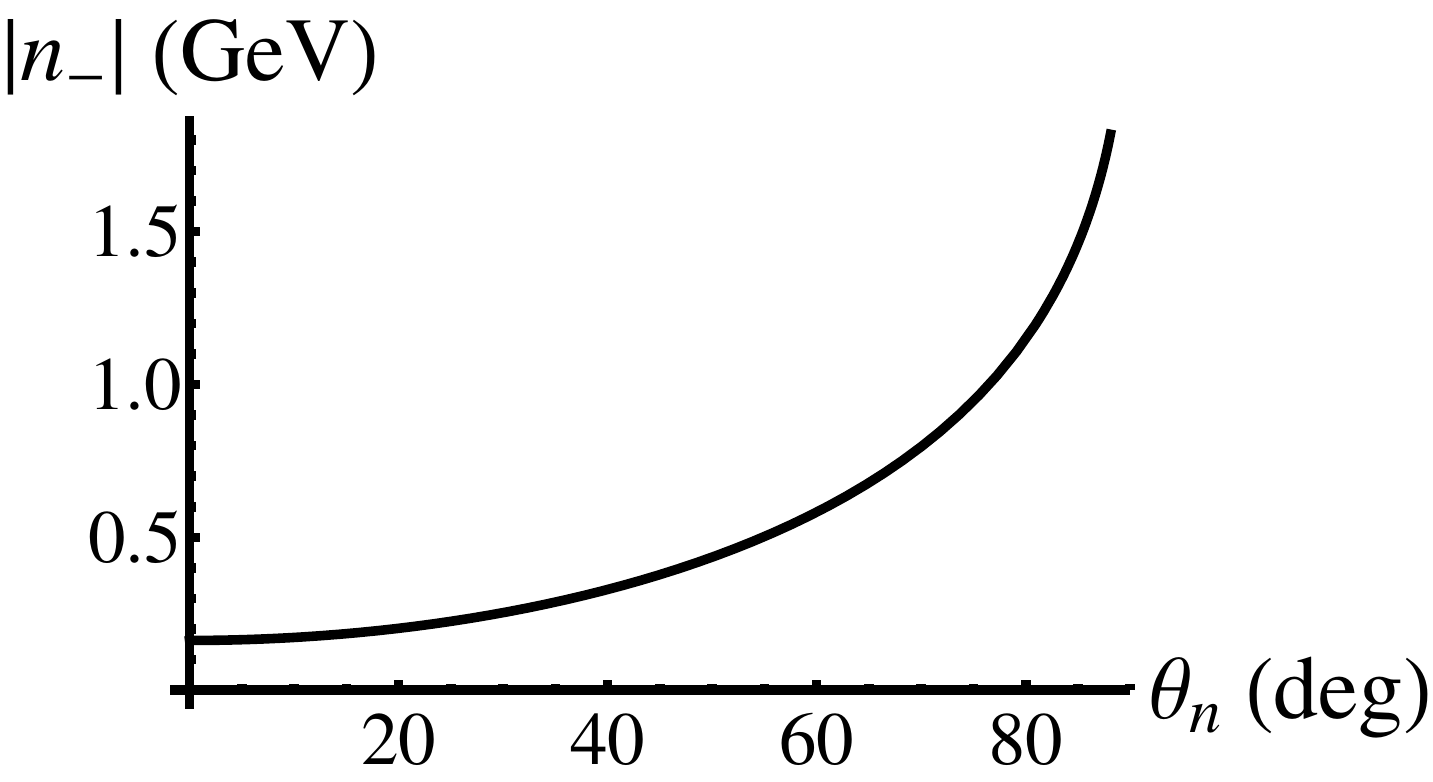}
        }%
        \hspace{0.5in}
         \subfigure[$J/\psi$-p rescattering]{%
           \label{fig:nminvnmin}
           \includegraphics[width=0.4\textwidth]{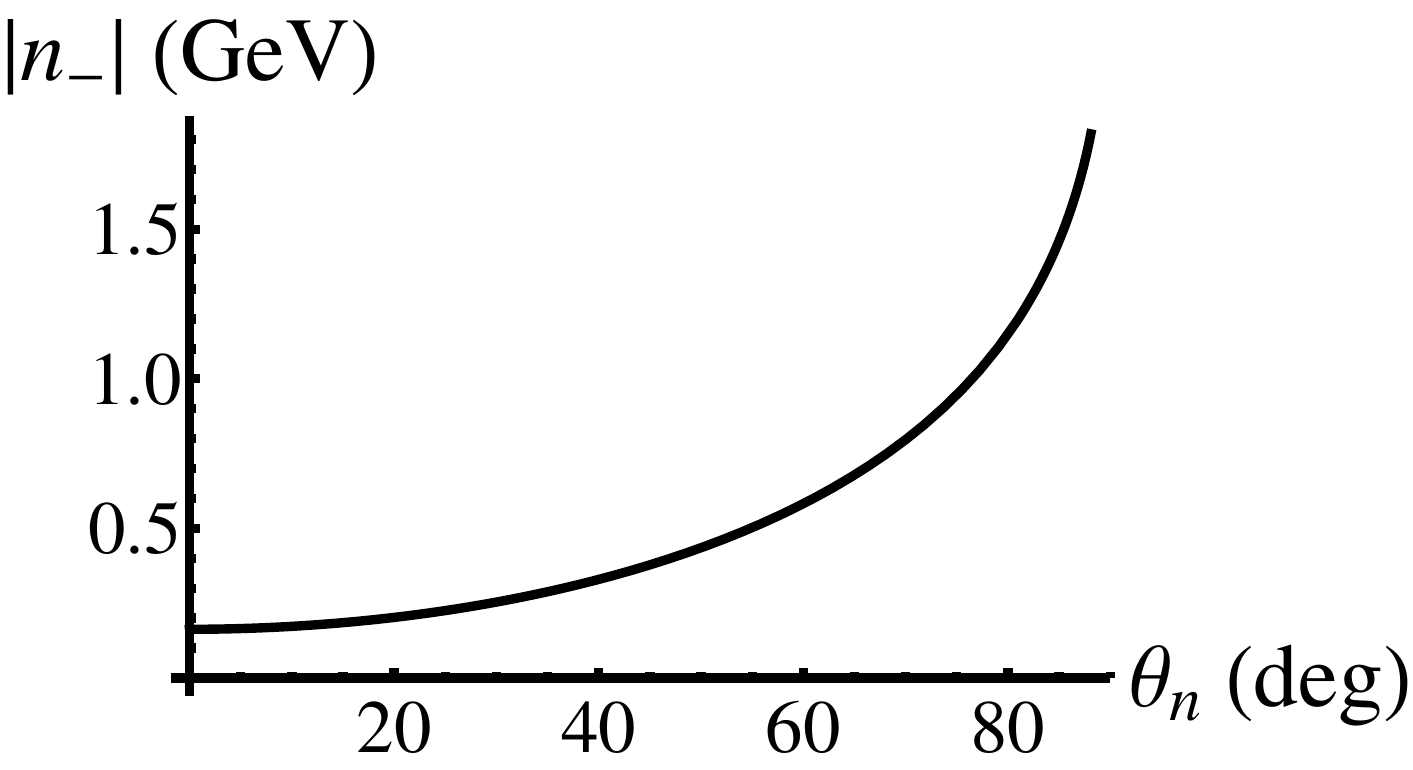}
        }\\ 


    \end{center}
    \caption{%
       $\vert n_-\vert$ vs. $\theta_n$ for $J/\psi$-p rescattering, for photon energy $\nu=10\;GeV$, and $t=-2\;GeV^2$.  (a) is for the ``minus" kinematics, (b) is for the ``plus" kinematics.
     }%
   \label{fig:nminvp}
\end{figure}

\begin{figure}[tbp]
     \begin{center}
        \subfigure[$p-n$ rescatter]{%
            \label{fig:int2aplus}
            \includegraphics[width=0.4\textwidth]{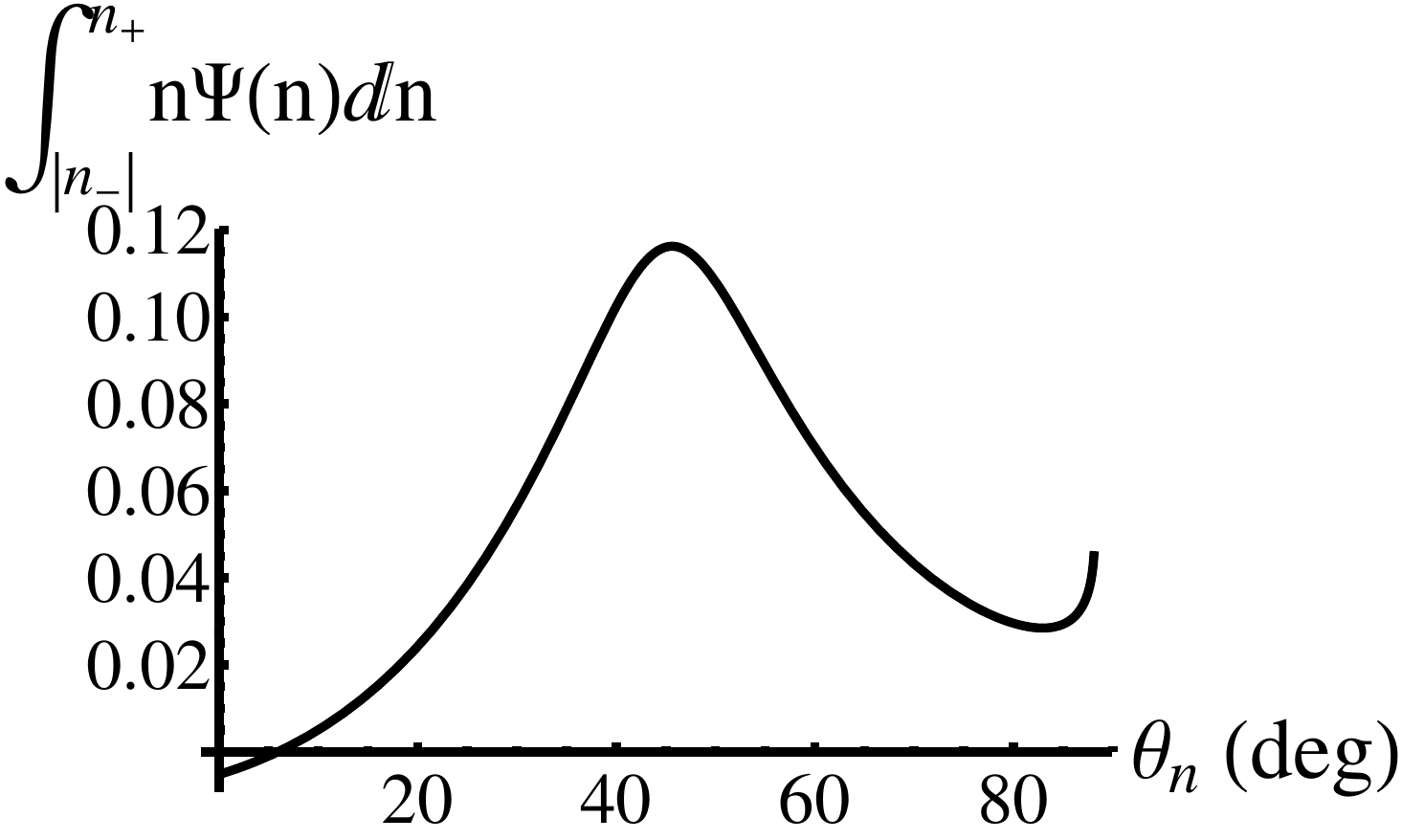}
        }%
        \hspace{0.5in}
         \subfigure[$V-n$ rescatter]{%
           \label{fig:int3aplus}
           \includegraphics[width=0.4\textwidth]{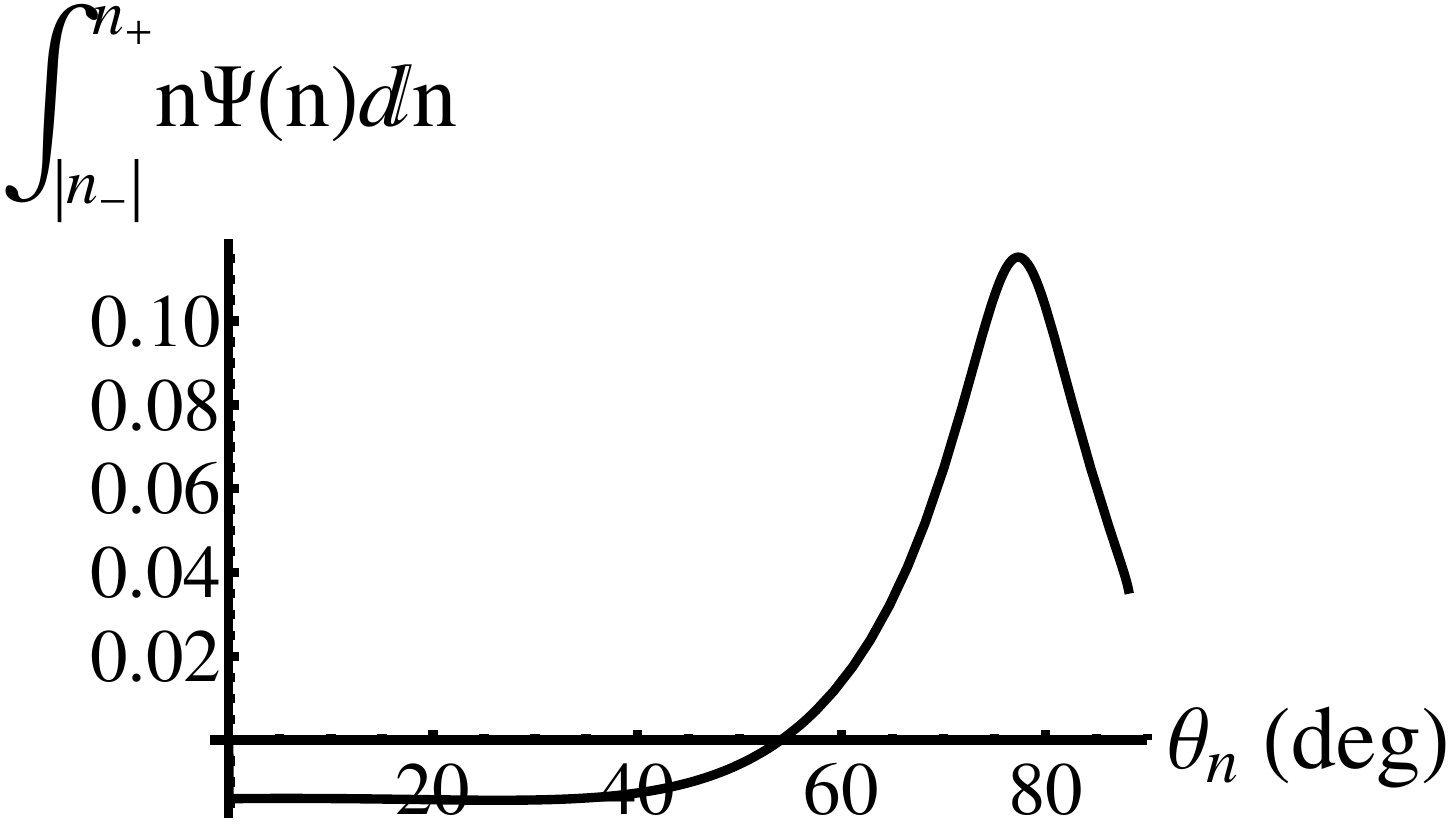}
        }\\ 

        \subfigure[$p-n$ rescatter]{%
            \label{fig:int2amin}
            \includegraphics[width=0.4\textwidth]{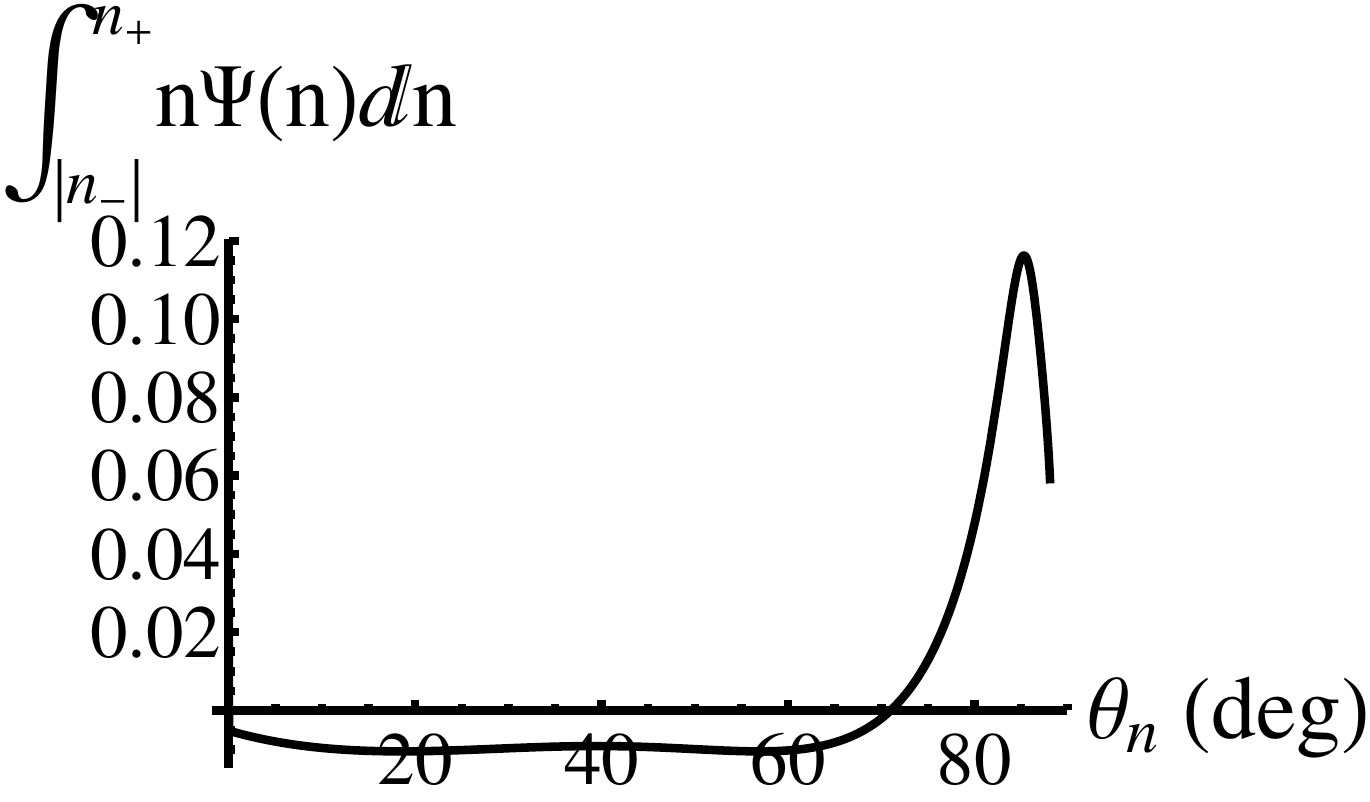}
        }%
         \hspace{0.5in} 
        \subfigure[$V-n$ rescatter]{%
            \label{fig:int3amin}
            \includegraphics[width=0.4\textwidth]{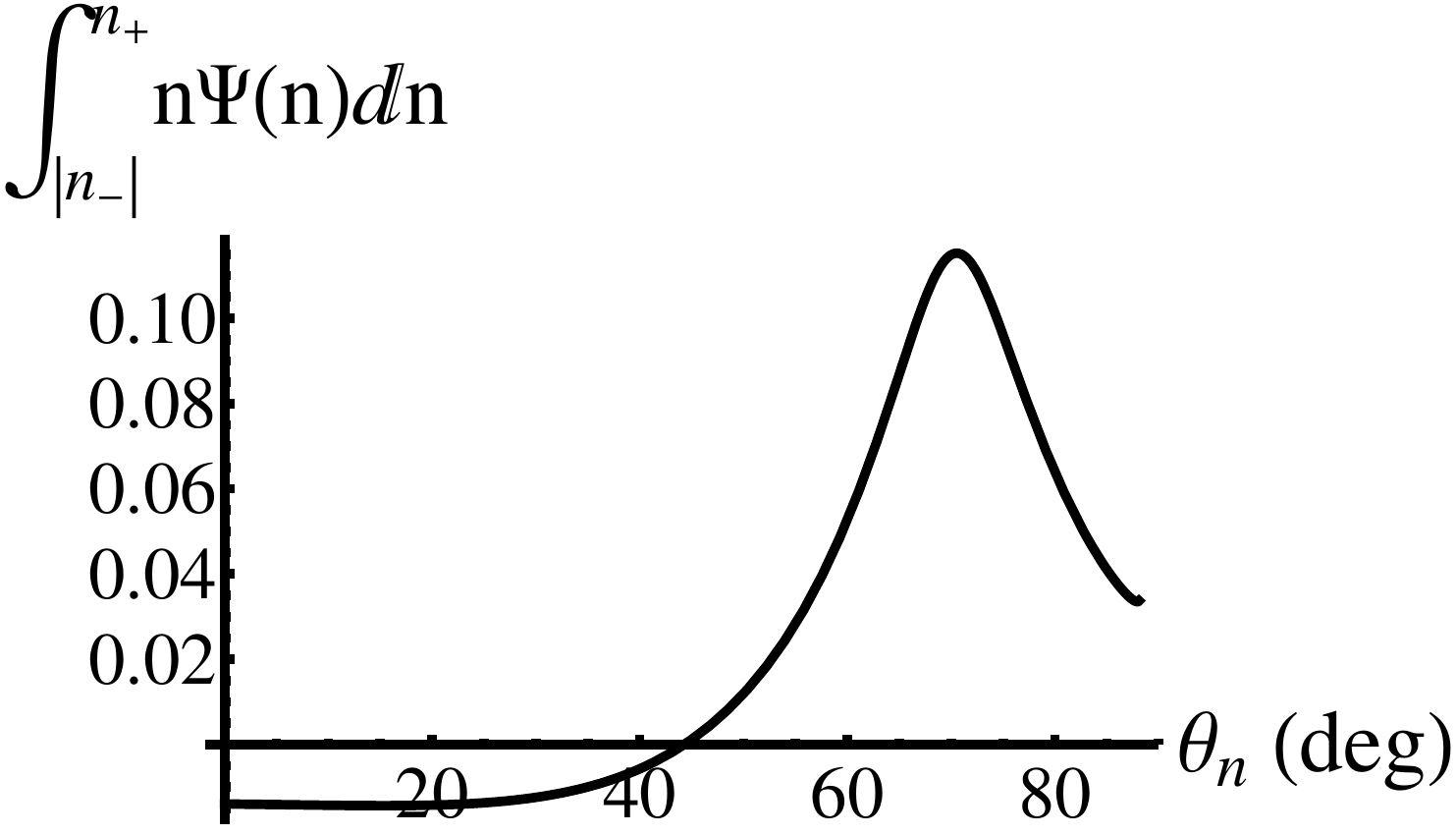}
        }%

    \end{center}
    \caption{%
      The integral from \eq{constMonshell}, for photon energy $\nu=10\;GeV$, and $t=-2\;GeV^2$.  (a) and (b) are for the ``plus" kinematics, (c) and (d) are for the ``minus" kinematics.
     }%
   \label{fig:onshellint}
\end{figure}
Figs. \ref{fig:nmin} and  Fig. \ref{fig:nminvp} show graphs of $\vert n_-\vert$ vs. $\theta_n$ for the 3 different pairs of outgoing particles.  Since the value of $\vert n_-\vert$ varies greatly with $\theta_n$, the value of the integral $\int_{\vert n_-\vert}^{n_+}dn\; n\Psi(n) $ varies greatly also, and has a prominent peak at the value of $\theta_n$ for which $\vert n_-\vert =0$.  This can be seen from Fig. \ref{fig:onshellint}, which shows the integral vs. $\theta_n$.  The general shape of these graphs is retained in the calculation of the exact amplitudes (i.e., including all dependence of the elementary amplitudes  $ {\cal M}^{\gamma V}$ and ${\cal M}$ on the internal momentum $\mathbf{n}$), including the location of the peaks.

\subsubsection{Calculation of Amplitudes}
For the calculation of the  amplitudes, the elementary amplitudes $ {\cal M}^{\gamma V}$ and ${\cal M}$ are taken to be of the diffractive form $Ae^{\frac{1}{2}B t}$ with parameters determined from existing experimental data.  For the $J/\psi$-nucleon rescattering diagrams, the only available data is from the experiment at SLAC~\cite{psidata77} discussed in Sec. \ref{sec:parameterssubsec}.  They determined the total $J/\psi$-nucleon cross-section to be $\sigma_{tot}^{J/\psi\; N}=3.5\pm0.8\;mb$, which gives via the optical theorem $A_{Vn}=1.61\pm0.4\;GeV^{-4}$.  The energy of the $J/\psi$ in this experiment was $\sim20$ GeV in the Lab frame (nucleon at rest).  However, for our kinematics the rescattering of the $J/\psi$ on the nucleon takes place at an energy in the outgoing neutron's rest frame of from 6 to 10 GeV, which is significantly smaller than in the SLAC experiment; thus the value of $A_{Vn}$ at our energy may be significantly different.  Since the entire reason for measuring the cross-section for this process is to extract the $J/\psi$-nucleon scattering amplitude in an energy region where it has not been measured before, I've used several different values of the parameter $A_{Vn}$ in the calculations, from the value measured at SLAC up to 10 times the SLAC value.  Since the total cross-section $\sigma_{tot}$ for $J/\psi$-nucleon scattering goes like $\sqrt{A_{Vn}}$ (Eq. \ref{Aandsigma}), this corresponds to a range of $\sigma_{tot}$ (which is what was actually measured in the SLAC experiment) of from 1 to $\sim3$ times the SLAC value.  In~\cite{brodsky97}, theoretical calculation of the $J/\psi$-nucleon elastic scattering cross-section at threshold yielded $7\;mb$, which is twice the value measured at higher energy at SLAC.

The full calculation of the amplitudes must of course include the off-shell parts.  If we use the same parametrization of the elementary amplitudes as in the on-shell part, then the off-shell parts are very small compared to the on-shell parts.
Fig. \ref{fig:2a3atot} shows $\vert F_{2a}+F_{3a}\vert^2$ (which are the amplitudes for production on the proton and rescattering on the neutron) as a function of $\theta_n$.  The negative values of $\theta_n$ are for the ``minus" kinematics, and the positive values are for the ``plus" kinematics.  It can be seen that the difference between the on-shell amplitude and the total (on- plus off-shell) amplitude is negligible.  Three different values of the $J/\psi$-nucleon amplitude coefficient $A_{Vn}$ are used.  It is seen that only if $A_{Vn}$ is of the order of 10 times as large as the previously measured value is there a noticeable peak due to the $J/\psi$-neutron rescattering.  The $p-n$ rescattering peak is much larger than, and close enough to, the $J/\psi$-neutron rescattering peak that it obscures the $J/\psi$ peak.  Note also that the position of each of the peaks is simply given by the value of $\theta_n$ where the value of the corresponding $\vert n_-\vert$ is zero (see Fig. \ref{fig:nmin}).

\begin{figure}[tbp]
     \begin{center}
        \subfigure[$A_{Vn}=1.6\;GeV^{-4}$]{%
            \label{fig:2a3a}
            \includegraphics[width=0.4\textwidth]{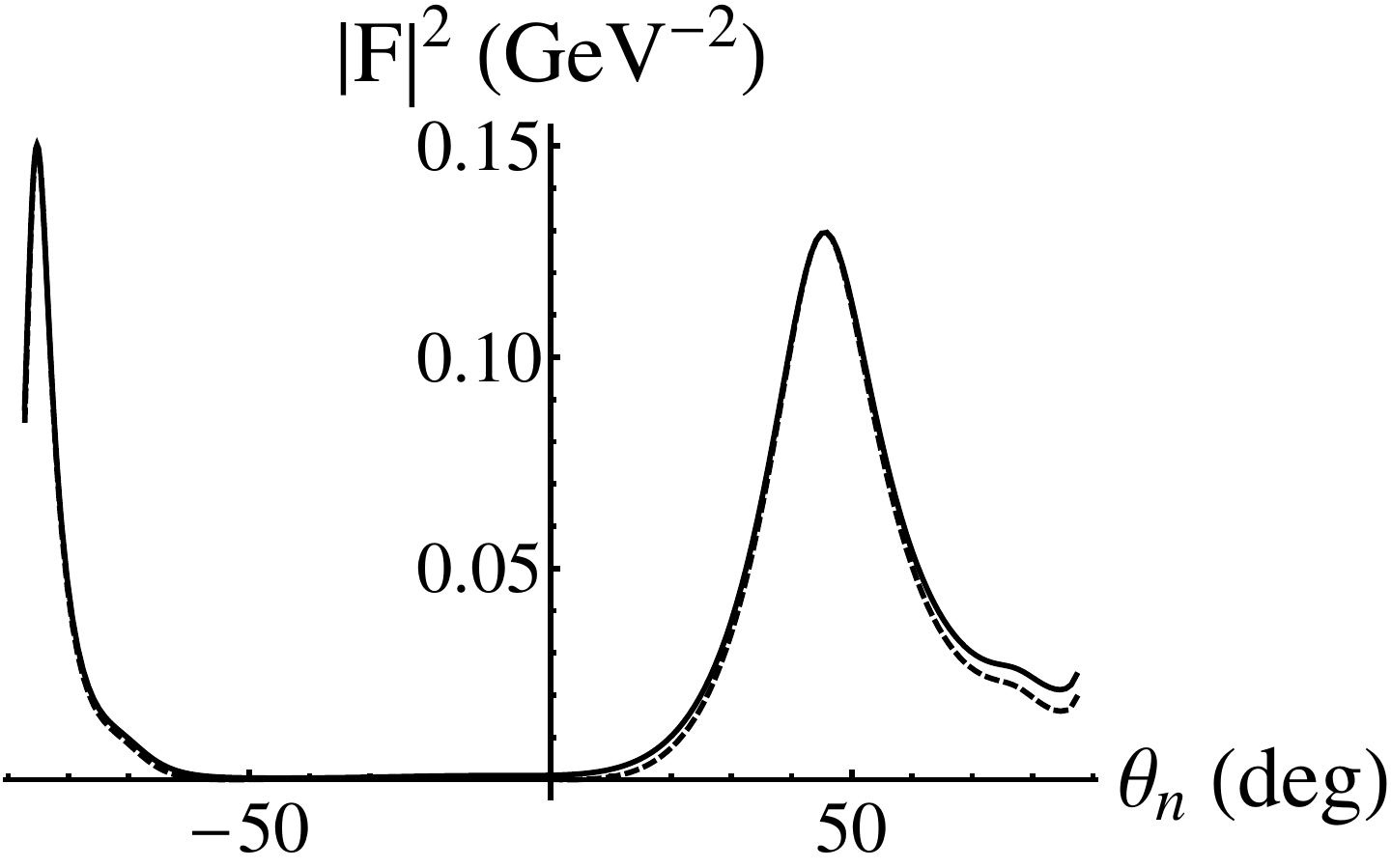}
        }%
        \hspace{0.5in}
         \subfigure[$A_{Vn}=5\times1.6\;GeV^{-4}$]{%
           \label{fig:2a3a5A}
           \includegraphics[width=0.4\textwidth]{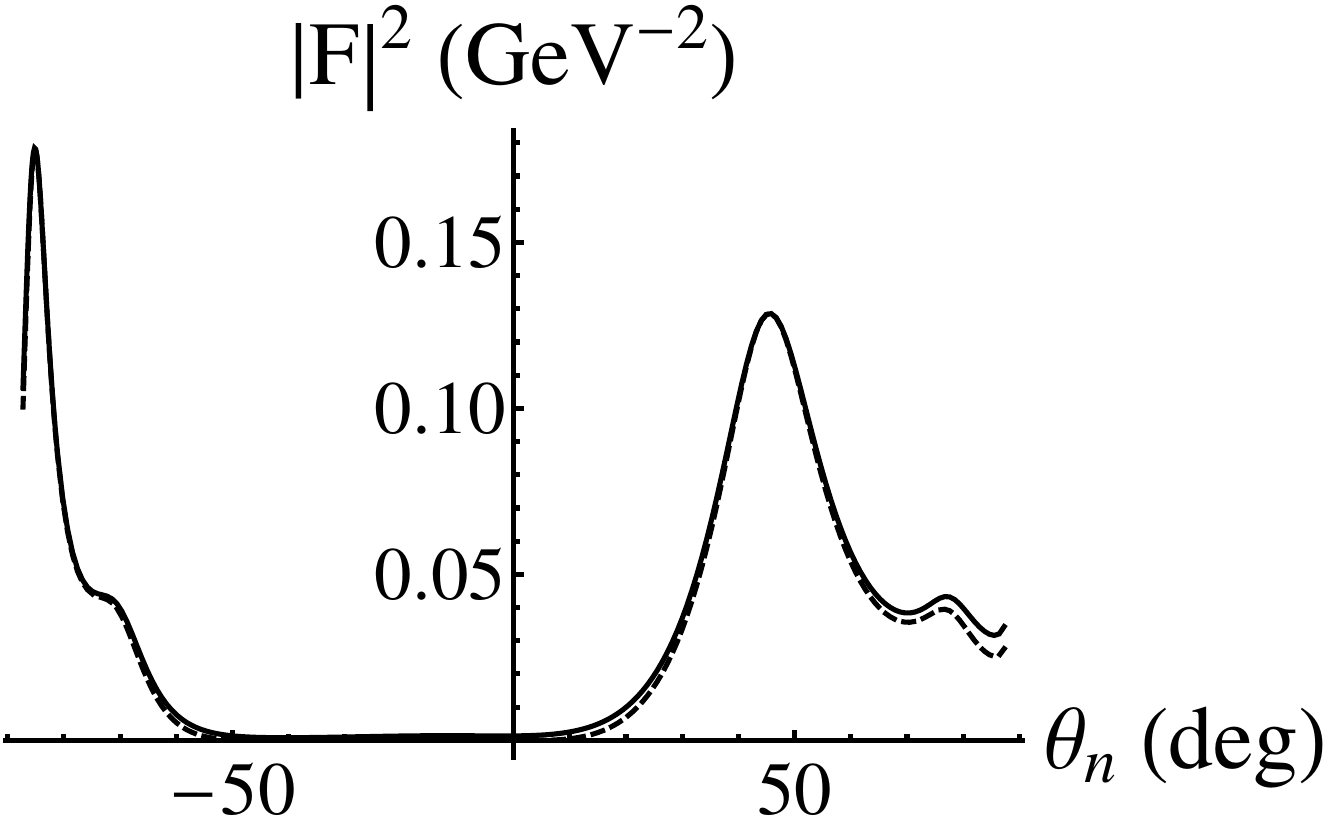}
        }\\ 

        \subfigure[$A_{Vn}=10\times1.6\;GeV^{-4}$]{%
            \label{fig:2a3a10A}
            \includegraphics[width=0.4\textwidth]{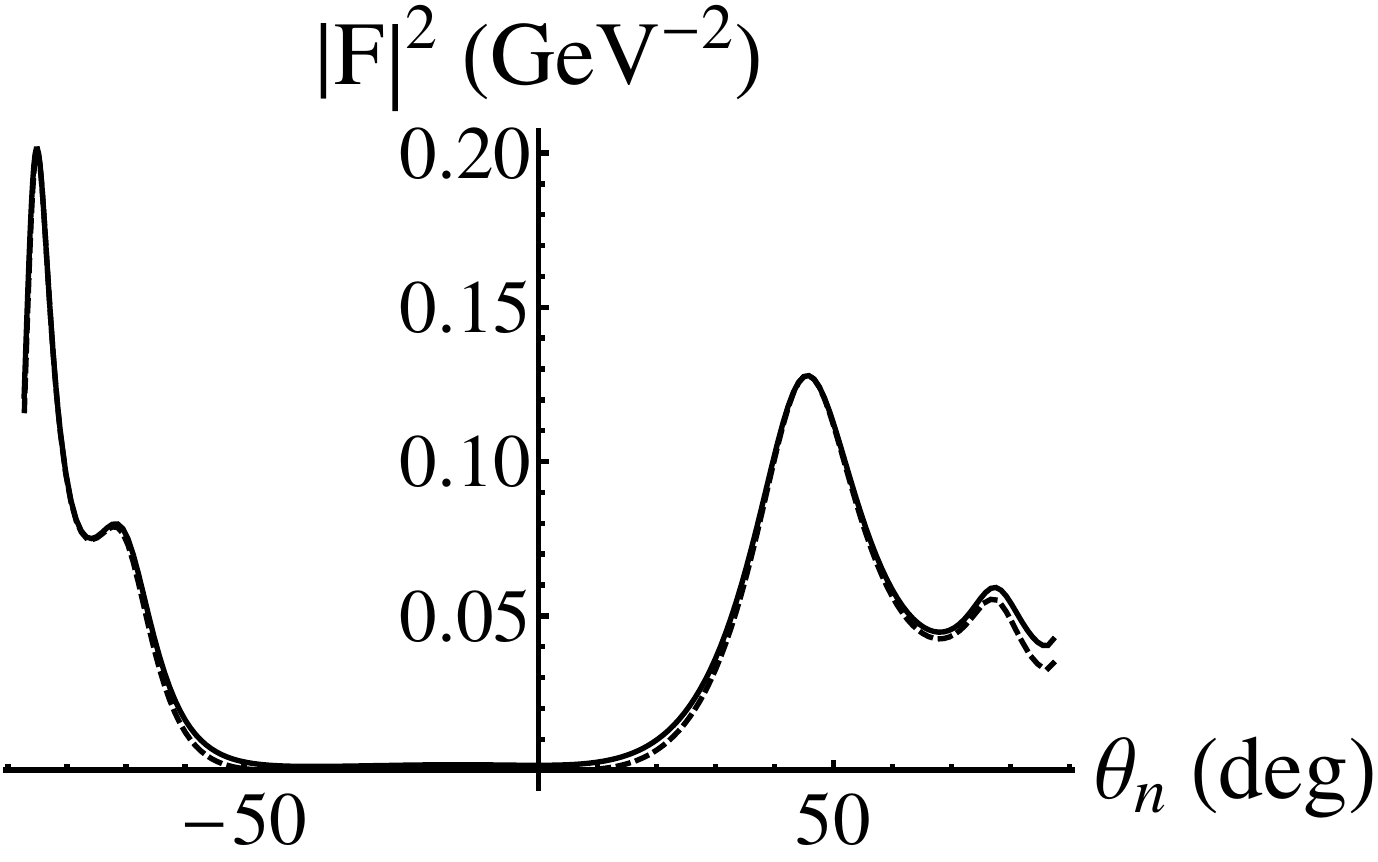}
        }%

    \end{center}
    \caption{%
      $\vert F_{1a}+F_{2a}+F_{3a}\vert^2$ vs. $\theta_n$, for 3 values of $A_{Vn}$, for $\nu=10\;GeV$, $t=-2\;GeV^2$.  The large peak on either side is due to $p-n$ rescattering, while the small peak (or bump) on either side is due to $J/\psi$-neutron rescattering.  The dashed curve is only including the on-shell amplitude, while the solid curve includes both the on-shell and the off-shell amplitude.  Here we show the ``plus" and ``minus" kinematics on the same graph by defining $\theta_n$ to be negative for the ``minus" kinematics. 
     }%
   \label{fig:2a3atot}
\end{figure}

The amplitudes $F_{2b}$ and $F_{3b}$, where the $J/\psi$ is produced on the neutron and then rescattering (of the neutron or $J/\psi$, respectively) occurs on the proton, are much smaller than $F_{2a}$ and $F_{3b}$, and do not exhibit the well-defined peaks that $F_{2b}$ and $F_{3b}$ do.  Fig. \ref{fig:electrocross} shows the 8-fold electroproduction differential cross-section, \eq{electrocross}, versus $\theta_n$.   Graphs are shown for 3 different values of the $J/\psi$-neutron elastic scattering parameter $A_{Vn}$:  $A_{Vn}=1.6\;GeV^{-4}$ (which is the value determined in the experiment at SLAC),  $A_{Vn}=8.0\;GeV^{-4}$, and $A_{Vn}=16\;GeV^{-4}$.  It is seen that only if $A_{Vn}$ is of the order of 10 times as large as the previously measured value is there a noticeable peak due to the $J/\psi$-neutron rescattering, for the ``plus" kinematics.  The $p-n$ rescattering peak is much larger than, and close enough to, the $J/\psi$-neutron rescattering peak that it obscures the $J/\psi$ peak.  For the ``minus" kinematics, the same statement holds; in addition, however, the size of the $p-n$ rescattering peak varies (by $\sim 40\%$) as the value of $A_{Vn}$ is varied.

It's important to note that the peak due to the $J/\psi$-neutron rescattering isn't observable at lower energies.  Fig. \ref{fig:nu9} shows the square of the total amplitude for photon energy of $\nu=9\;GeV$ and $t=-3$, for $A_{Vn}=10\times 1.6\;GeV^{-4}$.  On this graph the peak due to $p-n$ rescattering is visible, but there's no visible peak due to $J/\psi$-neutron rescattering.

\begin{figure}[tbp]
     \begin{center}
        \subfigure[]{%
            \label{fig:2a3a}
            \includegraphics[width=0.6\textwidth]{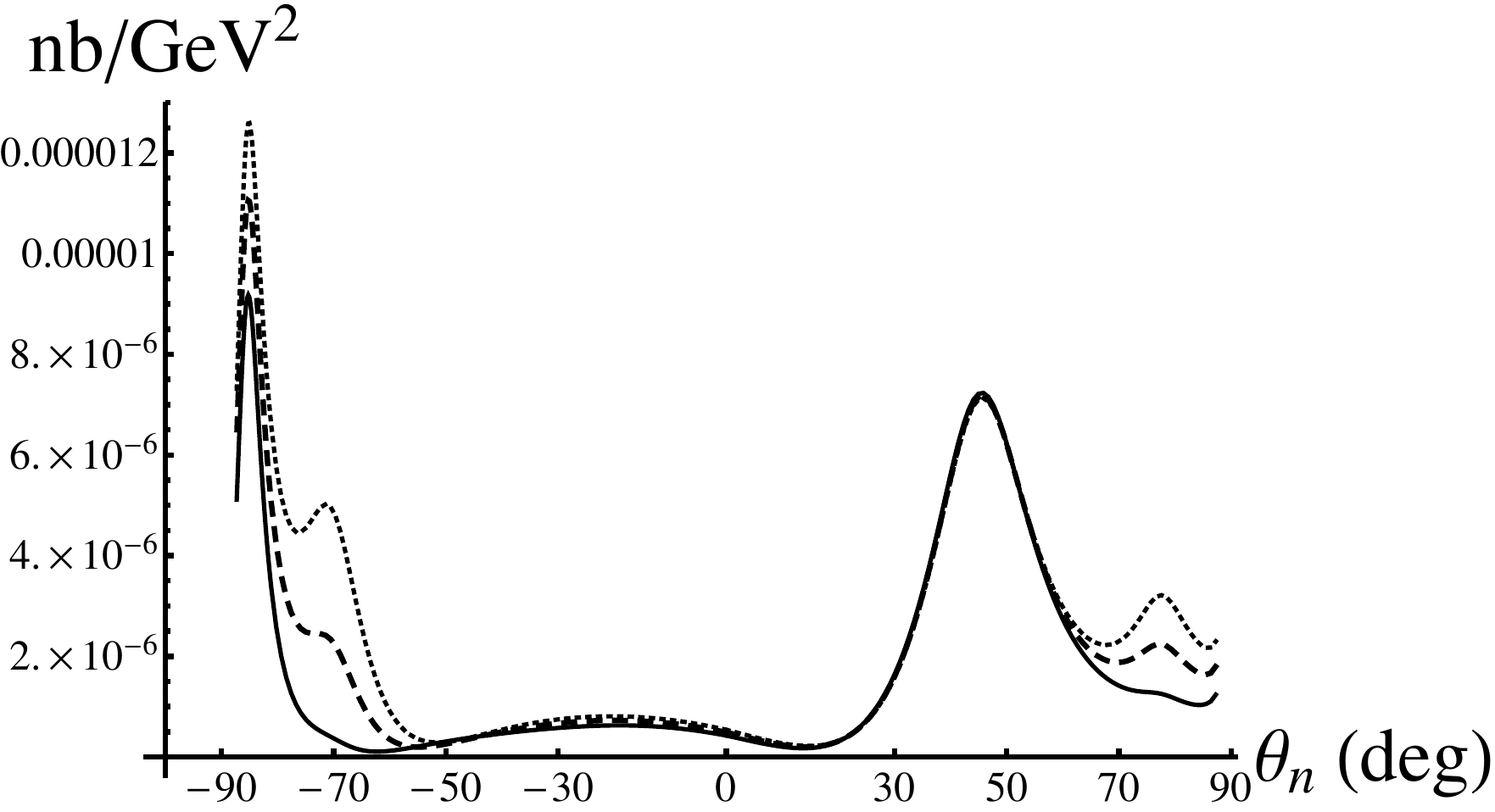}
        }%
        \hspace{0.5in}
         \subfigure[]{%
           \label{fig:2a3a5A}
           \includegraphics[width=0.6\textwidth]{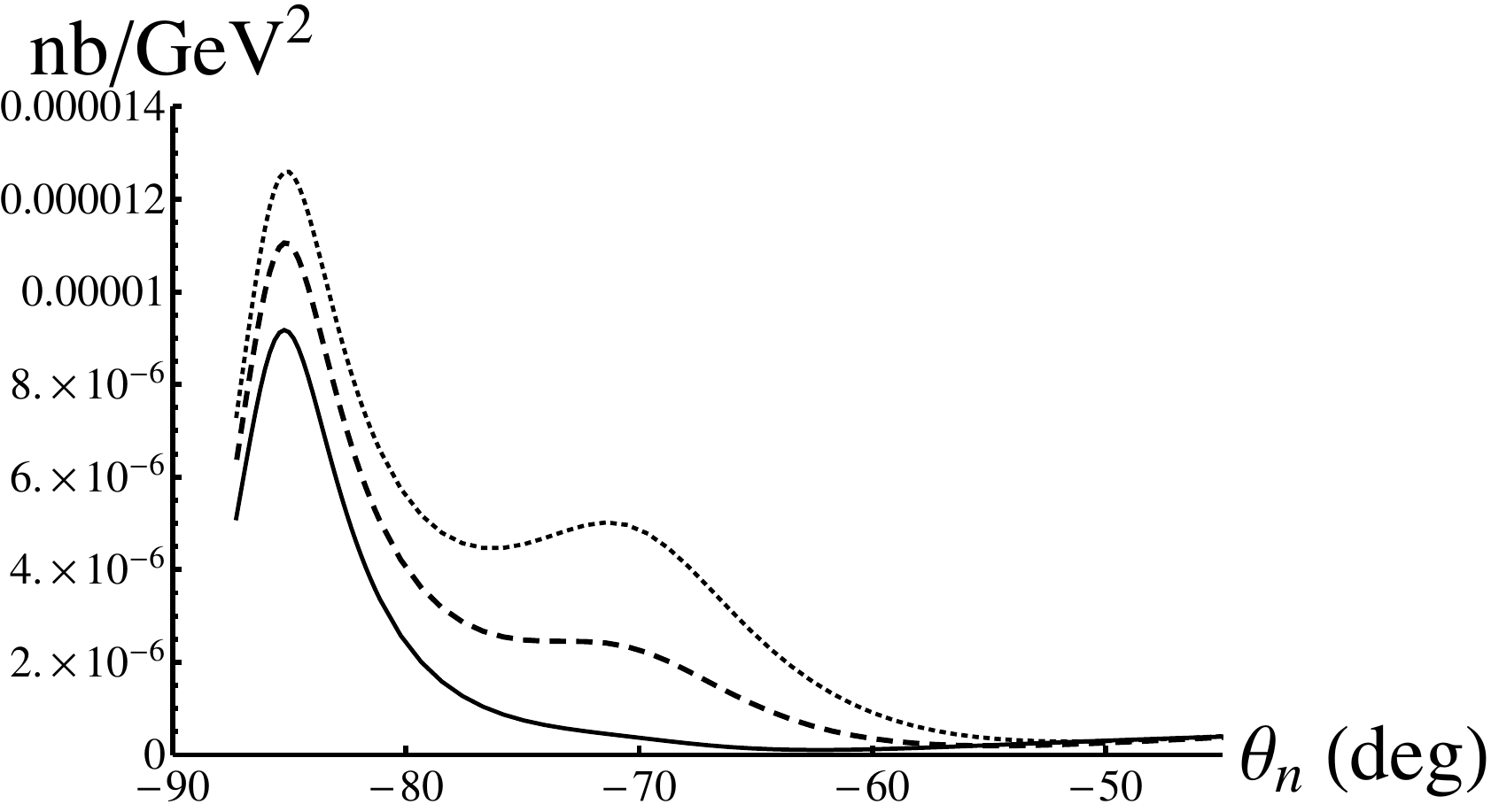}
        }\\ 
%

    \end{center}
    \caption{%
       Electroproduction differential cross-section vs. $\theta_n$, including all diagrams, for photon energy $\nu=10\;GeV$, $Q^2=0.5\;GeV^2$ and $t=-2\;GeV^2$, for 3 values of $A_{Vn}$:  Solid curve:  $A_{Vn}=1.6\;GeV^{-4}$.  Dashed curve:  $A_{Vn}=8.0\;GeV^{-4}$.  Dotted curve:  $A_{Vn}=16\;GeV^{-4}$.  In (a), the large peaks at $\theta_n\simeq 45^{\circ}$ and $\theta_n\simeq-85^{\circ}$ are due to $p-n$ rescattering, while the  small peaks (or bumps) at $\theta_n\simeq 80^{\circ}$ and $\theta_n\simeq-70^{\circ}$ are due to $J/\psi$-neutron rescattering.  (b) shows detail of left half of (a).  
     }%
   \label{fig:electrocross}
\end{figure}

\begin{figure}[tbp]
     \begin{center}

            \includegraphics[width=0.6\textwidth]{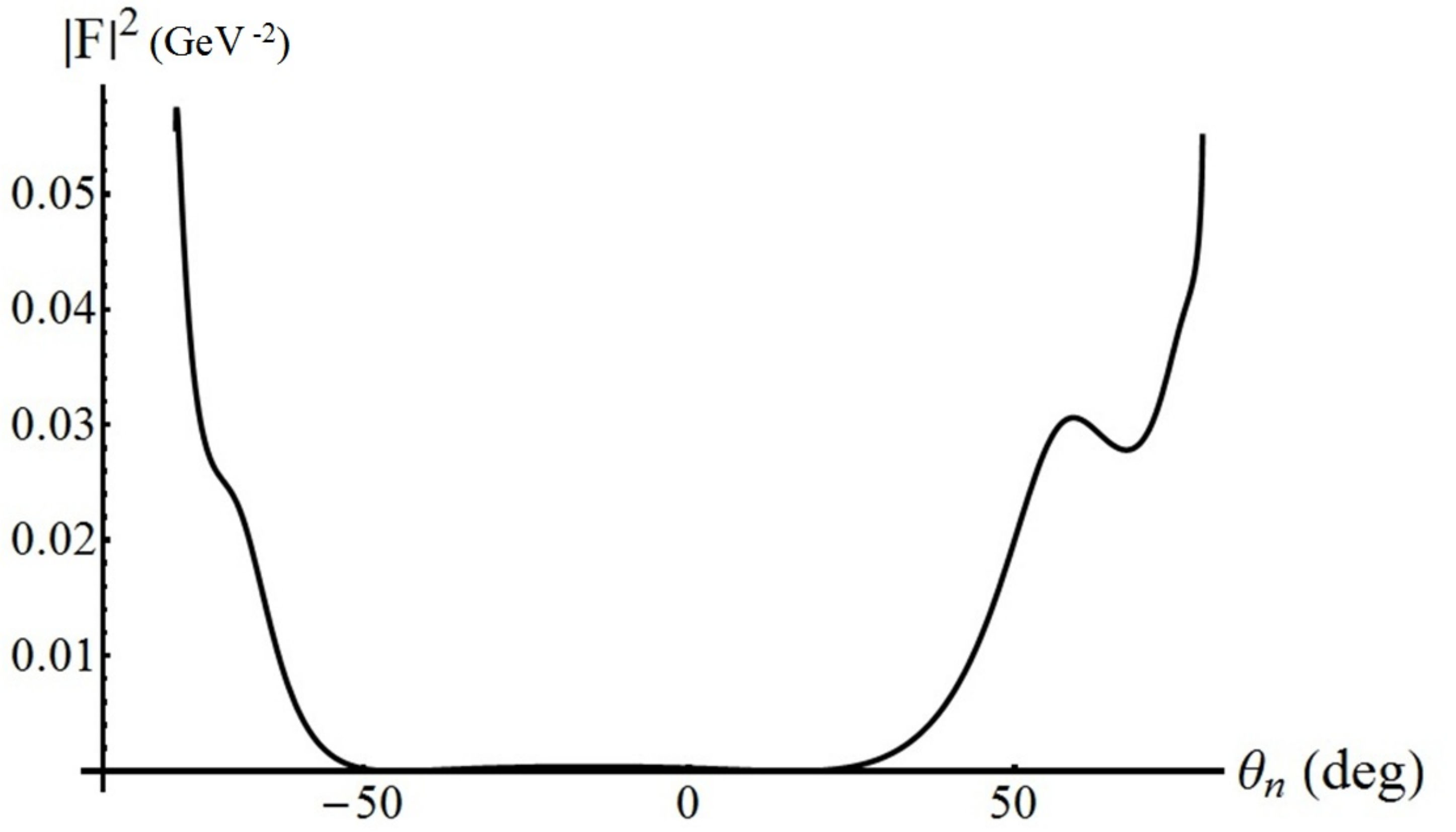}

    \end{center}
    \caption{%
         Amplitude squared vs. $\theta_n$, including all diagrams, for  $A_{Vn}=16\;GeV^{-4}$, for photon energy $\nu=9\;GeV$ and $t=-3\;GeV^2$.  The peak or bump on either side is due to $p-n$ rescattering, while the peak due to $J/\psi$-nucleon rescattering is not visible. 
     }%
   \label{fig:nu9}
\end{figure}

\subsection{Conclusion}

We have shown here the possibility of measuring the $J/\psi$-nucleon scattering amplitude for energies significantly smaller than the energy of the only existing data.  If the total $J/\psi$-nucleon cross-section $\sigma_{tot}^{J/\psi N}$ at these energies is of the order of $2-3$ times the previously measured value, then the differential cross-section as a function of $\theta_n$ should exhibit well-defined peaks corresponding to on-mass-shell $p-n$ and $J/\psi - n$ rescattering, for virtual photon energy of $\nu=10\;GeV$ and 4-momentum-transfer-squared $t=(q-p_V)^2=-2\;GeV^2$.  However, at lower photon energy (below $9\;GeV$) the $J/\psi - n$ rescattering peak would not be distinguishable.  As it is expected~\cite{brodsky97} that $\sigma_{tot}^{J/\psi N}$ should increase as the energy decreases, it is not impossible that the lower energy cross-section could be larger than the measured value by a factor of $\sim 2$.

%
%

\bibliographystyle{plain}


%
%
\appendix
\raggedbottom\sloppy
 

\chapter{Glauber theory relations for 2-body scattering}
\section{Scattering amplitude and profile function}
Under the conditions for which the Glauber model is valid (high incident energy, small scattering angle, neglect of longitudinal momentum transfer and energy transfer) the elastic scattering amplitude $f(\mathbf{q})$, where $\mathbf{q}=\mathbf{k}-\mathbf{k}'$ is the momentum transfer, can be written as the 2-dimensional Fourier transform of a function $\Gamma(\mathbf{b})$ known as the ``profile function":
\be
\label{fandgamma}
f(\mathbf{q})=\frac{ik}{2\pi}\int d^2b \;e^{i\mathbf{q}\cdot\mathbf{b}}\;\Gamma(\mathbf{b})
\ee
where the vector $\mathbf{b}$ is perpendicular to the incident momentum $\mathbf{k}$.  If the scattering amplitude is known, then the profile function may be obtained by the inverse Fourier transform:
\be
\Gamma(\mathbf{b})=\frac{1}{2\pi i k}\int d^2q \;e^{-i\mathbf{q}\cdot\mathbf{b}}\;f(\mathbf{q}).
\ee
The profile function $\Gamma$ can be written in terms of a phase shift function $\chi(\mathbf{b})$ as 
\be
\Gamma(\mathbf{b})=1-e^{i\chi(\mathbf{b})}
\ee
and the scattering amplitude in terms of $\chi$ is
\be
\label{fwithchi}
f(\mathbf{q})=\frac{ik}{2\pi}\int d^2b \;e^{i\mathbf{q}\cdot\mathbf{b}}(1-e^{i\chi(\mathbf{b})})
\ee
This shows the relation to the eikonal approximation in potential scattering:  in the eikonal approximation, the scattering amplitude is given by \eq{fwithchi} where the phase shift function is related to the potential $V(\mathbf{r})$ by
\be
\chi(\mathbf{b})=-\frac{m}{k}\int_{-\infty}^{\infty} V(\mathbf{b},z)\;dz.
\ee

\section{Cross-sections}

The elastic, total, and inelastic cross-sections can be written in terms of the profile function.    The differential cross-section is given by 
\be
\frac{d\sigma}{d\Omega}=\vert f(\mathbf{q})\vert^2
\ee
and so the cross-section for elastic scattering is 
\be
\sigma_{el}=\int d\Omega \vert f(\mathbf{q})\vert^2.
\ee
For small angle scattering, the solid angle element $d\Omega$ can be written as $d\Omega=\frac{d^2q}{k^2}$, and so
\be
\sigma_{el}=\frac{1}{k^2}\int d^2q \vert f(\mathbf{q})\vert^2.
\ee
which by using \eq{fandgamma} is

\be
\sigma_{el}=\frac{1}{k^2}\frac{k^2}{(2\pi)^2}\int d^2q \int d^2b d^2b' \;e^{i\mathbf{q}\cdot(\mathbf{b}-\mathbf{b}')}\;\Gamma(\mathbf{b})\Gamma^*(\mathbf{b}')
\ee
which gives
\be \boxed{
\sigma_{el}=\int d^2b\; \vert\Gamma(\mathbf{b})\vert^2}\;.
\ee

The optical theorem gives the total cross-section in terms of the forward scattering amplitude as
\be
\sigma_{tot}=\frac{4\pi}{k} Im f(0)=\frac{4\pi}{k} \frac{k}{2\pi} Re\;\int d^2b \;\Gamma(\mathbf{b})
\ee
\be \boxed{
\sigma_{tot}=2\; Re\;\int d^2b \;\Gamma(\mathbf{b})}\;.
\ee
Note that for $f(0)$ pure imaginary (which holds to good approximation for high-energy hadron-hadron scattering) we have
\be
\int d^2b \;\Gamma(\mathbf{b})=\frac{1}{2}\sigma_{tot}.
\ee

\chapter{Glauber model cross-section for $\pi+A\to\pi+A^*$}

The Glauber model calculation of the incoherent cross-section for $h+A\to h+A^*$ proceeds as follows~\cite{glaub67}.  Using the closure approximation (which neglects the dependence of the final states $f$ of $A$ on energy) the sum of the cross-sections for each final state is given by summing the squares of the amplitudes; then using closure on $f$ we obtain   
\be
\begin{split}
& \Biggl(\frac{d\sigma}{d\Omega}\Biggr)_{incoh}  =\sum_{f\neq i}\vert F_{fi}(\mathbf{q})\vert^2\\
&=(\frac{k}{2\pi})^2 \int d^2b\;d^2b'\;e^{i\mathbf{q}\cdot (\mathbf{b}-\mathbf{b}')}\;\bigl[<i\vert\Gamma_{tot}(\mathbf{b}) \Gamma_{tot}^*(\mathbf{b}')\vert i>
-<i\vert\Gamma_{tot}(\mathbf{b})\vert i><i\vert \Gamma_{tot}^*(\mathbf{b}')\vert i>\bigr ]\end{split}
\ee

Assuming an independent particle model for the ground-state wavefunction $\Phi_i$, i.e. $\vert \Phi_i \vert^2=\prod_{j=1}^A\rho_1(\mathbf{r}_j)$, with  $\int d^3r \rho_1(\mathbf{r})=1$ and  $\rho(\mathbf{r})=A\rho_1(\mathbf{r})$           we have
\be
<i\vert\Gamma_{tot}(\mathbf{b}) \Gamma_{tot}^*(\mathbf{b}')\vert i>  =1-\bigl[1-\overline{\Gamma}(\mathbf{b})\bigr]^A-\bigl[1-\overline{\Gamma}^*(\mathbf{b}')]^A+\bigl[1-\overline{\Gamma}(\mathbf{b}) -\overline{\Gamma}^*(\mathbf{b}')+\beta(\mathbf{b},\mathbf{b}')  \bigr]^A
\ee
where $\overline{\Gamma}(\mathbf{b})\equiv\int d^3r \rho_1(\mathbf{r})\Gamma(\mathbf{b}-\mathbf{s})$ and  $\beta(\mathbf{b},\mathbf{b}')\equiv\int d^3r \rho_1(\mathbf{r})\Gamma(\mathbf{b}-\mathbf{s})\Gamma^*(\mathbf{b}'-\mathbf{s})$.  And also
\be
 <i\vert\Gamma_{tot}(\mathbf{b})\vert i> <i\vert  \Gamma_{tot}^*(\mathbf{b}')\vert i>=1-\bigl[1-\overline{\Gamma}(\mathbf{b})\bigr]^A-\bigl[1-\overline{\Gamma}^*(\mathbf{b}')]^A+\bigl[1-\overline{\Gamma}(\mathbf{b})\bigr]^A\bigl[1-\overline{\Gamma}^*(\mathbf{b}')]^A.
\ee
Thus
\be
\begin{split}
&<i\vert\Gamma_{tot}(\mathbf{b}) \Gamma_{tot}^*(\mathbf{b}')\vert i>
-<i\vert\Gamma_{tot}(\mathbf{b})\vert i><i\vert \Gamma_{tot}^*(\mathbf{b}')\vert i>\\
&=\bigl[1-\overline{\Gamma}(\mathbf{b}) -\overline{\Gamma}^*(\mathbf{b}')+\beta(\mathbf{b},\mathbf{b}')  \bigr]^A-\bigl[1-\overline{\Gamma}(\mathbf{b})\bigr]^A\bigl[1-\overline{\Gamma}^*(\mathbf{b}')]^A\\
&=\bigl[(1-\overline{\Gamma}(\mathbf{b}))(1 -\overline{\Gamma}^*(\mathbf{b}'))+\beta(\mathbf{b},\mathbf{b}') -\overline{\Gamma}(\mathbf{b})\overline{\Gamma}^*(\mathbf{b}') \bigr]^A-\bigl[(1-\overline{\Gamma}(\mathbf{b}))(1-\overline{\Gamma}^*(\mathbf{b}'))]^A\\
&=\bigl[(1-\overline{\Gamma}(\mathbf{b}))(1-\overline{\Gamma}^*(\mathbf{b}'))]^A\\
&+ A \bigl[(1-\overline{\Gamma}(\mathbf{b}))(1-\overline{\Gamma}^*(\mathbf{b}'))]^{A-1}\bigl(\beta(\mathbf{b},\mathbf{b}')-\overline{\Gamma}(\mathbf{b}) \overline{\Gamma}^*(\mathbf{b}')\bigr)+\ldots -\bigl[(1-\overline{\Gamma}(\mathbf{b}))(1-\overline{\Gamma}^*(\mathbf{b}'))]^A\\
&= A \bigl[(1-\overline{\Gamma}(\mathbf{b}))(1-\overline{\Gamma}^*(\mathbf{b}'))]^{A-1}\bigl(\beta(\mathbf{b},\mathbf{b}')-\overline{\Gamma}(\mathbf{b}) \overline{\Gamma}^*(\mathbf{b}')\bigr)+\ldots 
\end{split}
\ee
where the $\ldots$ represents higher powers of $\bigl(\beta(\mathbf{b},\mathbf{b}')-\overline{\Gamma}(\mathbf{b}) \overline{\Gamma}^*(\mathbf{b}')\bigr)$.

The term proportional to $\overline{\Gamma}(\mathbf{b}) \overline{\Gamma}^*(\mathbf{b}')$ above only contributes significantly for $\mathbf{q}\simeq 0$.  It's contribution to $\sum_{f\neq i}\vert F_{fi}(\mathbf{q})\vert^2$ is
\be
(\frac{k}{2\pi})^2\Biggl\vert\int d^2b e^{i\mathbf{q}\cdot\mathbf{b}}\overline{\Gamma}(\mathbf{b})  \bigl[1-\overline{\Gamma}(\mathbf{b})\bigr]^{A-1}\Biggr\vert^2\simeq(\frac{k}{2\pi})^2\Biggl\vert\int d^2b e^{i\mathbf{q}\cdot\mathbf{b}}\overline{\Gamma}(\mathbf{b}) e^{-\frac{1}{2}\sigma T(\mathbf{b})} \Biggr\vert^2
\ee 
The integral over $\mathbf{b}$ here is
\be
\int d^2b e^{i\mathbf{q}\cdot\mathbf{b}}\int d^2s dz \rho_1(\mathbf{r})\Gamma(\mathbf{b}-\mathbf{s}) e^{-\frac{1}{2}\sigma T(\mathbf{b})}= \int d^2s dz \rho_1(\mathbf{r}) e^{-\frac{1}{2}\sigma T(\mathbf{s})}e^{i\mathbf{q}\cdot\mathbf{s}}\frac{2\pi}{ik} f(\mathbf{q})
\ee
which gives
\be
\vert f(\mathbf{q})\vert^2\frac{1}{A}\int d^2s dz e^{i\mathbf{q}\cdot\mathbf{r}} \rho(\mathbf{r}) e^{-\frac{1}{2}\sigma T(\mathbf{s})}.
\ee
Without the factor $e^{-\frac{1}{2}\sigma T(\mathbf{s})}$ this would be proportional to the form factor of the nucleon density.  Hence it is negligible for large $\mathbf{q}$.

So neglecting that term, we are left with the term proportional to $\beta$:
\be
\begin{split}
&(\frac{k}{2\pi})^2\int d^2b d^2b' e^{i\mathbf{q}\cdot(\mathbf{b}-\mathbf{b}')}\bigl[1-\overline{\Gamma}(\mathbf{b})\bigr]^{A-1}\bigl[1-\overline{\Gamma}^*(\mathbf{b}')\bigr]^{A-1}\;A\beta(\mathbf{b},\mathbf{b}')\\
&=\int d^2s dz \rho(\mathbf{s},z)   (\frac{k}{2\pi})^2\int d^2b d^2b' e^{i\mathbf{q}\cdot(\mathbf{b}-\mathbf{b}')} e^{-\frac{1}{2}\sigma T(\mathbf{b})} e^{-\frac{1}{2}\sigma T(\mathbf{b}')}\Gamma(\mathbf{b}-\mathbf{s})\Gamma(\mathbf{b}'-\mathbf{s})\\
&\simeq\int d^2s dz \rho(\mathbf{s},z) e^{-\sigma T(\mathbf{s})}  (\frac{k}{2\pi})^2\int d^2b d^2b' e^{i\mathbf{q}\cdot(\mathbf{b}-\mathbf{b}')}\Gamma(\mathbf{b}-\mathbf{s})\Gamma(\mathbf{b}'-\mathbf{s})\\
&=\vert f(\mathbf{q})\vert^2\int d^2s dz \rho(\mathbf{s},z) e^{-\sigma T(\mathbf{s})}  \\
\end{split}
\ee
where in the third line we've used the fact that $\Gamma(\mathbf{b}-\mathbf{s})$ and $\Gamma^*(\mathbf{b}'-\mathbf{s})$ are sharply peaked at $\mathbf{b}=\mathbf{b}'=\mathbf{s}$, to evaluate $e^{-\frac{1}{2}\sigma T(\mathbf{b})} e^{-\frac{1}{2}\sigma T(\mathbf{b}')}$ at $\mathbf{b}=\mathbf{b}'=\mathbf{s}$ and pull it out of the integral over $\mathbf{b}$.  Thus we have for the single-scattering result for the incoherent cross-section:
\be
\label{elasticincoherent}
\boxed{
 \Biggl(\frac{d\sigma}{d\Omega}\Biggr)_{incoh}  =\vert f(\mathbf{q})\vert^2\int d^2s dz \rho(\mathbf{s},z) e^{-\sigma T(\mathbf{s})}  }
\ee
The physical interpretation of this is that the projectile travels through the nucleus to the point $(\mathbf{s},z)$ where it scatters elastically from a nucleon at that point, with momentum transfer $\mathbf{q}$.  It then continues along approximately in the z-direction out of the nucleus.  The attenuation factor  $e^{-\sigma T(\mathbf{s})}$ represents the probability that the projectile is not absorbed by the other nucleons, on its incident path and on its outgoing path.

In ~\cite{glaub67}, Glauber derives an expression for the incoherent cross-section that includes all powers of $\beta$.  When this is expanded to first order in $\beta$ it gives \eq{elasticincoherent}.

\chapter{Summing over all final states of residual nucleus}

In this appendix I derive the Glauber result, for $\pi+A\to\pi+p+(A-1)^*$, where we sum over all final states of the residual nucleus (not just one-hole states).  The result is a multiple-scattering series, the terms representing one elastic scatter of the pion, two elastic scatters of the pion, etc., with the overall momentum transfer $\mathbf{q}$ shared between the individual scatterings.  The first term (the single-scattering term) gives exactly the result obtained by summing over only one-hole final states, \eq{sumsquared} which is as it should be.

Starting from the general expression for the scattering amplitude for a particular final state $f$ of the residual nucleus
\be
\begin{split}
F_{fi}^{}=\frac{ik}{2\pi}\int d^2b e^{i\mathbf{q}\cdot\mathbf{b}}\int & d^3r_1\ldots d^3r_A \chi_p^*(\mathbf{r}_1)\\
&\times \phi_{A-1}^{f*}(\mathbf{r}_2,\ldots,\mathbf{r}_A) \phi_{A}^{}(\mathbf{r}_1,\mathbf{r}_2,\ldots,\mathbf{r}_A)  \Gamma_{tot}(\mathbf{b},\lbrace \mathbf{r}_j \rbrace)
\end{split}
\ee
and then squaring and summing over all $f$ using the closure approximation gives:
\be
\sum_{f}\vert F_{fi}\vert^2=(\frac{k}{2\pi})^2\int d^2b e^{i\mathbf{q}\cdot\mathbf{b}} d^2b' e^{-i\mathbf{q}\cdot\mathbf{b}'}   \int d^3r_1   d^3r_1' d^3r_2\;d^3r_3  \ldots d^3r_A \chi_p^*(\mathbf{r}_1)    \chi_p(\mathbf{r}_1') 
\ee
\be
\times\phi_{A}^{*}(\mathbf{r}_1',\mathbf{r}_2,\ldots,\mathbf{r}_A) \phi_{A}^{}(\mathbf{r}_1,\mathbf{r}_2,\ldots,\mathbf{r}_A) \Gamma_{tot}^*(\mathbf{b}',\mathbf{r}_1',\mathbf{r}_2,\ldots, \mathbf{r}_A) \Gamma_{tot}(\mathbf{b},\mathbf{r}_1,\mathbf{r}_2,\ldots, \mathbf{r}_A).
\ee
Now substituting
\be
\phi_{A}^{*}(\mathbf{r}_1',\mathbf{r}_2,\ldots,\mathbf{r}_A) \phi_{A}^{}(\mathbf{r}_1,\mathbf{r}_2,\ldots,\mathbf{r}_A) =\rho_1(\mathbf{r}_1',\mathbf{r}_1)\prod_{j=2}^A\rho_1(\mathbf{r}_j),
\ee
where in the shell model $\rho_1(\mathbf{r}_1',\mathbf{r}_1)=\frac{1}{A}\sum_{n=1}^A\phi_n^*(\mathbf{r}_1')\phi_n(\mathbf{r}_1)$ with $\rho_1(\mathbf{r},\mathbf{r})=\rho_1(\mathbf{r})$, and using orthogonality of the single-particle wavefunctions $\chi$ and $\phi_n$, gives:
\be
\sum_{f}\vert F_{fi}\vert^2=(\frac{k}{2\pi})^2\int d^2b d^2b'  e^{i\mathbf{q}\cdot(\mathbf{b}-\mathbf{b}')}     \int d^3r_1   d^3r_1' \chi_p^*(\mathbf{r}_1)    \chi_p(\mathbf{r}_1')  \rho_1(\mathbf{r}_1',\mathbf{r}_1) \Gamma_{b1}\;\Gamma_{b'1'} 
\ee
\be
\times \int d^3r_2\;d^3r_3  \ldots d^3r_A  \rho_1(\mathbf{r}_2)\ldots \rho_1(\mathbf{r}_A)\prod_{j=2}^A(1-\Gamma_{bj}-\Gamma_{b'j}^*+\Gamma_{bj}\Gamma_{b'j}^*)
\ee
\be
\label{sumall}
\begin{split}
=(\frac{k}{2\pi})^2\int d^2b d^2b'  e^{i\mathbf{q}\cdot(\mathbf{b}-\mathbf{b}')}     \int d^3r_1 &  d^3r_1' \chi_p^*(\mathbf{r}_1)    \chi_p(\mathbf{r}_1')  \rho_1(\mathbf{r}_1',\mathbf{r}_1) \Gamma_{b1}\;\Gamma_{b'1'}\\ &\times[1-\bar{\Gamma}(\mathbf{b})-\bar{\Gamma}^*(\mathbf{b}')+\beta(\mathbf{b},\mathbf{b}')]^{A-1}
\end{split}
\ee
where $\bar{\Gamma}(\mathbf{b})\equiv\int d^2s dz\rho_1(\mathbf{s},z)\Gamma(\mathbf{b}-\mathbf{s})$ and $\beta(\mathbf{b},\mathbf{b}')\equiv\int d^2s dz\rho_1(\mathbf{s},z)\Gamma(\mathbf{b}-\mathbf{s})\Gamma^*(\mathbf{b}'-\mathbf{s})$.
This is to be compared to the expression obtained by squaring and only summing over one-hole states of the residual nucleus, which is \eq{amp1} squared and summed:
\be
\label{sumonehole}
=(\frac{k}{2\pi})^2\int d^2b d^2b'  e^{i\mathbf{q}\cdot(\mathbf{b}-\mathbf{b}')}     \int d^3r_1   d^3r_1' \chi_p^*(\mathbf{r}_1)    \chi_p(\mathbf{r}_1')  \rho_1(\mathbf{r}_1',\mathbf{r}_1) \Gamma_{b1}\;\Gamma_{b'1'}g(\mathbf{b}) g^*(\mathbf{b}')
\ee
where
\be
g(\mathbf{b}) g^*(\mathbf{b}')=[1-\bar{\Gamma}(\mathbf{b})]^{A-1}[1-\bar{\Gamma}^*(\mathbf{b}')]^{A-1}=[1-\bar{\Gamma}(\mathbf{b})-\bar{\Gamma}^*(\mathbf{b}')+\bar{\Gamma}(\mathbf{b})\bar{\Gamma}^*(\mathbf{b}')]^{A-1}.
\ee
Comparing the two expressions \eq{sumall} and \eq{sumonehole} we see that the difference lies in the last term inside the $[\ldots]^{A-1}$.  For the one-hole final states, it factorizes in $\mathbf{b}$ and $\mathbf{b}'$, while for the sum over all final states it does not.  

The result \eq{sumall} can be expanded in powers of $\beta(\mathbf{b},\mathbf{b}')$ to give a multiple scattering series:
\be
\begin{split}
[1-\bar{\Gamma}(\mathbf{b})-\bar{\Gamma}^*(\mathbf{b}')+\beta(\mathbf{b},\mathbf{b}')]^{A-1}&=[1-\bar{\Gamma}(\mathbf{b})-\bar{\Gamma}^*(\mathbf{b}')]^{A-1}\\
& + [1-\bar{\Gamma}(\mathbf{b})-\bar{\Gamma}^*(\mathbf{b}')]^{A-2} (A-1)\beta(\mathbf{b},\mathbf{b}') +\ldots
\end{split}
\ee
\be 
\simeq e^{-\frac{1}{2}\sigma_{tot}^{\pi N}T(\mathbf{b})} e^{-\frac{1}{2}\sigma_{tot}^{\pi N}T(\mathbf{b}')}(1+A\beta(\mathbf{b},\mathbf{b}') +\ldots)
\ee
\be
=g(\mathbf{b}) g^*(\mathbf{b}')(1+A\beta(\mathbf{b},\mathbf{b}') +\ldots)
\ee
using the approximation \eq{gofb1} (and in the large-$A$ limit).  The first term here thus contributes to $\sum_{f}\vert F_{fi}\vert^2$ :
\be
\label{exact2}
(\frac{k}{2\pi})^2\int d^2b d^2b'  e^{i\mathbf{q}\cdot(\mathbf{b}-\mathbf{b}')}     \int d^3r_1   d^3r_1' \chi_p^*(\mathbf{r}_1)    \chi_p(\mathbf{r}_1')  \rho(\mathbf{r}_1',\mathbf{r}_1) \Gamma_{b1}\;\Gamma_{b'1'}g(\mathbf{b}) g^*(\mathbf{b}')
\ee
which is exactly the same as the result obtained by only summing over one-hole states, \eq{sumonehole}.
Keeping one power of $\beta$ gives a term which corresponds to two scatterings of the pion, with the momentum transfer $\mathbf{q}$ shared between the two scatterings.

%
%
%

\chapter{Low-energy scattering wavefunctions for $J/\psi$-nucleon scattering}
\label{wavefuncappend}

The Schrodinger equation for the $J/\psi$-neutron system is:
\begin{equation}
\Bigl[\frac{1}{r^2}\frac{d}{dr}(r^2\frac{d}{dr}) +k_2^2 \Bigr]\Psi_{k_2}(r) =2\mu V(r)\Psi_{k_2}(r)    \end{equation}
for $l=0$ (S-wave scattering).  Normalization conventions:  the wavefunction $\Psi$ is related to $u(r)$ and the phase shift $\delta(k)$ by 
\begin{equation}
\Psi(r)=\frac{1}{\sqrt{(2\pi)^3}}\;e^{i\delta(k)}\frac{u(r)}{r}  \end{equation}
This also defines our normalization of plane wave states:  $\frac{1}{\sqrt{(2\pi)^3}}e^{i\mathbf{k}\cdot\mathbf{r}}$. 

The radial wavefunction $u(r)$ satisfies
\begin{equation}
\label{schrodu}
\frac{d^2u}{dr^2}+k_2^2 u=2\mu V(r)u(r)\equiv U(r)u(r)   \end{equation}

We assume a  finite range potential, so that $V(r)$ is zero for $r$ larger than some distance $R$.   For $r>R$, where the potential is zero, the general solution to Eq. \ref{schrodu} is 
\begin{equation}
u_{out}(r)=\frac{1}{k_2}\sin{\bigl(k_2 r+\delta(k_2)\bigr)}  \end{equation}
which defines the $S$-wave phase shift $\delta(k_2)$.

The phase-shift $\delta(k)$ satisfies the following well-known properties~\cite{joachain75} as $k\to 0$ :  
\begin{enumerate}
\item for a repulsive potential, or an attractive potential that doesn't admit a bound state:  $\delta\to -ak$ as $k\to 0$;
\item for an attractive potential which admits a single bound state:  $\delta\to \pi-ak$ as $k\to 0$
\end{enumerate}
where $a$ is a constant (the scattering length).
Thus in the first case the zero-energy radial wavefunction for $r>R$ is
\be
u_{out}^0(r)=r-a
\ee
and in the second case it is
\be
u_{out}^0(r)=-(r-a),
\ee
while the zero-energy wavefunction $\Psi^0_{out}$ is given in both cases by
\be
\Psi_{out}^0(r)=\frac{1}{\sqrt{(2\pi)^3}}\;e^{i\delta(k)}\frac{u(r)}{r}=\frac{1}{\sqrt{(2\pi)^3}}\;\frac{r-a}{r} 
\ee

With these conventions, $a>0$ for either a repulsive potential or an attractive potential with a bound state, and $a<0$ for an attractive potential that doesn't admit a bound state (see Fig. \ref{fig:aForPotentials}).

\end{document}